\documentclass[prb,superscriptaddress,showpacs,twocolumn]{revtex4}
\usepackage{graphicx}
\usepackage{graphicx,amsfonts}
\usepackage{epsfig,amsmath}
\usepackage{verbatim}

\begin{document}

\newcommand{\atanh}
{\operatorname{atanh}}
\newcommand{\ArcTan}
{\operatorname{ArcTan}}
\newcommand{\ArcCoth}
{\operatorname{ArcCoth}}
\newcommand{\Erf}
{\operatorname{Erf}}
\newcommand{\Erfi}
{\operatorname{Erfi}}
\newcommand{\Ei}
{\operatorname{Ei}}

\title{Exact results and open questions in first principle
functional RG}
\author{Pierre Le Doussal}
\address{LPTENS CNRS UMR 8549 24, Rue Lhomond 75231 Paris
Cedex 05, France}
\date{\today}

\begin{abstract}
Some aspects of the functional RG (FRG) approach to pinned elastic
manifolds (of internal dimension $d$) at finite temperature $T>0$
are reviewed and reexamined in this much expanded version of 
[Europhys. Lett. {\bf 76} 457 (2006)]. The particle limit $d=0$ provides a test for
the theory: there the FRG is equivalent to the decaying Burgers
equation, with viscosity $\nu \sim T$ - both being formally
irrelevant. An outstanding question in FRG, i.e. how temperature
regularizes the otherwise singular flow of $T=0$ FRG, maps to the
viscous layer regularization of inertial range Burgers turbulence
(i.e. to the construction of the inviscid limit). Analogy between
Kolmogorov scaling and FRG cumulant scaling is discussed. First,
multi-loop FRG corrections are examined and the direct loop
expansion at $T>0$ is shown to fail already in $d=0$, a hierarchy of ERG
equations being then required (introduced in
[L. Balents and P. Le Doussal, Annals of Physics {\bf 315} 213 (2005)]).
Next we prove
that the FRG function $R(u)$ and higher cumulants defined from the
field theory can be obtained {\it for any $d$} from moments of a
renormalized potential defined in an sliding harmonic well. This allows to measure the fixed point function $R(u)$ in numerics and
experiments. In $d=0$ the beta function (of the inviscid limit) is
obtained from first principles to four loop. For Sinai
model (uncorrelated Burgers initial velocities) the ERG hierarchy
can be solved and the exact function $R(u)$ is obtained.
Connections to exact solutions for the statistics of shocks
in Burgers and to ballistic aggregation is detailed. A relation is established 
between the size distribution of shocks and the one for droplets.
A droplet solution
to the ERG functional hierarchy is found for any $d$, and the form of
$R(u)$ in the thermal boundary layer is related to droplet
probabilities. These being known for the $d=0$ Sinai model the
function $R(u)$ is obtained there at any $T$. Consistency of the 
$\epsilon=4-d$ expansion in one and two loop FRG is studied
from first principles, and connected to shock and droplet relations
which could be tested in numerics. 
\end{abstract}
\maketitle

\begin{widetext}

\section{Introduction}

\subsection{overview}

Success of field theory and renormalization methods for pure systems
usually stems from being able to identify a few relevant
operators, usually from symmetry, and control them
in some e.g. dimensional expansion in order
to obtain universal quantities \cite{Zinn} (independent of short scale details,
i.e. a continuum renormalizable limit). Extending
these methods to systems with quenched disorder presents
new challenges. Often those exhibit strong
disorder regimes, or glass phases, signaled by a
flow away from weak coupling of some operators
associated to disorder. However the question of what is really
a relevant quantity in a random system is quite delicate.
Often these fast growing operators represent high moments of
the probability distribution of some observables over samples.
Thus they may be associated with far tails of these
distributions i.e. to ultra rare events, and may not be
that important in the end (if their feedback
into typical events is negligible). A better strategy is then to
focus on the RG flow of probability
distributions which contain information about
typical events. This is what functional RG methods
aim to do, and why they are needed in disordered
systems. 2D disordered Coulomb gases (DCG) for instance
exhibit a frozen phase which at first appears non perturbative,
since the {\it average} charge fugacity grows with the scale.
The RG flow for the {\it typical} fugacity however becomes
different from the average one in that phase, and turns out to
be manageable \cite{CarpentierLeDoussal,CarpentierLeDoussal2001,disloc}
through control of the probability distribution.
The same mechanism is at play in related 2D fermion models
with quenched disorder and leads to freezing transitions
\cite{castillo,HorovitzLe Doussal2002,MudryRyuFurusaki2003}.
Another example of a strong disorder phase which can
be controled by RG methods arises in one dimensional quantum
and classical random chains. There,
a (stochastic) real space RG (RSRG) which also decimates in energy space is
well adapted to the structure of the strong disorder fixed point (FP).
It requires consideration of probability distributions and
yields a host of exact results
\cite{dsfspin,LeDoussalMonthusFisher1999,LeDoussalMonthusFisher,Monthus}.
An outstanding question is how general this is, whether other systems
can be treated in that way, and whether
a unified (functional) RG approach to strong disorder problems
can be found.

Here we focus on pinned elastic objects, one of the simplest
class of systems which form glass phases
where temperature (thermal fluctuations) is
formally irrelevant. We investigate some
properties of the functional RG method to describe
these systems, and the issues to be solved. The standard model
involves an elastic manifold of internal dimension
$d$ parameterized by a displacement field noted
$u_x \equiv u(x)$ (which may have $N$-components $\vec u_x$). The energy in
a given sample (i.e. one realization of $V(u,x)$) is given by:
\begin{eqnarray}
H_V[u] = \int d^d x \frac{1}{2} ( (\nabla_x u)^2 + m^2 u_x^2 ) + V(u_x,x)
\end{eqnarray}
where the random potential lives in a $d+N$ dimensional space. Its
distribution can be chosen gaussian with second cumulant:
\begin{eqnarray}
&& \overline{V(u,x) V(u',x')} = \delta^d(x-x') R(u-u')
\end{eqnarray}
and $\overline{V(u,x)}=0$ where $\overline{..}$ describes average
over samples. Being the bare disorder one usually denotes this
function as $R_0(u)$, while $R(u)$ denotes the renormalized, or
coarse grained disorder, precisely defined in this paper. However in
this introduction section we will not keep the distinction. A small
confining parabolic potential (the mass term) is added for
convenience and provides an infrared cutoff at large length scale
$L_m \sim 1/m$. We do not review here the numerous numerical and
analytical studies of this problem, nor the many applications to
physical systems (see e.g. Ref.
\cite{blatter,bglass,natter,rosso,allemands} for references).
Pedagogical notes about the FRG method, following the plan
of this introduction can be found in \cite{windsor,begrohu}. An
introductory review is given in \cite{kayreview,WieseLeDoussalReview} where many
references can be found. 

Here we just recall a few important facts.
There are three main classes depending on whether the function
$R(u)$ is (i) periodic, then it describes periodic objects
such as lattices with substrate impurities, (ii) short range,
typically describing Ising domain walls with random bonds,
(iii) long range, typically Ising domain walls with random
field disorder. The minimal energy ground state configuration (unique for finite size and
continuous distribution), here denoted $u_1(x) \equiv u_{1x}$, is known
to be rough with roughness exponent $u_1(x) \sim x^\zeta$.
This is mostly from numerics and from very few exact \cite{henley,kardar,derrida1d}
or rigorous \cite{imbrie,johansson} results, i.e $\zeta=2/3$
for the directed polymer $d=N=1$. Universality classes are expected to
depend on $N,d$ and the symmetries and boundary conditions on
$R(u)$ (mentionned qbove). The minimum energy $E=H_V[u_1]$
fluctuates wildly from sample to sample with a width $\delta E \sim L^\theta$,
with $\theta =d-2+2 \zeta$ (a relation guaranteed
by the statistical translational invariance of the disorder,
the so called STS symmetry \cite{SchulzVillainBrezinOrland1988,HwaFisher1994b}). Here and below
$L$ is system size.
On a qualitative level, since the manifold
sees an energy landscape with roughness $L^\theta$,
equilibrium at non zero temperature $T>0$
does not alter the $T=0$ picture,
i.e. the manifold remains localized near $u_1(x)$. The
proper rescaled temperature $T_L \sim T L^{-\theta}$ flows to zero,
and the system is in a glass phase controlled
by a zero temperature FP (whenever $\theta >0$).
It is important to stress that the long range case (iii) has
a non trivial and quite useful $d=0$ limit, the so-called toy
model, one instance being the famous Sinai model
of a particle in a Brownian landscape (which is the
$d=0$ limit of random field disorder \cite{toy}).

Temperature is irrelevant, however even at very low $T$ there are
some accidental quasi-degeneracies where the energy difference
between two (or more) low lying configurations
$u_1(x)$ and $u_2(x)$ happen to be of order $T$.
Then the (normalized) Gibbs measure $e^{- H_V[u]/T}/Z$
is splitted between two (or more) states. How rarely it happens
is not settled yet. The droplet picture (or scenario) assumes
\cite{droplets0,droplets,dpfisher} that quasi-degeneracies
of two states differing only below scale $L$ happen more and more rarely,
with probability $p_L \sim T L^{-\theta}$ as $L$ grows large.
It also assumes a kind of statistical independence of
these rare events so that quasi-degeneracies between
more than two states are negligibly rare. By contrast, the many pure states picture (based
e.g. on the replica symmetry breaking (RSB) saddle point
for infinite $N$ \cite{mezpar}) finds that these
quasi-degeneracies (within $T$) always occur (i.e. at any $T>0$). For
manifolds, numerical studies \cite{dropevid} seem to disfavor
the many pure state picture at least for small
$N$ and $d$. The marginal case $\theta=0$ is
of particular interest, and may be intermediate
\cite{CarpentierLeDoussal2001} (the DCG
mentionned above falls in that class). In any case, even within the droplet
scenario these rare events produce large effects. The exact identity
(from STS) for thermal (i.e. connected) correlation (in Fourier):
\begin{eqnarray}
\overline{\langle u_q u_{-q} \rangle - \langle u_q \rangle \langle u_{-q} \rangle} =
\frac{T}{q^2 + m^2}
\end{eqnarray}
where $\langle .. \rangle$ denotes thermal average over the Gibbs measure
in a given sample,
suggests that although the typical $\delta u = u - \langle u \rangle$
is small, $\delta u_{typ} \sim O(1)$, $\delta u$ can be large,
$\delta u \sim L^\zeta$, with probability $T L^{-\theta}$, yielding a net result
after disorder (i.e. sample) averaging
$\overline{(\delta u)^{2 n}} \sim T L^{2(n-1) \zeta + 2 -d}$.
Thus certain correlation functions (thermal correlations) are dominated by rare events.
As emphasized in \cite{BalentsLeDoussal2004a,BalentsLeDoussal2005}
a challenge for the field theory is
to be able to describe both (i) typical
events, i.e. zero temperature correlations encoding
the full ground state statistics of $u_1(x)$
(believed, surely not yet proved, to be critical and universal),
and (ii) significant rare events, i.e. the ones which would
lead to universal behaviour in thermal
correlations. An important question is
whether these are two decoupled sectors or how much
mutual feedback they enjoy.

These correlations can be computed from
the replicated partition function and
action, $\overline{Z^p}= \int Du_a e^{- S[u]}$, $a=1,..p$, with:
\begin{eqnarray}
S[u] = \frac{1}{T} \sum_a \int d^d x (\nabla_x u_a)^2 + m^2 u_{ax}^2
- \frac{1}{2 T^2} \sum_{ab} \int d^d x R(u_{ax} -u_{bx}) \label{rep}
\end{eqnarray}
and the fields $u_a(x) \equiv u_{ax}$ have been introduced
(not to be confused with the $u_{1x}$ introduced above).
Various thermal and disorder averages are encoded through
averages with distinct replicas
$\langle u_{a_1}^{n_1} .. u_{a_k}^{n_k} \rangle_S
= \overline{ \langle u \rangle^{n_1} .. \langle u \rangle^{n_k} }$
and here and below the $p=0$ limit is implicit (space indices suppressed).
The disorder average
couples the replica, and their interaction
is precisely $R(u)$. This is a convenient tool to perform
perturbative calculations, although equivalent to
direct perturbation theory in given sample, averaging later.

It was observed long ago that weak disorder becomes relevant, and
the manifold rough, for $d \leq d_c=4$. The bare disorder being
smooth, perturbation theory can be performed in the derivatives
$R^{(2n)}(0)$. If only $R''(0)$ were non zero, i.e. keeping only the
quadratic part of $S[u]$ in (\ref{rep}), one obtains (upon replica
inversion):
\begin{eqnarray}
\langle u^a_{-q} u^b_q \rangle = \frac{T}{q^2+m^2} \delta_{ab} +
R''(0) \frac{1}{(q^2+m^2)^2} \label{2pt}
\end{eqnarray}
i.e. $d_c=4$ appears through the $1/q^4$ propagator and $u \sim
x^{\zeta_L}$ with $\zeta_L=(4-d)/2$, the so-called Larkin model
\cite{LarkinOvchinnikov}. Since simple power counting at the Larkin
fixed point (FP) shows that the coefficient of the $u^{2n}$ interaction, i.e. $R^{(2n)}(0)$  grows with scale as $L^{4-d + (2n-4)
\zeta}$, one expects that adding all these $n>1$ non-linearities
would yield flow away from the Larkin FP and produce a non trivial
$\zeta$. Amazingly, and quite distinctly from more standard field
theory (FT), one finds that at $T=0$ the corrections due to these
non linearities all cancel in any (smooth) observable, i.e.
(\ref{2pt}) is exact. This is the celebrated dimensional reduction
(DR) property \cite{EfetovLarkin1977} which also occurs in more
complicated, random field models
\cite{AharonyImryShangkeng1976,Grinstein1976}. It is in general a
(all-order) perturbative statement \cite{f1}, assuming that $R(u)$ is analytic at
$u=0$.

It was shown by Fisher \cite{fisher_functional_rg} that already at the one loop level
the functional RG equation for the function $R(u)$ at $T=0$ yields flow outside
the space of the analytic functions. Fixed point
functions with a linear cusp, $R^{* \prime \prime}(u) \sim |u|$,
were found \cite{fisher_functional_rg,bglass}. They are uniformly $O(\epsilon)$
with $\epsilon=4-d$
and thus yield non trivial $\zeta=0(\epsilon)$, evading DR. The physics of the
cusp was argued \cite{BalentsBouchaudMezard1996}
to be related to existence of many metastable states (see
below). One loop FRG has been applied
to numerous models and physical systems for two decades,
including the depinning
transition \cite{NattermanStepanowTangLeschhorn1992,NarayanDSFisher1993a},
moving systems
\cite{bglass}, quantum systems and correlated disorder
\cite{balents_loc,mottglass,mboseglass}. Amazingly then and for almost fifteen years,
it was not been extended to higher loop, nor its consistency
(even at one loop) really checked!
Given its usefulness, this was an uncomfortable situation. Claims that
it could not yield a consistent $\epsilon$
expansion beyond leading order \cite{fisher_functional_rg} may have discouraged efforts.
However, these were based on a toy model for which we
now know that FRG works to all orders (see below).
It is true however that handling multiple minima intuitively
appear as a non perturbative problem, so it remains to
be understood precisely how it can ever be controlled.

\subsection{review of our previous work}

In a series of works we attempted to make progress
\cite{frg2loop,twolooplong,frgdep2,threeloop,2loopN,ergchauve,scheidl}. First we identified the problem. Because of the
cusp, $R'''(0^+) = - R'''(0^-) \neq 0$, ambiguities arise
in $T=0$ calculations of the beta function
at two loop level, and even already in the one loop corrections
to correlations. Schematically there is no obvious way to
interpret a $R'''(0)$ in a graph, or, phrased otherwise,
perturbation theory requires evaluations of derivatives
{\it at} $u=0$, while only the left and right derivatives are known.
Thus it appears that {\it some additional information} about the
physics of the system is necessary to continue the calculation.
Furthermore it is crucial information since DR suggests,
and calculations confirm, that all the physics is contained in
these singularities.

We have followed two routes in parallel. The first one
is heuristics. One physical requirement is
the existence of a continuum limit, i.e. a renormalizable
field theory (RFT). Since one loop counterterms are found to be unambiguous,
this requirement strongly points to a simple prescription \cite{twolooplong} to lift
ambiguities at two loop in the case $N=1$. It then yields a
candidate field theory with nice properties: renormalizable to two loop,
preserving the linear cusp, and yielding reasonable predictions for $\zeta$
to next order in $\epsilon$. It also suggests more
powerful ''prescriptions'' to lift ambiguities and
produces a three loop RFT \cite{threeloop}. This route is more
delicate for $N$-component manifolds and random field sigma models,
but has also been attempted there
\cite{2loopN,rf,rfautres}. These theories
are in a sense the most natural candidates RFT.

Another route is to attempt to construct the
field theory {\it from first principles}.
At zero temperature, this was possible in two cases:
depinning and large $N$. The
depinning transition is continuously related to statics, since the system
remains pinned in presence of an additional force $f$ until the threshold is
reached $f=f_c$ (in fact, to one loop
\cite{NattermanStepanowTangLeschhorn1992,NarayanDSFisher1993a} the depinning
FP (and beta function) is {\it identical} to the statics one).
However at $f=f_c^+$, and for $N=1$, the famous Middleton theorem \cite{Middletontheorem}
shows that $\partial_t u_{xt}>0$ (in the stationary state, or at any time if the
initial condition satisfies it).
With this {\it additional information}, one can compute all graphs
without ambiguities (within the dynamical formalism) and check that the theory is
indeed renormalizable to two loop \cite{frgdep2,frg2loop}. This provides an instructive
and controlled higher loop extension of the theory. Since
the same theorem indicates also a unique state, one may
wonder whether $N=1$ depinning is, in a sense, simpler, the challenge
in the statics being to treat multiple quasi-degenerate minima.
Progress in the statics was possible in the large $N$ limit, the
second ''solvable'' case. There, it was shown \cite{frgN}
that the FRG yields
the contribution of the most distant states in the RSB solution of
Mezard and Parisi
\cite{mezpar}. Further aspects of this connection were 
elucidated more recently \cite{markus}. It still remains to
be understood though 
whether this is not a rather peculiar limit (i.e.
whether large but finite $N$ be different from
infinite $N$).

The most natural idea for a first principle understanding
of FRG is to perform it at non zero temperature $T>0$.
Then one finds (in some cases) that the effective action remains
smooth and there are no ambiguities. The price to pay is that
it is far more complicated to implement since one must
keep track of an irrelevant variable, the temperature.
On the other hand, the physics is then more accessible.
Let us summarize our recent results in that direction.

First we have argued for
the existence of a ''thermal boundary layer'' (TBL) in
the effective action \cite{balents_loc,chauve_creep,BalentsLeDoussal2004a}
which describes how the temperature
rounds the cusp, e.g. in $R''(u)$, on scales $u \sim T_L$.
For fixed $u = O(1)$, $R(u)$ converges to
the non analytic fixed point as $L \to \infty$, but
for $u \sim T_L$, $R(u)$ is an analytic function
(which is novel and is not encoded in the $T=0$ fixed point). We have
shown \cite{BalentsLeDoussal2004a,BalentsLeDoussal2005}
that not just the second cumulant, but all
cumulants possess TBL analytic scaling forms and derived the FRG equations
which couple all of them. This was done within the Wilson
one loop approach. To be more systematic we have
used exact RG schemes (ERG) \cite{BalentsLeDoussal2004a,BalentsLeDoussal2005}
extending previous studies \cite{ergchauve,scheidl,schehr_co_pre}.
In particular we could
analyze in a non perturbative manner the exact
FRG equation for the useful limit $d=0$. Second,
we have shown that these TBL forms reproduce the scaling
expected for correlations from the droplet picture \cite{ergchauve,
BalentsLeDoussal2005}.
The TBL thus appears to encode for droplet probabilities.

The ultimate goal is to understand how a critical theory
emerges in the large system size limit $L \to \infty$, i.e.
how, starting from a non zero bare temperature, the
limit $T_L = T L^{- \theta} \to 0$ can produce an asymptotic
''zero temperature theory'' with all ambiguities resolved.
It requires solving the so-called {\it matching problem}:
one wants to connect information
about e.g. $R(u)$ for $u \sim O(1)$ (which is what the
$\epsilon$ expansion a priori computes), to
derivatives of $R(u)$ at $u=0$, needed to compute correlations,
which a priori requires knowledge of the function
within the TBL $u \sim T_L$. We have progressed
towards that goal in the toy model $d=0$. By considering
partial boundary layers, i.e.
how renormalized disorder cumulants behave
when multiple points are brought within $u \sim T_L$,
we have been able to show how to derive a
beta function for the $T=0$ theory. Extensions
to higher $d$ and $N$ have been attempted
but are not yet conclusive \cite{BalentsLeDoussal2005,unpublished}.

\subsection{aim of the present paper and outline}

The aim of the present paper is to investigate
further the first principle approach to
FRG at non zero temperature. It is partially a
review, as far as explaining what has been already achieved.
We also derive some new results, clarify some points
from our previous works and detail some connections to other
problems and models such as the decaying Burgers equation
and ballistic agregation. Some of the results have already appeared in \cite{pldepl}. The
present paper is a much expanded version of Ref. \cite{pldepl}, giving all detailed
derivations, and presenting all the tools, which we hope will be useful. 
Since it took much time to be completed, some companion works, sometimes 
using some of the results presented here, have already appeared, for instance the
measurement of $R(u)$ using the method proposed here and in \cite{pldepl}
was successfully performed in \cite{MiddletonLeDoussalWiese2006}. Its extension to the dynamics 
was presented in \cite{LeDoussalWiese2006a,RossoLeDoussalWiese2006a,LeDoussalWiese2008}. Interesting developments about shocks and avalanches, quite complementary of the present work, have appeared in \cite{avalanches,markus}. 

The outline of the paper is as follows.

We start by showing (section \ref{sec:loopexpansion}) what
are the difficulties encountered
when trying to develop renormalization at $T>0$ along
the standard FT method, i.e. within the loop expansion.
It is illustrated in the text on the
$d=0$ model (up to four loop), the $d>0$ case
(up to three loop) is summarized in Appendix
\ref{app:3loop}. It is a pedagogical way of showing why more
systematic methods, such as ERG are needed here.

In our previous study Ref. \cite{BalentsLeDoussal2005,unpublished},
the thermal boundary layer scaling
was presented as an ''ansatz''. A hierarchy of ERG equations
for all cumulant's scaling functions was derived,
and the existence of a solution was assumed.
Here we obtain an {\it exact solution} for
the full hierarchy of these TBL equations. It makes
the connection to droplets simple and transparent.

This is achieved by first asking whether the function $R(u)$, which
was introduced and defined in the FRG somewhat abstractly (from the
effective action) could be related to an observable. In particular,
what precise physical information is contained in the derivatives
$R^{(2n)}(0)$ ? In our previous work Ref.
\cite{ergchauve,BalentsLeDoussal2005} we had shown how some low
order thermal correlations are related to some of these derivatives,
as well as to derivatives of higher cumulants. Here, and in Ref. \cite{pldepl}, we obtain the
full solution of the problem which turns out to be (a posteriori)
very simple. For that purpose we reexamine (section
\ref{sec:basictools}), in any dimension, various generating
functions for correlations, the effective action $\Gamma[u]$ and the
functional $W[j]$. We prove that they take similar dual forms and
that they define the same second cumulant functional $R[u]$, while
they differ in higher cumulants: $R[u]$ simply encodes the two point
correlation of a renormalized potential. One way to define this
potential is to add a quadratic well i.e. an external harmonic
potential, with a center which can be translated. It can thus be
measured in numerical simulations. Its definition in $d=0$ is
reminiscent of the ''toy RG'' model (with only two degrees of
freedom) introduced heuristically in
\cite{BalentsBouchaudMezard1996}, with somewhat different
interpretations. Here the correspondence with the field theoretic
definition is demonstrated in $d=0$ and we obtain the proper
generalization to any $d$. As a result, the renormalized force is
found to satisfy a {\it decaying functional Burgers equation}.

With this knowledge we then go back to the $d=0$ toy
model (Section \ref{sec:zerodim}). We write the
ERG hierarchy (Section \ref{subsec:erg}) and
then work out the full TBL solution corresponding to
droplets (Section \ref{subsec:dropletsolu}). The matching
property can then be checked explicitly. Next we derive
the correct beta function in $d=0$ up to four loop
with all ambiguities removed (Section \ref{subsec:4loop}):
the method was sketched in our previous work Ref. \cite{BalentsLeDoussal2005},
but not fully explicited. In the
case of the Sinai model (Brownian landscape, random
field disorder) using our previous result on the toy
model statistics (obtained through the strong disorder
RSRG \cite{toy}) we obtain
the exact result for the function $R(u)$ (Section \ref{subsec:sinai}),
in terms of multiple integrals of Airy functions,
and implicit forms for higher cumulants as well as
full and partial boundary layer functions.

In Section IV E and F we discuss the connection between FRG and 
Burgers turbulence. In $d=0$ the renormalized force in the FRG satisfies the usual $N$-dimensional
Burgers equation. Although it is a simplified version of the Navier Stokes equation
they share some common properties (see e.g. Ref. \cite{turbBernard,FrischBec01,khaninreview,BouchaudMezardParisi95}
for reviews). At large Reynolds number (small viscosity $\nu$)
both exhibit (i) a dissipative regime at small scale where viscosity dominates,
and (ii) an inertial range at larger scales with (multi)scaling and intermittency
(where viscosity can formally be set to zero). Understanding how the two
regimes connect is an outstanding problem in (decaying or stirred) turbulence
(resp. Burgulence). Here the FRG exactly
describes ''decaying Burgers'' with viscosity $\nu=2 T$, where the renormalization
scale plays the role of time. Although this connection is
not new, it has not been pushed too far. Here we study in detail the
mapping between Burgers and the FRG and find an exact relation between shocks
and droplet size distributions. In particular we emphasize
the general correspondence:
\begin{eqnarray}
&& - R''(0) \equiv \overline{v(x)^2} \\
&&  \frac{T}{2} R''''(0) \equiv \nu \overline{(\nabla v(x))^2} = \bar \epsilon
\end{eqnarray}
where $v(x)$ is the velocity field in Burgers. In the second line
the r.h.s. has a finite limit as $\nu \to 0$ called the
{\it dissipative anomaly}: also present in Navier Stokes, in
Burgers it is due to shocks. The (equivalent) finite limit of the
l.h.s. implies the existence of a
thermal boundary layer in the FRG. It is related to
droplets if shocks are dilute. The issue of the construction of the
inviscid limit $\nu \to 0$ in Burgers equation was
recently addressed using distributions
\cite{BernardGawedzki98,BauerBernard99}. We discuss
the relations to our work on the existence of
an ambiguity free $T_L \to 0$ critical limit in the
FRG. Indeed, the above mentionned matching problem in the FRG
maps onto the question of how the inertial range in Burgers turbulence
matches onto the dissipative scale. In particular, the celebrated Kolmogorov scaling
in the inertial range:
\begin{eqnarray}
&& \frac{1}{2} \bar S_{111}(0,0,u) \sim \bar \epsilon u \leftrightarrow
\overline{(v(x)-v(0))^3 } \sim - \bar \epsilon x
\end{eqnarray}
corresponds to the non analytic third cumulant behaviour
in the critical zero temperature theory, related
via the FRG hierarchy to the cusp in $R''(u)$:
\begin{eqnarray}
&& R'''(\pm 0) \equiv \overline{v(\pm 0) \nabla v(0)}
\end{eqnarray}
These relations hold for any $d=0$ model,
and can be fully explicited in the case
of the Sinai (random field) landscape $d=0,N=1$.
In that case many results about shock statistics are known, some since
Burgers \cite{Burgers,FrachebourgMartin99}. Here we derive a more
general result for the joint distribution of the renormalized potentials and
forces at several points. We can then relate it to the droplet solution to the
exact FRG hierarchy mentionned above, and prove matching in that case. 
The dynamics of the shocks
of also of high interest and given exactly by
a balistic aggregation model which can also
be solved exactly \cite{FrachebourgMartinPiasecki99}.
This can in turn be interpreted as a Markov property
in the merging of
droplets which holds in that case as we detail in Section IV F. These
relations can be pushed quite far in $d=0$ and
we hope they will help progress in $d>0$.

While the $d=0$ pinning problem maps onto
decaying Burgers, one should mention that
$d=1$ pinning problem (the so called directed
polymer) has connections to {\it noisy} Burgers.
Although the mapping is quite different, there
are some common issues. In particular, the
shocks and inviscid limit construction are
not expected to be too different. Attempts to
solve a hierarchy similar to the FRG was done
in \cite{Polyakov95}. 

Finally we extend our analysis to higher $d$ in Section
\ref{sec:higherd}. The corresponding ERG equation now relate
second (and higher) cumulant functionals ($R[u]$ is the
second cumulant functional and $R(u)$ its local part).
Again one identifies a ''zero temperature'' region $u = O(1)$
and a thermal boundary layer (TBL), $u\sim \tilde T_L$, in the functionals.
This hierarchy appears formidable, and a previous attempt to solve it in an
expansion in powers of $R$ was not successful \cite{unpublished}.
Here however, we obtain an {\it exact solution to all orders}
within the TBL. It is based on (and inspired by)
a simple droplet scenario. We then discuss the 
consistency and closure of the ERG hierarchy at $T_L \to 0$ and address the
ambiguity issue. We identify the
assumptions underlying the $\epsilon$ expansion in terms of continuity properties of the force functional $R''[u]$. 
Under these assumptions we perform the one and two loop derivation, from first principle, of the
anomalous terms in the FRG beta function.

\section{finite temperature FRG: preliminaries}

\label{sec:loopexpansion}

The aim of this Section is to examine the direct approach to the
FRG at finite temperature, i.e. computing the beta function for
the function $R(u)$ in a loop expansion, and understand how it
fails. It is an instructive exercise which allows to introduce a
few basic facts, and to motivate the use of more systematic exact
RG (ERG) methods. Since this method is more economical to use than
ERG, it is worth checking thoroughly.

\subsection{finite temperature beta function}

Let us consider model (\ref{rep}) calling $R_0(u)$ the bare
disorder appearing there, and consider perturbation theory in
$R_0(u)$. $R_0(u)$ plays the role of the coupling constant of the
theory, as $g_0 \phi^4$ in the $\phi^4$ theory. The graphical
rules have been described in \cite{twolooplong}. The interaction
is represented by a splitted (i.e. double) vertex with two free
replica indices $\sum_{cd} R_0(u_c-u_d)/2 T^2$. Each free
propagator line $T \delta_{ab} g_{q}$, $g_q=1/(q^2+m^2)$, gives a
factor of $T$ and identifies replica indices. Thus a graph with
$p$ connected components corrects a $p$-replica term. Each line
drawn from a vertex results in a derivative of the vertex with
respect to $u$. Vertices connected to a single replica component
are called saturated, i.e. evaluated at $u=0$.

One then computes the renormalized second cumulant $R(u)$, defined
from the effective action at zero momentum. Its detailed
definition is given in Section \ref{sec:basictools}. It plays the
role of the renormalized coupling $g \phi^4$ in the $\phi^4$
theory \footnote{As in massless $\phi^4$, one could set $m=0$ and
define instead renormalization conditions at non zero momentum.
Closing the FRG then becomes quite difficult. We found it more
convenient to add the mass $m$ and define the renormalized
disorder from the zero momentum limit.}. To compute it one writes
all one particle irreducible graphs \footnote{i.e. such that the
set of vertices cannot be disconnected by cutting a line} with two
replica connected components (since it is a two replica term), in
an expansion in a number of loops:
\begin{eqnarray} \label{higherloop}
&& R = R_0 + \delta^{(1)} R_0 + \delta^{(2)} R_0 + \delta^{(3)}
R_0 +..  \label{deltasum}
\end{eqnarray}
Including only the one loop diagrams one finds:
\begin{eqnarray}
&& \delta^{(1)} R_0 = T J_1 R_0'' + J_2 [ \frac{1}{2} (R_0'')^2 -
R_0''(0) R_0'' ] \label{delta1}
\end{eqnarray}
(here and below the $u$-dependence is often implicit, and primes
denote derivatives) while the two loop corrections read
\cite{twolooplong}:
\begin{eqnarray}
&& \delta^{(2)} R_0 = \frac{1}{2} T^2 R_0'''' J_1^2
+ \frac{1}{2} T (R_0''')^2 J_3 + T (R_0'''' (R_0'' - R_0''(0)) - R_0'' R_0''''(0))
J_2 J_1  \label{delta2} \\
&& + (R_0''')^2 (R_0'' - R_0''(0)) I_A
+ \frac{1}{2} R_0'''' (R_0'' - R_0''(0))^2 J_2^2 \nonumber
\end{eqnarray}
with the integrals:
\begin{eqnarray}
&& J_n=\int_{k_1+..+k_n=0} g_{k_1}..g_{k_n} \quad , \quad I_A=
\int_{k_1+k_2+k_3=0} g_{k_1} g_{k_2} g^2_{k_3} \label{int1}
\end{eqnarray}
The three loop corrections at $T>0$ are given in Appendix
\ref{app:3loop} (at $T=0$ they were computed in \cite{threeloop}
). Note that $n$-loop corrections are of the form $T^p
R_0^{n+1-p}$ with $p=0,..n$. The loop expansion is thus a double
expansion in $T$ and $R_0$ treated on the same footing (both are
considered ''small'' of the same order).

The standard method to extract the RG beta function is to first
compute $- m \partial_m R$ at fixed bare disorder $R_0$. This
results in similar expressions as
(\ref{deltasum},\ref{delta1},\ref{delta2}) where the integrals,
which are $m$-dependent, are differentiated. Next one reexpresses
$R_0$ as a function of $R$, i.e. one inverts (\ref{deltasum})
order by order in $R_0$, treating $R_0$ and $T$ to be of the same
order, i.e. one writes (schematically) $R_0 = R - T R_0'' +
(R_0'')^2 + .. = R - T R'' + (R'')^2 + ..$. Inserting in $- m
\partial_m R$ one finally obtains the beta function:
\begin{eqnarray}
&& - m \partial_m R|_{R_0} = \beta[R,T] \label{rgen}
\end{eqnarray}
which is also a polynomial expansion of the form $T^p R^{n+1-p}$.
Here STS guarantees that there are no corrections to the one
replica part of the effective action, hence to $T$, which can thus
be treated as a constant number. Note that in this calculation we
have taken both $R_0$ and $R$ analytic at $u=0$, which is natural
since we work at $T>0$. If one sets $T$ to zero in the result, one
can check order by order in $R$ that it yields a function
$\beta[R,T=0]$ whose coefficients are finite (no poles in
$\epsilon=4-d$). This would be the zero temperature beta function
for an analytic $R(u)$. Its finiteness would mean that the theory
is renormalizable at $T=0$, i.e. a continuum limit exists.
Unfortunately, it is not the correct zero temperature beta
function, since the assumption of $R(u)$ being analytic is not
self consistent at $T=0$ (it develops a cusp in finite
renormalization time). Dealing with the resulting ambiguities
directly at $T=0$ was studied in \cite{twolooplong}, but it is not
our aim in this Section. We want to keep $T>0$ and follow the RG flow: since
temperature is irrelevant, we hope that it may lead us to the
correct, ambiguity-free beta function.

\subsection{toy model}
\label{sectoyexp} 

We will now continue in $d=0$, for illustration, simplicity, and
also because there we know the answer by other means (see below).
Hence the model describes a particle in a $N=1$ dimensional random
potential (i.e. on a line):
\begin{eqnarray}
&& Z_V = \int du e^{- \frac{1}{T} H_V(u)} \quad , \quad H_V(u) = \frac{1}{2} m^2 u^2 + V(u) \\
&& S(\{u_a\}) = \frac{1}{2 T} \sum_a m^2 u_a^2 - \frac{1}{2 T^2} \sum_{ab}
R(u_a-u_b)
\end{eqnarray}
Although it is a simple integral, it is still a non trivial model.
It exhibits a glass phase when correlations of the potential grow
with $u$, $\overline{(V(u)-V(0))^2} = 2 (R(0)-R(u)) \sim |u|^{2
\alpha}$ with $\alpha>0$. The position $u_1$ of the minimum of the
energy $H_V(u)$ fluctuates from sample to sample as $u_1 \sim m^{-
\zeta}$ and $H_V(u_1) \sim m^{- \theta}$ and one has
$\alpha=\theta/\zeta$ and $\theta=2(\zeta-1)$. The case
$\alpha=1/2$ corresponds to a Brownian landscape and to the $d=0$
limit of the random field universality class for the manifold
($\zeta=(4-d)/3$). We stress that no bifurcation (such as a lower
critical dimension) is expected to occur between $d=4$ and $d=0$
so the toy model should be {\it continuously related} to the RF
manifold problem. In particular if the latter has a well defined
finite beta function, it should have a good limit in $d=0$. It is
this limit that we are studying here (for arbitrary $\alpha$).

We now compute $\beta[R,T]$, using that $J_n=m^{-2n}$ and $I_A=m^{-8}$ in $d=0$
and write the flow equation (\ref{rgen}). To obtain a fixed point it is
more convenient to replace $R$ in (\ref{rgen}) by the {\it rescaled} disorder $\tilde R$:
\begin{eqnarray}
&& R(u) = \frac{1}{4} m^{\epsilon-4 \zeta} \tilde R(u m^{\zeta})
\end{eqnarray}
One easily sees that this does not change the beta function, apart from
an additional linear rescaling term, and
replacing $T$ with the rescaled temperature $\tilde T = 2 T m^\theta$:
\begin{eqnarray}
&& -m \partial_m \tilde R(u) = (\epsilon-4 \zeta) \tilde R(u) +
\zeta u \tilde R'(u) + \tilde T \tilde R''(u) + \frac{1}{2} \tilde
R''(u)^2 - \tilde R''(0) \tilde R''(u) + \beta_{2 loop}[\tilde
R,\tilde T](u) + .. \label{betatot}
\end{eqnarray}
Here of course $\epsilon=4$, $\zeta=2/(2-\alpha)$, $\theta = 2
\alpha/(2-\alpha)$ and the rescaled temperature $\tilde T$ flows
to zero as $m \to 0$:
\begin{eqnarray}
&& \tilde T = 2 T m^\theta \to_{m \to 0} 0
\end{eqnarray}
and the main problem is to understand the limit, as $m \to 0$ of
(\ref{betatot}). Before studying higher loops let us recall the
physics of the one loop truncation (thus setting all $\beta_{n>1
loop}$ to zero).

\subsubsection{one loop}

For a fixed $u = O(1)$, as $m \to 0$ one has $\tilde R(u) \to
\tilde R^*_{1loop}(u)$ the non-analytic fixed point solution of
the naive zero temperature equation \footnote{where $\epsilon$ can
be scaled away setting $\zeta=\epsilon \zeta_1$ and $\tilde R =
\epsilon \tilde R_1$}:
\begin{eqnarray}
&& 0 = (\epsilon-4 \zeta) \tilde R(u) + \zeta u \tilde R'(u) +
\frac{1}{2} \tilde R''(u)^2 - \tilde R''(0) \tilde R''(u)
\label{oneloopFP}
\end{eqnarray}
which exhibits a non analytic expansion in $|u|$:
\begin{eqnarray}
&& \tilde R''(u) = \tilde R^{* \prime \prime}(0^+) + \tilde R^{*
\prime \prime \prime}(0^+) |u| + O(u^2) \quad , \quad u = O(1)
\end{eqnarray}
with the relation:
\begin{eqnarray}
&& - (\epsilon- 2 \zeta) \tilde R_{1loop}^{* \prime \prime}(0^+) =
\tilde R_{1loop}^{* \prime \prime \prime}(0^+)^2 \label{rel}
\end{eqnarray}
obtained from taking $u \to 0^+$ in the second derivative of
(\ref{oneloopFP}).

However, for any non zero $m$, $\tilde R(u)$ is analytic and
rounded in the thermal boundary layer (TBL) region $u \sim \tilde
T$, with a scaling form:
\begin{eqnarray}
&& \tilde R(u) = \tilde R(0) + \frac{1}{2} u^2 \tilde R^{* \prime \prime}(0)
+ \tilde T^3 r(u/\tilde T)
+ O(\tilde T^4) \quad , \quad u = O(\tilde T) \label{tblR}
\end{eqnarray}
Consistency is checked by plugging this form in the one loop
truncation of (\ref{betatot}). Leading terms are of order $\tilde
T^2$ (the term $ m \partial_m R = O(\tilde T^3)$ and only the
quadratic piece in the rescaling term contributes) and one finds
that the TBL function $r(x)$ satisfies
\begin{eqnarray}
(\epsilon-2 \zeta) \tilde R^{* \prime \prime}(0) x^2 + r''(x) +
\frac{1}{2} (r''(x)-r''(0))^2 = r''(0) \label{tblfunct}
\end{eqnarray}
yielding $r''(x)-r''(0) =\sqrt{1 - (\epsilon-2 \zeta) \tilde R^{*
\prime \prime}(0) x^2} -1$. Using the relation (\ref{rel}) one
sees that the large $x$ behaviour of $r''(x)$ exactly matches the
cusp of the $T=0$ fixed point. We also see that:
\begin{eqnarray}
\tilde R''(0) = \tilde R^{* \prime \prime}(0^+) \label{continuity}
\end{eqnarray}
Thus this one loop truncation of the FRG equation (for $\tilde R$
alone) yields a simple consistent answer (which has been used in
several studies at non zero temperature \cite{chauve_creep}). Let us
now see if this carries to higher loop.

\subsubsection{two loop}

\label{subsec2loop}

To two loop we find:
\begin{eqnarray}
&& \beta_{2 loop}[\tilde R,\tilde T](u)
= \frac{1}{8} \tilde T \tilde R'''(u)^2
- \frac{1}{4} \tilde T \tilde R''''(0) \tilde R''(u)
+ \frac{1}{4} \tilde R'''(u)^2 (\tilde R''(u) - \tilde R''(0)) \label{beta2loop}
\end{eqnarray}
and to check whether the one loop analysis carries through, we should study
this equation both in the so called {\it outer region} $u=O(1)$ and in the TBL
$u \sim \tilde T$. However one must first ask whether the same TBL
scaling to holds to two loop. Let us make an important observation. One can show
the exact result, displayed here both in rescaled and unrescaled version:
\begin{eqnarray}
&& - m \partial_m R''(0) = 2 T R^{(4)}(0) \\
&& - m \partial_m \tilde R''(0) = (\epsilon - 2 \zeta) \tilde R''(0)
+ \tilde T \tilde R^{(4)}(0) \label{exactrel}
\end{eqnarray}
We emphasizes that it holds {\it to all orders} and
yields a check on the expressions (\ref{betatot}), (\ref{beta2loop}),
(\ref{beta3loop}) and (\ref{beta4loop}) - a rather non trivial one
as it is satisfied for each combination
of $R$ derivatives corresponding to each power of $T$. To explain its origin
let us recall the physical content of the quantities which appear
in (\ref{exactrel}). First one has:
\begin{eqnarray}
&& \langle u_a^2 \rangle = \overline{ \langle u \rangle^2 } = \frac{- R''(0)}{m^4} =
\frac{- \tilde R^{\prime \prime}(0)}{4 m^{2 \zeta}}
\end{eqnarray}
a quantity expected to be continuous as $T \to 0$, hence $\tilde R''(0)$
must flows to a well defined limit. Then (\ref{exactrel}) strongly constrains
TBL scaling, i.e. $R^{(4)}(0) \sim 1/\tilde T$. To leading order in temperature,
the relation (\ref{exactrel}) connects the sample to sample fluctuations
$\overline{u_1^2}$ of the absolute minimum (i.e. a zero temperature quantity),
to the sample to sample fluctuations of the
thermal width $\chi_s = \langle (u - \langle u \rangle)^2 \rangle$
of the Gibbs measure in a given sample. Indeed one has:
\begin{eqnarray}
&& \overline{\chi_s^2} - \overline{\chi_s}^2 = \frac{T^2
R^{(4)}(0)}{m^8} = \frac{\tilde T^2 \tilde R^{(4)}(0) }{4 m^4}
\label{suscept2}
\end{eqnarray}
$\chi_s$ is a ''droplet quantity'', i.e. it is $O(1)$ in a typical sample, but
can be large, $\chi_s \sim m^{- 2 \zeta}$, with probability $p \sim \tilde T =
2 T m^{\theta}$ in the rare samples where almost denenerate minima
exist. There is an infinite set of relations
such as (\ref{exactrel}), consequence of STS and ERG.
We will recall and discuss them later.
We can now come back to the two loop beta function.

The first observation about the two loop contribution (\ref{beta2loop})
is that if we evaluate it in the TBL region, i.e. for $u \sim \tilde T$
a catastrophy happens. Indeed, using TBL scaling (\ref{tblR})
we find that all terms are as $\sim \tilde T$. But we should remember
that the one loop terms in (\ref{betatot}) are $\sim \tilde T^2$, thus the two loop
terms are huge, uniformly of order $1/\tilde T$ compared to the one loop
ones. One can check that no cancellation occur. These would be
unlikely anyway since the situation gets worse at three and
higher loops, each new loop yields a factor $1/\tilde T$.
The direct loop expansion thus does not appear to describe the
TBL and we see the need for the ERG method. The
ERG equations introduced in Ref. \cite{schehr_co_pre}
and analyzed in \cite{BalentsLeDoussal2004a,BalentsLeDoussal2005}
allow to understand what goes wrong in the above procedure,
which is subtle. We explain it here schematically,
for more details about the TBL structure see Ref. \cite{BalentsLeDoussal2005} and Section \ref{sec:zerodim} below. 

The first ERG equation
states that the sum of all loop corrections (two and higher)
for $R(u)$ is exactly $S_{110}(0,0,u)$, i.e. a second derivative of
the (renormalized) third cumulant $S(u_1,u_2,u_3)$ of the
disorder. Next, $S$ itself obeys an ERG equation of the type
(schematically) $\partial S = T S'' + T (R'')^2 + R''^3 + ..$ where the
$..$ contains a feedback from the fourth cumulant (and the
hierarchy goes on). To get the above loop expansion
one truncates the ERG equations in powers of $R$ (which is fine) and
in powers of $T$. For instance, the two loop term (\ref{beta2loop})
can be obtained solving only $\partial S = T (R'')^2 + R''^3$,
and computing the ensuing feedback of $S$ in the $R$ equation.
To three loop the term $T S''$ (and others)
will be added perturbatively to this equation, and so on.
Unfortunately, this double expansion in $T$ and $R$  {\it incorrect}
in the TBL. One can show that in the TBL equation for $S$ the term $\partial S$
is {\it negligible} compared to $T S''$ (very much
like the term $\partial R$ is also negligible in the
TBL for $R$). Thus to obtain the correct result in
the TBL one must rather equate $T S''$ with the feeding terms
$T (R'')^2 + R''^3 +..$, which is the opposite of
what is done here! It then amounts to a non trivial resummation
of the above loop expansion, necessary to recover consistency with TBL
scaling. One may wonder whether a resummation of
all orders in $T$, to a fixed order in $R$ may provide
a meaningful result in the TBL. This question is examined
in Appendix \ref{app:resumm}.

At this stage we must renounce to use this direct loop expansion method
in the TBL (i.e. for $u \sim \tilde T$). The next question is then
whether it can be used in the outer region $u \sim O(1)$, i.e.
whether there is a good limit as $m \to 0$ for fixed $u$.

That question is more subtle. In fact to two loop
it works! (at least for $d=0$). Consider (\ref{beta2loop}) for a $u = O(1)$
and $m \to 0$. The first term in the r.h.s. flows to zero. The only
problematic one is the second term. But we know that
it has a finite limit since, from (\ref{exactrel})
\begin{eqnarray}
 \tilde T R^{(4)}(0) &\to& - (\epsilon - 2 \zeta) \tilde R^{* \prime \prime}(0)
\quad , \quad \text{(always)}  \\
& =& \tilde R^{* \prime \prime \prime}(0^+)^2 \quad , \quad \text{(to one loop)}
\label{tooneloop}
\end{eqnarray}
Thus it seems that the two loop contribution flows to a well defined limit:
\begin{eqnarray}
&& \beta_{2 loop}(u) \to \frac{1}{4} (\tilde R'''(u)^2 - \tilde R'''(0^+)^2)
(\tilde R''(u) - \tilde R''(0))
\end{eqnarray}
where we have inserted a one loop relation into a
two loop one. This is allowed, to this order, if the end result is an
expansion in powers of $R$. This turns out to coincide with
the result from the ''correct'' method (see Section \ref{subsec:4loop}).
It obeys the requirement that the cusp remains linear, the constraint
of ''no supercusp'': expanding the r.h.s. in $|u|$ the term linear in $|u|$
cancels (i.e. $R'(0^+) \neq 0$).

\subsubsection{three loops and beyond}

Embolded by this success, we compute the three loop contribution:
\begin{eqnarray}
&& \beta_{3 loop}[\tilde R,\tilde T](u)
= \frac{1}{48} \tilde T^2 \tilde R'''' ( \tilde R'''' - 2 \tilde R''''(0))
+ \frac{1}{8} \tilde T (\tilde R''')^2 (\tilde R'''' - \frac{5}{4}  \tilde R''''(0))
+  \frac{1}{16} \tilde T {\sf R}'' (\tilde R'''')^2 \\
&& + \frac{1}{16} ({\sf R}'')^2 (\tilde R'''')^2 +
\frac{3}{32} (\tilde R''')^4 + \frac{1}{4} {\sf R}''
(\tilde R''')^2 \tilde R'''' \nonumber \\ \label{beta3loop}
\end{eqnarray}
where here and below we often use the shorthand notation:
\begin{eqnarray}
&& {\sf R}''(u) = \tilde R''(u) - \tilde R''(0)
\end{eqnarray}
We see that again it yields a well defined limit in the outer region $u = O(1)$.
Let us define the limit:
\begin{eqnarray}
&& \tilde T \tilde R''''(0) = r^{(4)}(0) + O(\tilde T)
\label{magic}
\end{eqnarray}
and keeping only terms with a finite limit as $m \to 0$
(discarding terms which are $O(\tilde T)$ we get:
\begin{eqnarray}
&& \partial \tilde R = (\epsilon - 4 \zeta) \tilde R + \zeta u \tilde R'
+ [ \frac{1}{2} (\tilde R'')^2 - \tilde R''(0) \tilde R'' ]
+ \frac{1}{4} (\tilde R''')^2 (\tilde R'' - \tilde R''(0))
- \frac{1}{4} r^{(4)}(0) \tilde R''  \nonumber \\
&& + \frac{1}{16} (\tilde R''- \tilde R''(0))^2 (\tilde R'''')^2 +
\frac{3}{32} (\tilde R''')^4 + \frac{1}{4} (\tilde R''- \tilde
R''(0)) (\tilde R''')^2 \tilde R''''
- \frac{5}{32} r^{(4)}(0) (\tilde R''')^2 \label{3loop0}
\end{eqnarray}
The question is now to fix the value of the number $r^{(4)}(0)$.
One requirement is the absence of supercusp, i.e. $- m \partial_m \tilde R'(0^+) =0$.
This yields:
\begin{eqnarray}
&& r^{(4)}(0) = \tilde R'''(0^+)^2 + \frac{5}{4} \tilde R'''(0^+)^2 \tilde R''''(0^+)
+ O(R^4)
\end{eqnarray}
Although it looks reasonable, there is something puzzling. The
flow of $\tilde R''(0^+)$ (in the outer region) reads, to two
loop:
\begin{eqnarray}
&& \partial \tilde R''(0^+) = (\epsilon - 2 \zeta) \tilde R''(0^+)
+ \tilde R'''(0^+)^2 + \frac{5}{4} \tilde R'''(0^+)^2 \tilde
R''''(0^+) - \frac{1}{4} r^{(4)}(0) R''''(0^+)
\end{eqnarray}
which can be compared with the exact identity (\ref{exactrel}):
\begin{eqnarray}
&& \partial \tilde R''(0) = (\epsilon - 2 \zeta) \tilde R''(0)
+ r^{(4)}(0)
\end{eqnarray}
Thus if we want to enforce (\ref{continuity}) i.e. the continuity
relation $\tilde R''(0) = \tilde R''(0^+)$ at two
loop it yields $r^{(4)}(0) = \tilde R'''(0^+)^2 +  \tilde R'''(0^+)^2 \tilde R''''(0^+)$
and is {\it incompatible} with the absence of supercusp at three loop.

Thus this procedure has produced a nice limit at three loop, but this limit
turns out to be incorrect. The correct beta function is derived below in Section \ref{subsec:4loop}
and is incompatible with (\ref{3loop0}) for any value of $r^{(4)}(0)$.
The reason why this procedure fails has to do with the existence of partial boundary layer forms,
discussed in Section \ref{subsec:4loop} and in Appendix  \ref{app:resumm}.
However, we cannot exclude that a clever way may exist to resum consistently,
equivalent to what is done with the ERG. Thus, to close this section
on a challenge, we display the result for the four loop contribution
to the beta function:
\begin{eqnarray}
&& \beta_{4 loop}[\tilde R,\tilde T](u) =
\frac{1}{384}
\tilde T^3 (\tilde R^{(5)})^2  - \frac{1}{192} \tilde T^3
R^{(6)}(0) R'''' - \frac{1}{64} \tilde T^2 \tilde
R''''(0)^2 \tilde R'''' - \frac{1}{32} \tilde T^2 \tilde R''''(0)
(\tilde R'''')^2 + \frac{1}{64} \tilde T^2 (\tilde R'''')^3 \nonumber \\
&& -
\frac{1}{48} \tilde T^2\tilde R''''(0) \tilde R''' \tilde R^{(5)}
+ \frac{5}{96} \tilde T^2 \tilde R'''  \tilde R'''' \tilde
R^{(5)} + \frac{1}{96} \tilde T^2(\tilde R'' -  \tilde R''(0))
(\tilde R^{(5)})^2 - \frac{1}{64} \tilde T^2(\tilde R''')^2 \tilde
R^{(6)}(0) \nonumber \\
&& - \frac{3}{64} \tilde T \tilde R''''(0)^2
(\tilde R''')^2 - \frac{13}{64} \tilde T \tilde R''''(0) (\tilde
R''')^2 \tilde R'''' + \frac{33}{128} \tilde T (\tilde R''')^2
(\tilde R'''')^2 - \frac{5}{64} \tilde T
\tilde R''''(0) (\tilde R''- \tilde R''(0)) (\tilde R'''')^2 \nonumber \\
&& + \frac{1}{16} \tilde T  (\tilde R''- \tilde R''(0)) (\tilde
R'''')^3 + \frac{1}{16} \tilde T  (\tilde R''')^3 \tilde R^{(5)} +
\frac{3}{16} \tilde T  (\tilde R''- \tilde R''(0)) \tilde R'''
\tilde R'''' \tilde R^{(5)} + \frac{1}{64} \tilde T (\tilde R''-
\tilde R''(0))^2 (\tilde R^{(5)})^2 \nonumber \\
 &&   + \frac{5}{16} (\tilde
R''')^4 \tilde R'''' +  \frac{9}{16} (\tilde R''- \tilde R''(0))
(\tilde R''')^2 (\tilde R'''')^2 +
\frac{1}{16} (\tilde R''-\tilde R''(0))^2 (\tilde R'''')^3 \nonumber \\
&& + \frac{1}{8} (\tilde R''-\tilde R''(0)) (\tilde R''')^3 \tilde
R^{(5)} + \frac{3}{16} (\tilde R''-\tilde R''(0))^2 \tilde R'''
\tilde R'''' \tilde R^{(5)} + \frac{1}{96} (\tilde R''-\tilde
R''(0))^3 (\tilde R^{(5)})^2 \label{beta4loop}
\end{eqnarray}
One notes that a new feature arises to four loop. All terms have a nice limit except for
the combination:
\begin{eqnarray}
&& - \frac{1}{64} \tilde T^2(\tilde R''')^2 \tilde
R^{(6)}(0) - \frac{3}{64} \tilde T \tilde R''''(0)^2
(\tilde R''')^2 \sim - \frac{1}{64 \tilde T} (r^{(6)}(0) + 3 r^{(4)}(0)^2) (\tilde R''')^2
\end{eqnarray}
Thus for the well defined limit in
the outer region to exist one must have:
\begin{eqnarray}
&& r^{(6)}(0) + 3 r^{(4)}(0)^2 = c \tilde T  + O(\tilde T)
\label{condition}
\end{eqnarray}
which also implies the cancellation of the anomalous term
linear in  $\tilde R''''(0)$. This condition happens to be correct
for the one loop truncation for the TBL function $r(x)$ given
above in (\ref{tblfunct}). However we also know (from ERG see
\cite{BalentsLeDoussal2005}) that the exact equation
\begin{eqnarray}
&& r^{(6)}(0) + 3 r^{(4)}(0)^2 + s^{(3)}_{114}(0,0,0) = 0
\end{eqnarray}
involves some derivative of the third cumulant in the TBL,
which has no good reason to vanish.

Thus we have shown in this Section some of the issues and difficulties
facing FRG at finite temperature, using $d=0$ as a test.
The direct loop expansion does show
some features which are qualitatively correct, such as
the TBL scaling and its matching with zero temperature
solution. However, unless it can be extended in a clever way,
it appears to fail beyond two loop (in $d=0$) as a quantitative method.
Of course we can be certain of that only if we know a
correct method. Before we get to a more promising approach
we need to understand better the physical meaning of the
tools used in the FRG, which is the aim of the next Section.

\section{Basic tools and functionals}
\label{sec:basictools}

In this Section we analyze the physical information contained in the
generating functionals in the replica formulation. We recall the
definition of the renormalized disorder in term of the effective
action. Then we show that this definition is equivalent to a much
more physical one directly related to an observable. For pedagogical
purpose we start by establishing the correspondence on the $d=0$
model. Then it is extended to higher $d$ and $N$.

\subsection{Renormalized disorder in $d=0$}

\subsubsection{connected correlations}

There are two basic generating functions in the replica formulation. The first
one is:
\begin{eqnarray}
&& W(j)= \ln Z(j) \\
&& Z(j) = \int \prod_a du_a \exp( \sum_a j_a u_a - S(u) ) \\
&& S(u) = \frac{1}{2 T} \sum_a m^2 u_a^2 - \frac{1}{2 T^2} \sum_{ab} R_0(u_a-u_b)
\label{wjd0}
\end{eqnarray}
where $R_0(u)$ is the second cumulant of the bare disorder. Here and
below we often use for simplicity the same notation for a function
of $p$ replica variables, e.g. $W(j) \equiv W(\{j_a\})$, and
functions of a single real variable (i.e. $H_V(u)$ below). $W(j)$ is
the generating function, via polynomial expansion, of the connected
correlations:
\begin{eqnarray}
W(j)= W(0) + \frac{1}{2} \sum_{ab} G_{ab} j_a j_b + \frac{1}{4!}
\sum_{abcd} G_{abcd} j_a j_b j_c j_d + .. \label{expW}
\end{eqnarray}
where $G_{a_1..a_n}=\langle u_{a_1}..u_{a_n} \rangle_S^c$ where $c$
here means connected with respect to the replica measure $S$, e.g.:
\begin{eqnarray}
&& G_{ab} = \langle u_a u_b \rangle = \frac{T}{m^2} \delta_{ab} - \frac{R''(0)}{m^4} \label{Gconn0} \\
&& G_{abcd} = \langle u_a u_b u_c u_d \rangle - (G_{ab} G_{cd} + 2
\text{perm}) \label{Gconn}
\end{eqnarray}
(sixth order correlations are studied in the Appendix
\ref{app:highercorrelations}). The last equality in the first
formula anticipates on the definition of $R(u)$ given below. Let us
examine the physical content of $W(j)$ which, at least within its
polynomial expansion, seems clear. There are a priori two distinct
two-point correlation $\langle u_a^2 \rangle = \overline{ \langle
u^2 \rangle}$ and $\langle u_a u_b \rangle = \overline{ \langle u
\rangle^2}$, where in these type of formula we will assume distinct
replicas, $a\neq b$. Similarly \footnote{technically replica
symmetry is assumed here. This can be enforced by restricting to a
system with a finite number of degree of freedom, and use continuous
distributions of disorder to avoid exact degeneracies. It turns out
that the formalism and results of this Section can be extended to
situations with spontaneous replica symmetry \cite{markus}. The
reason is that replica symmetry is broken explicitly when defining
the effective action} there are a priori five distinct four-point
correlations, each related (for $p=0$) to a particular combination
of thermal and disorder averages in the usual way: $\langle u_a^4
\rangle=\overline{\langle u^4 \rangle}$, $\langle u_a^3 u_b
\rangle=\overline{\langle u^3 \rangle \langle u \rangle}$, $\langle
u_a^2 u_b^2 \rangle =\overline{\langle u^2 \rangle \langle u^2
\rangle} $, $\langle u_a^2 u_b u_c \rangle=\overline{\langle u^2
\rangle \langle u \rangle^2}$ and $\langle u_a u_b u_c u_d
\rangle=\overline{\langle u \rangle^4}$. More generally one expects
$N_n$ distinct elements $G_{a_1..a_n}$, where $\sum_n N_n z^n =
\prod_{k=1}^\infty (1-z^k)^{-1}$ is the boson partition function.

In fact there are less. This is because of the STS symmetry, namely
that $W(j_a+\tilde j)=W(j_a)+\frac{T}{m^2} \tilde j \sum_a j_a + p
\frac{T}{2 m^2} \tilde j^2$ for any (replica independent) $\tilde
j$, as can be seen from shifting the integration in (\ref{wjd0}).
One easily sees that it implies (for arbitrary $p$):
\begin{eqnarray}
&& \sum_a G_{ab}= \frac{T}{m^2} \quad , \quad \sum_a G_{abcd}=0
\quad \cdots \label{STSW}
\end{eqnarray}
To order $u^n$, these imply $N_{n-1}$ STS relations linking the $N_n$ variables.
For $n=4$ there are $5$ variables and $3$ STS relations, which
leaves $2$ independent correlations. For $n=6$ one finds $11-7=4$
independent correlations. More insight into the physical meaning of
these relations will be given below. For now they are
an infinite set of constraints on the authorized form
for the correlations.

\subsubsection{effective action and renormalized disorder}

The second important generating function is the effective action
defined as the Legendre transform. Define:
\begin{eqnarray}
&& \partial_{j_a} W(j) = u_a \nonumber \\
&& \Gamma(u)=j_a u_a - W(j)  \label{legendre0}
\end{eqnarray}
Note that here we are a priori not trying to minimize anything, we
can construct this function e.g. perturbatively. As we will see
below it is somehow a formal definition. It also admits a polynomial
expansion \footnote{note that the constants $\Gamma[0]=W[0]=p
\overline{\ln Z(0)}$ are proportional to $p$ and to the quenched
free energy and play little role in the following}:
\begin{eqnarray}
&& \Gamma(u) = \Gamma(0) + \frac{1}{2} \sum_{ab} \Gamma_{ab} u_a u_b
+ \frac{1}{4!} \sum_{abcd} \Gamma_{abcd} u_a u_b u_c u_d + ..
\label{expG}
\end{eqnarray}
with:
\begin{eqnarray}
&& \Gamma_{ab} = G^{-1}_{ab} = \frac{m^2}{T} \delta_{ab} +
\frac{R''(0)}{T^2} \label{gammapoly0} \\
&& \Gamma_{abcd} = - (\frac{m^2}{T})^4 G_{abcd} \label{gammapoly}
\end{eqnarray}
and so on (see Appendix \ref{app:highercorrelations}). Note that we
somehow assume here that both $\Gamma$ and $W$ admit a polynomial
expansion, i.e. that they are smooth. This should pose no problem at
non zero temperature \footnote{if necessary, assuming also a finite
number of degree of freedom}. At low $T$ this smoothness region may
be limited to a boundary layer $u \sim T$ around $u=0$ (see below).
The more general approach presented below does not rely on
smoothness and can be used directly to handle the $T=0$ problem
\footnote{note that in the replica formulation $T$ appears in all
formula, even if in the end some observables have a well defined
$T=0$ limit. It is not too surprising since an infinitesimal
temperature is necessary to select the true minimum among all the
extrema which satisfy the zero force condition. The definition of
the $T=0$ problem, i.e. minimization of the energy, is to take first
the limit $T=0^+$ at fixed size (finite number of degree of freedom)
then take the thermodynamic limit}.

Let us point out that the effective action $\Gamma(u)$ is useful in
the field theory and renormalization because it is the generating function of
one particle irreducible graphs in perturbation theory (of the
interaction, here $R_0(u)$). It is often used to define renormalized
vertices and renormalization conditions. Physically, $\Gamma(u)$
contains all the coarse grained information. Since it sums all
fluctuations (all loops) the connected correlation functions are
then obtained as the sum of all connected tree graphs contructed
from the vertices of $\Gamma(u)$.

The STS symmetry also constrains the form of $\Gamma(u)$. From
(\ref{legendre0}) above one finds $\Gamma(u_a + \tilde u) =
\Gamma(u_a) + \frac{m^2}{T} \tilde u \sum_a u_a + p \frac{m^2}{2 T}
\tilde u^2$ and on the polynomial expansion it again implies
\begin{eqnarray}
&& \sum_a \Gamma_{ab}= \frac{m^2}{T} \quad , \quad \sum_a
\Gamma_{abcd}=0 \quad \cdots \label{STSG}
\end{eqnarray}
again valid for any $p$. The discussion of the number of independent
components within the polynomial expansion is thus identical to the
one above.

More global constraints can also be deduced from STS. If one assumes
that $\Gamma(u)$ can be expanded in number of replica sums, then the
STS implies the form:
\begin{eqnarray}
\Gamma(u) = \Gamma_0 + \frac{m^2}{2 T} \sum_a u_a^2 - \frac{1}{2
T^2} \sum_{ab} R(u_a-u_b) - \frac{1}{3! T^3} \sum_{abc}
S(u_a,u_b,u_c) - \frac{1}{4! T^4} \sum_{abcd} Q(u_a,u_b,u_c,u_d) +
\cdots \label{repsumG}
\end{eqnarray}
where $\Gamma_0$ is a constant (see below) and the functions $R$,
$S$, $Q$ and so on can be thought of ''renormalized disorder
cumulants'', second, third and fourth, respectively. More generally
they will be denoted $S^{(n)}$ with $S^{(3)}=S$, $S^{(4)}=Q$ etc..
The $n$-th cumulant then correspond to the $n$ replica terms, and
always come with the factors $T^{-n}$. Since they are defined from
$\Gamma(u)$ we call them $\Gamma$-cumulants. The STS imply that they
satisfy $S(u_1+v,u_2+v,u_3+v) = S(u_1,u_2,u_3)$ as if they were true
cumulants of some ''renormalized'' statistically invariant random
potential \footnote{it is not clear whether there are, if yes it
would be interesting to find its probability measure.}

Indeed since these $\Gamma$-cumulants are usually the output of the
FRG, it would be quite useful to relate them to observables. In
particular one wants to know to which physical quantity the
derivatives $R^{(2 n)}(0)$ correspond to. It should be possible to
answer in principle since, as mentionned above, the correlations can
be obtained as tree graphs from $\Gamma$ vertices. The Legendre
transform being involutive one can also express each $\Gamma$ vertex
as a sum of tree graphs constructed from $W(j)$, i.e. connected
correlations. One can do it systematically in the polynomial
expansion. To lowest order, $u^2$, the meaning of $R''(0)$ is simple
and transparent from (\ref{gammapoly0}, \ref{Gconn0}). To order
$u^4$, it is also clear from (\ref{gammapoly}, \ref{Gconn}), but
needs more calculation as one also needs to expand (\ref{repsumG})
to order $u^4$:
\begin{eqnarray}
&& \Gamma_{abcd}= - \frac{1}{T^4} F_4  - \frac{1}{T^2} R^{(4)}(0) (
(\delta_{ab} \delta_{cd} + 2 \text{perm}) - (\delta_{abc} + 3
\text{perm}) ) \label{G4}
\end{eqnarray}
thus it contains not only $R^{(4)}(0)$ but also the fourth cumulant
$F_4 = Q_{1111}(0,0,0,0)$ (this notation means one derivative w.r.t. each argument, see below). All (connected) four points correlations
can be expressed from only these two quantities, one finds:
\begin{eqnarray}
&& \langle u_a^4 \rangle_c =  \langle u_a^3 u_b \rangle_c
= \frac{-T^2}{m^8} R^{(4)}(0) + \frac{F_4}{m^8} \nonumber \\
&& \langle u_a^2 u_b^2  \rangle_c = \frac{T^2}{m^8} R^{(4)}(0)
+ \frac{F_4}{m^8} \label{fourthorder} \\
&& \langle u_a^2 u_b u_c  \rangle_c = \langle u_a u_b u_c u_d
\rangle_c = \frac{F_4}{m^8}  \nonumber
\end{eqnarray}
Thus $R^{(4)}(0)$ corresponds to various combinations (one of them
was given in (\ref{suscept2}) above) which are all thermal
correlations. As can be seen from (\ref{fourthorder}) the zero
temperature information (i.e. $\overline{ u_1^4 }$ where $u_1$ here
denotes the position of the absolute minimum, see below) is strictly
contained in the fourth cumulant $F_4$ (it can be interpreted as describing a renormalized
random force in a non gaussian Larkin model). 
One can go on and the order $u^6$ is given in the Appendix
\ref{app:highercorrelations}. It contains $R^{(6)}(0)$ together with
three other derivatives of higher cumulants. This is the approach
that was studied in Ref. \cite{BalentsLeDoussal2005} and is
summarized here. Unfortunately, this polynomial route does not lead
too far because the complexity increases very fast with the order.
We now turn to a more powerful approach. It will give us not only
the $R^{(2p)}(0)$ which contain information about thermal
excitations, but also the zero temperature part $R(u)$.

Before doing so let us note that the formulae
(\ref{wjd0},\ref{expW},\ref{STSW},\ref{legendre0},\ref{expG},
\ref{repsumG}) as well as the STS transformations for $W$ and
$\Gamma$ given in the text all hold for any number of replica $p$.
Similarly for the formula
(\ref{defWj},\ref{repsumW},\ref{legendredual}) of the next section.
The resulting definitions for the cumulant functions $R$,$S$, etc..
are such that they are independent of $p$, as will be obvious from
the next section. By contrast the explicit forms in
(\ref{gammapoly0}, \ref{G4}, \ref{fourthorder}) and in  the first
formula in (\ref{Gconn}) have been given for $p=0$ \footnote{for
arbitrary $p$ one has $\Gamma_{ab} = A \delta_{ab}+B$ with
$A=\frac{m^2}{T} + p R''(0)$ and $B=-R''(0)$, and
$G_{ab}=\frac{1}{A} \delta_{ab} + \frac{1}{p} ( \frac{1}{A+p B} -
\frac{1}A)$}.

\subsubsection{back to the $W(j)$ functional}

\label{sec:back}

We go back to $W(j)$ and ask what is the physical
information contained there beyond the polynomial
expansion. One has:
\begin{eqnarray}
&& e^{W(j) } = Z(j) = \overline{ \prod_a \int du_a e^{j_a u_a -
\frac{H_V(u_a)}{T}  } }
= \overline{ \prod_a \langle e^{j_a u_a} \rangle_{H_V} Z_V^p} \label{defWj} \\
&& H_V(u) = \frac{1}{2} m^2 u^2 + V(u) \label{defHV}
\end{eqnarray}
where $Z_V= \int du e^{-H_V(u)/T}$ and at $p=0$ the $Z(0)^p$ term in
the first formula can be set to unity. At low temperature, expansion
at small $j$ at fixed $T$ contains information about the minimum of
$H_V(u)$ and thermal flips between quasi-degenerate minima whenever
they exist. There is much more information in $W(j)$, hence in
$\Gamma(u)$. Indeed if one instead rescale $j=J/T$ the particle sees
a fixed force as $T \to 0$ and hence the position of the minimum is
shifted. Thus $W(j)$ and $\Gamma(u)$ also contain, in an essential
way \footnote{A too naive $p=0$, $T=0$ limit seems to fail: keeping
only the contribution of the absolute minimum $u_1$, and its
distribution $P(u_1)$, would yield $Z(j) = \int du_1 P(u_1) e^{u_1
\sum_a j_a}$. The LT condition then yield $u_a= u = \langle u_1
e^{u_1 \sum_a j_a} \rangle_P/\langle e^{u_1 \sum_a j_a} \rangle_P$,
independent of $a$. Evaluating $\Gamma[u^a=u]$ formally as the LT of
$W(\hat j=\sum_a j_a)$ gives $\Gamma[u^a=u] = u^2/(2 \overline{u^2})
+ \sum_{n \geq 3} c_n u^n$ where $c_n$ are some combinations of
connected correlations. However $\Gamma$ should contain $1/T^n$
factors, and there are none to be found here. Presumably this
Legendre transform is ill-defined as hinted by the fact that it
cannot be evaluated away from the point $u^a=u$.} information about
correlations of minima shifted by a force.

We will thus define $j = m^2 v/T$ so that $v$ has the same dimension
as $u$. To economize notation we will write $W(v) \equiv W(j)$. The
important observation is that the STS constraints are exactly the
same on $W(v)$ and $\Gamma(u)$, i.e. (\ref{STSW}) and (\ref{STSG})
are identical, thus if $W(v)$ has an expansion in replica sums the
STS implies the form:
\begin{eqnarray}
&& W(v) = W_0 + \frac{m^2}{2 T} \sum_a v_a^2 + \frac{1}{2 T^2}
\sum_{ab} \hat R(v_a-v_b) + \frac{1}{3! T^3} \sum_{abc} \hat
S(v_a,v_b,v_c) + \frac{1}{4! T^4} \sum_{abcd} \hat
Q(v_a,v_b,v_c,v_d) + \cdots \label{repsumW}
\end{eqnarray}
where $\hat S(v_a+w,v_b+w,v_c+w)=\hat S(v_a,v_b,v_c)$ etc.. and
$W_0$ a constant (see below). The general term in the above
expansion will be denoted $\hat S^{(n)}$ with $\hat S^{(3)}=\hat S$,
$\hat S^{(4)}=\hat Q$ etc.. The relation between $\Gamma(u)$ and
$W(v)$ is now very symmetric:
\begin{eqnarray}
&& \Gamma(u) + W(v) = \frac{m^2}{T} \sum_a v_a u_a \label{legendredual}
\end{eqnarray}
and there is a duality between the W-cumulants, denoted $\hat R$,
$\hat S$, $\hat Q$ and so on, and the $\Gamma$-cumulants., $R$, $S$, $Q$
etc..

It is easy to find which random potential has $\hat R$, $\hat S$, as
its cumulants and corresponds to $W(v)$. For each realization of the
random potential $V(u)$ one defines the ($W$-)renormalized potential
$\hat V(v) \equiv \hat V(V,v)$ in a given sample as:
\begin{eqnarray}
&& \exp( - \frac{1}{T} \hat V(v) )  = \int du \exp \big(
- \frac{1}{T} ( \frac{1}{2} m^2 u^2 + V(u+v) ) \big) \nonumber \\
&& = \int du \exp \big(
- \frac{1}{T} ( \frac{1}{2} m^2 (u-v)^2 + V(u) ) \big) \label{defVhat}
\end{eqnarray}
then one sees that:
\begin{eqnarray}
&& e^{W(v)} = Z(j=m^2 v/T) = \overline{ \int \prod_a du_a e^{
\frac{1}{T} m^2 \sum_a v_a u_a
- \frac{1}{T} ( \frac{1}{2} m^2 u_a^2 + V(u_a) ) } } \nonumber \\
&& = \overline{ \int \prod_a e^{\frac{1}{2 T} m^2 v_a^2 }
\int du_a e^{- \frac{1}{T} ( \frac{1}{2} m^2 (u_a-v_a)^2 + V(u_a) ) } } \label{harm} \\
&& = \overline{ \prod_a e^{\frac{1}{2 T} m^2 \sum_a v_a^2  -
\frac{1}{T} \hat V(v_a) } } \nonumber
\end{eqnarray}
Averaging over disorder (i.e. over $V$) reproduces the above
expansion (\ref{repsumW}) in terms of cumulants, i.e. {\it connected
moments}, provided:
\begin{eqnarray}
&& \overline{\hat V(v_1) \hat V(v_2) }^c = \hat R(v_1-v_2) \\
&& \overline{\hat V(v_1) \hat V(v_2) ..\hat V(v_n)}^c = (-1)^n
\hat S^{(n)}(v_1,..v_n)
\end{eqnarray}
Of course the overline here denote averaging over the measure on the
$\hat V(v) \equiv \hat V(V,v)$ induced by the bare measure (of
cumulants $R_0$ etc..) on the $V(u)$. Note that the formula
(\ref{harm}) can be written for any number of replica and together
with (\ref{defVhat}) shows quite clearly \cite{gauge} that the
functions $\hat R$, $\hat S$ in (\ref{repsumW}) and, through
Legendre transfoms $R$, $S$ etc.. are independent of the number of
replica $p$. Finally note that $\overline{\hat V(v)} =
\overline{\hat V(0)} := \overline{F_V} = - T \overline{\ln Z_V} = -
T W_0/p$ independent of $v$ from STS (i.e. translational invariance
of the measure on $V$) and equal to the averaged free energy. More
generally, since $W(u_a=0)=W_0+ \frac{p^2}{2 T^2} \hat R(0)+
\frac{p^3}{6 T^3} \hat S(0,0,0)+..$ the cumulants of the free energy
are given by $\overline{F_V^n}^c = (-)^n \hat S^{(n)}(0,..0) $.
Anticipating a bit \footnote{Using the Legendre Transform presented
in Appendix \ref{app:legendre}, it relies on $v_a(u=0)=0$ which is a
consequence of $R'(0)=0$, $S'(0,0,0)=0$ etc.. always true at non
zero $T$ (at $T=0$ it is equivalent to the absence of super-cusp).}
, Eq (\ref{legendredual}) implies that $\Gamma(0)=-W(0)$ hence
$\Gamma_0= - p \overline{\ln Z_V}$, and also $\overline{F_V^n}^c =
(-)^n S^{(n)}(0,..0)$.

Thus the information contained in $W(v)$ is exactly the statistics
of the W-renormalized random potential $\hat V(v)$. We can now
perform explicitly the Legendre transform (\ref{legendredual}), i.e.
relate (\ref{repsumW}) and (\ref{repsumG}). This is done in Appendix
\ref{app:legendre}, and we only quote the result. The most
interesting property that we find is that {\it the second
$\Gamma$-cumulant is the same as the second $W$-cumulant}, i.e. one
has:
\begin{eqnarray}
&& R(v) = \hat R(v)
\end{eqnarray}
the two functions are the same! Hence from now on we will use the same symbol,
i.e. note $R(v)$ both. This is remarquable, since, as we will see these two
function obey quite different RG equations. The difference of course appears
at the level of third and higher cumulants. One finds that:
\begin{eqnarray}
&& S(u_a,u_b,u_c) = \hat S(u_a,u_b,u_c) - \frac{1}{m^2}
\big( R'(u_{ab}) R'(u_{ac}) + R'(u_{ba}) R'(u_{bc}) + R'(u_{ca}) R'(u_{cb}) \big) \nonumber \\
&& Q(u_{abcd}) = \hat  Q(u_{abcd})  + \frac{6}{m^4}
\text{sym}_{abcd}[ R''(u_{ab})
(R'(u_{ac})-R'(u_{bc})) (R'(u_{ad})-R'(u_{bd}))] \nonumber \\
&& - \frac{12}{m^2} \text{sym}_{abcd}[\hat S_{100}(u_{abc})
R'(u_{ad})] \label{relGW}
\end{eqnarray}
where $\text{sym}_{abcd}$ is $1/4!$ times the sum of all
permutations of $abcd$ and here and below:
\begin{eqnarray}
&& u_{ab} : = u_a - u_b \\
&& u_{abc} : = u_a,u_b,u_c    \label{defarg}
\end{eqnarray}
and so on. We recall the notations used in this paper everywhere for
derivatives:
\begin{eqnarray}
S_{nmp}(u_{abc}) = \partial_{u_a}^n \partial_{u_b}^m \partial_{u_c}^p \label{defder} S(u_{abc})
\end{eqnarray}
and so on.
A simple graphical interpretation allows to recover from tree diagrams the
combinatoric factors in the $S^{(n)}$ cumulants from the $\hat
S^{(n)}$.

This solves, in $d=0$, the question of finding an operational way to
compute $R(u)$ and higher cumulants and relate it to observable. It
is thus:
\begin{eqnarray}
&& \overline{\hat V(u) \hat V(u')}^c = R(u-u')  \label{renru}
\end{eqnarray}
where the ''renormalized random potential'' $\hat V(u)$, defined
from (\ref{defVhat}), has a number of nice properties. At $T=0$ it
is, up to a constant piece, the Legendre transform (with a true
minimization) of $V(u)$, namely:
\begin{eqnarray}
&& \hat V(v) = min_u H_{V,v}(u) \quad , \quad H_{V,v}(u) =
\frac{1}{2} m^2 (u-v)^2 + V(u)
\end{eqnarray}
The role of the mass is important as one minimizes in presence of an
harmonic well centered on point $v$. $\tilde V(v)$ is the resulting
free energy. As $v$ is moved the absolute minimum will also move and
it will result in a non trivial renormalized energy landscape. At
$T=0$, in situations discussed below, these moves become discrete
jumps. At any temperature, the derivative (minus the renormalized
force) satisfies:
\begin{eqnarray}
&& \hat V'(v) = m^2 (v -  \langle u \rangle_v) \label{forced0} \\
&& \langle u \rangle_v \equiv \langle u \rangle_{H_{V,v}} =
\frac{1}{Z_{V,v}} \int du u e^{- \frac{1}{T} (\frac{1}{2} m^2
(u-v)^2 + V(u))}
\end{eqnarray}
in terms of the thermal averaged position in a given sample. At
$T=0$ it exhibits jump discontinuities at some discrete set of
values $v=v_s$, so called shocks (the above formula (\ref{forced0})
still holds then for left and right derivatives at the shock
positions $v_s$). These result in a non-analyticity in the force
correlator $-R''(u)$ computed from (\ref{renru}). One expects the
switch between minima to be abrupt at $T=0$ and smooth at finite $T$
as the Gibbs measure gradually shifts from one minimum to the other
as $v$ increases. This results in a thermal boundary layer in
$R(u)$. These issues are studied quantitatively in the following
Sections.

The random potential satisfies satisfies a Kardar-Parisi-Zhang (KPZ) \cite{kpz}
type of FRG equation:
\begin{eqnarray}
&& - m \partial_m \hat V = \frac{T}{m^2} \partial^2_v \hat V -
\frac{1}{m^2} (\partial_v \hat V)^2
\end{eqnarray}
with initial condition $\hat V=V$ for $m=\infty$. Its derivative
yields the {\it decaying Burgers} equation (in dimension $N$). It is
indeed well known that shock singularities appear in this equation.
There is no noise term and the equation describes the merging and
coarsening of shocks. Their physical connection to the FRG hierarchy
for cumulants is studied in the coming Sections. Note that the
reverse problem, often known as Polchinski RG, i.e. how to evolve
$V(u)$ so that the $\hat V(v)$ at fixed $m$ is fixed, satisfies the
same (reversed) equation.

Note that a version of the present renormalized potential was
proposed in $d=0$ in Ref. \cite{BalentsBouchaudMezard1996}, as a
''toy RG'' for the random manifold problem. The interpretation was
different, in terms of elimination of fast modes: the role of $m$,
which is here the infrared cutoff, was played by $\Lambda$, the UV
cutoff. Iterative minimization was discussed and somewhat
qualitative arguments were given as to relevance to the FRG. Here we
have {\it established} that $\hat V(v)$ has a precise content: it
yields the second cumulant defined from the replicated effective
action. We have also shown how to obtain the higher cumulants. In
Appendix we check explicitly, on the formulas established previously
that the (sixth) derivatives of $R(u)$ at $u=0$ agree with formulas
obtained by the polynomial method, explained above.

The main advantage of the present analysis is that it is now
easily extended to any $d$ (and $N$).

\subsection{Renormalized disorder functionals, any $d$}
\label{sec:rendisfunanyd}

\subsubsection{Functionals and their relations}

To generalize the previous Section we consider the model, defined on
a discrete $d$-dimensional lattice:
\begin{eqnarray}
&& H_V[u] = \frac{1}{2} \sum_{xy} g^{-1}_{xy} u_x u_y + \sum_x V(u_x,x) \\
&& S[u] = \frac{1}{2 T} \sum_{xy a} g^{-1}_{xy} u^a_x u^a_y -
\frac{1}{2 T^2} \sum_{x a b} R_0(u^a_x - u^b_x)  \label{modeld}
\end{eqnarray}
respectively the energy functional in a given sample and the
replicated action functional. We consider for now $g^{-1}_{xy}$ an
arbitrary matrix ($d=0$ is recovered suppressing space indices and
setting $g^{-1}_{xy}=m^2$). The disorder is chosen to be
uncorrelated from site to site on the lattice, i.e.
$\overline{V(u,x)V(u',x')}^c=\delta_{xx'} R_0(u)$, so that
(\ref{modeld}) enjoys the exact STS property at the level of the
lattice model. A continuum model can also be considered $\sum_x \to
\int d^d x$ and here we denote $u^a_x\equiv u^a(x)$. The $W[j]$
functional is defined as:
\begin{eqnarray}
&& W[j]= \ln \int \prod_{a x} d u^a_x e^{ \sum_{xa} j^a_x u^a_x -
S[u] } \label{defwjd}
\end{eqnarray}
The connected correlations $G^{a_1..a_p}_{x_1..x_n}=\langle
u^{a_1}_{x_1}..u^{a_n}_{x_n} \rangle_S^c$ are generated upon
polynomial multilocal expansion:
\begin{eqnarray}
W[j]=W[0]+\frac{1}{2} \sum_{xyab} G^{ab}_{xy} j^a_x j^b_y +
\frac{1}{4!} \sum_{xyztabcd} G^{abcd}_{xyzt} j^a_x j^b_y j^c_z j^d_t
+ \cdots
\end{eqnarray}
It satisfies the STS identity $W[\{j^a_x + j(x)\}] = W[\{j^a_x\}] +
T \sum_{xy} g_{xy} j(y) \sum_a j^a_x + p \frac{T}{2} \sum_{xy}
g_{xy} j(x) j(y)$ which is obtained performing the shift $u_x^a \to
u_x^a + \phi_x$ with $T j(x) = g^{-1}_{xy} \phi_y$. This implies:
\begin{eqnarray}
&& \sum_a G^{ab}_{xy} = T g_{xy} \quad , \quad \sum_a
G^{abcd}_{xyzt} = 0 \quad , \quad \cdots
\end{eqnarray}
In the latter we have used the simultaneous permutation symmetry $G^{abcd}_{xyzt} =
G^{bacd}_{yxzt}$ etc.. The proper change of variable is now:
\begin{eqnarray}
&& T \sum_y g_{xy} j_y^a = v^a_x \label{jtov}
\end{eqnarray}
so that $v$ has the same dimension as $u$. Then:
\begin{eqnarray}
W[v]= W[0]+ \frac{1}{2} \sum_{xy} \overline{G}^{ab}_{xy} v^a_x v^b_y
+ \frac{1}{4!} \sum_{xyzt} \overline{G}^{abcd}_{xyzt} v^a_x v^b_y
v^c_z v^d_t + \cdots
\end{eqnarray}
where:
\begin{eqnarray}
&& \overline{G}^{ab}_{xy} = T^{-2} \sum_{x'y'} g^{-1}_{x x'} g^{-1}_{y y'} G^{ab}_{x'y'} \\
&& \overline{G}^{abcd}_{xyzt} = T^{-4} \sum_{x'y'z't'} g^{-1}_{x x'} g^{-1}_{y y'}
g^{-1}_{z z'} g^{-1}_{t t'} G^{abcd}_{x'y'z't'}
\end{eqnarray}
and so on. Thus one still has:
\begin{eqnarray}
&& \sum_a \overline{G}^{ab}_{xy} = T^{-1} g^{-1}_{xy} \quad ,
\quad \sum_a  \overline{G}^{abcd}_{xyzt} = 0
\end{eqnarray}
which is exactly the symmetry obeyed by the polynomial expansion of
$\Gamma$. Because of STS one thus has:
\begin{eqnarray}
&& W[v] = W_0 +\frac{1}{2 T} \sum_{axy} g^{-1}_{xy} v^a_x v^a_y +
\frac{1}{2 T^2} \sum_{ab} \hat R[v^{ab}] + \frac{1}{3! T^3}
\sum_{abc} \hat S[v^{a},v^{b},v^c] + \cdots \label{wvd}
\end{eqnarray}
(for any $p$, $W_0=- p \overline{F_V}/T$ being a constant
proportional to the averaged free energy - see discussion in
previous Section) where $\hat R[v^{ab}]$ is a two replica functional
which only depends on the field $v^{ab}_x \equiv v^{a}_x - v^{b}_x$
with $a$, $b$ given. It can itself be decomposed into a local part,
the usual function $R(v^{a}_x - v^{b}_x)$, bilocal and higher. This
will be discussed in Section \ref{sec:higherd}. Similarly the third
cumulant functional $\hat S[v^{a},v^{b},v^c]$ only depends on the
fields $v^{a,b,c}_x$ with $a$, $b$, $c$ given. It satisfies
statistical translational invariance $\hat S[\{ v^{a}_x + \phi_x
,v^{b}_x + \phi_x,v^c_x + \phi_x \}] = \hat S[v^{a},v^{b},v^c]$.
This form is dual to the one for its Legendre transform $\Gamma(u)$:
\begin{eqnarray}
&& \Gamma[u] = \Gamma_0 + \frac{1}{2 T} \sum_{axy} g^{-1}_{xy} u^a_x
u^a_y - \frac{1}{2 T^2} \sum_{ab} R[u^{ab}] - \frac{1}{3! T^3}
\sum_{abc} S[u^{a},u^{b},u^c]
+ \cdots \label{gammaud} \\
&& \Gamma[u] = \sum_{ax} u^a_x j^a_x - W[j] =  \frac{1}{T}
\sum_{axy} u^a_x g^{-1}_{xy} v^a_y - W[v]  \label{gammaud2}
\end{eqnarray}
The calculation of the Legendre transform is performed in the
Appendix \ref{app:legendre}. One finds the functional identities:
\begin{eqnarray}
&& \hat R[v^{ab}] = R[v^{ab}] \\
&& S[u_a,u_b,u_c] = \hat S[u_a,u_b,u_c] - 3  ~ \text{sym}_{abc}
\sum_{xy} g_{xy} \frac{\delta R[u^{ab}] }{\delta u^a_x}
\frac{\delta R[u^{ac}] }{\delta u^a_y} \label{corresp}
\end{eqnarray}
which generalize the $d=0$ result. One also finds $\Gamma_0=-W_0= p
\overline{F_V}/T$.

We are now in position to define the renormalized random potential $\hat V[\{v_x\}]$
{\it functional} in a given sample whose connected cumulants will
reproduce $\hat R$, $\hat S$ and so on. Some equivalent definitions are:
\begin{eqnarray}
&& e^{- \frac{1}{T} \hat V[\{v_x\}] } = \int \prod_{x} d u_x exp
\bigg( - \frac{1}{T} ( \frac{1}{2} \sum_{xy} g^{-1}_{xy} u_x u_y
+ \sum_{x} V(u_x + v_x,x) ) \bigg) = \int \prod_{x} d u_x e^{ - \frac{1}{T} H_{V,v}[u] } \\
&& H_{V,v}[u] = \frac{1}{2} \sum_{xy} g^{-1}_{xy} (u_x-v_x)
(u_y-v_y) + \sum_{x} V(u_x,x)
\end{eqnarray}
Thus one has:
\begin{eqnarray}
&& \overline{ \hat V[\{v_x\}] \hat V[\{v'_x\}] }^c = R[\{v_x-v_x'\}]
\end{eqnarray}
note that this is now true {\it as a functional}. The simpler form
$\sum_x V(u_x,x)$ which defines the bare disorder is of course not
preserved under coarse graining as higher multilocal components
develop. Thus in general the functional $R[\{v_x\}]$ is not local.
However one can still define a {\it function} $R(v)$ from the {\it
local part} of the renormalized random potential functional. This is
sketched below and detailed in Section \ref{sec:higherd}.

There are again some nice properties. One can also define the
renormalized force functional, which satisfies:
\begin{eqnarray}
&& - {\cal F}_x[v] = \frac{\delta \hat V[\{v_z\}]}{\delta v_x} =
g^{-1}_{xy} ( v_y - \langle u_y \rangle_{H_{V,v}} )
\label{forcefunct}
\end{eqnarray}
The renormalized potential in a given sample obeys a RG functional
equation as $g_{xy}$ is varied (its variation is noted $\partial
g_{xy}$), also called Polchinski equation in the ERG context:
\begin{eqnarray}
&& \partial \hat Z_g[v] = \frac{T}{2} tr [ \partial g \frac{\delta^2
\hat Z_g[v]}{\delta v \delta v} ] \quad , \quad \hat Z_g[v] = e^{-
\frac{1}{T} \hat V[\{v_x\}] }
\end{eqnarray}
hence $\hat V[v]$ satisfies a functional KPZ type \cite{kpz} equation:
\begin{eqnarray}
&& \partial \hat V[v] = \frac{1}{2} tr[  \partial g ( T  \frac{\delta^2
\hat V[v]}{\delta v \delta v} -  \frac{\delta \hat V[v]}{\delta v }
\frac{\delta \hat V[v]}{\delta v }) ]
\end{eqnarray}
The ''initial condition'' at $g=0$ (analogous to $m=\infty$) is
again $\hat V[\{v_x\}] = V[\{v_x\}]$. Its (functional) derivative is
a functional decaying Burgers equation. One can then also expect
''functional shocks'' as a generalization of the standard Burgers
equation.

Finally we note that all formula of the present Section are
straightforwardly extended to the case of a $N$-component vector
$u \equiv u^i$, $i=1,..N$ for arbitrary $N$.

\subsubsection{how to measure $R(u)$}

The present study has opened the way to {\it measuring} the FRG
fixed point function(al) in numerical simulations (or in
experiments). The simplest procedure is to confine the manifold in a
harmonic well centered on a given configuration $v_x$. The simplest
model is then $g^{-1}(q)=q^2+m^2$ in Fourier space. As $v_x$ is
varied at $T=0$ shocks will occur as the manifold switches from one
ground state to another. A difference with $d=0$ is that these
switches occur now on various scales (in $x$). Very little is known
at present on these functional shocks and their statistics. They are
reminiscent of avalanches in the driven dynamics but they truly are
''static'' shocks where the manifold is at equilibrium (or in the
minimum energy position) for each $v$. The simplest choice is a
uniform $v_x=v$. Then $\tilde V[v_x=v]$ is the ground state
(free)-energy, which is proportional to the volume $L^d$. As shown
in Section \ref{sec:higherd} this allows to measure the local part
$R(v)$. One has simply:
\begin{eqnarray}
&& \overline{\hat V[\{v_x=v\}] \hat V[\{v_x=v'\}]}^c = R(v-v') L^d
\end{eqnarray}
which holds at any $T$. It can also be obtained from the force. At
$T=0$ is is particularly simple. Denote $u_x(v)$ the minimum energy
configuration for a fixed $v_x=v$ and $\bar u=L^{-d} \sum_x u_x(v)$
the center of mass position. Then:
\begin{eqnarray}
&& \overline{(v-\bar u(v))(v'-\bar u(v'))} = \Delta(v-v') L^{-d}
m^{-4} \label{measurede}
\end{eqnarray}
where from (\ref{forcefunct}) $\Delta(v)=-R''(v)$ is the local part
of the correlator of the pinning force (one has $\overline{v-\bar
u(v)}=0$). One expects the manifold to behave as roughly $(L/L_m)^d$
independent pieces, with $L_m \sim m^{-1}$, hence the inverse volume
factor $L^{-d}$ in (\ref{measurede}) simply expresses the central
limit theorem. Hence $\Delta(v)$ should have a limit as $L \to
\infty$ proportional to the fluctuations of the force density in a
correlation volume $L_m^d$. Since force density scale as $m^2 u \sim
m^{2-\zeta}$ this in turns suggests that, as a function of $m$,
$\Delta(u) \sim m^{-d+4-2 \zeta} \tilde \Delta(u m^\zeta)$ where
$\tilde \Delta$ should have a fixed point form as $m \to 0$. This is
indeed what is predicted by the FRG (see e.g. Section
\ref{sec:loopexpansion}). The above formula (\ref{measurede}) is
exact however for any $m$ and allows in principle to measure in
numerics all earlier stages of the RG, e.g. (i) the Larkin mass
where $\Delta(u)$ suddenly acquires a cusp \footnote{a finite $m_c <
+ \infty$ should exist for smooth and bounded bare disorder} and the
Larkin regime for $m>m_c$ where $\zeta=(4-d)/2$, (ii) the
convergence to the fixed point, (iii) crossovers between distinct
universality classes. At non-zero temperature one simply replaces
$u_x(v)$ by the thermal average. The effect of temperature is
discussed in the next Section where a ''droplet'' solution is
obtained. The predicted rounding of the cusp can also be measured in
numerics. Finally, it is also possible to measure the non-local part
of $R[v]$ using a non uniform $v_x$.

These formula can be generalized to a number of situations. First
they extend straightforwardly to any $N$. Next, they can be modified
in the case of a model which does not possess exact STS symmetry, as
is often the case in numerical simulations. This extension is
discussed in Appendix \ref{app:general}. They also allow to study
(and define properly) the FRG for chaos discussed in
\cite{PierreChaos}. Considering two copies indexed by $1,2$ seeing
slightly different disorders, e.g. $V \pm \delta ~ W$ with small
$\delta$, $V$ and $W$ being two statistically independent random potentials, one can study the
cross-correlation:
\begin{eqnarray}
&& \overline{\hat V_1[v_x=v] \hat V_2[v_x=v']}^c = R_{12}(v-v') L^d
\end{eqnarray}
The mutual correlation of the two ground states in each copy:
\begin{equation} \label{deltadef}
m^4 \overline{ (\bar u^1(v) - v) (\bar u^2(v') - v')} = L^{-d}
\Delta_{12}(v-v')
\end{equation}
defines the renormalized pinning force cross-correlator
$\Delta_{12}(u)=-R''_{12}(u)$. At zero temperature these functions
measure the correlations between the shocks in the two copies.

Having defined the important functional and observables we now turn
to the derivation and analysis of the FRG equations, first in zero
dimension.

\section{zero dimension, Exact FRG and decaying Burgers turbulence}

\label{sec:zerodim}

\subsection{ERG for moments and cumulants}

\label{subsec:erg}

Here we write the exact RG equations satisfied by the ''renormalized disorder''
in $d=0$. As was shown in the previous Section there are three interesting set
of functions. First the $W$-moments, i.e. the {\it moments} of
the renormalized potential $\hat V(v)$, i.e. the
simple averages, and second the $W$-cumulants, i.e.
the connected correlations of $\hat V(v)$, denoted respectively:
\begin{eqnarray}
&& \overline{ \hat V(v_1) .. \hat V(v_n) } = (-1)^n \bar S^{(n)}(v_1,..v_n) \\
&& \overline{ \hat V(v_1) .. \hat V(v_n) }^c = (-1)^n \hat S^{(n)}(v_1,..v_n)
\end{eqnarray}
Finally the $\Gamma$-cumulants denote $S^{(n)}(v_1,..v_n)$. We now
give the ERG equation for each set. We will use the notation
$v_{1,2,..,n} = v_1,..v_n$ and often denote $[..]$ the
symmetrization with respect to the all variables (i.e. sum over all
permutation divided by number of permutations). There are some
relations between the lowest ones \cite{anyp}:
\begin{eqnarray}
&& R = \hat R = \bar R  \quad , \quad \bar S = \hat S \\
&& \bar Q(v_{1234}) = \hat Q(v_{1234}) + 3 \text{sym}_{1234} R(v_{12}) R(v_{34}) \label{QR}
\end{eqnarray}
together with (\ref{relGW}).

\subsubsection{ERG for the  $W$-moments}

Starting from the flow equation:
\begin{eqnarray}
&& - m \partial_m \hat V = \frac{T}{m^2} \partial^2_v \hat V -
\frac{1}{m^2} (\partial_v \hat V)^2
\end{eqnarray}
Computing e.g. $- m \partial_m \overline{ \hat V(v_1) .. \hat
V(v_n) }$ one obtains:
\begin{eqnarray}
&& - m \partial_m \bar S^{(n)}(v_{1,2,..,n}) = \frac{n T}{m^2} [
\bar S^{(n)}_{20..0}(v_{1,2,..n})] + \frac{n}{m^2} [\bar
S^{(n+1)}_{110..0}(v_{1,1,2..n})] \label{Wrg}
\end{eqnarray}
The lowest ones are:
\begin{eqnarray}
&& - m \partial_m \bar R(v) = \frac{2 T}{m^2} R''(v) + \frac{2}{m^2} \bar S_{110}(0,0,v) \label{mom2} \\
&& - m \partial_m \bar S(v_{abc}) = \frac{3 T}{m^2} [\bar
S_{200}(v_{abc})] + \frac{3}{m^2} [\bar Q_{1100}(v_{aabc})]  \label{mom3} \\
&& - m \partial_m \bar Q(v_{abcd}) = \frac{4 T}{m^2} [\bar
Q_{2000}(v_{abcd})]  + \frac{4}{m^2} [\bar P_{11000}(v_{aabcd})] \label{mom4}
\label{frgw2}
\end{eqnarray}
where $\bar P$ is the fifth moment. As will be studied below, this
set of equation generate the loop expansion. More precisely,
inserting only the $R^2$ part of (\ref{QR}) into (\ref{mom3}),
solving for $\bar S$ and reporting into (\ref{mom2}) yields the
order $R^2$ part of the beta function, a priori to all orders in
$T$. To get the two loop contribution one needs to go up to the
equation for the fifth cumulant (not written) and so on, so while
these equations are very simple (they form a linear system) they are
not very economical. Note the ERG equation for the flow of the
averaged free energy $\overline{\hat V(0)}=\overline{F_V}$:
\begin{eqnarray}
&& - m \partial_m \overline{F_V} = \frac{1}{m^2} R''(0) = m^2
\overline{\langle u \rangle^2} \label{flowfreeenergy}
\end{eqnarray}
one of the many ERG identities derived in
\cite{BalentsLeDoussal2005} (see section III. D there). It implies the universal behaviour for the
averaged free energy
$\overline{F_V} \sim - (A/\theta) m^{-\theta}$ where $A = \lim_{m \to 0} m^{2
\zeta} \overline{\langle u \rangle^2}$ valid at any temperature \footnote{in dimension $d>0$
this generalizes as $- m \partial_m \overline{F_V} =- m^2 \sum_x
\overline{\langle u_x \rangle^2} \sim - A L^d m^{d-\theta}$. For short range disorder since $\theta \leq d/2$ this always gives universal {\it corrections} to the free energy density $f=\overline{F_V}/L^d=f_0 + \frac{A}{d-\theta} m^{d-\theta}$, but for sufficiently long range disorder such that $\theta>d$, then the {\it dominant} contribution is universal, very much as in $d=0$.}

\subsubsection{ERG for the  $W$-cumulants}

\label{ergW}

Another way to proceed, equivalent but a bit faster, is to study the
cumulants. One starts from the general (Polchinski type) functional
ERG equation \cite{schehrcoreview}:
\begin{eqnarray}
&& \partial W[j] = - \frac{1}{2 T} Tr \partial g^{-1} (
\frac{\delta^2 W[j] }{\delta j \delta j} + \frac{\delta W[j]
}{\delta j} \frac{\delta W[j]}{\delta j})
\end{eqnarray}
We specialize to $d=0$, $g=1/m^2$, $j=m^2 v/T$. One defines:
\begin{eqnarray}
&& W(v) = \frac{m^2}{2 T} \sum_a v_a^2 + \hat U(v)
\end{eqnarray}
with $W(0)=\hat U(0)$. The ERG for $W$ can then be rewritten:
\begin{eqnarray}
&& -m \partial_m \hat U = \frac{T}{m^2} \sum_a \frac{\partial^2
\hat U}{\partial v_a \partial v_a} + \frac{T}{m^2} \sum_a (
\frac{\partial \hat U}{\partial v_a} )^2
\end{eqnarray}
where to obtain the above one should remember that $W(v)=W(j=m^2
v/T)$ when differentiating w.r.t. $m$. Expanding in replica sums,
along (\ref{repsumW}), this yield the general equation for the
cumulants:
\begin{eqnarray} \label{frgw}
&& -m \partial_m  \hat S^{(n)}(v_{1,2,..,n}) = \frac{n T}{m^2} [
\hat S^{(n)}_{20..0}(v_{1,2,..n})] + \frac{n}{m^2} [\hat
S^{(n+1)}_{110..0}(v_{1,1,2..n})] \\
&& + \frac{1}{m^2} \sum_{k=2}^{E[(n+1)/2]} c_{k,n} [ \hat
S^{(k)}_{10..0}(v_{1,2,..,k}) \hat S^{(n+1-k)}_{10..0}
(v_{1,k+1,..,n})] \nonumber
\end{eqnarray}
with $c_{k,n} = 2 \frac{n!}{(k-1)! (n-k)!} = 2 k C_n^k$ except for
$n$ odd and $k=(n+1)/2$ then $c_{k,n}=k C_n^k$. The lowest ones are:
\begin{eqnarray}
&&  -m \partial_m R(v) = \frac{2 T}{m^2} R''(v) + \frac{2}{m^2}
\hat S_{110}(0,0,v) \label{Wcum2} \\
&&  -m \partial_m \hat S(v_{abc}) = \frac{3 T}{m^2} [\hat
S_{200}(v_{abc})] + \frac{3}{m^2} [\hat Q_{1100}(v_{aabc})] +
\frac{6}{m^2} [R'(v_{ab})
R'(u_{ac})] \label{Wcum3} \\
&&  -m \partial_m \hat Q(v_{abcd}) = \frac{4 T}{m^2} [\hat
Q_{2000}(v_{abcd})]  + \frac{4}{m^2} [\hat P_{11000}(v_{aabcd})]
+ \frac{24}{m^2} [\hat S_{100}(v_{abc}) R'(v_{ad})]
\label{Wcum4}
\end{eqnarray}
These equations are equivalent \cite{anyp} to the above (though
derived very differently) and allow to recover the loop expansion a
bit faster recursively.

\subsubsection{ERG for the $\Gamma$-cumulants}

\label{erggamma}

We now turn to the effective action, and define:
\begin{eqnarray}
&& \Gamma(u) =  \frac{m^2}{2 T} \sum_a u_a^2 - U(u)
\end{eqnarray}
The general ERG for $\Gamma(u)$ instead obeys \cite{schehrcoreview}:
\begin{eqnarray}
&&  -m \partial_m U(u) = Tr ( \delta -  \frac{T}{m^2}
\frac{\partial^2 U(u) }{\partial u \partial u} )^{-1}
\end{eqnarray}
The expansion in replica sums is more tedious, and a general formula
to all orders (for the rescaled cumulants) was derived in Appendix A
of \cite{BalentsLeDoussal2005}. Let us recall the result for the
second and third cumulants (here given in the present, i.e.
unrescaled, notations):
\begin{eqnarray}
 &&  -m \partial_m R(u) = \frac{2 T}{m^2} R''(u) + \frac{2}{m^2}
S_{110}(0,0,u) + \frac{2}{m^4}  (R''(u)^2 - 2 R''(0) R''(u))
\label{Gcum2}
\end{eqnarray}
\begin{eqnarray}
 &&  -m \partial_m S(u_{abc}) = \frac{3 T}{m^2} [S_{200}(u_{abc})] + \frac{6 T}{m^4}
 [R''(u_{ab}) R''(u_{ac})] + \frac{12}{m^4} (
[ R''(u_{ac}) S_{110}(u_{aad})] - [ (R''(u_{ac})- R''(0))
S_{110}(u_{acd}) ]) \nonumber \\
&&  + \frac{3}{m^2} [Q_{1100}(u_{aabc})] + \frac{6}{m^6} ( 3
[R''(u_{ab}) R''(u_{ac}) (R''(u_{ac})-R''(0)) ] - R''(u_{ab})
R''(u_{bc}) R''(u_{ca}) )  \label{Gcum3}
\end{eqnarray}
where $[..]$ denotes again symmetrization. Note that if one is
interested only in the dependence in $u_{abc}$ one can replace all
unmatched $R(u)$ by $R(u)-R(0)$ since this produces only gauge terms
for $p=0$ (thanks to a cancellation for the last term). But the
above equation contains a bit more information (e.g. about free
energy cumulants). The advantage of this set of ERG equation is that
the one loop beta function can already be read off from
(\ref{Gcum2}), the two loop from (\ref{Gcum3}), it is thus well
suited to the loop expansion. Obtaining the order $R^2$ to all
orders in $T$ still requires (\ref{Gcum3}).

In Appendix \ref{app:WtoGERG} we show how the $\Gamma$-ERG equations
(\ref{Gcum2}) and (\ref{Gcum3}) can be derived from (\ref{Wcum2})
and (\ref{Wcum3}), being careful with the flow term $-m \partial_m$
(being unrescaled equations, this term is always important).

\subsection{Exact ''droplet'' solution in the TBL region for the ERG hierarchy}

\label{subsec:dropletsolu}

We now show how one can obtain an {\it exact solution} to the full
ERG hierarchies displayed in the previous Section
\ref{sec:loopexpansion}. It is restricted to the thermal boundary
layer (TBL) region, defined in Section as $u m^\zeta = O(\tilde T)$.
Via matching however, it does also provide some zero temperature
information.

The strategy is to first compute the low temperature behaviour for the
$W$-moments of the potential $\tilde V(v)$. Once this is done, one can
easily get iteratively all $W$-cumulants and all $\Gamma$-cumulants.

In this Section, to avoid inflation of notations, we do not work
with rescaled quantities (unlike in Section
\ref{sec:loopexpansion}). Of course, the statements made below are
expected to become exact in the (universal) limit $m \to 0$, i.e.
$\tilde T= 2 T m^\theta \to 0$ with displacements scaled
appropriately $u \sim m^{- \zeta}$. These are easily restored
afterwards. The true expansion is in powers of $\tilde T$, but it is
as efficient to use instead a formal expansion in $T$ at small but
fixed $m$.

\subsubsection{droplets: general considerations}
\label{dropgen}

Let us study the difference of the renormalized potential at two
closeby points, and we define $v = T \tilde v$. The TBL region is
defined as $\tilde v$ being constant as $T$ goes to zero. The
definition (\ref{defVhat}) can be written in the form of a thermal
average in a given sample:
\begin{eqnarray}
&& e^{- \frac{(\hat V(v) - \hat V(0))}{T}}  = e^{- \frac{1}{2} T m^2
\tilde v^2} \frac{\int du e^{m^2 \tilde v u - \frac{1}{T}
(\frac{1}{2} m^2 u^2 + V(u))} }{\int du e^{- \frac{1}{T}
(\frac{1}{2} m^2 u^2 + V(u))} } = e^{- \frac{1}{2} T m^2 \tilde v^2}
\langle e^{m^2 \tilde v u} \rangle_{H_V(u)}
\end{eqnarray}
For fixed $\tilde v$ as $T \to 0$ (and fixed $m$) and for LR type
disorder ($\theta>0$) in (almost) all samples there is a single
minimum of $H_V(u)$ (we assume continuous disorder distributions)
which gives the dominant contribution to this average. The droplet
assumption is that one can obtain the behaviour up to order $T$ by
considering no more than two quasi-degenerate minima (i.e. wells).
Restricting to these two wells one obtains:
\begin{eqnarray}
&& \hat V(v) - \hat V(0) = \frac{1}{2} T^2 m^2 \tilde v^2 - T \ln( p
e^{m^2 \tilde
v u_1} + (1-p) e^{m^2 \tilde v u_2} ) \\
&& p=\frac{e^{-E_1/T}}{e^{-E_1/T} + e^{-E_2/T}}= \frac{1}{1 + w} \label{pot}
\end{eqnarray}
with $w=e^{-E/T}$ with $E=E_2-E_1 \geq 0$. Here we call $u_1$ the
absolute minimum, $u_2$ the secondary minimum, and $P(u_1,u_2,E) du_1 du_2 dE$ the
joint probability density of position and energy difference
(normalized to unity). One usually denotes:
\begin{eqnarray}
&& P(u_1) = \int P(u_1,u_2,E) du_2 dE \\
&& D(u_1,u_2) dE = P(u_1,u_2,E=0) dE
\end{eqnarray}
the probability of the position of the absolute minimum, and the
(unnormalized) probability density that there are two
quasi-degenerate minima within a window $dE$ near $E=0$ (this window
will be small, of order $T$, and the function of $E$ there can be
assumed to be constant, provided it does not vanish). Since many
observables depend only on the relative position $y=u_{12}=u_1-u_2$
we also denote:
\begin{eqnarray}
&& D(y) = \int du_1 D(u_1,u_1-y) \label{D}
\end{eqnarray}

The droplet assumption allows to compute any average over disorder
as the sum of single well events, and quasi-degenerate rare events,
as \cite{BalentsLeDoussal2005,Monthusdroplets}:
\begin{eqnarray}
&& F(w,u_1,u_2) := \prod_i \langle O_i(u) \rangle =
\prod_i ( p O_i(u_1) + (1-p) O_i(u_2) ) \\
&& \overline{F(w,u_1,u_2)} = \int du_1 P(u_1) F(1,u_1)  + T \int
du_1 du_2 D(u_1,u_2) \int_0^1 \frac{dw}{w} (F(w,u_1,u_2) - F(1,u_1))
\label{dropletaverage}
\end{eqnarray}
note that $F(1,u_1,u_2)= F(1,u_1)$ is independent of $u_2$. We often
use the shorthand notation $\langle O \rangle_{u_1}$ or $\langle O
\rangle_{P}$ for the first average and $\langle O \rangle_{u_1 u_2}$
or $\langle O \rangle_{D}$ or $\langle O \rangle_{u_i}$ or even $\langle O \rangle_{y}$ (when it depends only on the difference) for the
second (unnormalized) average. One may worry that the lack of
normalization of the second average may make predictions weaker,
thanks to STS this is not the case. Indeed the STS relation:
\begin{eqnarray}
&& \frac{T}{m^2} = \overline{\langle u^2 \rangle - \langle u \rangle^2 } =
\overline{p(1-p) (u_1-u_2)^2 } = T \int_0^1 \frac{dw}{w}
\frac{w}{(1+w)^2} \langle  (u_1-u_2)^2 \rangle_{u_i} = \frac{T}{2}
\langle  (u_1-u_2)^2 \rangle_{u_i}
\end{eqnarray}
provides a normalization
\begin{eqnarray}
&& \langle  (u_1-u_2)^2 \rangle_{u_i}
= \langle  y^2 \rangle_{y} = \frac{2}{m^2} \label{STSnorm}
\end{eqnarray}
where here and below $\langle O(y) \rangle_{y} = \int dy O(y) D(y)$.

In practice it is easier to compute moments of the renormalized
force, i.e. take a derivative of (\ref{pot}):
\begin{eqnarray}
&& \frac{\hat V'(v)}{m^2} = T \tilde v - ( u_1 X_1
\partial_{X_1} + u_2 X_2 \partial_{X_2} ) \ln(\frac{X_1 + w
X_2}{1+w})|_{X_i=e^{m^2 u_i \tilde v}}  \\
&& = T \tilde v - \frac{u_1 X_1 + u_2 w X_2}{X_1 + w
X_2}|_{X_i=e^{m^2 u_i \tilde v}}
\end{eqnarray}
One can check that the STS also guarantees
that the average force is zero:
\begin{eqnarray}
&& \frac{ \overline{ \hat V'(v) } }{m^4} = 0
\end{eqnarray}
We have used (\ref{dropletaverage}) and:
\begin{eqnarray}
&& T \int_0^1 \frac{dw}{w} (\frac{u_1 X_1  + u_2 w X_2}{X_1 + w
X_2 }-u_1) = T (u_2 -
 u_1) \ln \frac{X_1+X_2}{X_1} \label{formdrop} \\
 && - T \langle u_{12} \ln \frac{X_1+X_2}{X_1} \rangle_D  =
 \frac{1}{2} T \langle u_{12}^2 \rangle_D m^2 \tilde v =
T \tilde v
\end{eqnarray}
and (\ref{STSnorm}). Here and below we repeatedly use two important
symmetry property of the droplet averages:
\begin{eqnarray}
&& D(u_1, u_2) = D(u_2, u_1) \\
&& D(- u_1, - u_2) = D(u_1, u_2) \label{sym}
\end{eqnarray}

The functions $P(u_1)$ and $D(u_1, u_2)$ also satisfy two important
relations. It is shown in Ref. \cite{BalentsLeDoussal2005} (Section
IV.B there) that:
\begin{eqnarray}
&& P'(u_1) = m^2 \int du_2 u_{21} D(u_1,u_2) \\
&& - m \partial_m P(u_1) = m^2 \int du_2 (u_1^2 - u_2^2) D(u_1,u_2)
\label{rgdrop2}
\end{eqnarray}
The first equation is a consequence of STS and implies (upon
integration) an infinite set of relations between moments. It
encodes for the low temperature limit of a subclass of all the STS
relations between moments. The second equation is a consequence of
the more general one:
\begin{eqnarray}
&& - m \partial_m P(u_1,u_2,E) = - m^2 (u_1^2 - u_2^2) \partial_E
P(u_1,u_2,E) \label{droprgeq}
\end{eqnarray}
for $E>0$, which arises from the facts that (i) the dependence of $E
= H_V(u_1)-H_V(u_2)$ in $m$ arises only from the explicit $m$
dependence (using the minimum condition) and (ii) the dependence of
the typical $u_1$ and $u_2$ in $m$ becomes subdominant (compared to
$m^{-\zeta}$) when $E>0$ in the universal limit $m \to 0$. Indeed
changes in $u_1$ and $u_2$ come only from switching from one low
lying state to another (i.e. at $E=0$). Integrating (\ref{droprgeq})
over $E$ yields (\ref{rgdrop2}). This point reexplained in Section
\ref{sec:landscape}. Integrating (\ref{droprgeq}) over $E$ yields
(\ref{rgdrop2}). This equation also implies an infinite set of
relations between moments, as discussed in
\cite{BalentsLeDoussal2005} (Section III.D, IV B, and especially
Appendice C there). Of course (\ref{droprgeq}) neglects an
additional feeding term from three wells, and as such is a
truncation of an infinite hierarchy of equations. These additional
contributions are however expected to lead to corrections to higher
order in $T$.

\subsubsection{second moment $R(u)$ in the thermal boundary layer}

We now compute:
\begin{eqnarray}
&& - \frac{R''(v-t)}{m^4} =  \frac{ \overline{ \hat V'(v) \hat V'(t)} }{m^4} = \overline
{ (- T \tilde v + \frac{u_1 X_1 + u_2 w X_2}{X_1 + w X_2})(- T
\tilde t + \frac{u_1 Y_1 + u_2 w Y_2}{Y_1 + w Y_2})|_{X_i=e^{m^2
u_i \tilde v},Y_i=e^{m^2 u_i \tilde t}}) }
\end{eqnarray}
Upon expanding the first term is subdominant of order $T^2$ and is discarded.
The cross term vanish since, as we have just shown $\overline{ \hat V'(v)} =0$.
Remains to be computed:
\begin{eqnarray}
\overline { \frac{u_1 X_1 + u_2 w X_2}{X_1 + w X_2} \frac{u_1
Y_1 + u_2 w Y_2}{Y_1 + w Y_2} }  & = &  \langle u_1^2 \rangle_P +  T
\langle h(X_1,Y_1,u_1,X_2,Y_2,u_2) \rangle_{u_i} + O(T^2) \\
& = & \langle u_1^2 \rangle_P +  T
\langle h_{ss}(X_1,Y_1,u_1,X_2,Y_2,u_2) \rangle_{u_i} + O(T^2)
\end{eqnarray}
where we have defined:
\begin{eqnarray}
h(X_1,Y_1,u_1,X_2,Y_2,u_2)= \int_0^1 \frac{d w}{w} (\frac{ u_1 X_1
+ u_2 X_2 w}{X_1 + X_2 w} \frac{u_1 Y_1 + u_2 Y_2 w }{Y_1 + Y_2 w
} - u_1^2)
\end{eqnarray}
This integral is easily done but is greatly simplified if one uses
the symmetries (\ref{sym}). For that purpose one defines:
\begin{eqnarray}
&& h_s(X_1,Y_1,u_1,X_2,Y_2,u_2)= \frac{1}{2} (
h(X_1,Y_1,u_1,X_2,Y_2,u_2) +
h(X_2,Y_2,u_2,X_1,Y_1,u_1) )  \\
&& h_{ss}(X_1,Y_1,u_1,X_2,Y_2,u_2) = \frac{1}{2} (
h_s(X_1,Y_1,u_1,X_2,Y_2,u_2) + h_s(X_2,Y_2,-u_1,X_1,Y_1,-u_2) )
\label{symmetriz}
\end{eqnarray}
A little calculation yields:
\begin{eqnarray}
&& h_{ss}(X_1,Y_1,u_1,X_2,Y_2,u_2)= \frac{1}{4} (u_1 - u_2)^2 \frac{1 + Z}{1 - Z} \ln Z \\
&& Z = \frac{X_1 Y_2}{X_2 Y_1} = e^{m^2 (u_1-u_2) (v-t)}
\end{eqnarray}
Denoting $y=u_2-u_1$ the final result is very simple:
\begin{eqnarray}
&& R''(v) = R''(0) + m^4 T \langle y^2 F_2(m^2 y \tilde v)
\rangle_y \label{tbl2} \\
&& F_2(z) = \frac{z}{4} \coth \frac{z}{2} - \frac{1}{2} = \frac{z}{4} \frac{e^z + 1}{e^z - 1}  - \frac{1}{2} =
\frac{z^2}{24} - \frac{z^4}{1440} + O(z^6)
\end{eqnarray}
Thus we have now an exact correspondence between the droplet
probability of two degenerate minima distant from $y$, $D(y)$ in
(\ref{D}), and the full function $R(u)$ in the TBL: they contain the
same information. The question posed in the beginning of this paper
is thus finally answered. Each higher derivative $R^{(2p)}(0)$ is
proportional to a moment $\langle y^{2 p + 2} \rangle_y$ of $D(y)$.
It may come as a surprise that only $y=u_1-u_2$ appears in these
formulae, since after all the system is in an harmonic well. We will
see below that this property extends to the third moment, things
change after the fourth. The value of $R''(u=0)$ can also obtained
more directly as:
\begin{eqnarray}
&& \overline{ \langle u \rangle^2 } = - \frac{R''(0)}{m^4} = \langle u_1^2 \rangle_{P} + \overline{ (\frac{u_1 +
u_2 w }{1 + w })^2 - u_1^2} = \langle u_1^2 \rangle_{P} - T \frac{1}{2} \langle (u_1-u_2)^2 \rangle_{u_i} =
\langle u_1^2 \rangle_{P} - \frac{T}{m^2} \label{u2}
\end{eqnarray}

It is instructive to display the integrated versions:
\begin{eqnarray}
&& R'(v) = R''(0) v + m^2 T^2 \langle y G_2(m^2 y \tilde v) \rangle_y \\
&& R(v) = R(0) + \frac{1}{2} R''(0) v^2 + T^3 \langle H_2(m^2 y \tilde v) \rangle_y \\
&& G_2(z) = \frac{\pi^2}{12} + \frac{z}{8} (z-4+ 4 \ln(1-e^{-z}))
- \frac{1}{2} Li_2(e^{-z}) \\
&& H_2(z) = - \zeta(3) + \frac{\pi^2}{12} z - \frac{1}{4} z^2 +
\frac{5}{24} z^3 + \frac{1}{2} z^2 \ln(1-e^{-z}) - \frac{1}{2} z^2
\ln(1-e^{z}) - \frac{1}{2} z Li_2(e^{-z}) - z Li_2(e^{z}) +
Li_3(e^{z})  \nonumber \label{defh2}
\end{eqnarray}
which contain polylogarithm functions $Li_n(z)=\sum_{k=1}^\infty z^k/k^n$.
Those would have appeared if we had computed disorder averages of the
$\ln$ in the potential (\ref{pot}), rather than the force.

As was recalled in Section \ref{sec:loopexpansion} we expect that
the large $\tilde v$ limit coming from the TBL should match the
small $v$ behaviour coming from the outer (zero temperature) region.
We find that for $\tilde v \to \infty$:
\begin{eqnarray}
&&  \frac{R''(0)-R''(v)}{m^4} \sim - \frac{T m^2}{4} |\tilde v|
\langle |y|^3   \rangle_{y} =  - \frac{1}{2} |v| \frac{\langle
|y|^3 \rangle_{y}}{\langle y^2 \rangle_{y}} \nonumber
\end{eqnarray}
The linear cusp of the zero temperature fixed point is thus
beautifully reproduced with the exact result for $d=0$:
\begin{eqnarray}
&& \frac{R'''(0^+)}{m^4} =  \frac{1}{2}  \frac{\langle |y|^3
\rangle_{y}}{\langle y^2 \rangle_{y}} \label{drop1}
\end{eqnarray}
both sides have dimension of length. Similar relations will be
derived in higher $d$ in Section \ref{sec:higherd}. In $d=0$ it will
be further confirmed below for the Sinai Random Field case from the
exact solution for the function $R(u)$ in its outer region.

The last question of course is the validity of the droplet
calculation. One should see the droplet model as one possible
solution of all the STS and ERG constraints on correlations at low
temperature. In $d=0$ and for LR disorder (i.e. when there is a
glass phase) it seems fairly inescapable, although a proof would be
welcome (beyond the Sinai case). It could fail in two ways. In the
first case, two wells are indeed sufficient to describe low $T$, and
failure would then require very peculiar correlations between
$E_2-E_1$, when it is of order $T$, and well positions (e.g. like
eigenvalue repulsion). Or, second case, more than two wells are
needed. The latter may happen as $\theta \to 0$ since then we know
(e.g. for a logarithmic $R_0(u)$) that many wells are important in
the low temperature phase, where some replica symmetry breaking
(RSB) phenomenon takes place \cite{CarpentierLeDoussal2001}. We also
know that for infinite $N$ in $d=0$ these many-well
quasi-degeneracies take place. Lastly, let us point out that these
results, described here within the disordered model, can also be
understood within the Burgers approach. We do not wish to anticipate
on this topic until Section \ref{subsec:burgers}, and pursue here
along the FRG path.

\subsubsection{higher moments}

The droplet calculation of {\it all} the higher $W$-moments $\bar
S^{(n)}(v_{1..n})$ turns out to be possible along the same line in
the full TBL, i.e. for all arguments $\tilde v_i$ fixed. From these,
one can of course obtain all the $W$-cumulants and
$\Gamma$-cumulants. Here we give only the result for
the third and fourth moments. The general result and calculation is
performed in Appendix \ref{app:dropletmoments}.

The third cumulant reads:
\begin{eqnarray}
&& \bar S_{111}(v_1,v_2,v_3) = \hat S_{111}(v_1,v_2,v_3) = -
\overline{ \hat V'(v_1) \hat V'(v_2) \hat V'(v_3) } = m^6 T \langle
y^3 F_3(m^2
\tilde v_1 y, m^2 \tilde v_2 y, m^2 \tilde v_3 y) \rangle_y \\
&& F_3[z_1,z_2,z_3]=\frac{1}{4} (z_1 (F[z_1-z_2,z_1-z_3]-\frac{2}{3}) + 2 \text{perm} ) \label{tbl3} \\
&& F(a,b) - \frac{2}{3} =\frac{1+ e^{a+b}}{(1-e^{a})(1-e^b)}-
\frac{2}{3} =
 \frac{1}{2} \frac{\cosh(\frac{a+b}{2})}{\sinh(\frac{a}{2})\sinh(\frac{b}{2})}
- \frac{2}{3} = \frac{1}{2} \coth(\frac{a}{2})\coth(\frac{b}{2}) -
\frac{1}{6}
\end{eqnarray}
where $y=u_1-u_2$. One can check (statistical) translational
invariance (STS), i.e. $\bar S_{111}(v_1+v,v_2+v,v_3+v)=\bar
S_{111}(v_1,v_2,v_3)$ for any $v$, S using $\coth(x+y)=(1+\coth x
\coth y)/(\coth x + \coth y)$ and $sym_{abc} (1-a b)(1- a
c)/(a-b)(a-c) = 1/3$. Since one can shift all the arguments by
$-v_1$, the equivalent information is contained in the function:
\begin{eqnarray}
&& F_3[0,z_2,z_3]= \frac{1}{12} \frac{(z_2-2 z_3 - (z_2+z_3) \cosh
z_3) \sinh z_2 - (z_3-2 z_2 - (z_2+z_3) \cosh z_2) \sinh z_3
}{\sinh z_2 - \sinh z_3 - \sinh(z_2-z_3)}
\end{eqnarray}

Starting from the fourth moment the structure changes slightly:
\begin{eqnarray}
&& \bar Q_{1111}(v_1,v_2,v_3,v_4) = \langle u_1^4 \rangle_P + m^8 T
\langle y^4 F_4[m^2 y \tilde v_1,m^2 y \tilde v_2,m^2 y \tilde
v_3,m^2 y \tilde v_4])
\rangle_y \label{tbl4} \\
&& + m^8 T \langle u_1 u_2 y^2 \tilde F_4[m^2 y \tilde v_1,m^2 y  \tilde v_2,m^2
y  \tilde v_3,m^2 y  \tilde v_4]) \rangle_{u_1,y}
+ O(T^2) \\
&& F_4[z_1,z_2,z_3,z_4] = \frac{1}{4} (z_1
F[z_1-z_2,z_1-z_3,z_1-z_4] + 3 \text{perm} ) \\
&& \tilde F_4[z_1,z_2,z_3,z_4] = \frac{1}{4} (z_1 \tilde F[z_1-z_2,z_1-z_3,z_1-z_4] + 3 \text{perm} ) \\
&& F[a,b,c]=\frac{1+e^{a+b+c}}{(1-e^a)(1-e^b)(1-e^c)} \\
&& \tilde F(a,b,c)= -3 + \frac{2}{1-e^a} + \frac{2}{1-e^b} +
\frac{2}{1-e^c}
\end{eqnarray}
i.e. the second term involves not only $y=u_{12}$ but also an explicit $u_1$, i.e. on the precise nature of the
harmonic well. Note that this term resembles a disconnected contribution with one second cumulant and one
temperature. The fact that the TBL of $R$ and $S$ do not depend on $u_1$ probably reflects some universality of $R$
and $S$ with respect to infrared cutoff procedure.

\subsubsection{check that the droplet solution obeys the FRG equations}

\label{checktbl}

The above solution should satisfies the FRG equation (within the TBL). There should thus be differential
relations between the functions $F_n$ introduced above. Let us examine the derivative of the ERG equation for
the second moment (\ref{mom2}):
\begin{eqnarray}
&& - m \partial_m R'(v)  = \frac{2 T}{m^2} R'''(v) + \frac{2}{m^2} \bar S_{111}(0,0,v) \label{deriv22}
\end{eqnarray}
and consider $v=T \tilde v$ with $\tilde v=O(1)$. From (\ref{tbl3}) the feedback from the third moment involves:
\begin{eqnarray}
&& F_3[0,0,z]= \frac{3 + z + 4 z e^z + (z-3) e^{2 z}}{12 (e^z -1)^2} = \frac{z^3}{360} + O(z^5)
\end{eqnarray}
Thanks to the exact relation:
\begin{eqnarray}
&& F_3[0,0,z] + F'_2[z] = \frac{z}{12}
\end{eqnarray}
where $F_2$ is the TBL function of the second cumulant (\ref{tbl2}) one finds that the r.h.s. of (\ref{deriv22})
simplifies into:
\begin{eqnarray}
&& \frac{2 T}{m^2} R'''(v) + \frac{2}{m^2} \bar S_{111}(0,0,v) = \frac{1}{6} m^6 T \langle y^4 \rangle_y \tilde
v + O(T^2)
\end{eqnarray}
a simple (linear) quantity of order $O(T)$ in the TBL. The l.h.s. of (\ref{deriv22}) is a priori of order
$O(T^2)$ apart from the single ''zero temperature'' piece containing $R''(0) v$ (since $v=T \tilde v$) thus if
FRG is obeyed one should have:
\begin{eqnarray}
&& - m \partial_m R''(0) v = \frac{1}{6} m^6  \langle y^4
\rangle_y v + O(T^2)
\end{eqnarray}
This can be checked using now an exact relation (not dependent on droplet assumption) which is obtained by
taking an additional derivative of (\ref{deriv22}) at $u=0$, and noting that the third moment $\bar S$ can only
start at order $u^6$ (i.e. $\bar S^{(3)}_{112}(0,0,0) =S^{(3)}_{112}(0,0,0)=0$, see e.g. Appendix B of Ref.
\onlinecite{BalentsLeDoussal2005}). Hence:
\begin{eqnarray}
&& - m \partial_m R''(0) = \frac{2 T}{m^2} R''''(0)
\end{eqnarray}
If we now use (\ref{tbl2}) to evaluate $R''''(0)$ we find consistency since:
\begin{eqnarray}
&& T R''''(0) = m^8 \langle y^4 \rangle_y F_2''(0) + O(T) = m^8
\frac{1}{12} \langle y^4 \rangle_y + O(T)
\end{eqnarray}
Thus the FRG equation for the second cumulant is verified by the droplet
solution in the TBL.

Let us now check the ERG equation for the third moment. From (\ref{frgw2}), taking three derivatives one finds:
\begin{eqnarray}
&& - m \partial_m  \bar S_{111}(v_1,v_2,v_3) = \frac{T}{m^2} (
\partial_1^2 +
\partial_2^2 + \partial_3^2 ) \bar S_{111}(v_1,v_2,v_3)
\\
&& + \frac{2}{m^2} ( \bar Q_{2111}(v_1,v_1,v_2,v_3) + \bar
Q_{2111}(v_2,v_2,v_1,v_3) + \bar Q_{2111}(v_3,v_3,v_1,v_2) )
\end{eqnarray}
In the TBL the r.h.s. of this equation should be of order $O(1)$, and one can in fact show that it exactly
vanishes. Indeed one can check that the above solution (\ref{tbl3}) and (\ref{tbl4}) satisfies:
\begin{eqnarray}
&& (\partial_{z_1}^2 + \partial_{z_2}^2 + \partial_{z_3}^2)
F_3[z_1,z_2,z_3] + \partial_{z_1} F_4[z_1,z_1,z_2,z_3] +
\partial_{z_2} F_4[z_2,z_2,z_1,z_3] + \partial_{z_3}
F_4[z_3,z_3,z_1,z_2] = 0 \\
&& \partial_{z_1} \tilde F_4[z_1,z_1,z_2,z_3] +
\partial_{z_2} \tilde F_4[z_2,z_2,z_1,z_3] + \partial_{z_3}
\tilde F_4[z_3,z_3,z_1,z_2] = 0
\end{eqnarray}
In addition, there is no $O(1)$ constant piece from the r.h.s. since $\bar S_{111}(0,0,0) =0$. Thus the FRG
equation is obeyed in the TBL.

To conclude we have obtained the solution of the FRG equation for all cumulant in the TBL. It is parameterized
by a single droplet distribution, i.e. a function $D(u_1,u_2)$. This function remains arbitrary, i.e. one needs
information outside the TBL to determine it. Let us point out that in the terminology of Ref
\cite{BalentsLeDoussal2005} what is obtained here is the full TBL, all $\tilde v_i = O(1)$. Cumulants higher
than $R(u)$ exhibit additional intermediate regimes between the full TBL and the outer region where all $v_i =
O(1)$. These are the so-called partial BL, where some $v_{ij}$ are of order one, other of order $T$. We have not
obtained the solution for these (except, see below in the Sinai RF case where they can in principle be
computed). They will be discussed below in the language of Burgers, where they are associated to correlations
between shocks, while the full TBL corresponds to a single shock quantity, rounded by viscosity (i.e.
temperature).

\subsubsection{Droplet observables in Sinai case (random field):}
\label{dropletobs}

For the RF Sinai case, thanks to the Markovian properties of $V(u)$ (which is a simple Brownian walk in ''time'' $u$) it is
possible to obtain analytically the droplet probabilities $P(u_1)$ and $D(u_1,u_2)$. This was performed in Ref.
\cite{toy} and we recall the results. Here we choose the parameters $m^2=1$ and $\overline{(V(u)-V(u'))^2} = 2
\sigma |u-u'|$ with $\sigma=1$, the general case is easily restored at the end performing the change of spatial
scale $u \to u \sigma^{-1/3} m^{4/3}$ and energy $V \to V \sigma^{-2/3} m^{2/3}$.

\begin{figure}[h]
\centerline{\includegraphics[width=7cm]{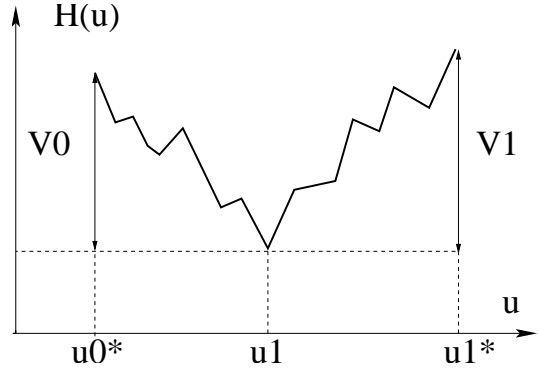}}
\caption{A random lansdscape such that the
minimum of $H(u) = V(u) + \frac{1}{2} u^2$ on a given interval
$u \in [u_0^*,u_1^*]$ is at position $u_1$ and
that the energy differences with the two edges are $V_{0}=H(u_0^*)-H(u_1)$
and $V_{1}=H(u_1^*)-H(u_1)$ \label{figtoy2}}
\end{figure}

\begin{figure}[h]
\centerline{\includegraphics[width=7cm]{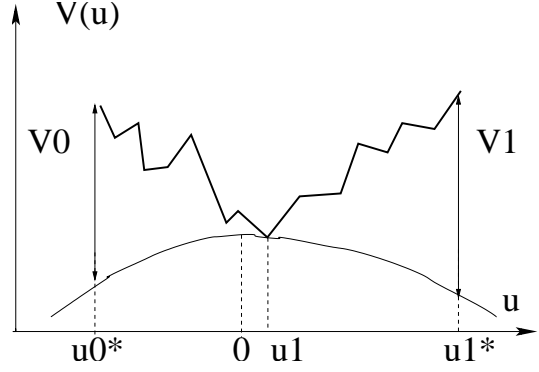}} \caption{The same as Fig \ref{figtoy2} where $V(u)$ is
plotted, together with the inverted parabola $y=-\frac{1}{2} u^2 +H(u_1)$.  \label{figtoy3}}
\end{figure}

Let us first recall a more general result obtained in Ref.~\onlinecite{toy} for the joint probability, denoted
\footnote{it is $N_\infty^{(2)}(x_L,u_L-u,x,u_R-u,x_R)$ in notations of Ref.~\onlinecite{toy}}
$N_\infty^{(2)}(u_0^*,V_0,u_1,V_1,u_1^*) du_1 dV_0 dV_1$ that the minimum of $H(u) = V(u) + \frac{1}{2} u^2$ on
a given interval $u \in [u_0^*,u_1^*]$ is at position $u_1$ (within $du_1$) and that the energy differences with
the two edges are $V_{0}=H(u_0^*)-H(u_1)$ and $V_{1}=H(u_1^*)-H(u_1)$ (within $dV_0$ and $dV_1$). This is
illustrated in Fig. \ref{figtoy2}. This is equivalent (see Fig. \ref{figtoy3}) to the condition that the
Brownian walk $V(u)$ must remain above the (inverted) parabola centered on $u=0$:
\begin{eqnarray}
&& V(u) > - \frac{u^2}{2} + E_1
\end{eqnarray}
with one contact point at $u_1$ where the equality holds (i.e. the walk is on the parabola), i.e. $E_1 = H(u_1)=
\frac{u_1^2}{2} + V(u_1)$, and start and finish at
\begin{eqnarray}
&& V(u_i^*) - V(u_1) = - \frac{(u_i^*)^2}{2} + \frac{u_1^2}{2} + V_i
\end{eqnarray}
It was found (formula (D.5) in Ref.~\onlinecite{toy}) that:
\begin{eqnarray}
&& N_\infty^{(2)}(u_0^*,V_{0},u_1,V_1,u_1^*)
= \tilde h(-u_1,-u_{0}^*,V_{0}) \tilde h(u_1,u_1^*,V_1) \label{N} \\
&& \tilde h(u,u_R,V_R) = e^{\frac{1}{12} (u^3 - u_R^3) + \frac{1}{2} u_R V_R}
h(u_R-u,V_R) \\
&& h(u,V) = \int_{-\infty}^{+\infty} \frac{d \lambda}{2 \pi} e^{i \lambda u} \frac{Ai(a V + i \lambda b)}{Ai(i
\lambda b)} \label{huV}
\end{eqnarray}
$b=1/a^2$ and $a=2^{-1/3}$ and the normalization identity:
\begin{eqnarray}
&& 1 = \int_0^\infty dV_1 \int_0^\infty dV_0 \int_{u_{0}^*}^{u_1^*} du_1
N_\infty^{(2)}(u_0^*,V_{0},u_1,V_1,u_1^*)
\end{eqnarray}
From the Markovian property it factors into two blocks, where the right block is $\tilde h(u,u_R,V_R)$ is the
probability that a walk which starts on the centered parabola at time $u'=u$ remains above it and ends up a
vertical distance $V_R$ above the parabola at $u'=u_R$ (the left block is its mirror image). These are obtained
as ''renormalized bonds'' \footnote{$h(u_R-u,V_R)=\tilde E_\infty^-(0,V_R,u_R-u)$ in the notations of 
Ref.~\onlinecite{toy}} in the RSRG method of Ref.~\onlinecite{toy}. The function $Ai(z+V)/Ai(z) \sim e^{- V
z^{1/2}}$ thus for $V>0$ it decays everywhere. If $u >0$ the contour in (\ref{huV}) can be closed on the side
$Re(z) <0$, and gives a strictly positive result. If $u < 0$ the contour can be closed on the $Re(z) >0$ side
and gives zero, i.e. $h(u,V) = 0$ for $u<0$. Note that $\int_{- \infty}^{+\infty} du h(u,V) = Ai(a V)/Ai(0)$ and
$h(u,0)= \delta(u)$.

From $N_\infty^{(2)}$ one can obtain both $P(u_1)$ and $D(u_1,u_2)$. Taking
the interval $[u_0^*,u_1^*]$ to become the whole real axis $]-\infty,+\infty[$
one obtains the distribution of the position of the absolute minimum:
\begin{eqnarray} \label{exactP}
&& P(u_1) = \tilde g(-u_1) \tilde g(u_1)  =  g(-u_1) g(u_1) \\
&& \tilde g(u) = \lim_{u_R \to \infty} \int_0^\infty dV_R \tilde h(u,u_R,V_R) = e^{u^3/12} g(u) \label{gtilde} \\
&& g(u) = \int_{- \infty}^{+ \infty} \frac{d \lambda}{2 \pi}
\frac{e^{- i \lambda u}}{a Ai(i \lambda b)} \label{g} \\
&& = \frac{1}{a b} \sum_k e^{s_k u/b}
\frac{1}{Ai'(-s_k)} \quad \text{for} u <0
\end{eqnarray}
The normalization condition $1 = \int_{-\infty}^{+\infty} du_1 P(u_1)$ follows from the Airy functions identity
$\pi (Bi(z)/Ai(z))' = 1/Ai(z)^2$ which implies $\int_{- \infty}^{+ \infty} \frac{d \lambda}{2 \pi} \frac{1}{a^2
Ai(i \lambda b)^2} =1$. Note also that $Ai(z) \sim z^{-1/4} e^{- 2 z^{3/2}/3}$ for large $|z|$. Thus $1/Ai(z)$
decays only for $|arg(z)| > \pi/3$ and the contour can be closed along negative $z$ only when $u<0$ resulting in
a sum over the zeroes $-s_k$ of the Airy function. Each factor $\tilde g(u)$ arises from the probability that a
walk which starts on the centered parabola at time $u'=u$ remains above it for all larger $u'$. One has the asymptotic behavior:
\begin{eqnarray} \label{asymp1} 
&& g(u) \approx_{u \to + \infty}  2 a^3 u e^{-a^6 u^3/3}  \\
&& P(u_1)  \approx_{|u| \to \infty} \frac{2 a^4}{Ai'(-|s_1|)} |u_1| e^{- a^2 |s_1| |u_1| -a^6 u_1^3/3 }
\end{eqnarray}
The probability of (quasi) degenerate minima (within $\epsilon$) can be obtained by considering two adjacent
blocks (see Fig \ref{figtoy5}), setting e.g. $V_1 = \epsilon$. One defines:
\begin{eqnarray}
&& \tilde d(u_1,u_2) = \partial_{V_R} \tilde h(u_1,u_2,V_R)|_{V_R=0}
\end{eqnarray}
such that the quantity:
\begin{eqnarray}
&& \epsilon \tilde d(u_1,u_2) = \epsilon e^{\frac{1}{12} (u_1^3
- u_2^3) } d(u_2-u_1) \nonumber \\
&& d(u)= \partial_{V} h(u,V)|_{V=0} = a \int_{- \infty}^{+ \infty} \frac{d \lambda}{2 \pi} e^{i \lambda u}
\frac{Ai'(i \lambda b) }{Ai(i \lambda b)} \label{d}
\end{eqnarray}
describes the probability that a walk which starts on the centered parabola at time $u'=u_1$ remains above it
and terminates within $\epsilon$ in energy of the parabola again at time $u'=u_2>u_1$. Note that the integral in (\ref{d}) should be taken in the sense that $u d(u)$ is the Fourier transform of the second derivative of $\ln Ai(z)$. Using the above asymptotics for Airy functions, this yields that at small $u$ one has:
\begin{eqnarray} \label{dsmallu} 
&& u d(u) \sim u^{-1/2}
\end{eqnarray}
as can be obtained from the return probability to the origin of a simple random walk (since on that scale the
curvature of the toy model energy landscape does not play any role). 
The total droplet probability
also takes into account the two outer intervals with the net result \cite{toy}:
\begin{eqnarray}
&& D(u_1,u_2) = \hat D(u_1,u_2) \theta(u_2-u_1) +
\hat D(u_2,u_1) \theta(u_1-u_2) \\
&& \hat D(u_1,u_2) =  \tilde g(- u_1) \tilde d(u_1,u_2) \tilde g(u_2) =  g(- u_1) d(u_2-u_1) g(u_2)
\label{exactD}
\end{eqnarray}
It is easy at this stage to restore the dependence in arbitrary $m$ and $\sigma$. One just needs to replace
everywhere here and below:
\begin{eqnarray}
a=2^{-1/3} \sigma^{-2/3} m^{2/3} \quad , \quad b=2^{2/3} \sigma^{1/3} m^{-4/3}
\end{eqnarray}
which still satisfies $a^2 b=1$. It was checked in Section IV-D of Ref. \cite{BalentsLeDoussal2005} that the STS
and ERG identities (\ref{rgdrop2}) are indeed satisfied \footnote{For general $m$, $\sigma$ one thus has
$P(u_1)=u_m^{-1} \tilde P(u_1/u_m)$, $D(u_1,u_2)=u_m^{-2} E_m^{-1} \tilde D(u_1,u_2)$ and $D(y)= u_m^{-1}
E_m^{-1} \tilde D(y/u_m)$ where $u_m= 2^{2/3} \sigma^{1/3} m^{-4/3}$ and $2^{-1/3} E_m=\sigma^{2/3} m^{-2/3}$
and $\tilde P$,$\tilde D$ are given by the same expressions as $P$ and $D$ here setting $a=b=1$. Hence the check
of STS and RG in Section IV-D of Ref. \cite{BalentsLeDoussal2005} was performed using $m=1$, $\sigma=1/4$, as
may not be clear from the explanation there.} by these exact results for $P(u_1)$
and $D(u_1,u_2)$.

\begin{figure}[h]
\centerline{\includegraphics[width=7cm]{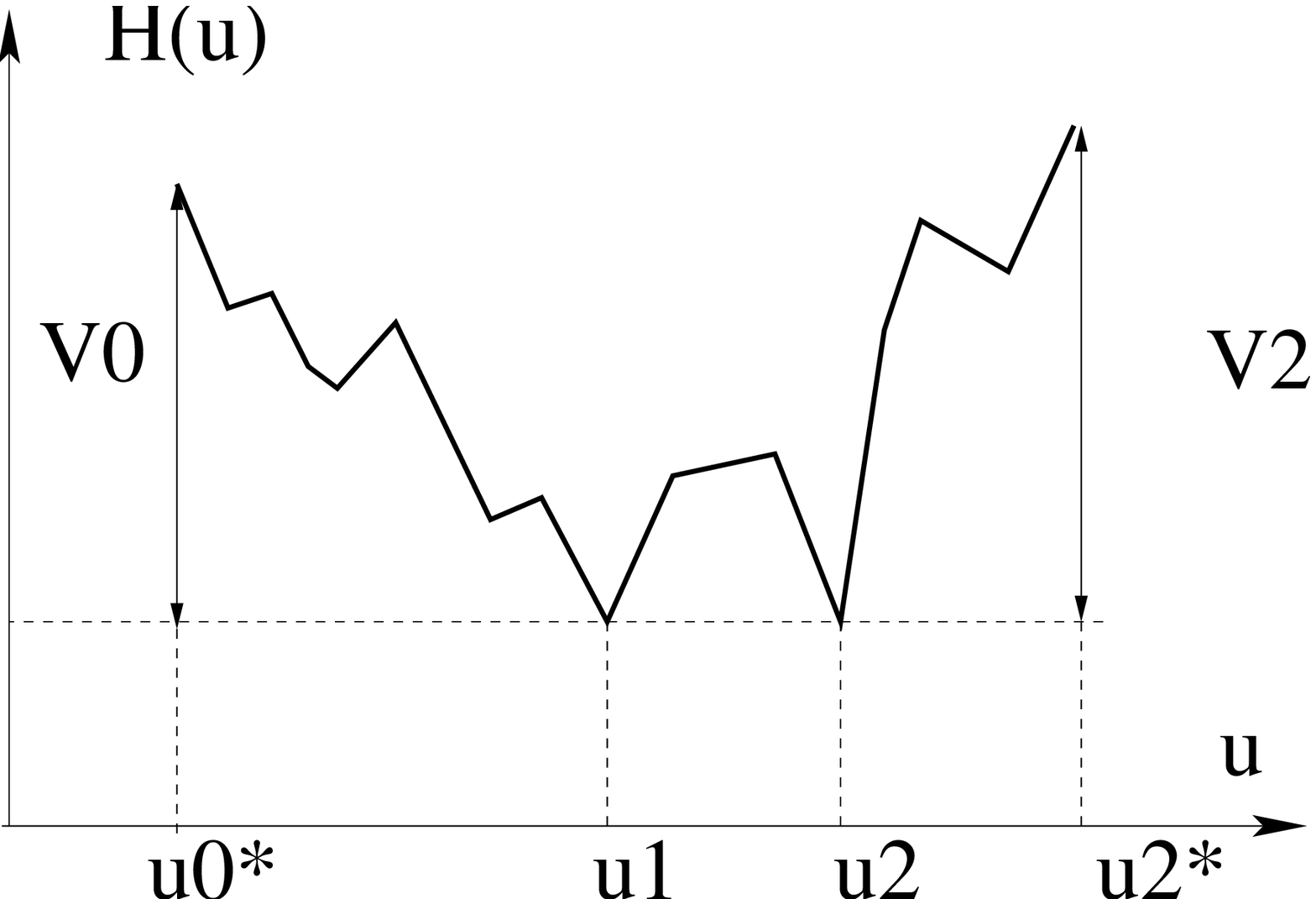}} \caption{The probability of a landscape with two
degenerate minima (within $\epsilon$) can be obtained from two adjacent blocks with $V_1 = \epsilon$
\label{figtoy5}}
\end{figure}

The total probability density (\ref{D}) that there are two quasi-degenerate minima separated by $y=u_1-u_2$ thus
reads
\begin{eqnarray} \label{dD} 
&& D(y) =  d(|y|) \int_{- \infty}^{+ \infty} du g(-u) g(u + |y|) = a b \int_{- \infty}^{+ \infty} \frac{d
\lambda_1}{2 \pi} \int_{- \infty}^{+ \infty} \frac{d \lambda_2}{2 \pi} e^{i (\lambda_1-\lambda_2) |y|}
\frac{Ai'(i
\lambda_1 b) }{Ai(i \lambda_1 b) Ai(i \lambda_2 b)^2} \\
&& = \frac{a}{b} \int_{- \infty}^{+ \infty} \frac{d \lambda_1}{2 \pi} \int_{- \infty}^{+ \infty} \frac{d
\lambda_2}{2 \pi} e^{i (\lambda_1-\lambda_2) |y|/b} \frac{Ai'(i \lambda_1) }{Ai(i \lambda_1) Ai(i \lambda_2)^2}
\label{sinaiD}
\end{eqnarray}
with $D(-y)=D(y)$ and which satisfies the normalization \footnote{for $y>0$ the function $D(y)$ is the same as
the one in Ref.~\onlinecite{toy}. However the definitions differ since here $y=u_1-u_2$ has arbitrary sign, and
$E_2-E_1>0$, while in Ref.~\onlinecite{toy} $y=|u_1-u_2|$ and $E_2-E_1$ has arbitrary sign. The end result for
expectation values are of course identical}:
\begin{eqnarray}
&& \int_{- \infty}^{+ \infty} dy y^2 D(y) = a b^2 = \frac{2}{m^2}  \label{drop2}
\end{eqnarray}

The final result for the renormalized disorder correlator in the TBL for the Sinai model is thus:
\begin{eqnarray}
&& R''(v) - R''(0) = m^4 T \int_{0}^{+ \infty} dy D(y) y^2 (\frac{m^2 y \tilde v}{4} \coth \frac{m^2 y \tilde
v}{2} - \frac{1}{2})
\end{eqnarray}
where $D(y)$ is given by (\ref{sinaiD}). Below we obtain $R(v)$ outside the TBL, and we check the matching
explicitly between the two regimes.

It is useful to recall the (half) generating function \cite{toy}:
\begin{eqnarray}
&& \hat D(p) = \int_0^\infty dy D(y) e^{- p y}  = a \int_{- i \infty}^{i \infty} \frac{dz_2}{2 i \pi}
\frac{1}{Ai(z_2)^2}
\int_{- i \infty}^{i \infty} \frac{dz_1}{2 i \pi} \frac{Ai'(z_1)}{Ai(z_1)} \frac{1}{b p + z_2-z_1} \\
&& = a \int_{- i \infty}^{i \infty} \frac{dz_2}{2 i \pi} \frac{1}{Ai(z_2)^2} \frac{Ai'(z_2+b p)}{Ai(z_2+b p)}
\end{eqnarray}
The second expression is obtained for $b p>0$ by closing the contour on the $Re(z_1)>0$ side and using Cauchy's
theorem, and is somewhat formal, but allows to obtain \footnote{by derivation w.r.t. $p$ the conditions for
convergence are restored} the moments \cite{toy}:
\begin{eqnarray}
&& \int_{- \infty}^{+ \infty} dy |y|^k D(y) = 2 \sigma^{(k-2)/3} m^{(2-4 k)/3} y_k \\
&& y_k = (-1)^k 2^{(2k-1)/3} \int_{- \infty}^{+ \infty} \frac{d \lambda}{2 \pi} (\frac{Ai'(z)}{Ai(z)})^{(k)}(z=i
\lambda) \frac{1}{Ai(i \lambda)^2}  \label{drop3}
\end{eqnarray}
One notes also, from formula (124) of \cite{toy}, an unexpected relation $\hat D(p) = a d/dE
P_{[0,\infty[}(E)|_{E=bp}$ with the probability of the minimum energy $-E$ on the half line. One has:
\begin{eqnarray}
&& y_2 = 1 \\
&& y_3 = 1.80258 \\
&& y_4 = 4.21695 \\
&& y_5 = 11.6187 \\
&& y_6 = 36.0516 \\
&& y_7 = 122.769
\end{eqnarray}

The droplet distribution immediately allows to obtain the (normalized) distribution $p(s)$ of shock sizes $s=u_{i+1}-u_i >0$ . This quantity will be defined in Section \ref{shockdroplet} where we demonstrate:
\begin{eqnarray}  \label{rel0}
&&  p(s) = \langle s \rangle_{\cal P} \frac{2}{\langle s^2 \rangle}_D s D(s) \theta(s)
\end{eqnarray}
From this we can compute the dimensionless universal ratios, independent of any parameters ($m$ or $\sigma$):
\begin{eqnarray}
&& \langle s^{k+1} \rangle \langle s^{k-1} \rangle \langle s^{k} \rangle^{-2} = y_{k+2} y_{k} y_{k-1}^{-2}
\end{eqnarray}
This yields:
\begin{eqnarray}
&& \langle s^{3} \rangle \langle s \rangle \langle s^{2} \rangle^{-2} = y_{4} y_{2} y_{3}^{-2} = 1.2978 \\
&& \langle s^{4} \rangle \langle s^2 \rangle \langle s^{3} \rangle^{-2} = y_{5} y_{3} y_{4}^{-2} = 1.17776 \\
&& \langle s^{4} \rangle \langle s \rangle^2 \langle s^{2} \rangle^{-3} = y_{5} y_{2}^2 y_{3}^{-3} = 1.98369
\end{eqnarray}

Note that the density of small droplets diverge as $D(y) \sim y^{- 3/2}$ at small $y$, hence the 
distribution of shock sizes has a $p(s) \sim s^{-1/2}$ at small $s$. At large $s$ they both decay with a faster than gaussian stretched exponential tail $\exp(-B s^3)$ as can be seen from (\ref{dD}) and (\ref{asymp1}). 

\subsection{Ambiguity-free zero temperature beta function to four loop via the ERG}

\label{subsec:4loop}

In this section we study the $\Gamma$-Exact RG defined in Section \ref{erggamma}, for $d=0$ and $N=1$. We show how it reproduces the $T>0$ beta function of Section \ref{sec:loopexpansion}. We then recall the method introduced in Ref. \cite{BalentsLeDoussal2005} to study the partial boundary layers (PBL) and obtain the correct large $l$, $\tilde T_l \to 0$, ''zero temperature'' limit of the beta function.
Next, we finish here the job started in Ref. \cite{BalentsLeDoussal2005}
to obtain the unambiguous beta function. There, only a simplified version was presented to three and four loop, in which one arbitrarily sets the roughness exponent $\zeta$ to zero. In the two loop part $\zeta$ was kept non trivial, but the version given there was unnecessarily heavy as it contained convolutions which, we show here, can be removed.

Let us recall the $\Gamma$-ERG equations for the rescaled cumulants of Section \ref{erggamma}:
\begin{eqnarray}
&& R(u) = \frac{1}{4} m^{\epsilon-4 \zeta} \tilde R(u m^{\zeta}) \quad, \quad
 S^{(3)} = \frac{1}{8} m^{2 \epsilon - 2 - 6 \zeta} \tilde S^{(3)} \quad, \quad
S^{(4)} = \frac{1}{16} m^{3 \epsilon - 4 - 8 \zeta} \tilde S^{(4)} \quad, \quad \tilde T = 2 T m^\theta
\end{eqnarray}
$u_i m^{\zeta}$ being implicit as arguments of all rescaled cumulants
and $\theta=d-2 + 2 \zeta=2 - \epsilon + 2 \zeta$. Everywhere here
$\epsilon=4$ ($d=0$). We denote $\partial_l=-m\partial_m$ everywhere
\begin{eqnarray}
  && \partial_l \tilde{R}(u) = (\epsilon - 4 \zeta + \zeta u \partial_{u} ) \tilde{R}(u) +
  \tilde {T}_l \tilde{R}''(u) +
  \frac{1}{2} \tilde{R}''(u)^2 - \tilde{R}''(0) \tilde{R}''(u) + \tilde{S}^{(3)}_{110}(0,0,u)
\label{Gcum2r}
\end{eqnarray}
\begin{eqnarray}
  && \partial_l \tilde{S}^{(3)}(u_{123}) = (2
  \epsilon - 2 - 6 \zeta + \zeta u_i \partial_{u_i}) \tilde{S}^{(3)}(u_{123}) +
   \, \big[ \frac{3\tilde{T} _l}{2}\tilde{S}^{(3)}_{200}(u_{123})
    + \frac{3\tilde{T} _l}{2}\tilde{\sf R}''(u_{13})
    \tilde{\sf R}''(u_{23}) \big]     \nonumber \\
  && + 3 \, \big[  \tilde{\sf R}''(u_{12}) (
    \tilde{S}^{(3)}_{110}(u_{113}) -
    \tilde{S}^{(3)}_{110}(u_{123}))
    +\gamma \tilde{\sf R}''(u_{12})\tilde{\sf R}''(u_{13})^2  -
    \frac{\gamma}{3} \tilde{\sf R}''(u_{12}) \tilde{\sf R}''(u_{23})
  \tilde{\sf R}''(u_{31}) \big]  , \nonumber \\
&& + \frac{3}{2} \big[ \tilde{S}^{(4)}_{1100}(u_{1123}) \big] \label{Gcum3r}
\end{eqnarray}
where $\gamma=3/4$ and $[..]$ denote symetrization over the three arguments. We have defined $\tilde {\sf R}''(u)=R''(u)-R''(0)$ and various other notations are
given in (\ref{defarg}) and (\ref{defder}). The fourth cumulant is needed only to three
loop. The equation (\ref{Gcum2r}) setting $\tilde S^{(3)} = 0$ already yields the
one loop beta function.

\subsubsection{two loop}

We first show how to recover the two loop contribution to the beta functions at non zero temperature,
displayed in Section \ref{sec:loopexpansion} and obtained there by a more standard field theoretic method. We recall, as detailed there, that the $n$-loop contribution is a sum of
terms of the form $T^p R^{n+1-p}$, $p=0,..n$. To two loop we need
the third cumulant equation (\ref{Gcum3r}) but we can discard
the fourth cumulant feeding term in (\ref{Gcum3r}) (which contains only $T R^3$ and $R^4$
terms), as well as the $R S$ term, and the $T S''$ term (since the expansion
is also in $T$), these terms only yield contributions at three loop.
Thus only the $T R^2$ and $R^3$ terms remain in the r.h.s. of (\ref{Gcum3r}),
apart from rescaling. It is then natural to look for the solution under
the form:
\begin{eqnarray}
  && \tilde{S}^{(3)}(u_{123}) = \alpha \tilde{T} \left[ \tilde{\sf R}''(u_{13})
    \tilde{\sf R}''(u_{23}) \right] + \beta
  \left[ \tilde{\sf R}''(u_{12})\tilde{\sf R}''(u_{13})^2  -
    \frac{1}{3} \tilde{\sf R}''(u_{12}) \tilde{\sf R}''(u_{23})
  \tilde{\sf R}''(u_{31})  \right] + O(T^2 R^2,T R^3,R^4) \nonumber \\
&& \label{2loopsolu}
\end{eqnarray}
where $[...]$ means symmetrization. One can compute its flow using (\ref{Gcum3r})
to lowest order (i.e. zero loop) and one finds, in schematic notations:
\begin{eqnarray}
&& {\cal L}_3 = \partial_l - \zeta u_i \partial_{u_i} - (2 \epsilon - 2 - 6 \zeta) \\
&& {\cal L}_3 [R'' R'' R''] = (2+\epsilon) [R'' R'' R''] +  O(R^4,T R^3) \\
&& {\cal L}_3 \tilde{T} [R'' R''] = \epsilon \tilde{T} [R'' R''] +  O(R^4,T R^3)
\end{eqnarray}
where we have used $\partial_l \tilde T_l
= (\epsilon - 2 - 2 \zeta) \tilde T_l$ and $(\partial_l - \zeta u_i \partial_{u_i})
[R'' R'' R''] = 3 [R'' R'' (\partial_l - \zeta u \partial_{u}) R''] + O(R^4,T R^3)
= 3 (\epsilon - 2 \zeta) [R'' R'' R''] + O(R^4,T R^3)$. This
implies
\begin{eqnarray}
&& \alpha = 3/8 \quad , \quad \beta = 3 \gamma/(\epsilon + 2)=3/8   \label{alphares}
\end{eqnarray}
Note that $\zeta$ plays no role here, hence it can be first set to zero and later restored in the beta function by global rescaling. In that case however one does not deal with a fixed point and it is essential to retain and compute to each number of loop the flow term $\partial_l$ (which provides order by order what is usually called counterterms).
The feeding term into (\ref{Gcum2r}):
\begin{eqnarray}
&& \tilde{S}^{(3)}_{110}(0,0,u) = \frac{3}{8} \tilde T [[R''
R'']''] + \frac{3}{8} [[R'' R'' R'']''] \\
&& = \frac{1}{8} \tilde T \tilde R'''(u)^2 - \frac{1}{4} \tilde T
(\tilde R''(u)- \tilde R''(0)) \tilde R''''(0) +  \frac{1}{4}
(\tilde R''(u)- \tilde R''(0)) \tilde R'''(u)^2
\end{eqnarray}
where we use schematic notations, reproduces correctly the two loop contribution (\ref{beta2loop}). As discussed
in Section (\ref{subsec2loop}) its large $l$ limit is:
\begin{eqnarray}
&& \tilde{S}^{(3)}_{110}(0,0,u) =  \frac{1}{4} (\tilde R''(u)-
\tilde R''(0)) (\tilde R'''(u)^2 - r^{(4)}(0) ) + O(\tilde T)
\end{eqnarray}

On the other hand one can evaluate (\ref{2loopsolu}) in the
outer region $u_{12}=0(1)=u_{13}$. It has a nice, but non-analytic
limit as $l \to \infty$, i.e. setting $\tilde T_l \to 0$.
We can now take the limit of the
resulting function when arguments become close:
\begin{eqnarray}
&& \tilde{S}^{(3)}_{110}(0,0^+,u) := \lim_{v \to 0^+}
\tilde{S}^{(3)}_{110}(0,v,u)  = \frac{1}{4} (\tilde R''(u)- \tilde
R''(0)) ( \tilde R'''(u)^2 - \tilde R'''(0^+)^2) \label{2loopbet} \\
&& = \frac{1}{2}
\tilde R'''(0^+)^2 \tilde R''''(0^+) u^2 + O(|u|^3)
\end{eqnarray}
where the limit $\lim_{v \to 0^+} (\tilde R''(v)-R''(0))
R''''(v)$ vanishes. Of course taking instead uniformly the limit $v \to 0^-$ yields the same result. The notation here means that $v$ is taken to zero
with $v=O(1)$, i.e. within the outer region, also called inertial range in Burgers turbulence (see Section \ref{subsec:burgers}). To the two loop
accuracy these two expressions are identical since
we have shown in (\ref{tooneloop}) that to one loop $r^{(4)}(0) = \tilde R'''(0^+)^2$
(in the turbulence context it is the Kolmogorov relation to one loop, see Section \ref{subsec:burgers}).
We will see in the Burgers Section \ref{subsec:burgers} below why we
expect very generally that:
\begin{eqnarray}
\tilde{S}^{(3)}_{110}(0,0^+,u) = \tilde{S}^{(3)}_{110}(0,0,u) \label{identity}
\end{eqnarray}
holds exactly (to any number of loop). This ''matching'' identity was in fact demonstrated in Ref \onlinecite{BalentsLeDoussal2005} using
a systematic analysis of partial boundary layers which we now recall.
Consider $u_{13} = O(1)$. Examination of the ERG equation (\ref{Gcum3r})
indicates that there are two
regions as $u_{12} \to 0$:
\begin{eqnarray}
&& \tilde S^{3}(u_{123}) = \tilde T_l^2 \tilde u_{12}^2 \phi(u_{13})
+ \tilde T_l^3 s^{(21)}(\tilde u_{12},u_{13}) + O(\tilde T_l^4) \quad \tilde u_{12} = O(1) \label{pbl21} \\
&& \tilde S^{3}(u_{123}) = u_{12}^2 \bar \phi(u_{13}) + |u_{12}|^3 \psi(u_{13}) + O(u_{12}^4)
\quad \quad \quad u_{12} = O(1) \label{pbl210}
\end{eqnarray}
with $\tilde u_{12}=u_{12}/\tilde T$,
the first region being the partial boundary layer (PBL21) and the second
the outer region. Plugging each form of (\ref{pbl210}) in
(\ref{Gcum3r}) one obtains two equations, one valid
for PBL21, the other, displayed below, is the
outer equation. As discussed in Ref \onlinecite{BalentsLeDoussal2005} they appear to
imply (as an exact relation to all orders) that $\bar \phi(u) = \phi(u)$, which we set from now on and is equivalent to
(\ref{identity}) since \footnote{note that we have not introduced the factor $\chi$ in the definition of
$\tilde u$ thus it can be set to one in Ref. \onlinecite{BalentsLeDoussal2005}} $\tilde{S}^{(3)}_{110}(0,0,u) = - 2 \phi(u)$. The (loop) expansion in powers of $R$ can be set up as:
\begin{eqnarray}
\tilde S^{(3)}_{110}(0,0,u)= - 2 \phi(u) = - 2 (\phi_0(u) + \phi_1(u) + ..)
\label{expandphi}
\end{eqnarray}
where $\phi_n(u)$ is $O(R^{n+3})$. At two loop order we can write
the equation for $\phi_0(u)$ from the one for $\tilde S^{(3)}$ in the
outer region:
\begin{eqnarray}
&& \partial_l \phi_0(u) = (2 \epsilon - 2 - 4 \zeta + \zeta u
\partial_u ) \phi_0(u) - \gamma  {\sf R}''(u) (\tilde R'''(u)^2
-  \tilde R'''(0^+)^2) \label{eqphi0}
\end{eqnarray}
the last term is the $[R'' R'' R'']$ feeding term in (\ref{Gcum3r}),
the only left to two loop order and $\tilde T_l \to 0$,
expanded to order $u_{12}^2$. It is by definition the same term
as (\ref{2loopbet}) above. Using that:
\begin{eqnarray}
&& (\partial_l - 3 \epsilon + 4 \zeta - \zeta u \partial_u) {\sf R}''(u) (\tilde R'''(u)^2 -  \tilde R'''(0^+)^2) = 0(R^4)
\end{eqnarray}
from the lowest order, rescaling part (i.e. zero loop) of the equation for $\tilde R$,
one can check that the solution obtained by equating $-2 \phi_0(u)$ with the right hand side of (\ref{2loopbet}) obeys indeed (\ref{eqphi0}). The coefficient $1/4$ in (\ref{2loopbet}) comes from the coefficient $2 \gamma/(\epsilon+2)=1/4$. Hence we recover the above two loop bewta function and the convolutions
in the two loop term of Ref. \onlinecite{BalentsLeDoussal2005} (Eq. (187)) are indeed unnecessary. Note finally that we would obtain the same
equation (\ref{eqphi0}) for $\phi_0(u)$ from the
large argument $\tilde u_{21}^2$ limit of the PBL21
for $\tilde S$, as shown in Ref. \onlinecite{BalentsLeDoussal2005}.

\subsubsection{Three loop}

To obtain the non zero temperature beta function to three loop
from the ERG one needs $R^4$,$T R^3$ or $T^2 R^2$, thus
one must now include the equation for the fourth cumulant
\begin{eqnarray}
&& \partial_l \tilde{S}^{(4)}(u_{1234}) = (3 \epsilon - 4 - 8 \zeta + \zeta u_i  \partial_{u_i})
\tilde{S}^{(4)}(u_{1234})
 + 4 \gamma \tilde{T} \Big[
\tilde {\sf R}''(u_{12}) \tilde {\sf R}''(u_{13}) \tilde {\sf
R}''(u_{14}) \Big] \label{Gcum4r}
\\
&& + 6 \gamma' \Big[ \tilde {\sf R}''(u_{12})^2 \tilde {\sf
R}''(u_{13}) (2 \tilde {\sf R}''(u_{14}) + \tilde {\sf
R}''(u_{24}) ) - 2 \tilde {\sf R}''(u_{31}) \tilde {\sf
R}''(u_{14}) \tilde {\sf R}''(u_{12}) \tilde {\sf R}''(u_{23}) +
\frac{1}{2} \tilde {\sf R}''(u_{41}) \tilde {\sf R}''(u_{12})
\tilde {\sf R}''(u_{23}) \tilde {\sf R}''(u_{34}) \Big]  \nonumber
\end{eqnarray}
with $\gamma=3/4$, $\gamma'=1/2$, keeping only the needed terms
(the complete one is displayed in Eq. A15 of Appendix A in Ref.\onlinecite{BalentsLeDoussal2005} ). $[...]$ denote symmetrization,
here over the four indices. One can solve (\ref{Gcum4r}) by extending the method
of the previous Section, plug in (\ref{Gcum3r}) and solve again using this
time the one loop beta function for $\tilde R$. It is somewhat tedious and
is summarized in Appendix \ref{app:3looperg}. One recovers the three loop finite
temperature beta function (\ref{beta3loop}).

We now recall and finish the derivation of the
correct three loop ''zero temperature'' beta function from
the method of Ref. \onlinecite{BalentsLeDoussal2005}.
Let us first write the equation (\ref{Gcum3r})
for $\tilde S^{(3)}$ in the outer region:
\begin{eqnarray}
&& \partial_l \tilde{S}^{(3)}(u_{123}) = (2
  \epsilon - 2 - 6 \zeta + \zeta u_i \partial_{u_i}) \tilde{S}^{(3)}(u_{123})
 - 3 \, \left[ \tilde{\sf R}''(u_{12}) (2 \phi(u_{13}) +
    \tilde{S}^{(3)}_{110}(u_{123})) \right] \\
&& + \gamma \left[ \tilde{\sf R}''(u_{12})\tilde{\sf R}''(u_{13})^2  -
    \frac{1}{3} \tilde{\sf R}''(u_{12}) \tilde{\sf R}''(u_{23})
  \tilde{\sf R}''(u_{31}) \right]
- 3 \left[ \phi^{(211)}(u_{12},u_{13}) \right]  \label{Gcum3rout}
\end{eqnarray}
where the last term is the fourth cumulant feeding. It requires two
points taken close together, and involves
a function $\phi^{(211)}$ which generalizes the function
$\phi$ of the third cumulant:
\begin{eqnarray}
&& \tilde S^{(4)}(u_{1234}) = \tilde T_l^2 \tilde u_{12}^2 \phi^{(211)}(u_{13},u_{14})
+ \tilde T_l^3 s^{(211)}(\tilde u_{12},u_{13}u_{14}) + O(\tilde T_l^4) \quad \tilde u_{12},
u_{13},u_{14},u_{34} = O(1) \\
&& \tilde S^{(4)}(u_{1234}) = u_{12}^2 \phi^{(211)}(u_{13},u_{14}) + O(|u_{12}|^3)
\quad \quad \quad u_{ij} = O(1) \label{pbl211}
\end{eqnarray}
the first region is called PBL211 and the second one is the outer region. The
identity of the two functions $\phi^{(211)}$ is again the continuity
statement of the zero temperature limits:
\begin{eqnarray}
\tilde{S}^{(4)}_{110}(u_1,u_1^+,u_3,u_4) = \tilde{S}^{(4)}_{110}
(u_1,u_1,u_3,u_4) \label{identity4}
\end{eqnarray}

Next one needs to expand (\ref{Gcum3rout}) itself to order $O(u_{12}^2)$. One finds:
\begin{eqnarray}
&& (\partial_l - (2
  \epsilon - 2 - 4 \zeta + \zeta u \partial_{u})) \phi(u) =
- \gamma  {\sf R}''(u) (\tilde R'''(u)^2
-  \tilde R'''(0^+)^2) \\
&& + (\phi''(u)-\phi''(0)) {\sf R}''(u) + 3 \phi'(u) \tilde R'''(u)
+ 6 \tilde R'''(0^+) \psi(u) - \eta(u)
 \label{eqphi}
\end{eqnarray}
with $u=u_{13}$. On the first line one recognizes the two loop terms,
the second line contains the contribution of the $R'' S''$ terms,
involving also the cubic expansion function $\psi$ in (\ref{pbl210}), and
$\eta$ represents the fourth cumulant feeding.

We now expand each function $\phi$, $\psi$ and $\eta$
as in (\ref{expandphi}). We recall from the previous Section:
\begin{eqnarray}
&& \phi_0(u) = - \frac1{2+\epsilon} \gamma
{\sf R}''(u) (\tilde R'''(u)^2 - \tilde R'''(0^+)^2)
\end{eqnarray}
Expanding (\ref{eqphi}) to next order in $\tilde R$ we obtain:
\begin{eqnarray}
&& (\partial_l - (2 \epsilon - 2 - 4 \zeta + \zeta u \partial_u ) ) \phi_1(u) =
 [(\phi_0''(u) - \phi_0''(0))
{\sf R}''(u) + 3 \phi'_0(u) \tilde R'''(u) + 6 \tilde R'''(0^+)
\psi_0(u) ] \\
&& + \eta_0(u) - \partial_l|_{R^2} \phi_0(u) \nonumber
\end{eqnarray}
where we must take also into account
\begin{eqnarray}
&& \partial_l|_{R^2} \phi_0(u) = - \frac1{2+\epsilon} \gamma  [
\delta {\sf R}''(u) (\tilde R'''(u)^2 - \tilde R'''(0^+)^2) + 2
{\sf R}''(u)( \tilde R'''(u) \delta \tilde R'''(u) - \tilde
R'''(0^+) \delta \tilde R'''(0^+) ] \nonumber \\
&& \delta {\sf R}''(u) = \tilde R'''(u)^2 - \tilde R'''(0^+)^2 +
{\sf R}''(u) R''''(u) \\
&& \delta R'''(u) = 3 \tilde R'''(u) \tilde R''''(u) + {\sf
R}''(u) R^{(5)}(u) \\
&& \delta R'''(0^+) = 3 \tilde R'''(0^+) \tilde R''''(0^+)
\end{eqnarray}
from the flow $\partial_l \tilde R$, i.e. the beta function of $\tilde R$ to order $\tilde R^2$ (since $\phi_0$
was found by neglecting those terms, and higher order ones).
The two other functions are obtained from:
\begin{eqnarray}
&& \partial_l \psi_0(u) = (2 \epsilon - 2 - 3 \zeta + \zeta u
\partial_u ) \psi_0(u) + \gamma   [\frac{1}{2} \tilde R'''(0^+)
\tilde R'''(u)^2 + {\sf R}''(u) \tilde R'''(0^+) \tilde
R''''(0^+)] \\
&& \partial_l \eta_0(u) = (3 \epsilon - 4 - 4 \zeta+ \zeta u
\partial_u ) \eta_0(u) + \gamma'   [\frac{7}{2} ( \tilde R'''(u)^2 - \tilde
R'''(0^+))^2 \\
&& + 12 {\sf R}''(u) ( \tilde R'''(u)^2 \tilde R''''(u) -
 \tilde R'''(0^+)^2 \tilde R''''(0^+) ) + {\sf R}''(u)^2  \tilde R''''(u)^2 ] \\
\end{eqnarray}

Proceeding as before one finds:
\begin{eqnarray}
 && \eta_0(u)= \frac1{4+\epsilon}
\gamma'   [\frac{7}{2} ( \tilde R'''(u)^2 - \tilde
R'''(0^+))^2 )  + 12 {\sf R}''(u) ( \tilde R'''(u)^2 \tilde
R''''(u) -
 \tilde R'''(0^+)^2 \tilde R''''(0^+) ) + {\sf R}''(u)^2  \tilde R''''(u)^2 ] \nonumber \\
 && \psi_0(u) =  \frac1{2+\epsilon}  \gamma   [\frac{1}{2} \tilde R'''(0^+) \tilde R'''(u)^2
+ {\sf R}''(u) \tilde R'''(0^+) \tilde R''''(0^+)]
\end{eqnarray}
All terms feeding $\phi_1$ have the same eigenvalue w.r.t. the linear
operator so we find:
\begin{eqnarray}
&&  \phi_1(u) = - \frac1{2+2 \epsilon} (\eta_0(u)  +
[(\phi_0''(0) - \phi_0''(u)) {\sf R}''(u) - 3 \phi'_0(u) \tilde
R'''(u) - 6 \tilde R'''(0^+) \psi_0(u) ] - \partial_l|_{R^2} \phi_0(u) )
\end{eqnarray}
It is instructive to compute each piece separately:
\begin{eqnarray}
&& (\phi_0''(0) - \phi_0''(u)) {\sf R}''(u) = \frac1{2+ \epsilon}
\gamma  {\sf R''}(u) ( 5 \tilde R''''(u) \tilde R'''(u)^2
- \tilde R'''(0^+)^2 (\tilde R''''(u) + 4 \tilde R''''(0^+)) \\
&& + 2 {\sf R''}(u) (\tilde R''''(u)^2 + \tilde R'''(u) \tilde
R^{(5)}(u)) )
 \\
&& - 3 \phi'_0(u) \tilde R'''(u) = \frac3{2+ \epsilon} \gamma
 ( \tilde R'''(u)^2 (\tilde R'''(u)^2 - \tilde
R'''(0^+)^2) + 2 {\sf R''}(u) \tilde R'''(u)^2 \tilde R''''(u))
\\
&& - 6 \tilde R'''(0^+) \psi_0(u) =  - \frac6{2+\epsilon} \gamma
 [\frac{1}{2} \tilde R'''(0^+)^2 \tilde R'''(u)^2 + {\sf
R}''(u) \tilde R'''(0^+)^2 \tilde R''''(0^+)]
\end{eqnarray}
while the contribution to the correction term is:
\begin{eqnarray}
&& \delta \beta_{3loop}[\tilde R](u) = \frac{2}{2+2 \epsilon}  \partial_l|_{R^2} \phi_0(u) = -
\frac{1}{40} [ \delta {\sf R}''(u) (\tilde R'''(u)^2 - \tilde
R'''(0^+)^2) + 2 {\sf R}''(u)( \tilde R'''(u) \delta \tilde
R'''(u) - \tilde R'''(0^+) \delta \tilde R'''(0^+) ] \nonumber
\end{eqnarray}

Putting all together we obtain the final result for the beta function to
three loop displayed in the next subsection.

\subsubsection{final result for the beta function to four loop}

Let us now display our final result for the unambiguous $\tilde T_l \to 0$ beta function to four loop. The three
loop term was obtained by three independent methods. The first two are based on the $\Gamma$-ERG: the first one
involves detailed considerations of the partial boundary layers and is described in the previous subsection. The
second one uses the continuity structure of the various cumulants and was described in the appendix G of Ref.
\onlinecite{BalentsLeDoussal2005}, hence we will not detail it here. It allowed us to obtain the four loop term
by going up to the $\Gamma$-ERG equation for the fifth cumulant. The third method is described in Appendix
\ref{app:threeloppwerg} and uses a closure of the $W$-moment hierarchy. Being a bit more memory consuming, we
could only use it to three loop. In all cases where they can be compared the results for the anomalous terms are
found to be non ambiguous and in agreement. One finds, up to a constant, with ${\sf R}''=\tilde R''-\tilde
R''(0)$):
\begin{eqnarray} 
&& -m \partial_m \tilde R = (\epsilon - 4 \zeta) \tilde R + \zeta u
\tilde R' + [ \frac{1}{2} (\tilde R'')^2 - \tilde R''(0) \tilde
R'' ] + \frac{1}{4} ((\tilde R''')^2- \tilde R'''(0^+)^2 ) {\sf R}'' \label{rg4b} \\
&& + \frac{1}{16} ({\sf R}'')^2 (\tilde R'''')^2 + \frac{3}{32}
((\tilde R''')^2- \tilde R'''(0^+)^2)^2 + \frac{1}{4} {\sf R}''
((\tilde R''')^2 \tilde R''''- \tilde R'''(0^+)^2 \tilde
R''''(0^+)) \nonumber \\
&& + \frac{1}{96} ({\sf R}'')^3 (\tilde R^{(5)})^2 + \frac{3}{16} ({\sf
R}'')^2 \tilde R''' \tilde R'''' \tilde R^{(5)} + \frac{1}{8}
{\sf R}'' ((\tilde R''')^3 \tilde R^{(5)} - \tilde R'''(0^+)^3 \tilde R^{(5)}(0^+)) \nonumber  \\
&&  + \frac{1}{16} ({\sf R}'')^2 (\tilde R'''')^3
+ \frac{9}{16} {\sf R}'' ( (\tilde R''')^2 (\tilde R'''')^2 - \frac{1}{6}
R'''(0^+)^2 (\tilde R'''')^2 - \frac{5}{6} \tilde R'''(0^+)^2 \tilde R''''(0^+)^2 ) \nonumber  \\
&& + \frac{5}{16} ((\tilde R''')^2-\tilde R'''(0^+)^2) ((\tilde R''')^2 \tilde R'''' +
\frac{1}{10} \tilde R'''' \tilde R'''(0^+)^2 - \frac{11}{10} \tilde R''''(0^+)
\tilde R'''(0^+)^2 ) + O(\tilde R^6) \nonumber
\end{eqnarray}
The first line are one and
2-loop terms, the second is 3-loop, the last three are 4-loop.
Normal terms (i.e. non vanishing for analytic
$R(u)$) are grouped with anomalous ''counterparts'' to show the
absence of $O(u)$ term, a strong constraint (linear cusp, no
supercusp): these combinations can hardly be guessed beyond 3 loop.
This shows the difficulty in constructing the FT, already in $d=0$.
We emphasize that (\ref{rg4b}) results from {\it a first principle derivation}.
This was the main point of this calculation, i.e. to show that it can be done.

We expect that a large class of scale invariant ''fixed point models'' in $d=0$ should be 
solution of this equation. That includes presumably a line of long range random 
potentials parameterized by a continuously varying $\zeta > 1$ (equivalently $\theta>0$). 
Unfortunately we do not, at this stage, have any 
such model which would be a perturbative test solution. Hence the interest of
this $\beta$-function is mostly as a $d=0$ limit of the one we hope can be
computed in higher $d$. Note that it would be interesting to derive the corresponding
beta function for depinning, since the two beta functions are expected to differ only
by anomalous terms.

\subsection{Sinai random field landscape: exact solution of the FRG hierarchy}

\label{subsec:sinai}

Here we describe the solution of the FRG in $d=0$ at zero temperature and
focus on the Sinai random field case. Modifications at non zero temperature
are discussed in the next Section, where the mapping to the Burgers equation
is further analyzed. Here we specialize to $N=1$ and whenever we deal
with the Sinai RF case we mention it.

\subsubsection{shape of the renormalized energy landscape at $T=0$ and shocks}
\label{sec:landscape}

At zero temperature one has:
\begin{eqnarray}
&& \hat V(v) = min_u H_V(u,v) =  min_u [ \frac{1}{2} m^2 (u-v)^2 + V(u) ]  \label{minimiz} \\
&& m^{-2} \hat V'(v)=v-u_1(v)=m^{-2}  V'(u_1(v))   \label{force}
\end{eqnarray}
where we denote $u_1(v)$ the value of $u$ which realizes the minimum in (\ref{minimiz}). There is an implicit
$m$ dependence everywhere, which plays the role of time in Burgers via $t=m^{-2}$, and we are interested in the
small $m$ regime (large Burgers time $t$). As can be seen integrating (\ref{force}) over $v$ if there is no
applied force (statics) $u_1(v)$ remains on average near $v$. Furthermore, the motion of $u_1(v)$ is always
forward as $v$ increases. It is either smooth and satisfies:
\begin{eqnarray}
&& \partial_v u_1(v) = \frac{1}{1 + m^{-2} V''(u_1(v))} \\
&& - \frac{1}{2} m \partial_m u_1(v) = - (v - u_1(v)) \partial_v u_1(v) \label{relu1}
\end{eqnarray}
Since $u_1(v)$ is the minimum of $H_V(u,v)$ the denominator is always positive and $\partial_v u_1(v) > 0$. If
the motion is smooth one easily show from (\ref{relu1}):
\begin{eqnarray}
&& - m \partial_m (u_1(v)-v)^2 = \frac{4}{3} \partial_v (u_1(v)-v)^3 + 4 (u_1(v)-v)^2
\end{eqnarray}
which, averaged over disorder produces the famous dimensional reduction result $\overline{(u_1(v)-v)^2} \sim
m^{-4}$. This cannot be correct, though since there are also discontinuous switches forward to another minimum.
This happens at special points $v_i$ called shocks where $H_V(u,v)$ has two minima. As $m$ decreases (RG time
increases) the quantity $m^{-2} V''(u_1(v))$ becomes larger and larger. This is because $V''$ is governed by the
typical curvature of the bare potential which we can take as smooth and of order unity. In the universal
asymptotic regime, when $u_1(v)$ is properly scaled by $m^{-\zeta}$, one then finds that in the smooth regions,
i.e. in between shocks, $u_1(v)$ is {\it both $v$-independent and $m$-independent}. The asymptotic motion
$u_1(v)$ thus becomes a staircase discontinuous forward motion. This can also be seen if one takes the
renormalized (rescaled) landscape to be e.g. the Brownian motion, which has $V''$ infinite. We expect it to
extend to any of the $\theta > 0$ landscape. We now study the asymptotic behaviour.

Thus for each (bare) disorder realization $V(u)$ there is a set of successive minima
$u_i$ and shock positions $v_i$, such that:
\begin{eqnarray}
&& m^{-2} \hat V'(v) = v - u_i  \quad v_{i-1} < v < v_{i}
\end{eqnarray}
such that $u(v)=u_i$ in this interval. At position $v=v_i$ minima $u_i$ and $u_{i+1}$ become degenerate and
switch. The discontinuity is $m^{-2} (\hat V'(v_i+) - \tilde V'(v_i-)) = u_i-u_{i+1}= V'(u_{i+1})-V'(u_i)$ (at
$T=0$ $\hat V$ is the Legendre transform of $V(u)$). This discontinuity will be rounded at small non zero $T$
since:
\begin{eqnarray}
&& m^{-2} \hat V'(v) = v - \langle u \rangle_{H_V(v)}
\end{eqnarray}
and the switch from one minimum to the next will occur
smoothly on scale $v \sim T$. This is related to
the internal structure of the shocks and studied in the
next Section.

There is a simple and well known geometric construction to
obtain $\hat V(v)$ illustrated in Fig \ref{figs1}. For
a given $v$, one writes the condition that the landscape $V(u)$
must remain, for all $u$, above a parabola centered on $u=v$:
\begin{eqnarray}
&& V(u) > E - \frac{(u-v)^2}{2}
\end{eqnarray}
with one contact point at $u=u_1$  where the equality holds.
The value of $E$ is fixed by the single contact point
condition. It corresponds to the apex of the parabola and
its value is precisely the minimum of $H_V(u,v)$,
i.e. $E=\hat V(v)$.

\begin{figure}[h]
\centerline{\includegraphics[width=7cm]{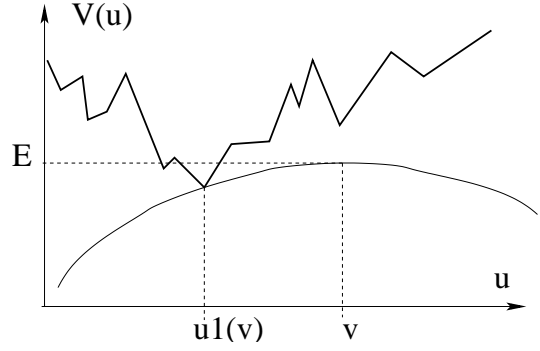}}
\caption{Geometrical construction of the renormalized landscape. The parabola
$y(u) = - \frac{m^2}{2} (u-v)^2 + E'$ centered on $u=v$ is raised ($E'$ increased
from $E' = -\infty$ to $E'=E$) until it touches the curve $y=V(u)$ at a single point
(for $E'=E$) $u=u_1(v)$, position of the minimum of $H_V(u,v)$. The
value at the minimum $\hat V(v) = E$ is obtained from the maximum of the
parabola. \label{figs1}}
\end{figure}

This construction is then repeated for increasing values of $v$.
Since $u_1(v)$ does not change, the touching parabola first rotates
around this point until, for a given $v=v_s$ a second contact
point appears. At this point $v=v_s$ there is a shock
(caracterized by $u_{21}=u_2-u_1$). The statistics is then
the one of degenerate minima in the toy model.

\begin{figure}[h]
\centerline{\includegraphics[width=7cm]{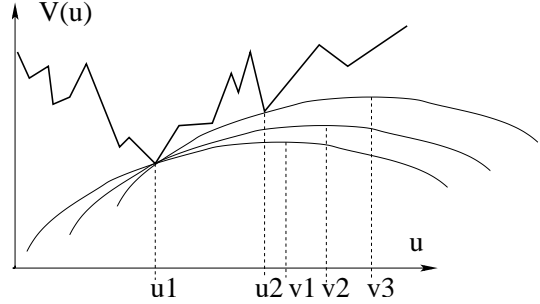}}
\caption{As the position $v$ of the center of the parabola is
increased (shift to the right with fixed curvature $m^2$), from
$v_1$ to $v_2$, the position of the minimum $u_1=u_1(v_1)=u_1(v_2)$
does not change: there is no shock between $v_1$ and $v_2$, the
parabola efectively rotates around the contact point. The
next shock is at $v=v_s=v_3$ when there are {\it two} contact
points at $u=u_1$ and $u=u_2$. Increasing $v$ further,
the parabola rotates again around $u_2$ until the next shock
and so on. \label{figs2}}
\end{figure}

There is another, equivalent, useful construction to find the shock positions
from the graph of the function $\Phi(u)=\frac{m^2}{2} u^2 + V(u)$. Since the (convex) function $\frac{m^2}{2} v^2 + \hat V(v)$ is the Legendre tranform of the function $\Phi(u)$, then it should also be the Legendre tranform of the convex enveloppe $\Phi_c(u)$ of $\Phi(u)$. The two functions coincide $\Phi_c(u)=\Phi(u)$ on regular points, while they differ on shock intervals $]u_i=u(v_i^-),u_{i+1}=u(v_i^+)[$. Note that this construction, as well the parabola construction generalize easily to Burgers in any dimension $N>1$. Finally note that there is yet another construction for shocks, known as the Maxwell rule \cite{khaninreview}, which does not seem to admit any known extension to $N>1$.

\subsubsection{statistics for the Random Field Sinai landscape case:
preliminaries}

\label{toy2}

We can now go back to the RF Sinai case, i.e. $V(u)$ a Brownian walk, where exact results can be obtained using
the Markov property. Extending the analysis of Section \ref{dropletobs} for droplet probabilities it is possible to obtain
analytically the full statistics of the renormalized landscape $\hat V(v)$. When convenient, we use the same
choice of parameters $m^2=1$ and $\overline{(V(u)-V(u'))^2} = 2 |u-u'|$ and emphazise connections between the
two set of results. Note that some of our results here are similar to those obtained in the context of Burgers equation \cite{FrachebourgMartin99}. Our method however is different (the real space RG of Ref. \cite{toy})
and more general, e.g we also obtain results about the distribution of renormalized potential $\hat V$ itself, and later compute the explicit form of the function $R(u)$.

Single point (i.e. $v$) correlations of the renormalized
landscape are related to the distribution $P(u_1)$ of the
position of the minimum in the toy model. From the discussion of Section,
the probability that the Brownian walk remains above the parabola centered on
$v$ and with one contact point at $u_1$ (see Fig \ref{figs1})
is:
\begin{eqnarray}
&& p(u_1,v) du_1 = \tilde g(v-u_1) \tilde g(u_1-v) du_1 = g(v-u_1) g(u_1-v) du_1
= P(u_1-v) du_1
\end{eqnarray}
where $g(u)$ and $\tilde g(u)$ are given in (\ref{g}) and (\ref{gtilde}), and there is a shift of $v$ due to the
position of the parabola. Each factor of $g$ represents the probability that $V(u)$ remains above the parabola
on the right, and on the left respectively. Since the force is $F(v)=\hat V'(v)= v-u_1$ this gives also the
distribution of the renormalized force at a single point:
\begin{eqnarray}
&& p_1(F,v) dF =  P(F) dF = g(-F) g(F) dF
\end{eqnarray}
which is normalized to unity. Its moments were computed in Ref.~\onlinecite{toy}.

It is easy to get the energy and force joint distribution in the case where there is no shock between $v_1$ and
$v_2$. The contact point $u_1$ remains the same and it is necessary and sufficient that the walk $V(u)$ remains
above the parabola centered on $v_1$ to the left of $u_1$ and above the parabola centered on $v_2$ to the right
of $u_1$. Then it is above both parabola everywhere. From the Markov property and the above result one sees that
the measure is:
\begin{eqnarray}
&& p(u_1,v_1,v_2)=\tilde g(v_1-u_1) \tilde g(u_1-v_2) du_1 \label{noshock1}
\end{eqnarray}
This implies that the joint probability that simultaneously (i) there is no shock in the interval $[v_1,v_2]$
{\it and} (ii) $E= \hat V(v_1) - \hat V(v_2)$, {\it and} (iii) $F_1=\hat V'(v_1)$, $F_2=\hat V'(v_2)$ is simply:
\begin{eqnarray}
&& e^{\frac{1}{12} (F_1^3 - F_2^3)} g(F_1) g(-F_2) \delta(F_2-(F_1 + v_{21}))
\delta(E + \frac{v_{21}}{2} (F_1 + F_2)) dF_1 dF_2 dE               \label{probnoshock}
\end{eqnarray}
since one has $F(v_1)=F_1=v_1-u_1$ and $F(v_2)=F_2=v_2-u_1$, also $u_1=1/2 (v_1+v_2) + E/v_{21}$ (here and below
$v_{21}=v_2-v_1$). From this measure one can extract one contribution to the function $R(u)$, which is done
below, i.e. the part corresponding to no shock (the other piece is more complicated). One can integrate over the
energy and obtain
\begin{eqnarray}
&& p_0(F_1,v_1;F_2,v_2) dF_1 dF_2 = \tilde g(F_1) \tilde g(-F_2)
\delta(F_2-(F_1 + v_{21})) dF_1 dF_2
\end{eqnarray}
i.e. the joint probability for $F_1$, $F_2$ and that there is no shock in the interval $v_{21}$. Integrating
further over $F_2$ and $F_1$ yields the probability that there is no shock in an interval of length $v_{21}$,
which varies between one and zero.

There is a direct connection between the shocks and
the degenerate minima. The statistics of the
shock is described by the the droplet
probability for a toy model whose parabola is centered
at the position of the shock. Let us call $v_2$ the point
where the first shock to the right of $v_1$
occurs (it is called $v_3$ in Fig. \ref{figs2}). The walk touches the
parabola centered on $v_1$ at $u_1$, and is above it to its left,
hence a first factor $\tilde g(v_1-u_1)$. The walk touches the
parabola centered at $v_2$ in two points, at $u_1$ and at $u_2$,
and is above it in between. Finally it must remain above the
parabola centered at $v_2$ for all points $u'>u_2$ (if it was
crossing there would be a shock at a smaller $v<v_2$). The total
probability is (taking into account for each parabola the shifted
position of its center):
\begin{eqnarray}
&& \epsilon \tilde g(v_1-u_1) \tilde d(u_1-v_2,u_2-v_2) \tilde
g(u_2-v_2) du_1 du_2 \label{firstshock}
\end{eqnarray}
where the function $\tilde d(u_1-v_2,u_2-v_2)$ was defined in (\ref{d}) and describes the probability of
degenerate minima in the toy model.

One finally obtains the joint probability of $u_1(v_1)=u_1$ and that the first
shock is at $v_2$ (within $dv_2$) with the new minimum at $u_2$:
\begin{eqnarray}
&& p_{ns}(u_1,v_1;v_2,u_2)= \tilde g(v_1-u_1) \tilde d(u_1-v_2,u_2-v_2) \tilde g(u_2-v_2) du_1 du_2 u_{21}
dv_{21} \\
&& = \tilde g(F_1) \tilde d(-F_2^-,-F_2^+) \tilde g(-F_2^+) \delta(F_2^{-}-F_1- v_{21}) (F_2^--F_2^+) dF_1
dF_2^- dF_2^+ dv_{21} \nonumber
\end{eqnarray}
where in the last line we have expressed the probability for the force variables $F_1=v_1-u_1$,
$F_2^{-}=v_2-u_1$, $F_2^+=v_2-u_2$. This is just (\ref{firstshock} ) taking into account that $\epsilon=u_{21}
dv_2$ is the vertical shift at $u_2$ of two parabola passing both through $u_1$ and corresponding to $v_2$ and
$v_2+dv_2$ respectively \footnote{The equations of the two parabolas are $-V=(u-u_1)^2 - v_2 (u-u_1)$ and
$-V=(u-u_1)^2 - (v_2+dv_2) (u-u_1)$. This means that $\epsilon=u_{21} dv_2$}. From (\ref{noshock1}) the
probability that the first shock is at $v_2$ (within $dv_2$) is $dv_2 \partial_{v_2}  p(u_1,v_1,v_2)$, thus
compatibility between the two results requires:
\begin{eqnarray}
&& \int_{u_1}^\infty du_2 \tilde d(u_1-v_2,u_2-v_2) \tilde
g(u_2-v_2) u_{21} = \tilde g'(u_1-v_2)
\end{eqnarray}
which is exactly the STS relation for the droplets.

On the other hand one can also study the probability density for a shock. There we have just a single parabola
at say, $v_2=0$ the position of the shock. Thus we get the one shock distribution function:
\begin{eqnarray}
&& p_{s}(u_1,u_2) du_1 du_2 = \tilde g(-u_1) \tilde d(u_1,u_2) \tilde g(u_2) u_{21} du_1 du_2 \\
&& = g(-u_1) d(u_2-u_1) g(u_2) u_{21} du_1 du_2 \nonumber
\end{eqnarray}
one can check that this result is consistent with the one obtained for $\rho_1(\mu,\eta) d\mu d\eta$ in
Ref.~\onlinecite{FrachebourgMartin99} in different variables such that $d\mu d\eta=du_{21} \frac{1}{2}
d(u_2^2-u_1^2) = \frac{1}{2} u_{21} du_{21} d(u_2+u_1) = u_{21} du_1 du_2$ ($u_2>u_1$ is assumed).

One method to describe the statitics of the full landscape is
to construct successive shocks, thus to write the probabilities of the
force at two points $v_L$ and $v_R$ and having $n$ shocks in
between:
\begin{eqnarray}
&& \tilde g(v_L-u_1) \tilde d(u_1-v_1,u_2-v_1) \tilde
d(u_2-v_2,u_3-v_2) .. d(u_n-v_{n},u_{n+1} - v_{n}) \tilde
g(u_{n+1}-v_R)  \nonumber \\
&& \times u_{21} u_{32} .. u_{n+1,n} du_1..du_{n+1} du_L du_R dv_1
.. dv_n \label{nshock2}
\end{eqnarray}
In principle summing this over $n$ should reproduce the probabilities
computed in the next section by different methods.

\subsubsection{multipoint statistics for the Random Field Sinai landscape}
\label{sec:shocks} 

Here we obtain the joint distribution of energies $E_i=\hat V(v_i)$ and forces $F_i=\hat V'(v_i)$
at multiple points. We need to impose the
condition that the
walk $V(u)$ remains above all inverted parabolas centered in $v_i$
of offset $E_i$:
\begin{eqnarray}
&& V(u) > - \frac{(u-v_i)^2}{2} + E_i
\end{eqnarray}
with one contact point at $u_i$ where the equality holds (i.e. the walk is
on the parabola). This is represented in Fig. \ref{shockR}.

\begin{figure}[h]
\centerline{\includegraphics[width=7cm]{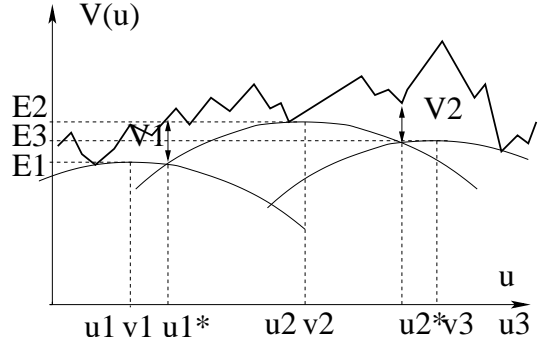}}
\caption{Construction of the joint probability that the
renormalized potentials are $\hat V(v_i)=E_i$ at points
$v_i$: the random walk $V(u)$ must remain above
all (inverted) parabola centered on the $v_i$ and
of apex $E_i$. \label{shockR}}
\end{figure}

We first assume that each interval $v_{i+1,i}$ contain at least
one shock. Neighboring parabolas intersect in:
\begin{eqnarray}
&& u^*_i-v_i = \frac{v_{i+1,i}}{2} - \frac{E_{i+1,i}}{v_{i+1,i}}
\label{energies}
\end{eqnarray}
where $E_{i+1,i}=E_{i+1}-E_i$. One must have $u_i < u^*_i < u_{i+1}$ (see Fig. \ref{shockR}). The case $u^*_i
\to u_i$ means that there is only one shock in the interval and it is in $v_{i+1}$ (whose parabola has then two
contact points $u_i$ and $u_{i+1}$). No shock in the interval corresponds to $u_{i+1}=u^*_i=u_i$ and is examined
separately. The intersection point of two neighboring parabola are at coordinate:
\begin{eqnarray}
&& y_i^* = \frac{(u_i^*-v_i)^2}{2} - E_i = \frac{(u_i^*-v_{i+1})^2}{2} - E_{i+1}
\end{eqnarray}
The random walk at $u=u^*_i$ must be above both parabola thus:
\begin{eqnarray}
&& - V(u_i^*) = y_i^* - V_i
\end{eqnarray}
where $V_i >0$ is the vertical distance, it is also:
\begin{eqnarray}
&& V_i = V(u_i^*) + \frac{(u_i^*-v_i)^2}{2} - E_i
\end{eqnarray}
i.e. the difference between the energy at $u_i^*$ and its
minimum value (i.e. its value at $u_i$).

The condition that the walk remain above all parabola with
a contact point on each is equivalent to the condition that
in each interval $[u_{i-1}^*,u_i]$
it remains above the corresponding parabola centered in $v_i$
with a contact point in each. For a single parabola
the corresponding probability was displayed in Section.
From the Markovian property, the total probability is
thus just a product of the same blocks, each shifted
by $v_i$. This gives for the probability in case of $n$ parabolas:
\begin{eqnarray}
&& \int_0^\infty dV_0 \int_0^\infty dV_1 .. \int_0^\infty dV_n N_\infty^{(2)}(-\infty,V_0,u_1-v_1,V_1,u_1^*-v_1)
\label{proba1}
\\
&& \times \prod_{i=2}^{n-1} N_\infty^{(2)}(u_{i-1}^*-v_i,V_{i-1},u_i-v_i,V_i,u_i^*-v_i)
N_\infty^{(2)}(u_{n-1}^*-v_n,V_{n-1},u_n-v_n,V_n,+\infty) \nonumber
\end{eqnarray}
for $n=2$ the central product is just suppressed. The functions $N_\infty^{(2)}$ were given in (\ref{N}).
Replacing $u_i^*-v_i$ by (\ref{energies}) and $v_i-u_i = \hat V'(v_i)=F_i$ (\ref{proba1}) becomes the joint
probability of the energies $E_i=\hat V(v_i)$ and the forces $F_i=\hat V'(v_i)$.

Using the explicit form for the block given in (\ref{N}), one finds, after some rearrangments and taking the
limits that (\ref{proba1}) takes the form:

\begin{eqnarray}
&& e^{ - \frac{1}{48} (v_{21}^3 + v_{32}^3 + .. +  v_{n,n-1}^3)
- \frac{v_{21}}{4} (u_1^* - \frac{v_1+v_2}{2})^2
- \frac{v_{32}}{4} (u_2^* - \frac{v_2+v_3}{2})^2 - ..
- \frac{v_{n,n-1}}{4} (u_{n-1}^* - \frac{v_{n-1}+v_{n}}{2})^2 } \nonumber  \\
&& \times \int_0^\infty dV_1 .. \int_0^\infty dV_{n-1}
e^{\frac{1}{2} (V_1 v_{21} + V_2 v_{32} +..+ V_{n-1} v_{n,n-1} )}
 g(v_1-u_1) h(u_1^*-u_1,V_1)  \nonumber
\\
&& \times [ \prod_{i=2}^{n-1} h(u_i - u_{i-1}^*,V_{i-1}) h(u_i^*-u_i,V_i) ]
h(u_n - u_{n-1}^*,V_{n-1}) g(u_n-v_n)        \label{joint1}
\end{eqnarray}
where the functions $h(u,V)$ and $g(u)$ are given in (\ref{g}) and (\ref{d}) in terms of Airy functions. Upon
integration over the $V_i$ this gives the joint distribution of the $E_{i+1,i}$ and $F_i$, via the replacements:
\begin{eqnarray}
&& u_i^*-u_i = \frac{v_{i+1,i}}{2} - \frac{E_{i+1,i}}{v_{i+1,i}} + F_i \label{joint2}  \\
&& u_{i+1}-u^*_i = \frac{v_{i+1,i}}{2} + \frac{E_{i+1,i}}{v_{i+1,i}} - F_{i+1}
\end{eqnarray}
and $v_i-u_i=F_i$.

Integrating over the forces amounts to integrate over the $u_i$. The distribution of the energies is then:
\begin{eqnarray}
&& e^{ - \frac{1}{48} (v_{21}^3 + v_{32}^3 + .. +  v_{n,n-1}^3) -
\frac{E^2_{21}}{4 v_{21}} - \frac{E^2_{32}}{4 v_{32}} - .. -
\frac{E^2_{n,n-1}}{4 v_{n,n-1}} }  \int_0^\infty dV_1 ..
\int_0^\infty dV_{n-1} \int_{-\infty}^{u_1^*} du_1
\int_{u_1^*}^{u_2^*} du_2 ..
\int^\infty_{u_{n-1}^*} du_n \\
&&
e^{\frac{1}{2} (V_1 v_{21} + V_2 v_{32} +..+ V_{n-1} v_{n,n-1} )}
 g(v_1-u_1) h(u_1^*-u_1,V_1)
[ \prod_{i=2}^{n-1} h(u_i - u_{i-1}^*,V_{i-1}) h(u_i^*-u_i,V_i) ]
h(u_n - u_{n-1}^*,V_{n-1}) g(u_n-v_n)  \nonumber
\end{eqnarray}
which can be simplified using convolutions. From this expression one can, in principle, compute all moments
$\bar S^{(n)}(v_1,..v_n)$. We now give an explicit expression for $n=2$.

\subsubsection{exact formula for the FRG function $R(u)$}

We now specialize to $n=2$ points $v_1$, $v_2$ and denote $v = v_{21}$. We obtain the distribution $p(E)$ of the
energy difference $E=E_1-E_2 = \hat V(v_1) - \hat V(v_2)$, and from there:
\begin{eqnarray}
&& 2 (R(0) - R(v)) = \int_{-\infty}^\infty dE E^2 p(E)   \label{RvsE}
\end{eqnarray}

The part with no shock is obtained by integrating (\ref{probnoshock}) over the forces $F_1$ and $F_2$ as:
\begin{eqnarray}
&& p^{(ns)}(E) dE = e^{- \frac{1}{48} v^3 - \frac{E^2}{4 v} } g(\frac{E}{v} - \frac{v}{2}) g(- \frac{E}{v} -
\frac{v}{2}) \frac{dE}{v}
\end{eqnarray}

The part with shocks can be read from the previous paragraph, we denote $V=V_1$ and
$u^*(E)=u_1^*=\frac{v_1+v_2}{2} + \frac{E}{v}$. One has:
\begin{eqnarray}
&& p^{(s)}(E)  = e^{- \frac{v^3}{48} - \frac{E^2}{4 v}} \int_{0}^\infty dV e^{\frac{1}{2} v V }
\int_{-\infty}^{u^*(E)} du_1 \int_{u^*(E)}^{+\infty} du_2
g(v_1-u_1) h(u^*(E)-u_1,V) h(u_2-u^*(E),V) g(u_2-v_2) \nonumber \\
&& = e^{- \frac{1}{48} v^3 - \frac{E^2}{4 v}} \int_{0}^\infty dV e^{\frac{1}{2} v V }
\phi(V,\frac{v}{2}-\frac{E}{v})
\phi(V,\frac{v}{2}+\frac{E}{v}) \quad , \quad  \phi(V,v)=\int_{0}^{+\infty} du h(u,V) g(u-v)
\end{eqnarray}
One notes that the part with shocks can be put in the
same form replacing $e^{\frac{1}{2} v V } \to \delta(v V)$.

The final formula for the distribution of the energy difference in terms of Airy functions is thus:
\begin{eqnarray}
&& p(E)  = (a b)^{-2} e^{- \frac{1}{48} v^3 - \frac{E^2}{4 v}} \int_{-\infty}^{+\infty} \frac{d z_1}{2 \pi i}
\int_{-\infty}^{+\infty} \frac{d z_2}{2 \pi i} \\
&& \times
 e^{\frac{v}{2 b} (z_1+z_2) + \frac{E}{v b}
 (z_2-z_1)} [ \frac{1}{v Ai(z_1) Ai(z_2)} +
\frac{\int_0^\infty dV e^{\frac{v}{2} V } Ai(a V + z_1)
 Ai(a V + z_2)}{Ai(z_1)^2 Ai(z_2)^2} ]
 \nonumber
\end{eqnarray}
where the integration contour is along the imaginary axis. Note that this form suggests $E \sim \sqrt{v}$, and
this is indeed true at large $E$. However, at small $E$ the landscape is more regular (random but finite first derivative) and one finds instead $E \sim F v$. This implies the absence of ''supercusp'' i.e.
$R'(0^+)=0$. Using (\ref{RvsE}) one finds finally:
\begin{eqnarray} \label{resru}
&& R(0)-R(v) = \frac{2}{b} \sqrt{\pi v} e^{- \frac{1}{48} v^3}
\int_{-\infty}^{+\infty} \frac{d \lambda_1}{2 \pi}
\int_{-\infty}^{+\infty} \frac{d \lambda_2}{2 \pi} [1 - \frac{2 (\lambda_2-\lambda_1)^2 }{b^2 v}]
 \frac{e^{i \frac{v}{2 b} (\lambda_1+\lambda_2) - \frac{(\lambda_2-\lambda_1)^2}{b^2 v}}}{Ai(i \lambda_1) Ai(i \lambda_2)} \\
&& \times [1 +
\frac{v \int_0^\infty dV e^{\frac{v}{2} V } Ai(a V + i \lambda_1)
 Ai(a V + i \lambda_2)}{Ai(i \lambda_1) Ai(i \lambda_2)} ]
 \nonumber
\end{eqnarray}
where we have chosen the contour $z_i=i \lambda_i$. In this formula the integral over $V$ converges very well and the double integral along the imaginary axis also converges. We recall that $ba^2=1$ and in this Section $a=2^{-1/3}$. The general case is obtained as $R_{m, \sigma}(v)=m^{-4/3} \sigma^{4/3} R_{1,1}(m^{4/3} \sigma^{-1/3} v)$, where $R(v)=R_{1,1}(v)$ is given by
(\ref{resru}).

Asymptotics and alternate formula are studied in Appendix \ref{appRu}. In the large $v$ limit it is found that:
\begin{eqnarray}
&& 2 (R(0)-R(v)) \sim 2 v + R_\infty + O(e^{- \frac{1}{48} v^3}) \\
&& R_\infty = \int_{-\infty}^{+\infty} \frac{d z_1}{2 \pi i} \int_{-\infty}^{+\infty} \frac{d z_2}{2 \pi i}
\frac{4}{b^2} \frac{(z_2-z_1)^2}{Ai(z_1)^2 Ai(z_2)^2} = \frac{8}{b^2} ( A_{2,2} - A_{1,2}^2 ) = - 2^{5/3}
0.510756 = - 1.62155
\end{eqnarray}
with $A_{p,n} = \int_{-\infty}^{+\infty} \frac{d u}{2 \pi} \frac{(i u)^p}{Ai(i u)^n}$, $A_{2,2}=1.06458$,
$A_{1,2}=-1.25512$ (and $A_{0,2}=1$), and $b=2^{2/3}$. This shows that the value of $\sigma$ (here chosen to be unity) is not renormalized, as expected from the long range nature of the random potential. Indeed, at large $v$ one finds, in a rescaled sense, $p(E) \approx \frac1{\sqrt{4 \pi v}} e^{- \frac{E^2}{4 v}}$.

In the small $v$ limit, one writes $E = \epsilon v$, define $\tilde p(\epsilon) d\epsilon=p(E) dE$
and expand in $v$:
\begin{eqnarray}
&& p(\epsilon) = p_0(\epsilon) + v p_1(\epsilon) + ..  \quad , \quad  p_0(\epsilon) = g(\epsilon) g(-\epsilon)
\end{eqnarray}
with $\int d \epsilon p_1(\epsilon)=0$ and where the first correction $p_1(\epsilon)$ is computed in the Appendix \ref{appRu}. Using that:
\begin{eqnarray}
&& R(v) = v^2 \int d \epsilon \epsilon^2 p_0(\epsilon) + v^3 \int d \epsilon \epsilon^2 p_1(\epsilon) + ..
\end{eqnarray}
one finds that there is indeed a linear cusp to the force correlator, $- R''(v) = - R''(0) - R'''(0^+) v + ..$, of amplitude:
\begin{eqnarray}
&& - R''(0) = \int d\epsilon \epsilon^2 g(\epsilon) g(-\epsilon) = b^2 \int_{-\infty}^{+\infty} \frac{du}{2 \pi} \frac{1}{Ai(i u)} \partial_{iu}^2 \frac{1}{Ai(i u)}
= 1.05423856 \\
&& - R'''(0^+) = (\frac{1}{4} \int d\epsilon \epsilon^4 g(\epsilon) g(-\epsilon) - 3 \int d\epsilon \epsilon^2
g'(\epsilon) g(-\epsilon) )    \label{teq0}
\end{eqnarray}

This cusp was obtained from the small $v=O(1)$ limit of the zero temperature function $R(v)$. As discussed in previous Sections it should match the large $\tilde v=v/T$ behaviour from the thermal boundary layer, and this provides a check for our droplet formula. The droplet formula, using (\ref{drop1},\ref{drop2},\ref{drop3}) predicts:
\begin{eqnarray}
&& R'''(0^+) = \frac{1}{4} \int_{-\infty}^{\infty} |y|^3 D(y) = \frac{1}{2} y_3
\end{eqnarray}
and it is checked in Appendix \ref{appRu} that this agrees with (\ref{teq0}) both expressions being equal to:
\begin{eqnarray}
&& R'''(0^+) = \frac{1}{a^2} \frac{8}{15} \int_{- \infty}^{\infty} \frac{d \lambda}{2 \pi} \frac{- \lambda^2}{Ai(i \lambda)^2} = 0.901289
\end{eqnarray}
which confirms matching and the exactness of the droplet hypothesis.

\subsubsection{exact formula for the FRG function $\Delta(v)$}

It is also useful to derive an independent formula for the correlator of the force $\Delta(v)=- R''(v)$
as:
\begin{eqnarray}
&& \Delta(v) = \int_{- \infty}^{+\infty} dF_1 \int_{- \infty}^{+\infty} dF_2 F_1 F_2 p(F_1,F_2,v)
\end{eqnarray}
where $p(F_1,F_2,v)$ is the two point force distribution. It can be obtained from our general formula for the joint distribution of forces and energies (\ref{joint1},\ref{joint2}). This yields:
\begin{eqnarray}
&& p(F_1,F_2,v)= g(F_1) g(-F_2) dF_1 dF_2 [ \delta(F_2-v-F_1)
e^{\frac{1}{12} (F_1^3 - F_2^3)} \\
&& + \theta(v+F_1-F_2) e^{ - \frac{1}{48} v^3} \int_0^\infty dV_1 e^{\frac{1}{2} V_1 v } \int_0^{v+F_1-F_2} du
e^{ - \frac{v}{4} (u -F_1 - \frac{v}{2})^2 }
 h(u,V_1) h(v+F_1-F_2-u,V_1)  ] \nonumber
\end{eqnarray}
After some manipulations summarized in Appendix \ref{appRu} one obtains:
\begin{eqnarray} \label{resdelta}
&& \Delta(v)  = - 2 \sqrt{\pi} v^{-1/2} b^2 a^{-2} e^{ - \frac{1}{48} v^3 } \int_{-\infty}^{+\infty} \frac{d
\lambda_1}{2 \pi} \int_{-\infty}^{+\infty}  \frac{d \lambda_2}{2 \pi}  e^{ - \frac{(\lambda_1-\lambda_2)^2}{v} +
i
\frac{v}{2} (\lambda_1+\lambda_2)} \\
&& \times \frac{Ai'(i b \lambda_1) }{Ai(i b \lambda_1)^2} \frac{Ai'(i b \lambda_2)}{ Ai(i b \lambda_2)^2} [ 1 +
\frac{\int_0^\infty dV e^{\frac{1}{2} v V }Ai(a V + i b \lambda_1) Ai(a V + i b \lambda_2) }{Ai(i b \lambda_1)
Ai(i b \lambda_2)} ] \nonumber
\end{eqnarray}

Equations (\ref{resru}) and (\ref{resdelta}) are thus the explicit form of the fixed point of the FRG in $d=0$ for the random field class. Up to a rescaling they should be a fixed point solution of the $d=0$ FRG equation, obtained to four loop in  (\ref{rg4b}), with the value $\zeta=4/3$ for the roughness exponent. These 
functions satisfy all the expected requirements (cusp, large $u$ behaviour, matching etc..) and confirm the validity of the FRG as a method to handle disordered systems with many metastable states leading to shock singularities.

\subsection{Decaying Burgers and FRG, inviscid limit}

\label{subsec:burgers}

\subsubsection{generalities}

Let us now detail the connection bewteen the FRG in $d=0$ (and $N$
components) and the decaying Burgers equation for a $N$-component
velocity field ${\sf u}(x,t)$ in $N$-dimension. We
focus on $N=1$, some aspects extend to any $N$. Let us recall that
the latter is a simplified version of the Navier Stokes equation
(without pressure) and reads, in standard notations (for $N=1$):
\begin{eqnarray}
\partial_t {\sf u}(x,t) + {\sf u}(x,t) \partial_x {\sf u}(x,t) = \nu \partial_x^2 {\sf u}(x,t)
\label{Burgers}
\end{eqnarray}
where ${\sf u}(x,t)$ is the velocity field and $\nu$ is the
viscosity. The decaying Burgers problem amounts to solve
(\ref{Burgers}) with an initial data ${\sf u}(x,t=0)={\sf u}_0(x)$. We are
interested in random initial data, a prominent example being ${\sf u}_0(x)$
gaussian with short range correlations. It corresponds to the random
field Sinai problem. More general initial (gaussian) velocity correlations are also studied
corresponding to initial (kinetic) {\it energy spectrum} $E_0(k)=k^{N-1} \overline{{\sf u}_0(-k) \cdot {\sf u}_0(k)}
\sim k^n$ at small $k$ (in Fourier), $n=0$ being the Sinai case.
Of high interest is the inviscid (i.e. large Reynolds number)
limit $\nu \to
0$. In that limit one recovers (formally) Euler equation whose
solutions ${\sf u}(x,t)={\sf u}_0(x-t {\sf u}(x,t))$ develop shock singularities at
finite time \cite{turbBernard}. These singularities are smoothed by a small non zero $\nu$, at
some scale called the dissipative length $L_d$, which must thus be kept
small but non zero, the question of the proper construction of the
inviscid limit $\nu \to 0$ being an outstanding problem, both for
Burgers and Navier Stokes turbulence.

One must also mention the driven Burgers problem, which has an
additional term $f(x,t) = \partial_x W(x,t)$ on the r.h.s. of
(\ref{Burgers}) where $W(x,t)$ is usually white noise in time and of correlation
scale $\xi$ in space. Under suitable boundary
conditions it is expected to reach a stationary measure (i.e.
suitable correlations become time independent). A lot of effort has
been devoted to study the statistics of the velocity field in that
case, as well as the stationary measure for shocks. It is important
to point out that the structure of shocks in stirred or decaying
Burgers is believed to be rather universal, the small distance (of
order and slightly above $L_d$) structure being analogous. The detailed time dependent
statistics of these shocks in the inertial range $L_d < x$ depend however on the model, with some universality classes.

In
decaying Burgers the evolution is expected to reach an asymptotic (statistically)
scale invariant form ${\sf} u(x,t) = t^{\frac{\zeta}{2}-1} u(w=x
t^{-\zeta/2})$ (in law), i.e. there is also a stationnary measure upon the
corresponding rescaling of lengths and time, while in stirred
Burgers no rescaling is necessary. For decaying Burgers the velocity
correlations in this stationnary measure identify with the FRG fixed
point, as discussed below, and the universality classes are parameterized by $\zeta$.
For uncorrelated initial velocities
$\zeta=4/3$, its random field value, other values of $\zeta$ correspond to
the other initial statistics, with $\zeta=4/(3+n)$ (see below).
The shocks are constantly merging and the typical scale (distance between shocks)
grows as $t^{\zeta/2}$. Although the width of an isolated shock grows as $L_d \sim \nu t$, one shows (see below) that because of merging of shocks the width of the {\it surviving shocks} actually grows as
$L'_d \sim \nu t^{1-\frac{\zeta}{2}}$. The important dimensionless parameter is
the ratio of (surviving) shock width to their separation $L'_d/t^{\zeta/2} \sim \nu
t^{1-\zeta}$. Hence shocks become effectively thinner for
$\theta=d-2+2 \zeta >0$ (here $d=0$), i.e. $\zeta >1$, which in the
FRG corresponds to an attractive zero temperature fixed point (FP). Equivalently
one can define an effective $\nu_{eff}$ flowing to zero as
$t^{-\theta/2}$. This picture holds for $n<1$, for which the (kinetic)
energy $\frac{1}{2} \overline{{\sf u}^2}$ asymptotic decay is $E(t) \sim t^{-2+\zeta} = t^{-2 (n+1)/(n+3)}$.
In the language of FRG it is called long range FP, and in Burgers it is
termed as being dominated by the persistence of large eddies \cite{khaninreview} . 

In addition there
is also a short range (SR) FP regime, called Kida regime in Burgers turbulence\cite{kida,khaninreview,BernardGawedzki98,jpkida}, which holds for
all $n>1$. There the decay is $E(t) \sim 1/(t (\ln t)^{1/2})$ (for gaussian statistics,
see below for generalization) and the scale 
\footnote{i.e. for a SR correlated gaussian random potential $V$, with non singular two point function at short scale $\lim_{\epsilon \to 0} |\ln \epsilon|^{1/2} {\sf u}(x/\epsilon, |\ln \epsilon|^{1/2}  t/\epsilon^2)$
exists} is $(t/\ln t)^{1/2}$. 
Since now the shock width grows faster than the typical distance (i.e. temperature is relevant) this regime
exists only strictly for $\nu \to 0$ before $t \to \infty$, and in that case the limits do not commute, while they do for the long range (LR) class $n<1$ ($\theta>0$). For $2>n>1$ one still has persistence of large eddies (persistence of tail of FRG function) but still the system flows to the SR (Kida) fixed point \footnote{the case $n=2$ corresponds to the LR Flory value $\zeta=4/5$, hence at short scale the SR correlator $\delta(x)$ of the random potential behaves effectively as $\sim 1/x$, while at large scale it remembers that it is short range and flows to the SR Kida FP.}. This is the so-called Gurbatov phenomenon, which states that the velocity statistics is not fully scale invariant,
and finds here a very natural interpretation in terms of the crossover from the LR to the SR FRG fixed point. The value $n=2$ corresponds to the Flory value $\zeta=4/5$, and at short scale the SR correlator of the random potential is behaves effectively as $\delta(x) \sim 1/x$, while at large scale it flows to the SR Kida FP.

Finally, the analogous of the random periodic class $\zeta=0$ correspond to Burgers in a periodic box which converges to a single random shock per period and $E(t) \sim t^{-2}$ ($n=\infty$). Note that although the above discussion was for $N=1$, the phenomenology of the LR FP holds for any $N$, but the precise exponent values for the crossover from LR to SR depend on $N$. 

\subsubsection{connection to FRG approach}

To be more specific, we now switch to the notations of this paper. The
renormalized potential $\hat V(v)$ satisfies a KPZ type \cite{kpz} equation:
\begin{eqnarray}
2 \partial_t \hat V = - m^3 \partial_m \hat V = T
\partial_v^2 \hat V - (\partial_v \hat V)^2
\end{eqnarray}
where the time is $t=m^{-2}$, large time corresponding to small mass and to the
universal region. Defining the renormalized force $F(v)=\hat V'(v)$,
it obeys the Burgers equation:
\begin{eqnarray}
\partial_t F(v) = \frac{T}{2} F''(v) -  F(v) F'(v)  \label{BurgersF}
\end{eqnarray}
here written for $N=1$, with the correspondence:
\begin{eqnarray}
&& {\sf u}(x) \equiv F(v) \quad , \quad t \equiv m^{-2} \\
&& \nu \equiv \frac{T}{2} \quad , \quad \text{viscous layer} \equiv \text{thermal layer}
\end{eqnarray}
and we will from now on switch freely between the two set of
notations for time (inverse mass) and viscosity (temperature). The
initial condition is precisely the bare potential (see Section
(\ref{sec:back}) $\hat V_{m=+\infty}(v)=V(v)$, hence the Sinai random potential
corresponds to random SR correlated initial force $F(v)$, the case $n=0$ defined above. In the
$T \to 0^+$ inviscid limit we recall that $t F(v)=v-u(v)$, where $u(v)$ is
position of minimum of $H_{V,v}(u)=(u-v)^2/(2 t) + V(u)$. Hence we
expect (equivalently for $n<1$ in the large $t$ limit upon rescaling of lengths) shock solutions of
the type (see Section \ref{sec:landscape} or below):
\begin{eqnarray}
t F'(v) = 1 - \sum_s u^{(s)}_{21} \delta(v-v_s)   \label{shocksolu1}
\end{eqnarray}
where $u^{(s)}_{21}=u_2-u_1=u(v_s^+)-u(v_s^-) >0 $ is the strength
of the shock, i.e. the force (velocity) discontinuity accross it
${\sf u}(0^-)-{\sf u}(0^+) \equiv F(v_s^-)-F(v_s^+) = u^{(s)}_{21}/t$.
Note that the term $F'(v) F(v)$ in (\ref{BurgersF}) becomes ill
defined in the $T=0$ inviscid limit. It does however possess a
distributional limit, i.e. as a distribution, as discussed below.
Note finally that the STS symmetry in the FRG corresponds to the
galilean invariance of the decaying Burgers equation, $F(v,t) \to
F(v+ a + b t,t)- b$ (the stirred Burgers has a larger invariance
which also involves the forcing).

The FRG approach consists in writing from (\ref{BurgersF})
the coupled RG flow (i.e. time evolution) equations for the
moments of the ''Burgers velocity field'':
\begin{eqnarray}
\overline{F(v_1) .. F(v_n)} = (-)^n \bar S^{(n)}_{1..1}(v_{1..n})
\end{eqnarray}
involving the derivatives of the $W$-moments of $\hat V$. In
particular the two point velocity correlation in Burgers corresponds
to the second cumulant of the renormalized force:
\begin{eqnarray}
\overline{F(v_1) F(v_2) } = - R''(v_1-v_2) = \bar
S^{(2)}_{11}(v_1,v_2)
\end{eqnarray}
i.e. $\overline{{\sf u}(x) {\sf u}(y)} \equiv - R''(x-y)$. They
satisfy the following hierarchy of dynamical equations:
\begin{eqnarray}
&& \partial_t \bar S^{(n)}_{1..1}(v_{12..n}) = n \frac{T}{2} [\bar
S^{(n)}_{31..1}(v_{12..n})] + n [\bar S^{(n+1)}_{21..1}(v_{112..n})]
\label{Wder}
\end{eqnarray}
where $[..]$ denotes symmetrization w.r.t. the $n$ arguments $v_i$.
These are well defined for $T>0$ and can be obtained directly from (\ref{BurgersF}) or by taking
derivatives of (\ref{Wrg}). As detailed in Section \ref{subsec:erg} one may also
study the hierarchy for connected $W$-cumulants, $\hat S^{(n)}$, or
$\Gamma$-cumulants $S^{(n)}$ of $\hat V$. In the FRG it is more
natural to study hierarchies of correlations of the potential, while
in Burgers one usually focus on the force (velocity field). The
former are usually less singular: as will be discussed below all
terms in (\ref{Wrg}) have a well defined limit for $T=0$ while the
term $\bar S^{(n+1)}_{21..1}(v_{112..n})=(-1)^{n+1}
\overline{F'(v_1)F(v_1)..F(v_n)}$ a priori is dominated by shocks
since it involves $F'(v_1)$, as discussed below. These FRG
functions can be obtained, whether at zero or non zero $T$, from an a priori more fundamental
object, namely the joint probabilities $P_n(v_1,F_1;..;v_n,F_n) dF_1 .. dF_n$ where
the $F_i=\hat V'(v_i)$ which at the fixed point should take the form:
\begin{eqnarray}
P_n(v_1,F_1;..;v_n,F_n) = m^{-(2-\zeta)n} p_n(m^\zeta v_1,
m^{-(2-\zeta)} F_1;.. ; m^\zeta v_n, m^{-(2-\zeta)} F_n )
\end{eqnarray}
leading to the moments:
\begin{eqnarray}
(-1)^n \bar S^{(n)}_{1..1}(v_1,..v_n) = \int dF_1.. dF_n F_1..F_n
P_n(v_1,F_1;..;v_n,F_n) = m^{(2-\zeta)n} \bar s_{1..1}^{(n)}(m^\zeta
v_1,..m^\zeta v_n)
\end{eqnarray}
In this formula the points are supposed to be all distinct, and
ordered. If some are repeated then the corresponding $F_i$ are
raised to the proper power. Thus there is no more information in the $P_n$ than in the $\bar
S^{(n)}$ provided one includes their values at coinciding points. These are also
the only quantities required for the
disordered model. For instance, at $T=0$ the one point probabilty of Burgers velocity
$P(F_1)=P_1(0,F_1)$ yields the distribution of the minimum $u_1$ of the toy model.
For the Sinai case, $\zeta=4/3$ and $n=0$ these probabilities can be
obtained in closed form (see Section \ref{toy2} and Ref. \cite{FrachebourgMartin99}) as they satisfy the Markov property:
\begin{eqnarray}
p_n(v_1,F_1;..;v_n,F_n) = p_1(F_1) \prod_{i=1}^{n-1}
\frac{p_2(v_{i+1}-v_{i},F_i,F_{i+1})}{p_1(F_i)}
\end{eqnarray}
In general it is of course quite difficult to solve this hierarchy.
Other hierarchies have been studied in
Burgers turbulence, usually for the generating functions $Z_N(v_i,\lambda_i) =
\langle e^{\sum_{i=1}^N \lambda_i F(v_i)} \rangle$, $G_N(F_i,v_i) =
\langle \prod_{i=1}^N \theta(F_i - F(v_i)) \rangle$, or
$P_N(v_i,F_i)=\partial_{F_1} .. \partial_{F_N} G_N(v_i,F_i)$,
with some (failed) attempts at ''exact'' closure \cite{Polyakov95}.

Another interesting case is the Kida model. This one is analyzed in the Appendix \ref{sec:kida} 
and it is recalled how to compute the two point force (or in Burgers, velocity) correlator. It also provides an interesting example of a fixed point function $R(u)$ which can be explicitly computed, i.e. formula (\ref{reskida}) providing an example of a short range fixed point in $d=0$.

\subsubsection{dissipation, viscous layer and inertial range}

Let us start with the first equation of the hierarchy and compare
the information it carries in Burgers and in the FRG for the disordered model.
\begin{equation}
\partial_t R''(v) = T R''''(v) + \bar S_{112}(0,0,v) \label{rel23}
\end{equation}
note that we have used the STS identity $- 2 \bar
S_{211}(0,0,v)=\bar S_{112}(0,0,v)$ to transform the quantity appearing in (\ref{Wder}) which a priori requires knowledge of
derivatives in the TBL (and is dominated by shocks) into one which is defined in the outer region
and has a $T=0$ limit. Eq. (\ref{rel23}) is the
usual dynamical equation which in Burgers relate two and three point
velocity correlations. At $v=0$ it yields:
\begin{equation}
\partial_t R''(0) = T R''''(0)
\end{equation}
an important identity encountered before (the Taylor expansion of a
third moment can only start as $\bar S \sim v^6$ at small $v$ as a
consequence of STS). In Burgers it expresses the decay of the energy
density $E(x)=\frac{1}{2} {\sf u}(x)^2$ on average:
\begin{equation}
- \partial_t \overline{E} = \bar \epsilon = \nu \overline{(\nabla
{\sf u})^2} \equiv \frac{T}{2} R''''(0) \label{energydecay}
\end{equation}
$\bar \epsilon$ being the ''dissipation rate'', i.e. the energy
dissipated from viscosity small scales in the shocks. Note that here, in decaying burgers this
rate is time dependent (i.e. $m$-dependent, see below). It is well
known in Burgers - decaying and stirred, as well as in Navier Stokes
- that this rate has a finite limit as $\nu \to 0$ (or $T \to 0$).
This is called the {\it dissipative anomaly}. It implies that
derivatives of velocity field must become very large at small
scales. As was discussed in Section \ref{sectoyexp}, the equivalent statement in
the FRG, i.e. that
\begin{equation}
T R''''(0) = 2 \bar \epsilon \label{dissrate}
\end{equation}
implies the existence of a non trivial thermal boundary layer (TBL). As detailed below the
correlations in the TBL region $v \sim T t$ are determined by the
fine structure of a shock (i.e. two points separated by $v_{21} \sim
T t$ will typically either both be inside a shock or both outside).
The dissipation occurs in the viscous layer and the dissipation
rate (\ref{dissrate}) is thus a TBL quantity.

Next, as discussed in Section \ref{checktbl}, one can consider (\ref{rel23}) for $v
=O(1)$ in the limit $T \to 0$, equivalently $T t \ll v$ (the so
called outer region). This corresponds to the {\it inertial range},
i.e. scales larger than $L_d$. On the small $v$ side of this region,
setting $v=0^+$ in (\ref{rel23}):
\begin{equation}
\partial_t R''(0^+) = \bar S_{112}(0,0,0^+) \label{rel234}
\end{equation}
since the term involving $T$ is smaller in that region. Now it
appears as a general property that:
\begin{equation}
- R''(0) = \overline{F(0)^2} = \overline{F(0) F(v=0^+)} = - R''(0^+)
\end{equation}
is continuous across the TBL, equivalently there is a well defined
limit for the joint probability distribution of the force (velocity
field) $p_{T=0}(F_1,0;F_2,v) \to p_{T=0}(F_1) \delta(F_2-F_1)$. This
is clear from (\ref{shocksolu1}), i.e. the force is discontinuous but
remains bounded in one shock, so
unless there is a strong accumulation of shocks near a point, the above
continuity should hold when averaging over a uniform density of shocks at random positions. Identifying (\ref{rel234}) and (\ref{rel23})
one finds that:
\begin{equation}
\bar S_{112}(0,0,0^+) = 2 \bar \epsilon
\end{equation}
This yields the celebrated Kolmogorov law for the third cumulant in
the inertial range:
\begin{equation}
\frac{1}{2} \bar S_{111}(0,0,u) \sim \bar \epsilon u \leftrightarrow
 \frac{1}{12} \overline{({\sf u}(x)-{\sf u}(0))^3 } \sim - \bar \epsilon x
\label{corr2}
\end{equation}
Identical coefficients in (\ref{energydecay}) and (\ref{corr2}) are a
consequence of matching across the TBL (i.e. viscous layer). Similar
relations exist in stirred Burgers (and Navier Stokes)
\cite{turbBernard}: there the dissipation rate $\bar \epsilon$ is
balanced by forcing instead of scale invariant time decay of
correlations, but small scale shock properties should be rather
similar.

Let us give some simple consequences for Burgers of the existence of fixed points in the FRG
in $d=0$. The correlations of the random potential, $\overline{V_k V_{-k}} \sim E(k)/k^2 \sim k^{n-2}$
associated to an initial (kinetic) energy distribution $E_0(k)$, are long
range for $n<1$ (and logarithmic for $n=1$). Using the disordered model notations, we know from FRG that there should be a LR fixed point where asymptotically $\overline{(\hat V(u)-\hat V(0))^2} \sim \overline{(V(u)-V(0))^2} \sim 2 \sigma |u|^{2 \theta/\zeta}$ with $\theta=2 (\zeta-1)$, i.e. renormalized and bare asymptotics should be the same, which implies that $\zeta=4/(3+n)$ and $\theta=2(1-n)/(3+n)$.
Hence one finds the law of energy decay:
\begin{equation}
R''(0) = \frac{1}{4} m^{4-2 \zeta} \tilde R^{* \prime \prime}(0)  \quad , \quad E(t) \approx \frac{1}{8} |\tilde R^{* \prime \prime}(0)| t^{-(2-\zeta)}
= \frac{1}{8} |\tilde R^{* \prime \prime}(0)| t^{-2(n+1)/(3+n)}
\end{equation}
where the prefactor is a universal function of $\sigma$, e.g. for
the Sinai case (uncorrelated initial velocities) we have that
$- \frac{1}{4} \tilde R^{* \prime \prime}(0) = 1.054238 \sigma^{2/3}$ using (\ref{teq0}).
Note that for the marginal case $n=1$ the disordered model
exhibits a freezing transition, hence the Burgers problem
will also exhibit an interesting phase transition
as a function of $\nu$, which can be studied using the results of
Ref. \onlinecite{CarpentierLeDoussal2001}.
As mentioned above the case $n>1$ corresponds to short range disorder, and $\theta <0$.
The corresponding $T=0$ FRG fixed point is thus unstable to temperature. It is dependent on the tail of
the distribution of disorder, and related to extreme value statistics, but for the so-called Gumbel class (which contains the gaussian) it seems fairly universal, up to non universal logarithmic corrections. This fixed point corresponds for a Gaussian disorder to the so-called Kida law in Burgers turbulence \cite{kida}.
Its explicit form is recalled in Appendix \ref{sec:kida} where the fixed point function $R(u)$ is given in (\ref{reskida}). Next, the {\it Gurbatov phenomenon} occurs when $R(u) \sim u^{1-n}$ for $1<n<2$. In that case there is still a memory of a LR fixed point, which is unstable towards the SR fixed point. However at any finite $t$, for $1<n<2$ the renormalized function will still decay as $\hat R(u) \sim u^{1-n} $ (conservation of tail in FRG, persistence of large eddies in Burgers). What happens is that this algebraic behaviour will hold only in the tail for $v>v_m$ where $v_m$ grows to infinity as $t=m^{-2} \to \infty$. Such crossover between LR and SR are well known and have been studied within the $\epsilon$ expansion (see e.g. \cite{fisher_functional_rg}).

Other quantities studied for Burgers, such as the dimensionless ''velocity flatness'' \cite{khaninreview}, have counterpart in the disordered model:
\begin{equation}
F = \lim_{t \to \infty} \frac{\overline{{\sf u}(x,t)^4}}{\overline{{\sf u}(x,t)^2}^2} \equiv \frac{\overline{u^4}}{\overline{u^2}^2}
\end{equation}
and indeed $F=2.83827..$ was derived \cite{toy} for the Sinai case $n=0$.

\subsubsection{shocks and droplets}
\label{shockdroplet}

As discussed in Section \ref{sec:landscape}, asymptotic (large time) solutions of the
Burgers equation in the $T=0$ inviscid limit are expected to be of
the form:
\begin{eqnarray}
&& t F(v,t) = v - u_i \quad v_{i-1}(t) < v < v_{i}(t)
\label{shocks}
\end{eqnarray}
where the $v_i(t)$ are the positions of the shocks, and the $u_i$
the minima $u_i=u(v_i^-)=u(v_{i-1}^+)$. This assumes dilute (i.e. well separated) shocks, i.e. $\theta/\zeta<1$, i.e. $\zeta<2$ (faster growing correlations
result in a function $\Phi(u)=V(u) + u^2/2m$ which differs almost everywhere from
its convex enveloppe). One can then show \cite{turbBernard} ballistic
motion of shocks:
\begin{eqnarray}
&& v_i(t) = b_i t + \frac{u_i+u_{i+1}}{2} + O(1/t) \label{shockvel}
\end{eqnarray}
$b_i$ being a constant discussed below.

At small $T>0$ one expects each shock to be smooth in a layer of size $t T$.
To find its shape one can directly look
for a solution of the Burgers equation (\ref{BurgersF}) of the form:
\begin{eqnarray}
&& F(v,t) = \frac{1}{t} \big( v - \frac{u_i+u_{i+1}}{2} -
\frac{u_{i+1} - u_{i}}{2} \phi(\frac{v-v_i(t)}{t T}) \big)
\end{eqnarray}
with $\phi(\pm \infty)=\pm 1$ to guarantee the boundary conditions
for a single shock (\ref{shocks}). One finds:
\begin{eqnarray}
&& (\frac{u_i+u_{i+1}}{2} + t \partial_t v_i - v_i) \phi' +
\frac{u_{i+1} - u_{i}}{2} \phi \phi' + \frac{1}{2} \phi'' = 0
\end{eqnarray}
The first term vanishes from (\ref{shockvel}) and one finds the
unique solution $\phi(x) = \tanh(\frac{u_{i+1} - u_{i}}{2} x)$. This
procedure can be pushed to any order in an expansion in $t T$ and to
the case of many shocks. In the regime where shocks are thin and dilute,
i.e. when their width $t T $ is much smaller than relative
distance, the velocity (renormalized force) can be written as:
\begin{eqnarray}
&& t F(v,t) = v - \hat u_s - \frac{u^{(s)}_{21}}{2}
\tanh(\frac{u^{(s)}_{21}}{2 t T} (v-v_s(t))) \label{profile}
\end{eqnarray}
where the shock parameters are denoted, from now on:
\begin{eqnarray}
&& \hat u_i = \frac{u_i+u_{i+1}}{2} \\
&& u^{(i)}_{21} = u_{i+1} - u_{i}
\end{eqnarray}
which makes contact with the droplet notations of Section
\ref{subsec:dropletsolu}. To make further connection, let us recall
that there we wrote a two well approximation:
\begin{eqnarray}
&& \tilde V(v) \approx - T \ln[ e^{- \frac{1}{T} (V(u_1) +
\frac{(v-u_1)^2}{2 t} )} + e^{- \frac{1}{T} (V(u_2) +
\frac{(v-u_2)^2}{2 t} )}] \nonumber \\
&& = \frac{(v-v_s)^2}{2 t} + \frac{V_{21}}{u_{21}} (v-v_s) - T \ln(2
\cosh(\frac{u_{21}(v-v_s)}{2 t T})) + C \label{pot2}
\end{eqnarray}
where $C$ is $v$ independent and $V_{21}=V(u_2)-V(u_1)$. The
position of the shock is given by equality of the two terms $v_s =
(u_1+u_2)/2 + t V_{21}/u_{21}$. This allows to identify the shock
velocity $b_i=V_{21}/u_{21}$. Hence shocks with zero velocity
corresponds to exact degenerate states in the bare disorder
potential. Taking a derivative of (\ref{pot2}) one recovers
(\ref{profile}).

Note from (\ref{profile}) that the width of the shock is really $L'_d = \Delta v \sim  t T/u_{21} \sim T/\Delta F$. Hence although an individual shock broadens with time as $\sim T t$, when there is a collection of (dilute) shocks (such as for random initial conditions with $\zeta<2$) they merge upon collision. As a result their typical separation $u_{21}$ (see below) grows as $u_{21} \sim t^{\zeta/2}$ and their width hence grows only as $T t^{1-\zeta/2}$. The dimensionless ratio
decays even faster $L_d/t^{\zeta/2} \sim  T t^{1-\zeta}$ as discussed above.
At the same time the shock amplitudes decay with time as $\Delta F \sim u_{21}/t \sim t^{-(1-\zeta/2)}$.

Let us now use the assumption of a (small) uniform density of shocks $\rho$, to compute the equal time velocity correlations at closeby points. Let us first consider:
\begin{eqnarray}
&& t (F(v_2)-F(v_1)) = v_{21} -  \frac{u_{21}}{2}
(\tanh(\frac{u_{21}}{2} \frac{v_2-v_s}{t T}) -
\tanh(\frac{u_{21}}{2} \frac{v_1-v_s}{t T}) ) \label{differ}
\end{eqnarray}
This expression holds if there is at most one shock in the common neighborhood of $v_1$
and $v_2$, and we will consider $v_{21} \ll 1/\rho$. We now average
over this shock position, with measure $\rho dv_s$ (or $\rho^2
\int_{-1/(2 \rho)}^{1/(2 \rho)} dv_s$ to be more specific). This yields:
\begin{eqnarray}
&& t \overline{F(v_2)-F(v_1)}= v_{21} (1 - \rho \int_0^\infty du_{21}
p(u_{21}) u_{21} ) \label{avdiffer}
\end{eqnarray}
where $p(x)$ denotes the normalized probability that for a given
shock the parameter $u_{21}=x$. We have used the identity
$\int_{-\infty}^{\infty} da (\tanh(\frac{z+a}{2})- \tanh
\frac{a}{2})=2 z$, and in averaging (\ref{differ}) we assumed $t T
\rho/u_{21} \ll 1$ (dilute shocks) hence we can push the integration to
infinity in the terms containing the $\tanh$. Since for statistically translational
invariant initial conditions the above
average (\ref{avdiffer}) must be zero (by the STS symmetry) it implies:
\begin{eqnarray}
&& \rho \int_0^\infty du_{21} p(u_{21}) u_{21} = 1     \label{stsshock}
\end{eqnarray}
hence the shock parameter $u_{21}$ is also the typical distance between shocks.

We now compute the single shock contribution to the correlator of the force.
Taking the square of (\ref{differ}) and averaging over the shock position,
we expand the square and use the above calculation to
evaluate the cross terms. Using
\begin{eqnarray}
&& \int_{-\infty}^{+\infty} da [ \tanh(\frac{z+a}{2})- \tanh
\frac{a}{2}]^2 = 16 F_2(z)
\end{eqnarray}
where $F_2(z)$ was introduced in (\ref{tbl2}), we finally obtain:
\begin{eqnarray}
&& R''(v_{21})-R''(0) =  \frac{1}{2} \overline{(F(v_2) - F(v_1))^2} =
\frac{2 T}{t} \rho \int du_{21} p(u_{21}) u_{21} F_2(\frac{u_{21}
v_{21}}{t T}) - \frac{v_{21}^2}{2 t^2}
\end{eqnarray}
The second term can be dropped since it is subdominant in the region
$v_{21} \sim t T/u_{21} \sim \rho t T$ that we are studying. This result, obtained here from Burgers, can be compared to the result (\ref{tbl2}) obtained from the
disordered model. It is consistent, provided
\begin{eqnarray}
\frac{p(y)}{\int_0^\infty y' p(y')} = \frac{y \tilde D(y)}{\int_0^\infty (y')^2 \tilde D(y')}  \quad , \quad y>0   \label{connect}
\end{eqnarray}
using $\langle y^2 \rangle_y = 2 t$ and (\ref{stsshock}) we have defined
$\tilde D(y)=D(y)+D(-y)=2 D(y)$ for $y=u_{21}>0$. This relates the shock size distribution $p(s=u_{21})$ to the droplet size distribution $D(y)$.
This relation was analyzed in the Sinai case $n=0$, $\zeta=4/3$ at the end of Section \ref{dropletobs}, where the universal ratios where computed. Although it makes sense that shocks statistics should be generally related to droplets since a shock is nothing but a droplet with exact degeneracy, it remains as a tantalazing question to generalize the relation (\ref{connect}) to higher $N$ and $d$.

\subsubsection{inviscid limit}
\label{sec:inviscid}

In Burgers (and Navier Stokes) one is particularly interested in the inviscid
limit $\nu \to 0$, equivalent to the limit $T\to 0$ in the FRG. As discussed above, for $\theta>0$, $n<1$, this limit is also the relevant one for the dynamics in Burgers and for the flow of the FRG. An important question is thus whether it is possible to construct {\it directly} this inviscid limit without solving the complete TBL or viscous layer problem, but keeping only the minimal information from its structure. Equivalently, in the FRG, whether one can compute directly the $T=0$ beta function, all ambiguities resolved.

Let us recall the analysis of Ref. \cite{BernardGawedzki98,BauerBernard99} for $N=1$. In the limit $\nu \to 0$ the derivative $F'(v)$ of the Burgers velocity field (\ref{shocksolu1}) becomes a distribution. Using the limit of the single shock
profile:
\begin{eqnarray}
&& F(v) = \frac{v - \hat u_s}{t} - \frac{u^{(s)}_{21}}{2 t} \epsilon(v-v_s(t))
 \label{profile0}
\end{eqnarray}
with $\epsilon(x)=\theta(x)-\theta(-x)$, one sees that quantities such as
$F(v) F'(v)$ are ill defined directly at $\nu=0$: they are not distributions
since the test functions for distributions should be infinitely differentiable
w.r.t $v$. However one can {\it define} them as:
\begin{eqnarray}
&& F(v)^{n-1} F'(v) := \frac{1}{n} \partial_v (F(v)^n) \quad , \quad
e^{\lambda F(v)} F(v) F'(v) := \partial_\lambda \frac{1}{\lambda} \partial_v  e^{\lambda F(v)}
\end{eqnarray}
In each case the r.h.s. is a perfectly legitimate distribution at $\nu=0$, and
the relations are evidently true for any $\nu>0$. One can then verify explicitly using (\ref{profile0}) and (\ref{shockvel}) that:
\begin{eqnarray}
&& \partial_t F(v) + F(v) F'(v) := \partial_t F(v) + \frac{1}{2} \partial_v F(v)^2 = 0
\label{limit}
\end{eqnarray}
holds in the sense of distributions at $\nu=0$.
It is not a fully trivial statement since by contrast:
\begin{eqnarray}
&& F(v) \partial_t F(v) + F(v)^2 F'(v) := \frac{1}{2} \partial_t F(v)^2 + \frac{1}{3} \partial_v F(v)^3  \neq 0
\end{eqnarray}
This is because of the neglected term $\nu F(v) F''(v) = \frac{1}{2} \nu \partial_v^2 F(v)^2 - \frac{1}{2} \nu F'(v)^2$. While the second term has a (zero) distributional limit as $\nu \to 0$ the second does not due to the dissipative anomaly field:
\begin{eqnarray}
&& \lim_{\nu \to 0} \nu F'(v)^2 = \frac{1}{12} \sum_s (\frac{u^{(s)}_{21}}{t})^3 \delta(v-v_s(t))
\end{eqnarray}
as found from direct integration of (\ref{profile}) around the shock. The following general equation was then shown to hold:
\begin{eqnarray}
&& (\partial_t  + \lambda \partial_\lambda \frac{1}{\lambda} \partial_v ) e^{\lambda F(v)} = - \frac{2}{\lambda} \sum_s e^{\lambda b_s} G(\frac{\lambda u^{(s)}_{21}}{2 t})
\delta(v-v_s(t))
\end{eqnarray}
with $G(x)=x \cosh x - \sinh x$ and $b_s$ the shock velocity. This can be checked directly using (\ref{profile}) and taking the limit $\nu \to 0$. Expanding in $\lambda$ one obtains all the anomalies, which contain information about the shock form factor, i.e. the distribution of shock sizes and velocities. It turns out that the dissipative anomaly field can be rewritten using left and right shock velocity, which leads to
a very simple and elegant form:
\begin{eqnarray}
&& (\partial_t F(v) + \frac{1}{2} (F(v+\delta)+F(v-\delta)) F'(v)) e^{\lambda F(v)} = 0
\label{final}
\end{eqnarray}
where the limit $\delta \to 0$ is implicit and selects at each shock position the left or right velocity. Note that to lowest order in $\lambda$ one recovers indeed (\ref{limit}) since obviously
$F(v+\delta) F'(v) + F'(v+\delta) F(v) = 2 \partial_v (F(v+\delta ) F(v))
= 2 \partial_v F(v)^2$ in the distribution sense (i.e. integrated with a test function). The form (\ref{final}) of the inviscid Burgers equation is physically very natural since it describes convection and that the true shock velocity is the half sum of left and right one. Note that these results invalidate the attempts at closure of Ref.
\cite{Polyakov95} since closures necessarily involve non trivial information about shocks as also discussed for
stirred Burgers \cite{WeinanE}.

Let us now come to the natural $T=0$ limit of the FRG hierarchy (\ref{Wder}). Given that it comes from
a KPZ like \cite{kpz} equation, it is thus natural to define it to be:
\begin{eqnarray}
&& \partial_t \overline{ F(v_1)..F(v_n) } = - \frac{n}{2} [ \partial_{v_n} \overline{ F(v_1)..F(v_n) F(v_n) }]
\label{Wder0}
\end{eqnarray}
If we assume that all force (i.e. Burgers velocity) correlation functions $\overline{ F(v_1)..F(v_n) }$ are {\it continuous} functions of their arguments, which is expected to hold at least for dilute shocks ($\zeta<2$) and can be checked explicitly from the exact solution in the case of Sinai landscape $\zeta=1/2$ ($n=0$) given in Section, then the r.h.s of (\ref{Wder0})
is well defined. As explained in Appendix \ref{app:threeloppwerg} iterative truncation of this hierarchy is one of the several methods used to obtain the $T=0$ beta function given in (\ref{rg4b}). For $N=1$ it does not generate any ambiguity, as was checked up to four loop. If we compare with (\ref{limit}) and (\ref{final}) we can now understand why the $T=0$ FRG based on (\ref{Wder0}) does work. Indeed this procedure is exactly the one performed in that case. The limits of $[\bar S^{(n)}_{1..12}(v_1,..v_n,v_n+0^+)]$ computed in the inertial range for $\nu >0$ coincide with the values $\frac{1}{2} [ \partial_{v_n} \overline{ F(v_1)..F(v_n) F(v_n) }]$ in the direct $\nu=0$ solution.

As a last application of the Burgers-FRG correspondence let us note that the shock form factor controls the small distance behaviour of the moments of the velocity difference in the inviscid limit (equivalently for $\nu>0$ in the inertial range).
Since the jump across the shock is $\Delta F=u_{21}/t$ a simple one shock calculation yields:
\begin{eqnarray}
&& \overline{(F(v)-F(0))^{p}} \sim \mu_{p} v \text{sign}(v)^{p+1} t^{-p} \\
&& \mu_{p} = \rho \int_0^\infty du_{21} u_{21}^p p(u_{21}) = \overline{\sum_s (u_{21}^{(s)})^p \delta(v-v_s)}
\end{eqnarray}
which can also be expressed in terms of droplet distribution, i.e. $\mu_p=\langle |y|^{p+1} \rangle/\langle y^{2}
\rangle$, e.g. consistent with $R'''(0^+) \equiv \overline{{\sf
u}(0^{+}) \nabla {\sf u}(0)}= \mu_{2}/(2 t^2)$ given above. One can check for instance, using the full
TBL form given in Section, that (setting $t=m^{-2}=1$:
\begin{eqnarray}
&& \overline{(F(0)-F(v))^4} = 2 \bar Q_{1111}(0,0,0,0)-8 \bar
Q_{1111}(0,0,0,v)+6 \bar Q_{1111}(0,0,v,v) \\
&& = \frac{1}{24} T y^4 \frac{1}{\sinh^3(\tilde v y/2)} (3 \tilde v
y (9 \cosh(\tilde v
y/2) + \cosh(3 \tilde v y/2)) - 27 \sinh(\tilde v y/2) - 11 \sinh(3 \tilde v y/2)) \\
&& = \frac{1}{280} T \langle y^8 \rangle_y  \tilde v^4 + O(\tilde v^6) \\
&& = \frac{1}{2} \langle |y|^5 \rangle_y |v| + O(v^2)
\end{eqnarray}
for small and large $\tilde v=v/T$ respectively.

\subsection{ballistic agregation of shocks and closures}
\label{subsec:ballistic}

As discussed in Section \ref{dropgen} there exist an exact RG equation (\ref{rgdrop2}) relating $P(u_1)$ the probability that in the disordered model the absolute minimum is at $u_1$ and $D(u_1,u_2)$ the (droplet) probability density that there are two degenerate absolute minima at $u_1$ and $u_2$. It is physically reasonable since as $m$ is varied the absolute minimum can change by abrupt switch events whose probability is governed by $D$. An exact solution for this equation (together with the additional STS relation) was given in
(\ref{exactP}), (\ref{exactD}) for the Sinai RF class ($\zeta=4/3$). In general however, these equations (FRG and STS) may not be sufficient to fully determine $D$ (and hence $P$), and one may need another FRG equation for $D$ itself. Such equation would then involve the probability of three degenerate absolute minima, and it is then not clear whether this set of equations would close. We now see that, at least in the Sinai case, one can close this hierarchy. This closure is related to the ballistic aggregation dynamics of shocks.

Let us first recall the expected scaling with time $t = m^{-2}$:
\begin{eqnarray}
&& P_t(u_1) = t^{- \zeta/2} P(u_1 t^{- \zeta/2}) \quad , \quad D_t(u_1,u_2) = P_t(u_1,u_2,0) = t^{1 - 2 \zeta} D(u_1 t^{- \zeta/2},u_2 t^{- \zeta/2}) \\
&& P_t(u_1) = \tilde g_t(- u_1) \tilde g_t(u_1) \quad , \quad \hat D_t(u_1,u_2) = \tilde g_t(- u_1) \tilde d_t(u_1,u_2) \tilde g_t(- u_1) \\
&& \tilde g_t(u_1) = t^{- \zeta/4} \tilde g(u_1 t^{- \zeta/2})
\quad , \quad  \tilde d_t(u_1,u_2) = t^{1 - \frac{3 \zeta}{2}} \tilde d(u_1 t^{- \zeta/2},u_1 t^{- \zeta/2})
\end{eqnarray}
where in the last two lines we have put the solution in the same form as in the Sinai case, for which $\tilde d(u_1,u_2)=e^{\frac{1}{12} u_1^3 -\frac{1}{12} u_2^3} d(u_{21})$ is proportional to the probability that if the absolute minimum is in $u_1$ then there is a second one in $u_2$. It turns out that its time dependent version satisfies the following remarkable FRG equation:
\begin{eqnarray}
&& t \partial_t \tilde d(u_1,u_3) = (1 - \frac{3 \zeta}{2}) \tilde d(u_1,u_3)
- \frac{\zeta}{2} (u_1 \partial_{u_1} + u_3 \partial_{u_3})
\tilde d(u_1,u_3) - \frac{3}{8} \zeta u_{31} (\frac{u_1+u_3}{2})^2 \tilde d(u_1,u_3) \nonumber  \\
&& - \frac{1}{2} \int du_2 u_{21} \tilde d(u_1,u_2) u_{32} \tilde d(u_2,u_3) \label{frgdtilde}
\end{eqnarray}
The first line is simple scaling, but the second represents the event of three degenerate minima, which amounts to switch between two valleys (i.e. two sets of two degenerate minima), and allows to close the equation. The weight factors $u_{21} \tilde d(u_1,u_2) u_{32} \tilde d(u_2,u_3)$ is the usual weight as discussed in Section \ref{toy2}, see e.g. Eq. (\ref{nshock2}). (\ref{frgdtilde}) is equivalent to the following equation for $d(u)$:
\begin{eqnarray}
&&  t \partial_t d(u_{31}) = (1 - \frac{3 \zeta}{2}) d(u_{31}) - \frac{\zeta}{2} u_{31} d'(u_{31})
+ \frac{\zeta}{32} u_{31}^3 d(u_{31})
- \frac{1}{2} \int du_2 u_{21} d(u_{21}) u_{32} d(u_{32}) \label{frgd}
\end{eqnarray}
and in both equations the term $t \partial_t$ vanishes at the fixed point, i.e. when scale invariance holds exactly. For the Sinai case $\zeta=4/3$ one explicitly check that it is equivalent to:
\begin{eqnarray}
&& - d(u) - \frac{2}{3} u d'(u) + \frac{1}{24} u^3 d(u) = \frac{1}{2} (u d(u)) * (u d(u))
\end{eqnarray}
where $*$ means convolution, which indeed holds because of the following identity of Airy
functions:
\begin{eqnarray}
&&  - \frac{Ai'(z)}{Ai(z)} + \frac{2}{3} (z \frac{Ai'(z)}{Ai(z)})' -
\frac{1}{6} (\frac{Ai'(z)}{Ai(z)})''' = [ (\frac{Ai'(z)}{Ai(z)})' ]^2
\end{eqnarray}
using that $d(u) = (a/b) \int_z e^{z u/b} Ai'(z)/Ai(z)$.

Three minima degeneracies in the Burgers setting is related to collision of shocks. It is well known that the dynamics of shocks for $N=1$ is simple ballistic agregation. It was studied in Ref. \cite{FrachebourgMartinPiasecki99} where some exact results where obtained, and we now make contact with notations of that paper. One denotes $M=\mu=u_{21}$ and $P=-\eta=-\frac{1}{2 t} ((u_1-v)^2-(u_2-v)^2)$ respectively, the mass and the momentum of a shock at position $v$. In between collisions the motion of a shock $v(t)$ is ballistic, i.e the shock velocity $V = P/M = \frac{1}{t} (\frac{u_1+u_2}{2} - v(t))$ is constant in time, where the $u_i$ are time independent. The collision process is related
to the three well droplet. Let us call $(u_1,u_2)$ the first shock at $v(t)$ and
$(u_2,u_3)$ the second shock at $v'(t)>v(t)$ with $u_1 < u_2 < u_3$. Neighboring
shocks share a minimum of the random potential $u_2$. Collision occurs at the time such that $v(t)=v'(t)$ and just amounts to erase $u_2$ and can hence be seen as a decimation process. It is characterized by three conservation laws (ballistic aggregation),
as detailed in Ref. \cite{FrachebourgMartinPiasecki99}, $M=M_1+M_2$,
$P=P_1+P_2$, $(V_1 - V_2) t = \frac{M}{2}$, which are
exactly equivalent to the operation of erasing $u_2$. Indeed:
\begin{eqnarray}
&& M_1 + M_2 = u_{21} + u_{32} = u_{31} = M \\
&& P_1 + P_2 = \frac{1}{2 t} (u_2^2 - u_1^2) + \frac{1}{2 t} (u_3^2 - u_2^2)
= \frac{1}{2 t} (u_3^2 - u_1^2) \\
&& (V_1 - V_2) t  = - \frac{1}{2} (u_1 + u_2) + \frac{1}{2} (u_2 + u_3)
= \frac{1}{2} (u_3 - u_1) = M/2
\end{eqnarray}
we set $v=0$ the shock collision position,
and $t$ is time of the collision. We can now look at formula (122) of Ref. \cite{FrachebourgMartinPiasecki99}
which describes the statistics of the ballistic aggregation process in terms of a probability weight $I(M,P,t)$ of aggregating particles. One can then check that
\begin{eqnarray}
&& \tilde d_t(u_1,u_3) = I_t(M=u_{31}, P=\frac{1}{t} u_{31} ( \frac{u_1+u_3}{2}
- v(t)) )
\end{eqnarray}
and that (122), which makes more explicit the conservation laws of the dynamics, is fully equivalent to  (\ref{frgdtilde}) if one also assumes scale invariance, leading to the same explicit solution in terms of Airy function (working out the jacobian in the collision integral in (122) simply replaces $M$ by $M_1$ $M_2$ and recover the measure as in (\ref{frgdtilde}). To conclude, one may wonder, as do the authors of Ref. \cite{FrachebourgMartinPiasecki99}, whether similar closing procedure could work for other values of $\zeta$, an open problem. There seem to be however a one to one correspondence between the problem of $N=1$ ballistic aggregation and the FRG, i.e. $m$ dependent solution of the disordered model. It would be of high interest to understand what type of aggregation process occur for $N>1$ and $d>0$ (functional shocks). 

\subsection{other solvable models}

Note that the FRG fixed point function can also be computed for the random periodic model in $d=0$. It produces some interesting results. Another case amenable to analytical results is the fully connected model 
(or the large $d$ limit). Both are studied in Appendix \ref{app:last}.

\section{higher dimension}

\label{sec:higherd}

We now study FRG for pinned manifolds in $d>0$. One outstanding
question is whether it can be controlled near $d=d_{uc}$, where
$d_{uc}=4$ for the standard type of elasticity studied here. This
being a difficult question we proceed by first trying to extend
what we have learned from $d=0$ in previous Sections. We study the
general non zero $T$ case but the important issue is whether a
$T=0$ limit can be constructed, i.e. an ''inviscid'' limit for the
decaying functional Burgers equation equivalent to the FRG.

In Section \ref{sec:rendisfunanyd} the $W[j]$ and $\Gamma[u]$ functionals and their associated $W$ and
$\Gamma$-cumulants ($R[u]$,..) where introduced and related. Here we start by giving the ERG equations that they
satisfy, and discussing a few other constraints coming from STS. Next we obtain a ''droplet'' solution to the
functional hierarchy valid in any $d$. Finally we discuss in more detail the $\epsilon$-expansion.

\subsection{ERG equations}
\label{ergeqd} 

The derivation of the ERG equation is a simple extension of the $d=0$ case presented in Section
\ref{subsec:erg}. Upon infinitesimal change $\partial g$ of the bare propagator matrix $g$ in (\ref{modeld},
\ref{defwjd}) the functional $W[j]$ satisfies the standard W-ERG equation:
\begin{eqnarray}
&& \partial W[j] = - \frac{1}{2 T} Tr \partial g^{-1} ( \frac{\delta^2 W[j]}{\delta j \delta j} + \frac{\delta
W[j]}{\delta j} \frac{\delta W[j]}{\delta j})
\end{eqnarray}
where $\partial g^{-1}=-g^{-1} \partial g g^{-1}$ and here and below $Tr M = \sum_{a} tr M_{aa}$ and $tr
M=\sum_{x} M_{xx}$. Upon the change of variable (\ref{jtov}) which implies $T g \frac{\delta}{\delta v} =
\frac{\delta}{\delta j}$ one obtains the W-ERG equation for $W[v]$. Separating the bare part from the
interacting one (i.e. due to disorder) in (\ref{wvd}):
\begin{eqnarray}
&& W[v] = \frac{1}{2 T} \sum_{axy} v^a_x g^{-1}_{xy} v^b_y + \hat W[v]
\end{eqnarray}
it can be written as:
\begin{eqnarray}
&& \partial \hat W[v] = \frac{T}{2} \sum_a tr [ \partial g ( \frac{\delta^2 \hat W}{\delta v_a \delta v_a} +
\frac{\delta \hat W}{\partial v_a} \frac{\delta \hat W}{\partial v_a} )]
\end{eqnarray}
up to a constant proportional to the number of replicas (the
linear term cancels upon the change from $j$ to $v$, i.e.
$W(v)=W(j=g^{-1} v/T)$).

In a similar fashion the effective action functional $\Gamma[u]$:
\begin{eqnarray}
&& \Gamma[u] = \frac{1}{2 T} \sum_{axy} u^a_x g^{-1}_{xy} u^b_y -
\hat \Gamma[u] \\
&& \hat \Gamma[u] = \frac{1}{2 T^2} \sum_{ab} R[u_{ab}] + \frac{1}{3! T^3} \sum_{abc} S[u_{abc}] + ..
\end{eqnarray}
satisfies the $\Gamma$-ERG equation:
\begin{eqnarray}
&&  \partial \hat \Gamma[v] = \frac{1}{2} Tr \partial g g^{-1} [1 - T g \frac{\delta^2 \hat \Gamma[v]}{\delta v
\delta v} ]^{-1}
\end{eqnarray}
up to a constant proportional to the number of replicas. Expanding $\hat W$ and $\hat \Gamma$ in number of
replica sums from (\ref{wvd}) and (\ref{gammaud}) yields ERG equations for the $p$-replica connected cumulants
functionals $\hat R$, $\hat S^{n}$ $n\geq 3$ and $R$, $S^{(n)}$, respectively. These are very similar to the
$d=0$ ones, except that all functions now become functionals, and all derivatives become functional derivatives.

Alternatively one can start from the RG equation obeyed by the
renormalized potential in a given sample:
\begin{eqnarray}
&& \partial \hat V[v] = \frac{1}{2} tr[ \partial g ( T
\frac{\delta^2 \hat V[v]}{\delta v \delta v} -  \frac{\delta \hat
V[v]}{\delta v } \frac{\delta \hat V[v]}{\delta v } ]
\end{eqnarray}
from which one easily obtains the linear functional equation for
the moments:
\begin{eqnarray}
&& \partial \bar S^{(n)}[v_1,..v_n] = \frac{T}{2} tr[ \partial g
\sum_{i=1}^n \frac{\delta^2 \bar S^{(n)}[v_1,..v_n]}{\delta v_i
\delta v_i} ] + \frac{n}{2} \text{sym}_{1..n} tr[ \partial g
\frac{\delta^2 \bar S^{(n+1)}[v_1,v_1',v_2..v_n]}{\delta v_1
\delta v_1'}|_{v_1'=v_1} ]
\end{eqnarray}
where, as before:
\begin{eqnarray}
&& \overline{\hat V[v_1] .. \hat V[v_n]} = (-1)^n \bar
S^{(n)}[v_1,..v_n]
\end{eqnarray}
The lowest order relates the second moment functional $\bar S^{(2)}[v_1,v_2]=\bar R[v_1-v_2]$ to the third:
\begin{eqnarray}
&& \partial \bar R[v] = T tr[ \partial g \frac{\delta^2  \bar
R[v]}{\delta v \delta v} ] +  tr[ \partial g \frac{\delta  \bar
S[v_1,v_2,v]}{\delta v_1 \delta v_2}|_{v_1=v_2=0} ]
\end{eqnarray}
the same equation being valid for the connected cumulants $\hat R$ and $\hat S$, see below. A standard choice is
$g^{-1}_k=k^2+m^2$ in Fourier, and $\partial=- m \partial_m$. Then one has $\partial g^{-1}_{xx'} = - 2 m^2
\delta_{xx'}$ and $\partial g= 2 m^2 g^2$. In $d=0$ setting $g=m^{-2}$ one then recovers the equations given in
Section \ref{subsec:erg}.

We now introduce more convenient notations for functional
derivatives:
\begin{eqnarray}
&& \frac{\delta R[v]}{\delta v_x} = R'_x[v] \quad , \quad  \frac{\delta^2 R[v]}{\delta v_x \delta v_y} =
R''_{xy}[v] \quad , \quad  \frac{\delta \bar S[v_1,v_2,v_3]}{\delta v_1 \delta v_2} = \bar
S^{110}_{xy}[v_1,v_2,v_3]
\end{eqnarray}
and so on. We now recall the relations between the $W$-moments (over-bars) $W$ cumulants (hat) and $\Gamma$
cumulants, up to the fourth cumulant, derived in Appendix \ref{app:legendre}:
\begin{eqnarray}
&& \bar R = \hat R = R \quad , \quad \bar S = \hat S \\
&& S[v_{123}] = \bar S[v_{123}] - g_{xy} ( R'_x[v_{12}]
R'_y[v_{13}] + R'_x[v_{23}] R'_y[v_{21}] + R'_x[v_{31}]
R'_y[v_{32}] ) \label{defofS} \\
&& \bar Q[v_{1234}]= \hat Q[v_{1234}] + R[v_{12}] R[v_{34}] +
R[v_{13}] R[v_{24}]  + R[v_{14}] R[v_{23}]  \\
&& Q[v_{1234}] = \hat Q[v_{1234}] - 12 \text{sym}_{1234} g_{xy} \hat S^{100}_x[v_{123}] R'_y[v_{14}] + 6
\text{sym}_{1234} g_{xy} g_{zt} R''_{xz}[v_{12}]
(R'_y[v_{13}]-R'_y[v_{23}])(R'_t[v_{14}]-R'_t[v_{24}]) \nonumber \\
&& = \hat Q[v_{1234}] - 12 \text{sym}_{1234} g_{xy} S^{100}_x[v_{123}] R'_y[v_{14}] - 6 \text{sym}_{1234} g_{xy}
g_{zt} R''_{xz}[v_{12}] (R'_y[v_{13}]-R'_y[v_{23}])(R'_t[v_{14}]-R'_t[v_{24}]) 
\label{defofQ}
\end{eqnarray}
Here and below repeated indices are contracted unless stated otherwise (or the trace notation is used).

Using these relations can now write the $W$-ERG equations for
second and third moments (and cumulant) as:
\begin{eqnarray}  \label{erg2momWd}
&& \partial R[v] = T \partial g_{xy} R''_{xy}[v] + \partial
g_{zz'} \bar S^{110}_{zz'}[0,0,v] \\
&& \partial \bar S[v_{123}] = \frac{3}{2} T \text{sym}_{123}
\partial g_{xy} \bar S^{200}_{xy}[v_{123}] + \frac{3}{2} \text{sym}_{123}
\partial g_{xy} \bar Q^{1100}_{xy}[v_{1123}] \\
&& = \frac{3}{2} T \text{sym}_{123}
\partial g_{xy} \bar S^{200}_{xy}[v_{123}] + \frac{3}{2} \text{sym}_{123}
\partial g_{xy} \hat   Q^{1100}_{xy}[v_{1123}] + 3 \text{sym}_{123} \partial g_{xy}
R'_x[v_{12}] R'_y[v_{13}]
\end{eqnarray}
and the corresponding $\Gamma$-ERG equations as:
\begin{eqnarray}
&& \partial R[v] = T \partial g_{xy} R''_{xy}[v] + \partial g_{zz'} g_{xy} ( R''_{xz}[v] R''_{yz'}[v] - 2
R''_{xz}[0] R''_{yz'}[v]  ) + \partial g_{zz'}
S^{110}_{zz'}[0,0,v] \label{erg2momGd} \\
 && \partial S[v_{123}] = \frac{3}{2} T
\text{sym}_{123}
\partial g_{xy} S^{200}_{xy}[v_{123}] + 3 T \text{sym}_{123}
\partial g_{xy} g_{zt} {\sf R}''_{xz}[v_{12}] {\sf
R}''_{yt}[v_{13}] \\
&& + 6 \text{sym}_{123}
\partial g_{xy} g_{zt} {\sf R}''_{xz}[v_{12}] (S^{110}_{yt}[v_{113}]
- S^{110}_{yt}[v_{123}]) + \frac{3}{2} \text{sym}_{123} \partial g_{xy}
Q^{1100}_{xy}[v_{1123}]  \nonumber  \\
&& + 3 \text{sym}_{123} \partial g_{xy} g_{zt} g_{rs} \big( {\sf R}''_{xz}[v_{12}] {\sf R}''_{yr}[v_{12}] {\sf
R}''_{st}[v_{13}] + 2 {\sf R}''_{xz}[v_{12}] {\sf R}''_{st}[v_{12}] {\sf R}''_{yr}[v_{13}] - {\sf
R}''_{xz}[v_{12}] {\sf R}''_{yr}[v_{23}] {\sf R}''_{st}[v_{13}] \big) \label{erg3momGd} \nonumber
\end{eqnarray}
where in the last formula we have introduced the notation:
\begin{eqnarray}
&& {\sf R}_{xy}''[u] =  R_{xy}''[u] - R_{xy}''[0]
\end{eqnarray}

It is important to note that up to now we have written the $W$ and $\Gamma$-ERG equations at non zero
temperature $T>0$, i.e. assuming analyticity of the functional at coinciding points (certainly correct for a
finite number of degrees of freedom, i.e. a finite size system). The quantity $R_{xy}''[0]$ is then well
defined.

Solving these functional equations seems hopeless beyond an
$\epsilon$-expansion. We show however in the following that an
exact solution can be found in the thermal boundary layer. Before
doing so let us give some further definitions and exact constraints on correlation
functions.

\subsection{local part and non local part of the functionals}

It is useful in the following to note that the functional $R[v]$ can be split unambiguously into a local
part and a non local one:
\begin{eqnarray}
&& R[v] = \int_x R(v_x) + \tilde R[v]
\end{eqnarray}
such that the non local part vanishes for a uniform configuration:
\begin{eqnarray}
&& \tilde R[\{v_z=v\}] = 0   \label{dec0}
\end{eqnarray}
this is in agreement with the definition of the local part $R(v)$ given in previous sections, $R[\{v_z=v\}] =L^d R(v)$. As a consequence:
\begin{eqnarray}
&& R''_{xy}[v] = R''(v_{x}) \delta_{xy} + \tilde R''_{xy}[v]  \\
&& \int_y \tilde R''_{xy}[v]|_{\{v_z=v\}} = 0   \label{dec2}
\end{eqnarray}
and the second line can also be used to specify the local part, i.e. the function $R(v)$ up to a constant. It was obtained taking two derivatives of (\ref{dec0}) and using translational invariance. A similar decomposition exists for all higher moments $\bar S^{(n)}$ and $S^{(n)}$ functionals. It may sometimes be useful to further split the non local part in multilocal components, and this is discussed in Appendix 
\ref{app:multilocal}.

\subsection{correlation functions}

\label{sec:corrd}

Here we give relations between two and four point correlations and
the renormalized disorder cumulants, as well as some exact
relations satified by these correlations. These are extensions of
the relations presented in Section in $d=0$ apart from subtleties
related to space arguments.

\subsubsection{two point functions}

There are only two distinct two point functions:
\begin{eqnarray}
&& G_{xy} = \langle u^a_x u^b_y \rangle = G_{xy} = - g_{xx'} g_{yy'} R''_{x'y'}[0] \\
&& \tilde G_{xy} = \langle u^a_x u^a_y \rangle = G_{xy} + T g_{xy}
\label{corr2d}
\end{eqnarray}
both related to the second derivative of the $R$ ($=\hat R=\bar R$) functional at zero. It implies that these
derivatives, here $R''_{xy}[0]$ must be well defined at $T>0$ and that the limit:
\begin{eqnarray}
&& \lim_{T \to 0^+} R''_{xy}[0]
\end{eqnarray}
should exist for finite system size and be related to the second moment of the configuration $u_1(x)$ of minimum
energy $G_{xy}^{T=0}=\overline{u_1(x) u_1(y)}$.

\subsubsection{four point functions}

There are five possible four point connected correlations. However
they depend on only two functions:
\begin{eqnarray}
&& \langle u^a_x u^a_y u^a_z u^a_t \rangle_c = \langle u^a_x u^a_y
u^a_z u^b_t \rangle_c = g_{xx'} g_{yy'}g_{zz'} g_{tt'} (
Q^{1111}_{x'y'z't'}[0] - T^2
R''''_{x'y'z't'}[0] ) \\
&& \langle u^a_x u^a_y u^b_z u^b_t \rangle_c = g_{xx'}
g_{yy'}g_{zz'} g_{tt'} ( Q^{1111}_{x'y'z't'}[0] + T^2
R''''_{x'y'z't'}[0] ) \\
&& \langle u^a_x u^a_y u^b_z u^c_t \rangle_c = \langle u^a_x u^b_y
u^c_z u^d_t \rangle_c =  g_{xx'} g_{yy'}g_{zz'} g_{tt'}
Q^{1111}_{x'y'z't'}[0]  \label{corr4d}
\end{eqnarray}
As for $d=0$ this is obtained using that connected correlations
are tree graphs from effective action, and that $\sum_b G_{ab}
\partial_{u_b}= T g \partial_{u_a}$ from the above form of the
exact two point function and the STS property. Notation here is again that different replica indices means
distinct replicas. Replica symmetry is assumed \footnote{this is automatically satisfied for a finite size
system and continuous distribution of disorder}. Note that these functions have higher permutation symmetry
(full with respect to the space points) than can naively be inferred from replica symmetry alone. This is
explained below. It is useful to give explicitly also the disconnected parts:
\begin{eqnarray}
&& \langle u^a_x u^a_y u^a_z u^a_t \rangle_{disc} = \tilde G_{xy}
\tilde  G_{zt} + \tilde G_{xz} \tilde G_{yt} +
\tilde G_{xt} \tilde G_{yz}  \quad , \quad 
\langle u^a_x u^a_y u^a_z u^b_t \rangle_{disc} = \tilde G_{xy}
G_{zt} + \tilde G_{xz} G_{yt} + G_{xt} \tilde G_{yz} \nonumber 
\\
&& \langle u^a_x u^a_y u^b_z u^b_t \rangle_{disc} = \tilde G_{xy}
\tilde  G_{zt} + G_{xz} G_{yt} + G_{xt} G_{yz} \quad, \quad  \langle u^a_x u^a_y u^b_z u^c_t \rangle_{disc} = \tilde G_{xy}
G_{zt} + G_{xz} G_{yt} + G_{xt} G_{yz} \nonumber \\
&& \langle u^a_x u^b_y u^c_z u^d_t \rangle_{disc} = G_{xy} G_{zt}
+ G_{xz} G_{yt} + G_{xt} G_{yz}
\end{eqnarray}

\subsubsection{STS identities}

One can prove \cite{BalentsLeDoussal2005} the general STS identity:
\begin{eqnarray}
&& T \sum_c g_{xy} \langle \frac{\delta O[u]}{\delta u_c^y  }
\rangle =  \sum_f \langle  O[u] u_f^x \rangle
\end{eqnarray}
for an arbitrary functional $O[u]$. Choosing respectively
$O[u]=u^a_y u^a_z u^a_t$, $O[u]=u^a_y u^a_z u^b_t$, $O[u]=u^a_y
u^b_z u^c_t$ one obtains:
\begin{eqnarray}
&& T (g_{xy} \tilde G_{zt} + g_{xz} \tilde G_{yt} + g_{xt} \tilde
G_{yz} ) = \langle u^a_x u^a_y u^a_z u^a_t \rangle - \langle
u^a_y u^a_z u^a_t u^b_x \rangle \\
&& T (g_{xy}  G_{zt} + g_{xz} G_{yt} + g_{xt} \tilde G_{yz} ) =
\langle u^a_x u^a_y u^a_z u^b_t \rangle + \langle u^a_y u^a_z
u^b_x u^b_t \rangle - 2 \langle u^a_y u^a_z u^b_x u^c_t \rangle \\
&& T (g_{xy}  G_{zt} + g_{xz} G_{yt} + g_{xt} G_{yz} ) = \langle
u^a_x u^a_y u^b_z u^c_t \rangle + \langle u^a_x u^a_z u^b_y u^c_t
\rangle + \langle u^a_x u^a_t u^b_y u^c_z \rangle  - 3 \langle
u^a_x u^b_y u^c_z u^d_t \rangle
\end{eqnarray}
We have used replica symmetry to relabel some indices. It turns out that the l.h.s of these equations (which do
not possess full symmetry with respect to spatial indices) are simply the disconnected parts of the r.h.s. .
Thus these equations are equivalent to the property that the connected parts of the r.h.s. must be zero. One can
check that they are indeed obeyed by the above parameterization in terms of $R$ and $Q$.

\subsubsection{ERG identities}

Similarly one can prove \cite{BalentsLeDoussal2005} ERG identities directly on correlations:
\begin{eqnarray}
&& \partial \langle O[u] \rangle = - \frac{1}{2 T} \sum_f (u_f
\partial g^{-1} u_f) O[u] \rangle
\end{eqnarray}
They yield:
\begin{eqnarray}
&& - 2 T \partial \langle u^a_x u^a_y \rangle  = \langle ( u^a
\partial g^{-1} u^a ) u^a_x u^a_y \rangle -  \langle ( u^a
\partial g^{-1} u^a ) u^b_x u^b_y \rangle  \\
&& - 2 T \partial \langle u^a_x u^b_y \rangle  = \langle ( u^a
\partial g^{-1} u^a ) u^a_x u^b_y \rangle +  \langle ( u^a
\partial g^{-1} u^a ) u^a_y u^b_x \rangle  - 2 \langle ( u^a
\partial g^{-1} u^a ) u^b_x u^c_y \rangle \label{ergcorr}
\end{eqnarray}
where here and below we denote $( u^a
\partial g^{-1} u^a )= u_z^a
\partial g^{-1}_{zt} u_t^a$. These equations are not independent, since, subtracting
the second to the first yields an identity always true (using the results of the previous paragraph, the
connected parts of the r.h.s. cancel and the rest simplifies). One can easily see that either correlation ERG
identity is equivalent, using (\ref{corr2d}) and (\ref{corr4d}) to the $\Gamma$-ERG identity:
\begin{eqnarray}
&& \partial R''_{zt}[0] = T \partial g_{xy} R''''_{xyzt}[0]
\label{identityd}
\end{eqnarray}
which generalizes the relation (\ref{exactrel}) to any $d$. It can be separately shown from (\ref{erg2momGd})
using the fact that $S^{110}_{xy}[0,0,v]$ starts at small $v$ only as $v^4$ (indeed the generic term
$S=u_x^a u_y^b u_z^c u_t^c$ can be excluded by STS, see Appendix of Ref. \cite{BalentsLeDoussal2005}, so that $S$
starts at $u^6$, as in $d=0$ - and a similar property for $\bar S$). The relation between (\ref{ergcorr}) and (\ref{identityd}) is obtained noting
that $\partial R''_{xy}[0] = -
\partial (g^{-1}_{xz} g^{-1}_{yt} G_{zt})$ and that, from (\ref{ergcorr}):
\begin{eqnarray}
&& \partial G_{zt} = - T \partial g_{xy} g_{zz'} g_{tt'}
R_{xyz't'}''''[0] -
\partial g^{-1}_{x'y'} (g_{x'z} G_{y't} + g_{x't} G_{y'z})
\end{eqnarray}

Since, as discussed above, the l.h.s. of (\ref{identityd}) should have a limit as $T \to 0$, it again suggests
some scaling $v \sim T$ for the {\it functional thermal boundary layer}. In the next section we present a
solution of the functional hierarchy which admits such a scaling. Taking the local part of (\ref{identityd}) one has:
\begin{eqnarray}
&& \partial R''(0) = T \int_{z} \partial g_{xy} R''''_{0xyz}[0]
\label{identitydloc}
\end{eqnarray}
hence the local part $R(v)$ should exhibit a TBL similar to $d=0$, but not identical since the r.h.s. of this equation cannot be expressed in terms of the local part alone.

\subsection{A droplet solution to the functional hierarchy}

We now obtain an exact solution of the full ERG functional hierarchy in the thermal boundary layer region $v
\sim T$, inspired by the droplet picture. We do not claim that this is necessarily the unique ''correct''
solution. The droplet picture serves as a heuristic method to find such a solution, and there are some
assumptions, detailed below, which go into this construction. It is quite possible that a more complex solution
based on a more complex (and realistic) picture can be constructed in the future. It is already very interesting
that an exact solution can be found to this highly non trivial hierarchy. In viewing the FRG in higher $d$ as a decaying functional Burgers equation, this droplet picture holds in what could be also called a ''dilute
functional shocks'' scenario. 

\subsubsection{structure of droplets}

Let us first give a qualitative description. Consider a sample of volume $L^d$ and keep the mass $m$ fixed. It
is simplest to think (and draw) the case of the directed polymer $d=1$, although we consider general $d$. Let us
call $u_{1x}$ the ground state configuration in a given sample (i.e. disorder environment). It is assumed to be
unique for continuous disorder probability distributions. We call a droplet a configuration $u_{2x}$ which is
close in energy from the ground state, the energy difference being $E=H_V[u_2]-H_V[u_1]$, i.e. a
quasi-degenerate state, where $H_V[u]$ is defined in (\ref{modeld}). To be qualified as an active droplet $E$
should be of order $T$, where $T$ is small and fixed. One calls the size (volume) of the droplet the distance
(volume) over which it differs from the ground state. There are of course many such droplet configurations,
especially of small sizes, and we label them $u^{(i)}_{2x}$ and their energy difference with the ground state
$E^{(i)}$. At any temperature $T$ the Gibbs measure is split between the ground state and the active droplets.

The main assumption within the droplet picture is that {\it large droplets are rare}. More precisely, the
probability $p$ to find a droplet with energy $E$ (fixed, of order $T$) of size $l < L_d < l+dl$ is $p \sim E
l^{-\theta} dl/l$. Thus in a sample of size $L$ will most often contain no (active) droplet of size of order $L$
(e.g. of size between $k L$ and $L$ where $k<1$ is a fixed number), and rarely will contain one, with
probability $\sim T L^{- \theta}$. When this occurs, the probability that one of the two quasi-degenerate states
contains another droplet of size of the same order is again vanishingly small. Thus there are no droplets within
droplets at large scale. This is a very important assumption. For $\theta>0$ it breaks down below some small
scale, and for marginal glasses, $\theta=0$, it breaks down at all scales. In that case there is indeed a finite
probability that a droplet contains another one of size one order of magnitude less (or a factor $k$ fixed),
resulting in a tree-like structure of droplet excitations. One may surmise based on results for Cayley trees
\cite{DerridaSpohn}, that in that case replica symmetry breaking occurs (for $\theta=0$, $d=0$ and any $N$
evidence for it was obtained in \cite{CarpentierLeDoussal2001}. In this section however we consider $\theta>0$
and assume rare, non overlapping, droplets.

Here we consider a geometry with fixed $m$ and $L^d$ very large
(i.e. $m L \gg 1$). Thus one can consider that the system is
roughly cut in $N \sim (L m)^d$ independent pieces (and samples)
of internal volume $m^{-d}$ where in each $u$ fluctuates (from
sample to sample) of order $m^{- \zeta}$ (i.e. the ground state
can be assumed to be uncorrelated over internal distances larger
that $1/m$). The droplets in each piece are thus also uncorrelated
(over distances larger than $1/m$). We will call ''elementary''
droplets the ones in each piece. In a given sample there are few
of them, i.e. only a few of the pieces contain an active droplet.
In the limit considered here, of small $T m^\theta$, their density
is small, of order $T m^\theta$. For these elementary droplets we
denote:
\begin{eqnarray}
u^{(i)}_{12x} = u_{1x} - u^{(i)}_{2x}
\end{eqnarray}
and consider that this quantity is non zero only over a region of size $m^{-d}$. The elementary droplets in a
given sample are assumed to be well separated along the directed polymer (or manifold). At this stage we
consider only droplets of volume of order $m^{-d}$. We ignore here questions arising from possible
accumulation of very small droplets.

\subsubsection{Droplet calculation and thermal boundary layer form}

We now implement these assumptions in a calculation. Extending the $d=0$ arguments presented in Section
\ref{dropgen}, the renormalized potential can be written:
\begin{eqnarray}
&& e^{- \frac{1}{T} (\hat V[v] - \hat V[0])} = e^{- \frac{1}{2 T}
\sum_{xy} g^{-1}_{xy} v_x v_y } \langle e^{ \frac{1}{T} \sum_{xy}
g^{-1}_{xy} u_x v_y } \rangle_{H_V[u]}
\end{eqnarray}
where $H_V[u]$ is defined in (\ref{modeld}). For $v = T \tilde v$ the average in the r.h.s. can be evaluated
from the droplet partition function:
\begin{eqnarray}
&& \langle e^{ \sum_{xy} g^{-1}_{xy} u_x \tilde v_y } \rangle_{H_V[u]} \approx Z_d^{-1} \sum_{n_i=0,1} e^{\tilde
v g^{-1} u_1 - \sum_i n_i (\tilde v g^{-1} u^{(i)}_{12} ) -
\frac{1}{T} \sum_i n_i E_i } \\
&& = Z_d^{-1} e^{(\tilde v g^{-1} u_1)} \prod_i (1 + e^{ \frac{- E_i}{T} - (\tilde v g^{-1} u^{(i)}_{12} )})
\end{eqnarray}
Here $Z_d=\sum_{n_i=0,1} e^{- \frac{1}{T} \sum_i n_i E_i }$ is the
partition sum of active elementary droplets and we use the
notation $(\tilde v g^{-1} u_1)=\sum_{xy} \tilde v_x g^{-1}_{xy}
u_{1y}$ when convenient. Thus one has:
\begin{eqnarray}
&& \tilde V[\{v_x\}] - \tilde V[\{v_x=0\}]  = \frac{1}{2} (v g^{-1} v) - T (\tilde v g^{-1} u_1) - \sum_i  T
\ln(1 + w_i e^{- (\tilde v g^{-1} u^{(i)}_{12x} )}) + \sum_i  T \ln(1 + w_i)
\end{eqnarray}
with $w_i=e^{-E_i/T}$. It is convenient to work with the
renormalized force:
\begin{eqnarray}
&& \hat V'_x[v] = g^{-1}_{xy} ( T \tilde v_y - (
u_{1,y} - \sum_i \frac{u^{(i)}_{12,y} w_i \tilde a^{(i)}  }{1 + w_i  \tilde a^{(i)}} ) ) \\
&& \tilde a^{(i)} = \exp(- (\tilde v g^{-1} u^{(i)}_{12}))
\end{eqnarray}

We now compute disorder averages. For that we need to make minimal assumptions which generalize the $d=0$
analysis. One first denotes $P[u_1]$ the (functional) probability for the ground state configuration. Next one
defines a droplet probability functional $D[u_1,u_2]$. Contrarily to $d=0$ there can be several elementary
droplet configurations $u^{(i)}_2=u_1-u^{(i)}_{12}$, thus one writes symbolically $D[u_1,u_2]=\sum_i
D_i[u_1,u_2^{(i)}]$. These functionals can be derived from a more general droplet functional $P[u_1,\{
u_2^{(i)},E^{(i)} \}]$. One calls $P_i[u_1,u_2^{(i)},E^{(i)}]$ the single droplet functional (all others
droplets variables $u_2^{(j)},E^{(j)}$ with $j \neq i$ integrated out). One has, as in $d=0$,
$D_i[u_1,u_2^{(i)}]=P_i[u_1,u_2^{(i)},E^{(i)}=0]$. Since elementary droplets are assumed to be independent, upon
computing disorder averages, only one active elementary droplet at a time need be considered. Terms involving
two simultaneously active elementary droplets give contributions of higher order $T^2$.

Hence to compute the disorder averaged force we can now use
formula (\ref{formdrop}) and obtain:
\begin{eqnarray}
&& \overline{ \sum_i \frac{u^{(i)}_{12,y} w_i a^{(i)}  }{1 + w_i a^{(i)}}  }
= T \langle  \sum_i u^{(i)}_{12,y} \ln(1 + e^{- (\tilde v g^{-1} u^{(i)}_{12}) }) \rangle_D \\
&& = - \frac{T}{2} \langle \sum_i u^{(i)}_{12,y} (\tilde v g^{-1} u^{(i)}_{12}) \rangle_D
\\
&& = - T \tilde v_y
\end{eqnarray}
Here and below the droplet average is defined as $\langle A[u_{12}] \rangle_D = \sum_i \langle A[u^{(i)}_{12}]
\rangle_{D_i}$ according to the previous paragraph. We will however usually drop the index $i$ on $D_i$. In the
second line we have assumed, and used the {\it local} symmetry $u^{(i)}_{12} \to - u^{(i)}_{12}$ of the droplet
probability distribution, which generalizes (\ref{sym}). In the last line we have used the STS identity:
\begin{eqnarray}
&& \sum_i \langle u^{(i)}_{12,x} u^{(i)}_{12,y} \rangle_D = 2 g_{xy}
\end{eqnarray}
which generalizes (\ref{STSnorm}). Since $\langle u_1 \rangle_P=0$ by parity, one correctly recovers (to lowest
order in $T$) that $\overline{ V'_x[v] } =0$.

We can now compute the second cumulant:
\begin{eqnarray}
&& \overline{ \hat V'_{x_1}[v_1] \hat V'_{x_2}[v_2] } = g^{-1}_{x_1 y_1} g^{-1}_{x_2 y_2} \big( \langle u_{1
y_1} u_{1 y_2} \rangle_P
\nonumber \\
&& + T \sum_i \int_0^1 \frac{dw_i}{w_i} \langle [ (u_{1 y_1} - \frac{u^{(i)}_{12,y_1} w_i a_1^{(i)}  }{1 + w_i
a_1^{(i)}} ) (u_{1 y_2} - \frac{u^{(i)}_{12,y_2} w_i a_2^{(i)}  }{1 + w_i a_2^{(i)}} ) - u_{1 y_1} u_{1 y_2} ]
\rangle_D \nonumber
\\
&& + \sum_{i \neq j} \overline{ \frac{u^{(i)}_{12,y_1} w_i a_1^{(i)}  }{1 + w_i  a_1^{(i)}}
\frac{u^{(j)}_{12,y_2} w_j a_2^{(j)}  }{1 + w_j  a_2^{(j)}} }  \big) + O(T^2)
\end{eqnarray}
Elementary droplets being independent averages can be decoupled in
the last term, which is thus of order $T^2$. Since it involves an
extensive double sum one could fear that it could contain an
additional factor of volume $L^d$ and be of order $T L^d$ as
compared to the dominant one, hence not negligible. This is not
the case, and in fact, up to adding and subtracting the $i=j$ term
(which is clearly subdominant), it exactly cancels against the $T
\tilde v$ terms. The algebra is now similar to the case $d=0$ and
we find:
\begin{eqnarray}
&& - \overline{ \hat V'_{x_1}[v_1] \hat V'_{x_2}[v_2] } = R''_{z_1 z_2}[v_1-v_2]= R''_{z_1 z_2}[0]  + T
g^{-1}_{z_1 x_1} g^{-1}_{z_2 x_2} \sum_i  \langle u^{(i)}_{12 x_1} u^{(i)}_{12 x_2} F_2 \big((u^{(i)}_{12}
g^{-1} \tilde v_{12}) \big) \rangle_D \label{drop2cumd}
\end{eqnarray}
where $\tilde v_{12}=\tilde v_1-\tilde v_2$ and the function $F_2(z)=\frac{z}{4} \text{coth} \frac{z}{2} -
\frac{1}{2}$ was defined in Section \ref{dropgen}. Thus the droplet picture yields a prediction for the exact
second cumulant functional directly related to the droplet probabilities. One can easily infer from
(\ref{drop2cumd}), or derive independently, that in the functional thermal boundary layer $\tilde v =v/T \sim
O(1)$ one has:
\begin{eqnarray}
&& R[v] = \frac{1}{2} v_x R_{xy}''[0] v_y + T^3 \sum_i \langle
H_2( (\tilde v g^{-1} u^{(i)}_{12} )) \rangle_D
\end{eqnarray}
where the function $H_2(z)$ was defined in (\ref{defh2}). Thus the second cumulant functional predicted from
this independent droplet picture has a simple structure. It is an average of a functional which is simply a {\it
function} of the quantity $(\tilde v g^{-1} u_{12})=\sum_{xy}\tilde v_x g^{-1}_{xy} u_{12y}$. It is nicely
proportional to $L^d$ for a uniform configuration, since each term contains a deformation $u^{(i)}_{12}$ non
zero only within a fixed volume $m^{-d}$. For a uniform configuration $\tilde v_x=\tilde v$ the argument in the
function $F_2$ (or $H_2$) is $(\tilde v g^{-1} u_{12})= m^2 \tilde v \int_x u_{12x}$.

For the usual massive propagator the result can be written, in
Fourier:
\begin{eqnarray}
&& - \frac{R''_q[v]}{(q^2+m^2)^2} = \langle u_{1q} u_{1,-q} \rangle_{u_1} - \frac{1}{4}  T \langle u_{12,q}
u_{12,-q} \int_k u_{12k} (k^2 + m^2) \tilde v_{-k}  \coth[ \frac{1}{2} \int_p u_{12p} (p^2 + m^2) \tilde v_{-p}
] \rangle_{D}
\end{eqnarray}
which includes a generalization of (\ref{u2}).

The calculation of all higher moments can be performed similarly. Some details are given in Appendix \ref{sec:2welld}. We
quote here only the result for the third moment:
\begin{eqnarray}
&& \bar S^{111}_{z_1 z_2 z_3}[v_1,v_2,v_3] = T g^{-1}_{z_1 x_1} g^{-1}_{z_2 x_2} g^{-1}_{z_3 x_3} \langle
u_{12x_1} u_{12x_2} u_{12x_3} F_3\big((u_{12} g^{-1} \tilde v_1),(u_{12}
g^{-1} \tilde  v_2),(u_{12} g^{-1} \tilde  v_3)\big) \rangle_{D} \nonumber \\
&& F_3[z_1,z_2,z_3] = \frac{1}{4} (z_1 (F[z_1-z_2,z_1-z_3]-\frac{2}{3}) + 2 p.c. ) \label{3mombl}
\end{eqnarray}
where we recall $F[a,b] = \frac{1+e^{a+b}}{(1-e^a)(1-e^b)}$. Note that the notion of a TBL in high $d$ may contain some momentum dependence if one chooses $v_x$ non uniform, a property which remains to be investigated.

\subsubsection{check that ERG equation are obeyed}

Since it is quite amazing to obtain a solution to a functional hierarchy, we will check explicitly at least the
first equation of the $W$-ERG hierarchy:
\begin{eqnarray} \label{ergg2}
&& \partial R[v] = T \partial g_{xy} R''_{xy}[v] + \partial g_{xy} \bar S^{110}_{xy}[0,0,v]
\end{eqnarray}
The two terms which must be added in the r.h.s. have the explicit form, using the droplet solution
(\ref{drop2cumd}) and (\ref{3mombl}):
\begin{eqnarray}
&& R''_{xy}[v] = R''_{xy}[0] + T \langle (g^{-1} u_{12})_x (g^{-1} u_{12})_y F_2\big((\tilde v g^{-1}
u_{12})\big) \rangle_D
\\
&& \bar S^{110}_{xy}[0,0,v] = T^2 \langle (g^{-1}  u_{12})_x (g^{-1} u_{12})_y G_3[(\tilde v g^{-1}  u_{12})]
\rangle_D \label{limS}
\end{eqnarray}
where one finds $G_3[z] = \frac{z^2}{24} - F_2[z] \sim \frac{z^4}{1440} + O(z^6)$. Hence the two term cancel and, up to a constant,
one is left with a single term, a quadratic term $\sim v^2$, of order $T^2$ in the TBL. Hence in the TBL the above equation becomes $ \partial R[v] = \frac{1}{2} \partial R''_{xy}[0] v_x v_y$ and the only thing
left to check is:
\begin{eqnarray}
&& \partial R''_{zt}[0] = \frac{1}{12} \partial g_{xy} \sum_i \langle  (g^{-1} u^{(i)}_{12})_x (g^{-1}
u^{(i)}_{12})_y (g^{-1} u^{(i)}_{12})_z (g^{-1} u^{(i)}_{12})_t   \rangle_D = \lim_{T \to 0} T \partial g_{xy}
R''''_{xyzt}[0] \label{limit1}
\end{eqnarray}
One can check that the second identity holds, as a consequence of (\ref{drop2cumd}). Hence the last equation to
check is the identity (\ref{identityd}) but this one can be proved exact independently of droplets, as
consequence of ERG and STS identities, as discussed in the previous Section (see also below). Note that Eq. (\ref{limit1}) is the generalization of the dissipation rate or anomaly equation in decaying Burgers turbulence.

Hence we have found that $\partial R''_{xy}[v]$ is constant in the whole TBL (i.e. from $v=0$ to large $\tilde v$). Hence if matching holds between small $v_x$ and large $\tilde v_x$ then the functional second derivative should be continuous in the whole range. 

It is useful to give the local version of (\ref{limit1}):
\begin{eqnarray}
&& \partial R''(0) = \frac{1}{12}
m^4  L^{-d} \partial g_{xy} \sum_i \langle (g^{-1} u^{(i)}_{12})_x (g^{-1} u^{(i)}_{12})_y (\int_t
u^{(i)}_{12 t})^2 \rangle_D \\
&& = \frac{1}{6} m^6  L^{-d} \sum_i \langle (\int_x
u^{(i)}_{12x} u^{(i)}_{12x})  (\int_t u^{(i)}_{12 t})^2 \rangle_D =
 \frac{1}{3} m^4  \frac{ \sum_i \langle (\int_x u^{(i)}_{12x} u^{(i)}_{12x})  (\int_t u^{(i)}_{12
t})^2 \rangle_D }{\sum_i \langle (\int dx u^{(i)}_{12}(x))^2 \rangle_D}
\end{eqnarray}
where in the second equality we have restricted to the form $g_k^{-1}=k^2+m^2$ and $\partial=-m\partial_m$. Again $R''(0)$ is continuous and equal to $R''(0^+)$.

\subsubsection{matching and relations between shocks and droplets}

We now study the limit $\tilde v \to \infty$. In $d=0$, for $R(v)$, this nicely matches to the small $v$ limit
of the outer region $v=O(1)$. For higher moments, such as $S(v_{123})$, as $\tilde v_{13}$ and $\tilde v_{23}$
go to infinity, it matches the partial boundary layer (here PBL21). Let us examine what happens here.

Let us consider first a uniform $\tilde v_x=\tilde v$. From (\ref{drop2cumd}) we find for large $\tilde v$:
\begin{eqnarray}
&& R(v) = \frac{1}{2} R''(0) v^2 + \frac{1}{24} L^{-d} m^6 |v|^3
\sum_i \langle |\int dx u^{(i)}_{12}(x)|^3 \rangle_D \\
&& = \frac{1}{2} R''(0) v^2 + \frac{1}{12} m^4 |v|^3 \frac{\sum_i \langle |\int dx u^{(i)}_{12}(x)|^3
\rangle_D}{\sum_i \langle (\int dx u^{(i)}_{12}(x))^2 \rangle_D}
\end{eqnarray}
where we have used the normalization of the droplet measure given by the STS symmetry:
\begin{eqnarray} \label{stsdrop} 
&& \sum_i \langle (\int dx u^{(i)}_{12}(x))^2 \rangle_D = 2 L^d m^{-2}
\end{eqnarray}
Hence we obtain the cusp as the following droplet average:
\begin{eqnarray}
 && R'''(0^+) = \frac{1}{2} m^4 \frac{\sum_i
\langle |\int dx u^{(i)}_{12}(x)|^3 \rangle_D}{\sum_i \langle (\int dx u^{(i)}_{12}(x))^2 \rangle_D} \label{cuspdrop}
\end{eqnarray}
On the other hand, within some assumptions, one can also show that \cite{avalanches}:
\begin{eqnarray}
 && R'''(0^+) = \frac{1}{2} m^4 \frac{\langle s^2 \rangle_P}{\langle s \rangle_P}
\end{eqnarray}
where $s=\int dx u_{12}(x)$ are the {\it shock sizes}. This is compatible with a relation between shock and
droplet size distributions which generalizes the $d=0$ relation (\ref{connect}), with, in $d$ dimension  the droplet size $y \to Y=\int dx
u^{(i)}_{12}(x)$.

Indeed let us consider the large $\tilde v=O(1)$ limit of (\ref{ergg2}). From (\ref{limS}) it behaves as
$O(\tilde v^2)$ as the term $T R''''[v]$ becomes negligible in that limit, and the coefficient should equal the
r.h.s. of (\ref{limit1}). Indeed in the TBL, from (\ref{3mombl}):
\begin{eqnarray}
\lim_{\tilde v \to \infty} \bar S^{112}_{xyzt}[0,0,v] = \frac{1}{12}  \sum_i \langle  (g^{-1} u^{(i)}_{12})_x (g^{-1}
u^{(i)}_{12})_y (g^{-1} u^{(i)}_{12})_z (g^{-1} u^{(i)}_{12})_t   \rangle_D
\end{eqnarray}
On the other hand, matching, i.e. considering the small $v=O(1)$ limit of (\ref{ergg2}) and making the usual
assumption that the term $T R''''$ can be neglected in that region, requires:
\begin{eqnarray}
 \partial g_{xy}  \bar S^{112}_{xyzt}[0,0,0^+] = \lim_{\tilde v \to \infty}
\partial g_{xy}  \bar S^{112}_{xyzt}[0,0,v] \label{connect2} 
\end{eqnarray}
a generalization of the anomaly equation in Burgers turbulence (matching of dissipation rate from dissipation range to inertial range).
Since one can also show that the third moment of the shock size distribution is given by \cite{avalanches}:
\begin{eqnarray}
 \int_{xzt} \bar S^{112}_{0xzt}[0,0,0^+] = \frac{1}{6}  m^6 \frac{\langle s^3 \rangle_P}{\langle s \rangle_P}
\end{eqnarray}
the continuity (\ref{connect2}) implies that the same quantity is related to the fourth moment of droplet sizes, hence it is again compatible with the generalization of the droplet-shock relation  (\ref{connect}). It would be quite interesting to investigate further the relation between droplets and shocks, in particular in the non-local, momentum-dependent aspects. While a shock always correspond to a droplet (with exact degeneracy), it is not fully clear how the reverse works, i.e. given a droplet at low $T$, does it correspond to a unique shock nearby in phase space, and which $v_x$ then to choose to find this underlying shock.

\subsubsection{STS and ERG droplet identities}

Let us close this Section by mentioning another useful check of the droplet solution. One can compute using
droplets all four point correlations given in Section \ref{sec:corrd} and check that all STS and ERG identities are indeed
obeyed. This is performed in Appendix \ref{app:sts}. Here let us just mention that the STS identities can be encoded in the
following functional RG equation which relate $P[u_1]$ and $D[u_1,u_2]$:
\begin{eqnarray}
&& - \frac{\delta P[u_1]}{\delta u_{1x} } = \int Du_2 g^{-1}_{xy} (u_{1y} - u_{2y}) D[u_1,u_2]
\end{eqnarray}
which, upon integration, generate an infinite set of identities between correlation functions, the first one
being the famous $ \langle u_{12,x} u_{12,y} \rangle_D = 2 g_{xy}$. The ERG identities yield:
\begin{eqnarray}
&& \partial P[u_1] = - \frac{1}{2} \int Du_2 ( (u_{1} \partial g^{-1} u_{1}) -  (u_{2} \partial g^{-1} u_{2}))
D[u_1,u_2]
\end{eqnarray}
These are the functional generalization of the equations given in $d=0$ \footnote{note that these equations, as
their $d=0$ counterpart, where found on the basis of all analytic correlation functions. If they hold they
should also imply relations for averages of non-analytic observables (such as $|u|$)}. It is interesting to
present these equations since one expects, if the $\epsilon$ expansion makes sense, that $P[u_1]$ be nearly
gaussian near $d=4$, and it may inspire other approaches to this expansion (in particular one wonders what
simplification occurs then, if any, in the shape of $D[u_1,u_2]$)

\subsection{$T=0$ limit and $\epsilon$-expansion}

\subsubsection{$T=0$ limit and continuity properties}

The ERG equations (\ref{erg2momWd}) and (\ref{erg2momGd}) in Section \ref{ergeqd} were derived for analytic moment and cumulant functions, and as such they are always exact for 
$T>0$. The zero temperature limit $T=0^+$ (alternatively the small $m$ limit
$T m^\theta \to 0$) of these equations necessitates a careful analysis. We will follow closely what was learned in $d=0$. 

Consider the W-ERG equation (\ref{erg2momWd}) and naively set $T=0$. One is left with the hierarchy:
\begin{eqnarray}
&& \partial R[v] = \partial g_{zz'} \bar S^{110}_{zz'}[0,0,v]  \label{ergWT0}  \\
&& \partial \bar S[v_{123}] =  \frac{3}{2} \text{sym}_{123}
\partial g_{xy} \bar Q^{1100}_{xy}[v_{1123}]
\end{eqnarray}
and so on. The questions are (i) what is the meaning of the functions evaluated at coinciding arguments on the right hand side (ii) can this hierarchy be closed (iii) can it be solved. The last question is about the $\epsilon$ expansion and is examined in the next section. 

To answer (i), we note that these equations are expected to be correct for small $T>0$ and $v$ in the outer region, i.e. $v = O(1) \gg T$, all $v_{ij}=0(1) \gg T$ for $i \neq j$. Indeed in that region one may assume that the terms proportional to $T$ in  (\ref{erg2momWd}) are negligible (they become of the same order only in the TBL region $v \sim T$, $v_{ij} \sim T$ discussed in the previous section). Of course this is an assumption but it is supported by the analysis of the previous section. It extends the $d=0$ analysis to the functional. As a consequence the meaning of the derivatives appearing in the r.h.s. of (\ref{ergWT0}) is:
\begin{eqnarray}
&&  \lim_{T \to 0} \bar S^{110}_{zz'}[0,0,v]
\end{eqnarray}
and similarly for all members of the hierarchy. These are perfectly well defined quantities, but defined in the TBL, since the derivatives are exactly at coinciding arguments.

The next question (ii) then is can the hierarchy be closed, i.e. is there really a hierarchy valid at $T=0^+$ ? Or, in the language of Burgers trubulence, is there a hierarchy defined solely in the inertial range (by ''closed'' here we do not mean truncated - this is the business of the $\epsilon$ expansion, we mean defined only in terms of quantities involving the outer region itself). For this one needs continuity properties, as was the case also in $d=0$. Since here we deal with a functional we may need to distinguish:

(a) the strong continuity property:
\begin{eqnarray}
&& \lim_{v_2 \to v_1} \bar S^{(n,T=0)1100..0}_{xy}[v_{1234..n}] \quad \text{exists for arbitrary $v_{21x} \to 0$} \\
&& \text{and} \nonumber \\
&& = \lim_{T \to 0} \bar S^{(n,T)1100..0}_{xy}[v_{1134..n}]
\end{eqnarray}

(b) the weak continuity \footnote{Note that we have termed these properties strong and weak for pure convenience and in any relation to a customary sense in which these terms are used in mathematics.} property:
\begin{eqnarray}
&& \lim_{v_{12} \to 0} \bar S^{(n,T=0)1100..0}_{xy}[v_{1234..n}] \quad  \text{exists for $v_{21x}=v_{12} \to 0$} \\
&& \text{and} \nonumber \\
&& = \lim_{T \to 0} \bar S^{(n,T)1100..0}_{xy}[v_{1134..n}]
\end{eqnarray}

Hence the limit may exist uniformly for arbitrary argument, or for a spatially uniform argument. Of course (a) implies (b) and is a stronger property. For $n=2$ it reads:
\begin{eqnarray} \label{Rxycont} 
&& \lim_{v_x \to 0} R''_{T=0,xy}[v]  =  \lim_{T \to 0} R''_{T,xy}[0]   \quad \text{strong continuity} \\
&&  \lim_{v_x=v \to 0} R''_{T=0,xy}[v]  =  \lim_{T \to 0} R''_{T,xy}[0]   \quad \text{weak continuity} 
\end{eqnarray}

These continuity properties are necessary for the ERG equation to be globally closed, i.e. to relate quantities computable in the outer region, hence defining a meaningful $T=0^+$ W-ERG hierarchy. The continuity of the second derivative of Eq. (\ref{Rxycont}) is necessary for the calculation of the two point correlation function $\overline{u_x u_y}$. In both cases, weak continuity is sufficient (b) but we believe that in fact strong continuity (a) usually holds. Indeed one has, directly at $T=0$:
\begin{eqnarray} \label{Rxycont} 
&& \overline{(v_{1x} - u_x[v_1])(v_{2y} - u_y[v_2])} = - g_{xz} g_{yt} R''_{zt}[v_{12}]
\end{eqnarray}
where $u_x[v]$ is the ground state configuration of the interface in an harmonic well centered on $v \equiv \{v_x\}$. Although we cannot prove it, it is hard to imagine from the picture of shocks that the left hand side is not a continuous function with a unique limit as $v_{12x} \to 0$, at least for a finite number of degree of freedom. This is basically the same argument that in $d=0$ (see section \ref{sec:inviscid}) that $\hat V'_x[v]$ and $u_x[v]$ have jumps at discrete locations as $v$ is varied, here in the space of configurations. Unless there is some accumulation of shocks it seems unlikely that continuity in arbitrary moments of $\hat V'_x[v]$ and $u_x[v]$ should fail. This implies that all $W$-moment functionals $ \bar S^{(n,T=0)11..1}_{xy}[v_{1234..n}] $ should be continuous in their arguments and equal to the $T=0^+$ limit of the same function in the TBL at exacty coinciding points.

What about the $\Gamma$-ERG hierarchy? From the definition (\ref{defofS}) one finds:
\begin{eqnarray}
&& \bar S^{110}_{zt}[v_{123}] = S^{110}_{zt}[v_{123}]  \\
&& +
g_{xy} ( R'''_{xzt}[v_{12}] (R'_y[v_{23}] - R'_y[v_{13}]) - R''_{xt}[v_{12}]
R''_{yz}[v_{13}] - R''_{xz}[v_{21}] R''_{yt}[v_{23}] + R''_{xz}[v_{31}] R''_{yt}[v_{32}]) ) \nonumber
\end{eqnarray}
thus to recover the first $\Gamma$-ERG equation in (\ref{erg2momGd}) setting $T=0$ one needs the regularity condition:
\begin{eqnarray}
&& \lim_{v_{12} \to 0} g_{xy} ( R'''_{xzt}[v_{12}] (R'_y[v_{23}] - R'_y[v_{13}]) = 0 \label{lim22}
\end{eqnarray}
since it vanishes by parity in the TBL,
and continuity of $R''_{xy}[v]$. Hence for the first $\Gamma$-ERG equation to be valid for all $v \equiv \{v_x\}$ weak continuity of $R''$ as well as the weak version of (\ref{lim22}) should be sufficient. There are very similar continuity, or regularity conditions to be satisfied for the second $\Gamma$-ERG equation in (\ref{erg2momGd}) to be valid. 

Hence if we make the minimal assumption of weak continuity and regularity we should be able to use the $\Gamma$-ERG equation directly at $T=0$.

\subsubsection{$\epsilon$ expansion}

Let us now investigate the $\epsilon=4-d$ expansion using the $\Gamma$-ERG hierarchy. A version using the W-ERG hierarchy is given in Appendix \ref{app:werg}.

Let consider the $\Gamma$-ERG hierarchy and set $T=0$. The two lowest order equations read:
\begin{eqnarray}
&& \partial R[v] =
\partial g_{zz'} g_{xy} ( R''_{xz}[v] R''_{yz'}[v] - 2 R''_{xz}[0] R''_{yz'}[v]  ) +
\partial g_{zz'} S^{110}_{zz'}[0,0,v]  \label{zeroT2}
\end{eqnarray}
and
\begin{eqnarray}
&& \partial S[v_{123}] = 6 \text{sym}_{123}
\partial g_{xy} g_{zt} {\sf R}''_{xz}[v_{12}] (S^{110}_{yt}[v_{113}]
- S^{110}_{yt}[v_{123}]) + \frac{3}{2} \text{sym}_{123}
Q^{1100}_{xy}[v_{1123}]  \nonumber  \\
&& + 3 \text{sym}_{123} \partial g_{xy} g_{zt} g_{rs} \big( {\sf R}''_{xz}[v_{12}] {\sf R}''_{yr}[v_{12}] {\sf
R}''_{st}[v_{13}] + 2 {\sf R}''_{xz}[v_{12}] {\sf R}''_{st}[v_{12}] {\sf R}''_{yr}[v_{13}] - {\sf
R}''_{xz}[v_{12}] {\sf R}''_{yr}[v_{23}] {\sf R}''_{st}[v_{13}] \big) \nonumber \\
&& \label{zeroT3}
\end{eqnarray}

The spirit of the $\epsilon$ expansion is that there is a solution of the ERG hierarchy which has the following structure. Let us recall the definition of the local part $R(v)$, and non local parts $\tilde R[v]$, of the functional $R[v]$ as defined in (\ref{dec0}) and (\ref{dec2}). 
 The $\epsilon$ expansion states that:
\begin{eqnarray}
&& R(v) \sim O(\epsilon) \quad , \quad \tilde R[v] \sim O(\epsilon^2)  \label{orders} \\
&& S[v_{123}] \sim O(\epsilon^3) \quad , \quad Q \sim O(\epsilon^4) \nonumber
\end{eqnarray}
and so on $S^{(n)} \sim O(\epsilon^n)$ for $n \geq 3$, these statements being valid for the properly rescaled
fixed point forms (see details below). Examination of the structure of the hierarchy shows superficially
compatibility with this counting. We must now check that it works and is unambiguous. 

Here we will distinguish again the {\it strong $\epsilon$ expansion} in which the counting (\ref{orders}) is obeyed for any $v=\{v_x\}$ and the {\it weak $\epsilon$ expansion} in which the counting (\ref{orders}) is obeyed for uniform configuration $v$ or infinitesimally close to it, i.e. by all derivatives evaluated at a uniform configuration. Note that the weak version is sufficient to compute all correlations of the $u_x$ field at zero temperature (or at the fixed point) hence it is perfectly respectable.

We now study the $\epsilon$ expansion. We focus on the hard question of ambiguities and anomalous terms, as in $d=0$. Once these are understood the rest, i.e. rescaling and derivation of the FRG equation is easy and similar to what was done in Ref.\cite{twolooplong}, hence we will not detail that part. Our aim here is to give a first principle derivation of the anomalous terms, or at the very least specify clearly what the assumptions are, which was not done in Ref \cite{twolooplong}. There a {\it candidate} field theory was proposed, based on some global 
consistency.

\subsubsection{one loop: order $O(\epsilon)$}

Following the strategy  (\ref{orders}) outlined above, to study $R[v]$ to one loop, i.e. to lowest order in $\epsilon$, we can discard the $S$ term in (\ref{zeroT2}) and obtain:
\begin{eqnarray}
&& \partial R[v] =
\partial g_{zz'} g_{xy} ( R''_{xz}[v] R''_{yz'}[v] - 2 R''_{xz}[0] R''_{yz'}[v]  ) + O(\epsilon^3) \label{1looptrunc}
\end{eqnarray}
Let us insert the decomposition (\ref{dec2}) and assume that $\tilde R[v] \sim O(\epsilon^2)$. Of course these assumptions (\ref{orders}) must be checked a posteriori self consistently. One finds:
\begin{eqnarray}
&& \partial R[v] =
\partial g_{xy} g_{xy} {\sf R}''(v_x) {\sf R}''(v_y) + O(\epsilon^3)
\end{eqnarray}
The local part gives:
\begin{eqnarray} \label{oneloopusual}
&& \partial R(v) = (\int_y \partial g_{y} g_{y}) {\sf R}''(v)^2  + O(\epsilon^3)
\end{eqnarray}
Upon evaluation of $\int_y \partial g_{y} g_{y}=\frac{1}{2} \partial J_2$ with $J_2=\int_k (k^2+m^2)^{-2}$ and rescaling 
 this yields the standard one loop FRG equation and fixed point for the rescaled $R$ of order $O(\epsilon)$. It is important to note that $J_2=\int_y g_y^2 \sim m^{-\epsilon}/\epsilon$ has a pole in $\epsilon$ which disappear as the derivative $\partial J_2$ is finite in $d=4$, producing a finite $\beta$-function as it should for a renormalizable theory. Using the fixed point value of $R''(0)$ and weak continuity yields the $T=0$ correlation function. Its general expression is:
 \begin{eqnarray} \label{zeroq}
&& \overline{u_{x} u_{y}}  = - g_{xz} g_{yt} R''_{yt}[0] = - g_{xz} g_{yz} R''(0) -  g_{xz} g_{yt} \tilde R''_{yt}[0] 
\label{corr0} 
\end{eqnarray}
and the knowledge of the local part $R''(0)$ only gives it at $q=0$, i.e. the center of mass fluctuations,
as $\int_y \overline{u_x u_y}=- R''(0)/m^4$. 

Let us now study the equation for the non local part which would result from (\ref{1looptrunc}). 
\begin{eqnarray}
&& \partial \tilde R[v] = (\partial g_{xy} g_{xy} - \delta_{xy} \partial g_{z} g_{z}) {\sf R}''(v_x) {\sf
R}''(v_y) + O(\epsilon^3)
\end{eqnarray}
which integrates into:
\begin{eqnarray}  \label{oneloopbiloc}
&& \tilde R[v] = \frac{1}{2} (g_{xy}^2 - \delta_{xy} g_{z}^2) {\sf R}''(v_x) {\sf R}''(v_y) + O(\epsilon^3)
\end{eqnarray}
here (\ref{oneloopusual}) has been used, and the fact that the term $\partial R$ terms should produce only higher order terms in $\epsilon$. The corresponding formula for the second derivative is:
\begin{eqnarray} \label{secder}
&& \tilde R''_{ts}[v] = (g_{ts}^2 - \delta_{ts} g_{z}^2) {\sf R}'''(v_t) {\sf R}'''(v_s) +  \delta_{st}
(g_{ty}^2 - \delta_{ty} g_{z}^2) {\sf R}''''(v_t) {\sf R}''(v_y) + O(\epsilon^3)
\end{eqnarray}
and no summation on $s,t$. Hence for a uniform configuration:
\begin{eqnarray} \label{nonlocunif} 
&& \tilde R''_{ts}[v] = (g_{ts}^2 - \delta_{ts} g_{z}^2) {\sf R}'''(v)^2  + O(\epsilon^3)
\end{eqnarray}
as the second term automatically vanishes. Since $R'''(0^+)^2=R'''(0^-)^2$ weak continuity then produces an {\it unambiguous limit}:
\begin{eqnarray}
&& \tilde R''_{ts}[0] = (g_{ts}^2 - \delta_{ts} g_{z}^2) {\sf R}'''(0^+)^2  + O(\epsilon^3) \label{resdiag}
\end{eqnarray}
and yields the $T=0$ correlation function at any momentum $q$ through (\ref{corr0}):
 \begin{eqnarray} \label{zeroq}
&& \overline{u_{q} u_{q'}}  =  - g_{q}^2 (R''(0) + \int_k g_k (g_{q+k} - g_k))  {\sf R}'''(0^+)^2) + O(\epsilon^3) 
\end{eqnarray}
which is the result displayed in \cite{twolooplong} where some (rough) discussion of the ambiguities was also given.
Note the important fact that $g_{ts}^2$ is in Fourier $\int_k g_k g_{q+k} \sim 1/\epsilon$ and that the divergence is removed by the local part (counterterm), i.e. only the difference $g_{ts}^2 - \delta_{ts} \int_k g_k^2$ is finite, and produces a finite result in $d=4$ for the correlation \footnote{as $d \to 4^-$ $g_x^2$ becomes proportional to a a delta function $\delta_x$ to leading order.}

There is however a feature of the result for the non-local part which at first sight may appear disturbing. 
Since ${\sf R}''(v) \sim |v|$ at small $v$ Eq. (\ref{oneloopbiloc}) appears {\it incompatible} with the existence of an unambiguous second derivative $\tilde R''_{xy}[0]$ for the non local part to the lowest order in $\epsilon$. More precisely, the limit of $\tilde R''_{xy}[v_{12}]$ as $v_{12} \to 0$ is unambiguous and equal to (\ref{resdiag}) {\it only} for {\it non crossing configurations} $v_{1x}$ and $v_{2x}$, i.e. such that $v_{12 x}$ remains of the same sign for all $x$. Since these are interface configurations ($N=1$) it also mean they are {\it partially ordered}, a concept familiar from depinning \cite{frgdep2}.

Hence there seem to be two mutually incompatible properties: (i) strong epsilon expansion (ii) strong continuity. At least one of them must fail. Faced with this dilemma we prefer to assume that strong continuity holds, as argued in the previous Section. Then it means that formula (\ref{oneloopbiloc}) and (\ref{secder}) cannot be correct in the counting in $\epsilon$ if evaluated for non uniform configurations $v_x$, more precisely for configurations with at least one sign change. In fact one can check that even as $v_x \to 0$ in a subspace of configurations with sign changes at fixed positions, the first term in the r.h.s of (\ref{secder}) retains a complicated momentum structure which may indeed be incompatible with the epsilon counting, while the second term exhibits delta function singularities at the points where $v_t$ vanishes (since $R''''(v) = 2 \delta(v) R'''(0^+) + R''''(0^+)$ as can be checked by integrating over a small region containing $v=0$). 

Note that this does {\it not} mean a failure of the $\epsilon$ expansion since as we found above to one loop the weak epsilon expansion is perfectly fine and is sufficient to give the correlation functions of $u_x$. It just means that the $T=0$ functional hierarchy can be easily solved using the usual $\epsilon$ counting {\it only} in a neighborhood of uniform configurations $v_x=v+\delta v_x$. This neighborhood can be infinitesimal, such that all derivatives of arbitrary order taken at a uniform configuration will satisfy the usual $\epsilon$ expansion counting, or one may try to extend it to partially ordered configurations $v_1..v_p$ with no intersections. In Appendix \ref{app:multilocal} exact equations are written using the multilocal expansion, and confirm that the weak epsilon expansion does work to one loop. We check below that it appears to work also to two loop.

A non trivial question is to understand exactly how the formula (\ref{oneloopbiloc}) and (\ref{secder}) fail for non uniform configurations. There must clearly be contributions of the same order in $\epsilon$ coming from higher order terms. There are then two possibilities. Either (\ref{1looptrunc}) remains true and corrections only come from the non local part itself, or it fails and corrections also come from the third moment term $S$  in (\ref{zeroT2}).

One reason for which we are confident that indeed (\ref{oneloopbiloc}) fails for (sign changing) non uniform configurations is that it contradicts the TBL solution which was shown in the previous sections to be an exact (and fully non perturbative in $\epsilon$) solution of the hierarchy (order by order in $T$). More precisely the large $\tilde v$ limit of (\ref{limit1}) and (\ref{ergg2}) is clearly incompatible with the naive small $v$ limit of  (\ref{oneloopbiloc}) for (sign changing) non uniform configurations. Hence terms of the same order at small $v$ must be hiding in the neglected terms of (\ref{zeroT2}).

\subsubsection{droplet relations to order $O(\epsilon)$}

It is interesting to note that if indeed $R'''(0^+)$ is of order $\epsilon$, then from (\ref{cuspdrop}) and (\ref{stsdrop}) one expects that:
\begin{eqnarray}
&& \int dx u^{(i)}_{12}(x) \sim O(\epsilon) \quad , \quad 
\sum_i \langle 1 \rangle_D  \sim O(\frac{1}{\epsilon^2})
\end{eqnarray}
and the typical value $\int dx
u^{(i)}_{12}(x) \sim \epsilon m^{-d+\zeta}$ with the TBL variable for a uniform $v$ being $z = m^2 v (\int dx u^{(i)}_{12}(x))/T \sim \epsilon (v m^\zeta) /\tilde T$.

In addition to the local relation (\ref{cuspdrop}), the assumption that (\ref{oneloopbiloc})  is correct for a uniform $v$, i.e. that (\ref{nonlocunif}) holds implies a few relations involving droplets which should be valid to leading order in $\epsilon$. Matching the flow of $R''_{xy}[0]$ inside and outside the TBL implies (for $x \neq y$:
\begin{eqnarray}
&& 2 \partial g_{x-y} g_{x-y} R'''(0^+)^2 =  \frac{1}{12} \langle (u_{12} \partial g^{-1} u_{12}) (g^{-1}
u_{12})_x (g^{-1} u_{12})_y \rangle_D + O(\epsilon^3)
\end{eqnarray}
to be valid to leading order in $\epsilon$, not only the amplitude, but also the spatial dependence - the local part reproduces the relation  (\ref{cuspdrop}). Matching of the term proportional to $O(|v|)$ from (\ref{nonlocunif}) and the large $\tilde v$ behavior from the non-local part of (\ref{drop2cumd}) yields:
\begin{eqnarray}
&& 2 (\partial g_{x} g_{x}- \delta(x) \int_y
\partial g_{y} g_{y}) R'''(0^+) R''''(0^+) \\
&& =  \frac{m^2}{4}  \langle \big( (g^{-1} u_{12})_0 (g^{-1} u_{12})_x  - \delta(x) \int_y  (g^{-1} u_{12})_0 (g^{-1} u_{12})_y \big) |\int_z u_{12z}| \rangle_D + O(\epsilon^3) \\
\end{eqnarray}
a relation which has no local analogous. Note that testing these relations would be as much a check of the structure of the droplet solution (after all, we have considered only the simplest minded one), than a check of the weak epsilon expansion.

\subsubsection{one loop: third moment}

Let us now compute the third moment to lowest order in the $\epsilon$ expansion. We need to solve to lowest order, keeping the dominant term in  (\ref{zeroT3}):
\begin{eqnarray}
&& \partial S[v_{123}] = 3 \text{sym}_{123} \partial g_{xy} g_{zt} g_{rs} \big( {\sf R}''_{xz}[v_{12}] {\sf
R}''_{yr}[v_{12}] {\sf R}''_{st}[v_{13}] + 2 {\sf R}''_{xz}[v_{12}] {\sf R}''_{st}[v_{12}] {\sf
R}''_{yr}[v_{13}] - {\sf R}''_{xz}[v_{12}] {\sf R}''_{yr}[v_{23}] {\sf R}''_{st}[v_{13}] \big) \nonumber
\end{eqnarray}
The solution is:
\begin{eqnarray}
&& S[v_{123}] =  \text{sym}_{123}  g_{xy} g_{zt} g_{rs} \big( 3 {\sf R}''_{xz}[v_{12}] {\sf R}''_{yr}[v_{12}]
{\sf R}''_{st}[v_{13}]  - {\sf R}''_{xz}[v_{12}] {\sf R}''_{yr}[v_{23}] {\sf R}''_{st}[v_{13}] \big) 
+ O(\epsilon^4) \nonumber
\end{eqnarray}
Here again we have assumed that $\partial {\sf R}$ is higher order. The leading behaviour is expected to be:
\begin{eqnarray} \label{solu3mom}
&& S[v_{123}] =  \text{sym}_{123}  g_{xy} g_{xt} g_{yt}  \big( 3 {\sf R}''(v_{12x}) {\sf R}''(v_{12y}) {\sf
R}''(v_{13t}) -  {\sf R}''(v_{12x}) {\sf R}''(v_{23y}) {\sf R}''(v_{13t}) \big) + O(\epsilon^4)
\end{eqnarray}
with validity a priori either for $v_x$ near a uniform configuration (weak epsilon expansion) or any $v_x$ (strong epsilon
expansion). As discussed below only the first one presumably holds. Since $\bar S=S$ one can compute, from (\ref{solu3mom}):
\begin{eqnarray}
&& S^{111}_{z z' z''}[v_{123}] = g_{z s}^{-1} g_{z' s'}^{-1} g_{z'' s''}^{-1} \overline{(v_{1s} - u_{s}[v_1])
(v_{2s'} - u_{s'}[v_2]) (v_{3s''} - u_{s''}[v_3])}
\end{eqnarray}
for $v_1$, $v_2$ and $v_3$ partially ordered (weak epsilon expansion).

\subsubsection{two loop: order $O(\epsilon^2)$ }

To go to next order in $\epsilon$ we must now study:
\begin{eqnarray}
&& \partial R[v] =
\partial g_{zz'} g_{xy} {\sf R}''_{xz}[v] {\sf R}''_{yz'}[v] +
\partial g_{zz'} S^{110}_{zz'}[0,0,v] \\
&& =
\partial g_{xy} g_{xy} {\sf R}''(v_x) {\sf R}''(v_y)
+ 2 \partial g_{zz'} g_{xz'} \tilde {\sf R}''_{xz}[v] {\sf R}''(v_{z'}) +
\partial g_{zz'} S^{110}_{zz'}[0,0,v]
\end{eqnarray}
at this stage there is no approximation. The second term is the feeding from the non-local part of the second moment. We will now insert
(\ref{oneloopbiloc}), but we need:
\begin{eqnarray}
&& \tilde {\sf R}''_{xz}[v] = \tilde R''_{xz}[v] - \tilde R''_{xz}[v=0]
\end{eqnarray}
we need to use only the weak continuity property:
\begin{eqnarray}
&& \tilde R''_{xz}[v=0] = \lim_{v \to 0^+} \tilde R''_{xz}[\{v_y\}=v]
\end{eqnarray}

One then obtains the contribution of the non-local term:
\begin{eqnarray}
&& 2 \partial g_{zz'} g_{xz'} \tilde {\sf R}''_{xz}[v] {\sf R}''(v_{z'}) = \\
&& 2 \partial g_{zz'} g_{xz'} (g_{xz}^2 - \delta_{xz} g_t^2) ({\sf R}'''(v_{x}) {\sf R}'''(v_{z}) - {\sf
R}'''(0^+)^2) {\sf R}''(v_{z'}) + 2
\partial g_{xz'} g_{xz'} (g_{xy}^2 - \delta_{xy} g_t^2) {\sf R}''''(v_{x}) {\sf R}''(v_{y}) {\sf R}''(v_{z'})
\nonumber
\end{eqnarray}

For the feeding from the third moment we use the result (\ref{solu3mom}). It is shown in the Appendix \ref{app:details} that:
\begin{eqnarray}
&& S[v_{123}] = v_{12z} v_{12y} g_{xy} g_{yz} g_{zx} \big( {\sf R}'''(v_{12y}) {\sf R}'''(v_{12z}) {\sf
R}''(v_{13x}) - {\sf R}'''(v_{13z}) {\sf R}'''(v_{13y}) {\sf R}''(v_{13x}) \big) + O(v_{12}^3)
\end{eqnarray}
where we have used that ${\sf R}''(v)={\sf R}'''(v) v + O(v^2)$. Hence:
\begin{eqnarray}
&& S^{110}_{zz'}[v_{123}] = - 2 g_{xz'} g_{z'z} g_{zx} \big( {\sf R}'''(v_{12z'}) {\sf R}'''(v_{12z}) {\sf
R}''(v_{13x}) - {\sf R}'''(v_{13z}) {\sf R}'''(v_{13z'}) {\sf R}''(v_{13x}) \big) + O(v_{12})
\end{eqnarray}
We use again the weak continuity:
\begin{eqnarray}
&& \partial g_{zz'} S^{110}_{zz'}[0,0,v] = \lim_{w \to 0^+} S^{110}_{zz'}[v_{123}]|_{v_{13x}=v_x,v_{12x}=w} \\
&& = - 2 \partial g_{zz'} g_{xz'} g_{z'z} g_{zx} ( {\sf R}'''(0^+)^2 - {\sf R}'''(v_z) {\sf R}'''(v_{z'}) ) {\sf
R}''(v_{x})
\end{eqnarray}

Putting all together we find the two loop FRG equation for the functional:
\begin{eqnarray}
&& \partial R[v] =
\partial g_{xy} g_{xy} {\sf R}''(v_x) {\sf R}''(v_y) \\
&& + 2 \partial g_{zz'} g_{xz'} (g_{xz}^2 - \delta_{xz} g_t^2) ({\sf R}'''(v_{x}) {\sf R}'''(v_{z}) - {\sf
R}'''(0^+)^2) {\sf R}''(v_{z'}) +
2 \partial g_{xz'} g_{xz'} (g_{xy}^2 - \delta_{xy} g_t^2) {\sf R}''''(v_{x}) {\sf R}''(v_{y}) {\sf R}''(v_{z'}) \nonumber \\
&& - 2 \partial g_{zz'} g_{xz'} g_{z'z} g_{zx} ( {\sf R}'''(0^+)^2 - {\sf R}'''(v_z) {\sf R}'''(v_{z'}) ) {\sf
R}''(v_{x})
\end{eqnarray}
This equation is certainly valid near a uniform configuration, i.e. within the weak epsilon expansion. Presumably it is valid again for non-sign changing $v_x$, although we have not checked it in details. As explained above, for it to be correct for arbitrary $v_x$ one needs to discard strong continuity, which we are not ready to do.

Let us now consider a uniform configuration $v_x=v$. One then gets the two loop FRG equation for the local part:
\begin{eqnarray}
&& \partial R(v) =
\partial g_{x} g_{x} {\sf R}''(v) {\sf R}''(v)  + [ 2 \partial g_{z0} g_{x0} (g_{xz}^2 - \delta_{xz} g_t^2) + 2 \partial g_{z0} g_{x0} g_{0z} g_{zx} ] ({\sf
R}'''(v)^2  - {\sf R}'''(0^+)^2) {\sf R}''(v)
\end{eqnarray}
which can also be written as:
\begin{eqnarray}
&& \partial R(v) = \frac{1}{2}
\partial J_2 {\sf R}''(v) {\sf R}''(v) + \partial( I_A - \frac{1}{2} J_2^2) ({\sf R}'''(v)^2  - {\sf R}'''(0^+)^2) {\sf R}''(v)
\end{eqnarray}
with $J_2=\int_x g_x^2$ and $I_A= \int_{xz} g_{0z}^2 g_{0x} g_{xz}$. This has now a very standard form for the $T=0$ FRG equation, and having carefully justified the anomalous terms we refer to  \cite{twolooplong} for the further simple steps leading to the rescaled two loop FRG equation. From the equation above we could also compute the two loop correction to the correlation function, a task left for the future.

\acknowledgments

My understanding of FRG has been greatly enhanced by vigorous, ongoing
and multiple collaborations with Kay Wiese whom I warmly thanks.
I am greatly indebted to collaborations with Leon Balents on the subtelties related
to finite temperature aspects, part of that work being reviewed
here. The ERG techniques where constructed with P. Chauve and G. Schehr
whom I also thank. I am also greatly
indebted to T. Giamarchi and C. Monthus, as well as A. Rosso, W. Krauth, A. Middleton and C. Marchetti
for numerous discussions and collaborations on related topics.

\appendix

\section{finite temperature beta function to three loop in any dimension}
\label{app:3loop}

Using the notations of the text one finds the following corrections
to the disorder to three loop:
\begin{eqnarray}
&& \delta^{(3)} R_0 =
\frac{1}{6} T^3 J_1 R_0^{(6)} + T^2 [
(\frac{1}{2} J_1^2 J_2 + \frac{7}{24} J_4) (R_0'''')^2
- (J_1^2 J_2 + \frac{1}{12} J_4) R_0''''(0) R_0''''
\\
&& + J_1 J_3 R_0''' R_0^{(5)}
+ \frac{1}{2} J_1^2 J_2 ( (R_0'' - R_0''(0)) R_0^{(6)} - R_0'' R_0^{(6)}(0)) ]
\nonumber \\
&& +
T [(2 I_r + I_A J_1 ) (R_0''')^2 R_0''''
- (\frac{1}{2} I_r + I_A J_1 ) (R_0''')^2 R_0''''(0)
- J_1 J_2^2 (R_0''-R_0''(0)) R_0''''(0) R_0''''
- I_{2A} R_0''''(0)^2 R_0''
\nonumber \\
&& + (J_1 J_2^2 + I_{2A})  (R_0''-R_0''(0)) (R_0'''')^2
+ (2 I_A J_1 + J_2 J_3) (R_0''-R_0''(0)) R_0''' R_0^{(5)}
+ \frac{1}{2} J_1 J_2^2 (R_0''-R_0''(0))^2 R_0^{(6)} ] \nonumber \\
&& +
\frac{1}{2} (I_i+I_m ) (R_0''')^4 +
(4 I_l + I_o + I_A J_2) (R_0''')^2 R_0'''' (R_0''-R_0''(0)) \nonumber \\
&& +
( I_j + \frac{1}{2} J_2^3) (R_0'''')^2 (R_0''-R_0''(0))^2
+ 2 I_A J_2 (R_0''-R_0''(0))^2 R_0''' R_0^{(5)}
- \frac{1}{6} J_2^3 (R_0''-R_0''(0))^3 R_0^{(6)} \nonumber
\end{eqnarray}
where some integrals are defined in (\ref{int1}),
the others are:
\begin{eqnarray}
&& I_m = \int_{1234} g_{13}^2 g_{14} g_{23} g_{24}^2  \quad , \quad
I_i = \int_{1234} g_{12} g_{13} g_{14} g_{23} g_{24} g_{34} \nonumber \\
&& I_l = \int_{1234} g_{12} g_{13}^2  g_{14} g_{23} g_{24}  \quad , \quad
I_o = \int_{1234} g_{12}^2 g_{13}^2 g_{24} g_{34} \nonumber \\
&& I_j = \int_{1234} g_{12}^2 g_{13} g_{14} g_{23} g_{24} \quad , \quad
I_k = \int_{1234} g_{12}^3 g_{13} g_{34} g_{24} \nonumber \\
&& I_r = \int_{1234} g_{12}^2 g_{13}^2 g_{23} \quad , \quad
I_{2A} = \int_{1234} g_{12}^3 g_{13} g_{23} \nonumber
\end{eqnarray}
where $\int_{1234}$ denotes the real space integral over the four points $\frac{1}{\Omega}
\int_{x_1,x_2,x_3,x_4}$, $\Omega$ the volume of the system. The corresponding diagrams are represented in Ref.
(one and two loop) and \cite{threeloop} (three loop at $T=0$). The last line contains finite $T$ integrals.
Denoting $\partial=- m \partial_m$, and $R=m^\epsilon \tilde R$, one obtains, via the procedure described in the
text, the beta function:
\begin{eqnarray}
&& \partial \tilde R = \epsilon \tilde R + \partial J_1 T \tilde R''
+ m^{\epsilon} \partial J_2 [ \frac{1}{2} (\tilde R'')^2
- \tilde R''(0) \tilde R'' ] \\
&& + m^{\epsilon} [\frac{1}{2} \partial J_3 - J_2 \partial J_1] T (\tilde R''')^2
- m^{\epsilon}  J_2 \partial J_1 T \tilde R''''(0) \tilde R''
+ m^{2 \epsilon} [ \partial  I_A  - J_2 \partial J_2 ]
(\tilde R''')^2 (\tilde R'' - \tilde R''(0)) \nonumber \\
&& + m^{\epsilon}
[\frac{7}{24} \partial J_4 - J_3 \partial J_1] T^2  (\tilde R'''')^2
- \frac{1}{12}  m^{\epsilon} \partial J_4 T^2 \tilde R''''(0) \tilde R''''
\nonumber \\
&& + m^{2 \epsilon} [ - \frac{1}{2} \partial I_r + (J_2^2 - I_A) \partial J_1 ]
T \tilde R''''(0) (\tilde R''')^2
+
m^{2 \epsilon} [ 2 \partial I_r + (5 J_2^2 - 4 I_A) \partial J_1
- 3 J_2 \partial J_3 ] \tilde T  (\tilde R''')^2 \tilde R''''
\nonumber \\
&& + m^{2 \epsilon}  [\partial I_{2A} - 2 I_A \partial J_1 - J_3 \partial J_2]
 T (\tilde R'' - \tilde R''(0)) (\tilde R'''')^2
+ m^{2 \epsilon} [- \partial I_{2A} + 3 J_2^2  \partial J_1 + J_3 \partial J_2]
T \tilde R''''(0)^2 \tilde R'' \nonumber \\
&& +
m^{3 \epsilon}
[\partial I_j - 2 I_A \partial J_2 ]
(\tilde R''- \tilde R''(0))^2 (\tilde R'''')^2
+ \frac{1}{2}
m^{3 \epsilon} [\partial I_i + \partial I_m - 2 J_2 \partial I_A +
 J_2^2 \partial J_2 ] (\tilde R''')^4
\nonumber \\
&& + m^{3 \epsilon} [ 4 \partial I_l + \partial I_o - 6 J_2 \partial I_A - 4 I_A \partial J_2 +
5 J_2^2 \partial J_2 ]
(\tilde R''- \tilde R''(0)) (\tilde R''')^2 \tilde R'''' \nonumber
\end{eqnarray}
and the flow of the second derivative is now:
\begin{eqnarray}
&& \partial \tilde R''(0) = (\epsilon - 2 \zeta) \tilde R''(0)
+ 2 \tilde J_2 \tilde T \tilde R''''(0)
+ 6 [\tilde I_A  - \tilde J_2^2] \tilde T (\tilde R''''(0))^2
+ m^{\epsilon} [\frac{1}{2} \partial J_4 - 2 J_3 \partial J_1] T^2
 \tilde R''''(0) \tilde R^{(6)}(0) \\
&& + 3 m^{ 2 \epsilon} [ \partial I_r - 4 I_A \partial J_1 + 5 J_2^2 \partial J_1
- 2 J_2 \partial J_3 ]
T (\tilde R''''(0))^3 \nonumber
\end{eqnarray}

One can now use special relations such as: $\partial J_1 = 2 m^2 J_2$,
$\partial J_3 = 6 m^2 I_A$, $\partial J_4 = 8 m^2 I_{2A}$,
$\partial I_{2A} = 2 m^2 (2 I_k + 3 I_j)$,
$\partial I_{r} = 2 m^2 (4 I_l +  I_o)$
or scaling relations
$m^{\epsilon} \partial J_2 = \epsilon \tilde J_2$,
$m^{2 \epsilon} \partial I_A = 2 \epsilon \tilde I_A$,
$m^{3 \epsilon} \partial I_l = 3 \epsilon \tilde I_l$ and
similar for any of the zero temperature three loop
integrals (the tilde denotes scaled integrals with $m=1$).
For $d<2$ one has $J_n = m^{2 n - 4 - (n-1) \epsilon} \tilde J_n$
and thus $\partial J_n = ((n-1) \epsilon + 4 - 2 n) J_n$.
This yields, upon introducing further rescaling of $u$ by $m^{-\zeta}$
and $R$ by $m^{4 \zeta}$, and
defining here $\tilde T=m^{d-2+ 2 \zeta} T$:
\begin{eqnarray}
&& \partial \tilde R = (\epsilon - 4 \zeta) \tilde R + \zeta u \tilde R'
+ 2 \tilde J_2 \tilde T \tilde R''
+ \epsilon \tilde J_2 [ \frac{1}{2} (\tilde R'')^2
- \tilde R''(0) \tilde R'' ] \\
&& + [3  \tilde I_A  - 2 \tilde J_2^2] \tilde T (\tilde R''')^2
- 2 \tilde J_2^2 \tilde T \tilde R''''(0) \tilde R''
+ \epsilon [  2 \tilde I_A  - \tilde J_2^2  ]
(\tilde R''')^2 (\tilde R'' - \tilde R''(0)) \nonumber \\
&& +
[\frac{7}{3} \tilde I_{2A} - 2 \tilde J_2 \tilde J_3] \tilde T^2  (\tilde R'''')^2
- \frac{2}{3} \tilde I_{2A} \tilde T^2 \tilde R''''(0) \tilde R''''
\nonumber \\
&& + [- 4 \tilde I_l -  \tilde I_o + 2 \tilde J_2^3
- 2 \tilde J_2 \tilde I_A ]
\tilde T \tilde R''''(0) (\tilde R''')^2
+
[ 16 \tilde I_l + 4 \tilde I_o + 10 \tilde J_2^3
- 26 \tilde J_2 \tilde I_A ] \tilde T  (\tilde R''')^2 \tilde R''''
\nonumber \\
&& + [ 4 \tilde I_k + 6 \tilde I_j - 4 \tilde I_A
\tilde J_2 - \epsilon \tilde J_2 \tilde J_3 ]
\tilde T (\tilde R'' - \tilde R''(0)) (\tilde R'''')^2
[- 4 \tilde I_k - 6 \tilde I_j
+ 6 \tilde J_2^3  + \epsilon \tilde J_2 \tilde J_3 ]
\tilde T \tilde R''''(0)^2 \tilde R'' \nonumber \\
&& +
\epsilon [3 \tilde I_j - 2 \tilde I_A \tilde J_2 ]
(\tilde R''- \tilde R''(0))^2 (\tilde R'''')^2
+ \frac{1}{2}
\epsilon [3 \tilde I_i + 3 \tilde I_m - 4 \tilde J_2 \tilde I_A +
 \tilde J_2^3  ] (\tilde R''')^4
\nonumber \\
&& + \epsilon [ 12 \tilde I_l + 3 \tilde I_o - 16 \tilde J_2 \tilde I_A +
5 \tilde J_2^3 ]
(\tilde R''- \tilde R''(0)) (\tilde R''')^2 \tilde R'''' \label{betagen}
\end{eqnarray}
One must be careful that the rescaled integrals $\tilde I_{2A}$,
$\tilde J_3$ are still functions of $\Lambda/m$ for
$d>10/3$ and $d>8/3$, respectively, while all
other integrals have a well defined UV limit.
All combinations of rescaled integrals
which appear as coefficients of the $T=0$ terms
are finite when multiplied by $\epsilon$, which
is a consequence of (formal) renormalizability
of the $T=0$ theory. The combinations entering the $T=0$
part of the three loop term is found to be identical to the
one obtained in Ref. \cite{threeloop}, where these integrals
are computed. Here we have in addition non zero temperature
terms up to three loop. In $d=0$, upon the rescaling
$\tilde T \to \tilde T/2$, $\tilde R \to \tilde R/4$
and setting all rescaled integrals to unity,
this equation yields the results given in the text.

The flow of the second derivative becomes, for $d<2$:
\begin{eqnarray}
&& \partial \tilde R''(0) = (\epsilon - 2 \zeta) \tilde R''(0) + 2 \tilde J_2 \tilde T \tilde R''''(0)
+ 6 [\tilde I_A  - \tilde J_2^2] \tilde T (\tilde R''''(0))^2
+ [(\frac{3}{2} \epsilon - 2) \tilde J_4 - 4 \tilde J_3 \tilde J_2] \tilde T^2
 \tilde R''''(0) \tilde R^{(6)}(0) \nonumber  \\
&& + 3 [ 8 \tilde I_l + 2 \tilde I_o  - 8 \tilde I_A \tilde J_2 + 10 \tilde J_2^3
+ 4 (1 - \epsilon) \tilde J_2 \tilde J_3 ] \tilde T
(\tilde R''''(0))^3 \label{newsecder}
\end{eqnarray}
Remarkably all but the first two terms vanish in
$d=0$, when all rescaled integrals are unity. As discussed in the text this is a consequence of an
exact identity which is local only in $d=0$. It holds also in $d>0$, see (\ref{identityd}), but involves also non local contributions of the $R[v]$ functional, producing the extra terms above. 

One can then analyze this beta function (\ref{betagen}) as in the text. The one
loop equation (first line) is identical (up to rescaling) to
the $d=0$ case and the analysis is identical, with a
well defined $\epsilon$ expansion. To one loop
one has a well defined TBL as for $d=0$, namely:
\begin{eqnarray}
&& 2 \tilde J_2 \tilde T \tilde R''''(0) \to - (\epsilon- 2 \zeta)
\tilde R''(0) = \epsilon \tilde J_2 \tilde R'''(0^+)^2
\end{eqnarray}
Now, however, one is not guaranteed that the TBL holds
without modification beyond one loop. Indeed inserting
TBL scaling in (\ref{newsecder}) yields additional
power of $1/\tilde T$ for each additional loop. This is
not necessarily a failure of TBL scaling, as we know
already from the analysis in $d=0$ that in the TBL
this loop expansion fails. If TBL still holds, resummations
of all loops higher than one in (\ref{newsecder}) should yield a
result $O(1)$. Assuming that this is the case one
can again examine whether (\ref{betagen}) is finite
in the outer region $u=O(1)$. Now trouble already
starts at two loop (second line). In the outer
region $u=O(1)$ the term $\tilde T (\tilde R''')^2$
flows to zero and the other term has a finite limit if one uses the one loop
result $2 \tilde J_2 \tilde T \tilde R''''(0) \to - (\epsilon- 2 \zeta)
\tilde R''(0) = \epsilon \tilde J_2 \tilde R'''(0^+)^2$, namely:
\begin{eqnarray}
&& - 2 \tilde J_2^2 \tilde T \tilde R''''(0) \tilde R''
\to -  \epsilon \tilde J_2^2 \tilde R'''(0^+)^2 \tilde R'' \label{twoloopd}
\end{eqnarray}
Unfortunately this limit cancels the supercusp
from the $T=0$ two loop term only for $d=0$. Worse,
close to $d=4$, $\epsilon [  2 \tilde I_A  - J_2^2  ]$
has a finite limit, while the coefficient of (\ref{twoloopd})
diverges as $1/\epsilon$.

Let us close by indicating the calculation of the correlation function to two loop at non zero temperature. We obtain:
\begin{eqnarray}
&& \langle  u_a^q u_b^{-q} \rangle
= \frac{1}{(q^2 + m^2)^2} [R_0''(0) + (\delta^{(1)} R_0)''(0) + (\delta^{(2)} R_0)''(0)
+ T R_0''''(0)^2 [J_3(q) - J_3(0)] \nonumber \\
&& = \frac{1}{(q^2 + m^2)^2} [m^\epsilon \tilde R''(0)
+ T m^{2 \epsilon} [J_3(q) - J_3(0)] \tilde R''''(0)^2 ]
\end{eqnarray}
where $\tilde R''(0)$ contains all $q=0$ contributions. The lowest non trivial $q$ dependent
diagram thus occurs at two loop and contained
in $J_3$.

\section{resummation in temperature at order $R^2$ in $d=0$}
\label{app:resumm}

In this Appendix we derive in $d=0$ the $\beta$-function resumming all
temperature loops to a fixed order $R^2$. We discuss
whether this $\beta$-function can be used at low temperature in the TBL, or
in the outer region. We show how it can be used in a high temperature expansion.

\subsection{low temperature analysis} 

To derive the FRG equation to order $R^2$ one can either truncate the
equations for the connected W-moments:
\begin{eqnarray}
&& - m \partial_m R(u) = 2 T m^{-2} R''(u) + 2 m^{-2} \hat S_{110}(0,0,u) \\
&& - m \partial_m \hat S(u_{123}) = T m^{-2}
(\partial_1^2 + \partial_2^2 + \partial_3^2)
\hat S(u_{123}) + 2 m^{-2} (R'(u_{12}) R'(u_{13}) + R'(u_{21}) R'(u_{23})
+ R'(u_{31}) R'(u_{32}) \nonumber \\
&& \label{truncW}
\end{eqnarray}
which amounts to set $\hat Q$ which is formally $O(R^3)$ to zero in (\ref{Wcum3}). Equivalently one can 
set $Q$ to zero in (\ref{Gcum3}) and obtain for the $\Gamma$-moments:
\begin{eqnarray}
&& - m \partial_m R(u) = 2 T m^{-2}  R''(u)
+ 2 m^{-4} (R''(u)^2 - 2 R''(0) R''(u))  + 2 m^{-2} S_{110}(0,0,u) \\
&& - m \partial_m  S(u_{123}) = T m^{-2}
(\partial_1^2 + \partial_2^2 + \partial_3^2)
S(u_{123}) + 2 m^{-4}  T (R''(u_{12}) R''(u_{13})
+ R''(u_{21}) R''(u_{23}) + R''(u_{31}) R''(u_{32})) \nonumber \\
&& \label{truncG}
\end{eqnarray}
This can be solved as an expansion in powers of $m^{-2}$, through the recursion:
\begin{eqnarray}
&& \hat S(u_{123}) = \sum_{n=1}^\infty m^{-2 n} s_n(u_{123}) = m^{-2} s_1(u_{123})
+ S(u_{123})  \\
&& s_1(u_{123}) = R'(u_{12}) R'(u_{13}) + R'(u_{21}) R'(u_{23}) + R'(u_{31}) R'(u_{32}) \\
&& 2 (n+1) T s_{n+1}(u_{123}) =
T (\partial_1^2 + \partial_2^2 + \partial_3^2) s_{n}(u_{123})
+ m^3 \partial_m s_{n}(u_{123})
\end{eqnarray}
where the evaluation of $m^3 \partial_m s_{n}$ simply amounts to use only the FRG to
linear order $m^3 \partial_m R = - 2 T R''$. The solution is:
\begin{eqnarray}
&& s_n(u_{123}) = \frac{3}{n!}
T^{n-1} \text{sym}_{123} R^{(n)}(u_{12}) R^{(n)}(u_{13}) \label{res3cum}
\end{eqnarray}
Using that $3 \partial_{u_1} \partial_{u_2} \text{sym}_{123} g(u_{12}) g(u_{13})|_{u_1=u_2=u,u_3=0}
= g'(-u)^2 - 2 g'(0) g'(u) - 2 g''(0) g(u)$ one finds the $\beta$ function:
\begin{eqnarray}
&& - m \partial_m R(u) = 2 T m^{-2} R''(u)
+ 2 \sum_{n=1}^\infty \frac{m^{-2 (n+1)}}{n!} T^{n-1}
[ R^{(n+1)}(u)^2 - 2 R^{(n+1)}(0) R^{(n+1)}(u) - 2 R^{(n+2)}(0) R^{(n)}(u) ]  \nonumber \\
&&
\end{eqnarray}
Since odd derivatives at zero of $R$ (supposed analytic) vanish,
for each $n$ only one of the last two terms is non zero. In rescaled
variables:
\begin{eqnarray}
&& - m \partial_m \tilde R(u) = (4 - 4 \zeta) \tilde R(u) + \zeta u \tilde R'(u) +
\tilde T  \tilde R''(u) \label{betaT}
\\
&& + \sum_{n=1}^\infty \frac{\tilde T^{n-1}}{2^n n!}
[ \tilde R^{(n+1)}(u)^2 - 2 \tilde R^{(n+1)}(0) \tilde R^{(n+1)}(u)
- 2 \tilde R^{(n+2)}(0) \tilde R^{(n)}(u) ]  \nonumber
\end{eqnarray}
As discussed in the text, this equation does not make sense for $u$ in the TBL,
since each new term is more divergent. Going to Fourier representation
one can formally resum it. However one checks that it does not lead to a meaningful
resummation which respects TBL scaling. TBL scaling for $R(u)$ should hold
since the first equation in (\ref{truncG}) implies (\ref{exactrel}) since
any third cumulant can start only at order $u^6$. A successful resummation would mean that
(\ref{truncG}) discarding the left hand side has a solution.
This does not appear to be the case.
In other words one cannot, in the full TBL equations, truncate as
was done here, i.e. discard the term $\hat Q$. This can
be seen from (\ref{truncW}), since there the feeding $R^2$
term is $\sim T^4$ it cannot balance the $T \hat S'' \sim T^3$ term (we recall that $R' \sim T^2$ and $\hat S'' \sim S'' \sim T^2$ in the TBL and that all $\bar S^{n} \sim \hat S^{n} \sim S^{n} \sim T^{1+n}$ in the TBL). 
It seems that {\it any} truncation at any order where
a connected cumulant will be set to zero will violate TBL scaling.
The reason is that in the TBL these connected cumulant are always dominant
with respect to disconnected parts.

Next one can check whether the above $\beta$-function (\ref{betaT}) makes sense
in the outer region $u=O(1)$. Since $\tilde R^{2p}(0) \sim \tilde T^{3-2p}$,
the term $\tilde T^{n-1} \tilde R^{(n+2)}(0) \sim r^{(n+2)}(0)$
has a well defined limit, while for $n > 1$ one can set the other
one $\tilde T^{n-1} \tilde R^{(n+1)}(0)$ to zero, as well as
the first term. Thus the $\beta$-function has a good limit.
Unfortunately this limit is incorrect. We can trace it to
an inconsistency of the present truncation with the partial boundary layer (PBL) scaling when two out of three arguments are brought close together. The PBL scaling (\ref{pbl21}) involves a
function $\phi$, which yields the terms beyond one
loop in the above $\beta$-function (\ref{betaT}), and
a PBL scaling function $s^{(21)}(\tilde u_{12},u_{13})$
which can grow at most as $s^{(21)} \sim \tilde u_{12}^3$
at large $u_{12}$ (otherwise, e.g. $\tilde T^3 s^{(21)} \sim
\tilde T^3 \tilde u_{12}^4$ would diverge as $\tilde T \to 0$).
Let us write the equation in the PBL, Eq. (175) of
Ref. \cite{BalentsLeDoussal2005}, keeping only the terms not discarded here. It reads:
\begin{eqnarray}
&& \tilde u_{12}^2 [ (2 \epsilon -2 - 4 \zeta) \partial \phi(u_{13})
+ \zeta u \phi'(u_{13}) ] =
{\sf R}''(u_{13}) r''(\tilde u_{12}) + s_{20}^{(21)}(\tilde u_{12},u_{13})
\end{eqnarray}
It implies that $s_{20}^{(21)}(\tilde u_{12},u_{13}) \sim \tilde u_{12}^2$
at large $\tilde u_{12}^2$, hence $s^{(21)} \sim \tilde u_{12}^4$, and violates finiteness of PBL.
Indeed the above solution (\ref{res3cum}) contains terms such as:
\begin{eqnarray}
&& \tilde S(u_{123}) \sim T^2 r''(\tilde u_{12}) {\sf R}''(u_{13}) + ..
\end{eqnarray}
which violate the TBL assumption that the only $T^2$ terms must be
of the form exactly $\tilde u_{12}^2$. A correct analysis of
the terms discarded shows that the present
truncation implies $\phi_{out}(u_{13})=0$ in the outer region.
Matching then implies that $\phi(u_{13})=0$, i.e. all
contributions found here are exactly cancelled in the PBL by
terms formally of higher order in $R$.

\subsection{high temperature expansion and logarithmic disorder}
\label{sec:log} 

The beta function (\ref{betaT}) is useful however in the opposite limit of
high temperature $\tilde T$. One example is the marginal case, $\theta=0$,
where $\tilde T=2 T$ does not flow. In $d=0$ this is achieved for
a bare disorder with logarithmic correlations:
\begin{eqnarray}
&& R_0(u) \sim - \sigma \ln |u| \\
&& R(u) \sim - \sigma \ln |u|
\end{eqnarray}
the renormalized disorder having the same large $u$ behaviour.
Logarithmic disorder correspond to $\zeta=\epsilon/4$ in any $d$, with $\theta=2-\frac{\epsilon}{2}$, hence
temperature is marginal only in $d=0$. The one loop ERG equation
for logarithmic disorder reads \footnote{this equation is
valid in any dimension with $\hat \sigma= \epsilon A_d \sigma$
with $A_d= \epsilon \int_q (q^2+1) = \epsilon (4 \pi)^{-d/2} \Gamma[2-(d/2)]$.}:
\begin{eqnarray}
&& \frac{\epsilon}{4} u  \tilde R'(u)
+ \frac{1}{2} \tilde R''(u)^2  - \tilde R''(0) \tilde R''(u)
+ \tilde T \tilde R''(u) = - \frac{\hat \sigma}{4}
\end{eqnarray}
i.e. it has a solution $\tilde R(u) \sim - (\hat \sigma/\epsilon) \ln |u|$ at large $u$, with $\hat \sigma=16 \sigma$.
The last term arises because $R(u) \sim - \sigma \ln u
+ \frac{\epsilon}{4} \sigma \ln m$ i.e., only $R(u) - R(0)$ has a $m$-independent limit. This fixed point equation for $\tilde R'(u)$ is quite simple to analyze, but does not seem to have a closed form solution. The second derivative at zero, however, is obvious to obtain as $- \tilde R''(0)= \sqrt{\frac{\hat \sigma}{2} + \tilde T^2}- \tilde T$, from which one knows $\overline{\langle u^2 \rangle}=- R''(0)/m^4 = - \tilde R''(0)/(4 m^2)$ in $d=0$ to a one loop approximation. 

To go beyond the one loop approximation in $d=0$ one notes that
a systematic high $T$ expansion of the ERG hierarchy can be constructed. One
easily finds that:
\begin{eqnarray}
&& \tilde R(u) = r(\frac{u}{\sqrt{\tilde T}}) + \frac{1}{\tilde T^2} r_2(
\frac{u}{\sqrt{\tilde T}})
+ .. \\
&& \tilde S(u_{123}) = \frac{1}{\tilde T} s_1(\frac{u_{123}}{\sqrt{\tilde T}})
\end{eqnarray}
with $\hat Q \sim Q \sim \tilde Q \sim \frac{1}{\tilde T^2}$ and
so on. Note that $\bar Q = 3 [R R] + \hat Q$ remains of order one, only its
connected part is subdominant in $1/\tilde T$. This extends to
all cumulants. To dominant order (with $\epsilon=4$) the equation and its solution are:
\begin{eqnarray}
&&  x r' + r'' = - 4 \sigma \\
&& r'(x) = - 4 \sigma e^{- x^2/2} \int_0^x dy e^{y^2/2} 
\end{eqnarray}
and one recovers the large $\tilde T$ limit for $\overline{\langle u^2 \rangle}= \sigma/( \tilde T m^2)$.

To next order one can check that all terms in (\ref{betaT})
are of the same order $1/\tilde T^2$, hence we obtain the equation which determine $r_2$:
\begin{eqnarray}
&& 0 = x r_2' + r_2'' +
 \sum_{n=1}^\infty \frac{1}{2^n n!} r_1^{(n+1)}(x)^2
- 2 r_1^{(n+1)}(0) r_1^{(n+1)}(u)
- 2 r_1^{(n+2)}(0) r_1^{(n)}(u)
\end{eqnarray}
We will not give the solution here, but it clearly can be done. It shows that the expansion in power of
$\tilde R$ at $T>0$ is more suited as a high temperature expansion, and that it fails at low temperature.

\section{higher correlations in $d=0$}

\label{app:highercorrelations}

Here we examine the polynomial expansion to sixth order. In
particular we check explicitly that $\hat R^{(6)}(0)=R^{(6)}(0)$.
Since six point correlations have been examined in detail in
Appendix C.2. of \cite{BalentsLeDoussal2005} we only give material
not presented there. The six point connected correlation reads:
\begin{eqnarray}
&&  G_{abcdef} = \langle u_a u_b u_c u_d u_e u_f \rangle - 15 [
\langle u_a u_b \rangle \langle u_c u_d u_e u_f \rangle ] + 30
\langle u_a u_b \rangle \langle u_c u_d \rangle \langle u_e u_f
\rangle \\
&& = \langle u_a u_b u_c u_d u_e u_f \rangle - 15 [ \langle u_a
u_b \rangle \langle u_c u_d u_e u_f \rangle_c ] - 15 [ \langle u_a
u_b \rangle \langle u_c u_d \rangle \langle u_e u_f \rangle ]
\end{eqnarray}
where $[..]$ denote full symmetrization upon the $p$ indices. It
is related to the $\Gamma$ vertices through:
\begin{eqnarray}
 && \Gamma_{abcdef} = - (\frac{m^2}{T})^6 G_{abcdef} + 10
(\frac{m^2}{T})^7 [ G_{abch} G_{defh} ] \\
&& G_{abcdef} = - (\frac{T}{m^2})^6 \Gamma_{abcdef} + 10
(\frac{T}{m^2})^7 [ \Gamma_{abch} \Gamma_{defh} ]
\end{eqnarray}
since Legendre transform is involutive $\Gamma$ is also made of tree
graphs of $W$, and there is a symmetry between these formulae. In
principle there is a $G^{-1}_{ab}$ on each external leg, but thanks
to the above STS property, only the Kronecker delta part remains
($R''(0)$ disappears). In Appendix C.2. of
\cite{BalentsLeDoussal2005} the following parameterization of the
$\Gamma$ vertex was used:
\begin{eqnarray}
&& \Gamma_{abcdef}= - \frac{1}{T^6} m_6 + \frac{6!}{4! 4  T^4} q_6
(8 [\delta_{abc}] - 6 [\delta_{ab} \delta_{cd}] ) - \frac{6!}{3! 8
T^3} s_6 ( [\delta_{abcd}] - 2 [\delta_{abc} \delta_{de}] +
[\delta_{ab} \delta_{cd} \delta_{ef}] \\
&& - \frac{1}{ T^2} R^{(6)}(0) (- (\delta_{abcde}+ 5 \text{perm}) +
(\delta_{abcd} \delta_{ef} + 14 \text{perm}) - (\delta_{abc}
\delta_{def} + 9 \text{perm}) ] )
\end{eqnarray}
where $s_6$, $q_6$ and $m_6$ are derivatives at zero of
respectively third, fourth and sixth cumulants given there. In all
above formula replica indices are arbitrary (they can be equal).
This form is imposed by STS since there are $11$ a priori distinct
elements $\Gamma_{aaaaaa},\Gamma_{aaaaab},..$ (distinct indices)
etc.. (same for the $G$) and the general STS relation $\sum_h
\Gamma_{abcdeh}=0$ (arbitrary indices) implies $7$ relations. It
thus remains $4$ independent variables at order $u^6$.

In Appendix C.2. of \cite{BalentsLeDoussal2005} all eleven distinct
connected correlations where given as a function of these parameters
(formula C15-C25 in the condmat version). We have checked that these
relations are correct. Taking linear combinations of those one can
hope to relate $R^{(6)}(0)$ to some simple observable. This is not
that simple, and in particular (see below) it involves correlations
with more that two distinct replica indices. Relation if any is
indirect \footnote{The quantity $s_6$ was found to be related to
third moment of susceptibility. Because of vanishing of $D_3$ to
leading order it means that in effect $r_6$ is related to third
moment of susceptibility.}

We now follow the route of the present paper and first compute
$\hat R^{(6)}(0)$ which does have a simple physical meaning from
the renormalized potential. Formula (\ref{forced0}) yields,
introducing two sets of replicas with $n_1$ and $n_2$ separately
going to zero:
\begin{eqnarray}
&& - m^{-4} \tilde R''(v_1-v_2) = \langle (u_{a_1} - v_1)
(u_{a_2}-v_2) e^{\frac{m^2}{T} ( v_1 \sum_{\alpha_1} u_{\alpha_1} +
v_2 \sum_{\alpha_2} u_{\alpha_2}  ) } \rangle
\end{eqnarray}
where $\langle .. \rangle$ is the standard replica average.
Indices $\alpha_1$ are always distinct from indices $\alpha_2$
(and $a_i$ belongs to $\alpha_i$). The above expression is clearly
a function of $v_1-v_2$ as can be verified by performing the shift
$u_c \to u_c + v_2$. Setting $v_2=0$ and $v_1=v$ and Taylor
expanding one finds:
\begin{eqnarray}
&& - m^{-4} \hat R''(v) = \langle u_{a_1} u_{a_2} \rangle +
\sum_{p=2}^\infty \frac{m^{2p} v^p}{T^p p!}  \langle u_{a_1}
(\sum_{\alpha_1} u_{\alpha_1} )^p u_{a_2} \rangle
\label{derivatives}
\end{eqnarray}
where a term proportional to $n_1$ has been dropped. Here one has
to be careful with the constraints. It yields for the sixth
derivative:
\begin{eqnarray}
&& - m^{-4} \hat R^{(6)}(0) = \frac{m^{8}}{T^4}  \langle u_{a_1}
(\sum_{\alpha_1} u_{\alpha_1} )^4 u_{a_2} \rangle
\end{eqnarray}
To perform the calculation the safest method is to add variables one
by one, recursively ($n_1=0$ being implicit everywhere):
\begin{eqnarray}
&& u_a  u_\alpha u_\beta u_\gamma u_\delta = ( u_a^2 - u_a u_b)
u_\beta u_\gamma u_\delta =
 ( u_a^3 - u_a^2 u_b  - 2 u_a^2 u_b + 2 u_a u_b u_c) u_\gamma
 u_\delta \\
 && = ( u_a^3 - 3 u_a^2 u_b  + 2 u_a u_b u_c) u_\gamma
 u_\delta
\end{eqnarray}
and so on.. the final result is:
\begin{eqnarray}
&& - m^{-4} \hat R^{(6)}(0) = \frac{m^{8}}{T^4} ( \langle u_a^5
u_b \rangle - 5 \langle u_a^4 u_b u_c \rangle - 10 \langle u_a^3
u_b^2 u_c \rangle \\
&& + 20 \langle u_a^3 u_b u_c u_d \rangle + 30 \langle u_a^2 u_b^2
u_c u_d \rangle - 60 \langle u_a^2 u_b u_c u_d u_e \rangle + 24
\langle u_a u_b u_c u_d u_e u_f \rangle )
\end{eqnarray}
Now let us give here explicitly the disconnected pieces not given in
Appendix C.2. of \cite{BalentsLeDoussal2005}:
\begin{eqnarray}
&&  \langle u_a^5 u_b \rangle_{disc}= 15 \langle u_a^2 \rangle^2
\langle u_a u_b \rangle + 5 \langle u_a u_b \rangle \langle u_a^4
\rangle_c + 10 \langle u_a^2 \rangle \langle u_a^3 u_b \rangle_c
\\
&& \langle u_a^4 u_b u_c \rangle_{disc}= 12 \langle u_a^2 \rangle
\langle u_a u_b \rangle^2 + 3 \langle u_a^2 \rangle^2 \langle u_a
u_b \rangle + \langle u_a u_b \rangle  \langle u_a^4 \rangle_c + 8
\langle u_a u_b \rangle \langle u_a^3 u_c \rangle_c  + 6 \langle
u_a^2 \rangle \langle u_a^2 u_b u_c \rangle_c \\
&& \langle u_a^3 u_b u_c u_d \rangle_{disc}= 9 \langle u_a^2
\rangle \langle u_a u_b \rangle^2 + 6  \langle u_a u_b \rangle^3 +
3 \langle u_a u_b \rangle  \langle u_a^3 u_b \rangle_c + 9 \langle
u_a u_b \rangle \langle u_a^2 u_b u_c \rangle_c + 3 \langle u_a^2
\rangle \langle u_a u_b u_c u_d \rangle_c \\
&& \langle u_a^2 u_b^2 u_c u_d \rangle_{disc}= \langle u_a^2
\rangle^2 \langle u_a u_b \rangle + 4  \langle u_a^2 \rangle
\langle u_a u_b \rangle^2 + 10  \langle u_a u_b \rangle^3 +
\langle u_a u_b \rangle  \langle u_a^2 u_b^2 \rangle_c + 2 \langle
u_a^2 \rangle \langle u_a^2 u_b u_c \rangle_c \\
&& + 8 \langle u_a u_b \rangle \langle u_a^2 u_b u_c \rangle_c + 4
\langle u_a u_b
\rangle \langle u_a u_b u_c u_d \rangle_c \\
&& \langle u_a^2 u_b u_c u_d u_e \rangle_{disc}= 3 \langle u_a^2
\rangle \langle u_a u_b \rangle^2 + 12  \langle u_a u_b \rangle^3
+ 6 \langle u_a u_b \rangle  \langle u_a^2 u_b u_c \rangle_c + 8
\langle u_a u_b \rangle \langle u_a u_b u_c u_d \rangle_c +
\langle u_a^2
\rangle \langle u_a u_b u_c u_d \rangle_c \nonumber \\
&& \langle u_a u_b u_c u_d u_e u_f \rangle_{disc}=  15  \langle
u_a u_b \rangle^3 + 15  \langle u_a u_b \rangle \langle u_a u_b
u_c u_d \rangle_c \nonumber \\
\end{eqnarray}
Amazingly, a tedious calculation shows that in the above
combination all disconnected parts cancel (as is already the case
for the fourth cumulant). One finally gets:
\begin{eqnarray}
&& - m^{-4} \tilde R^{(6)}(0) = \frac{m^{8}}{T^4} ( \langle u_a^5
u_b \rangle_c - 5 \langle u_a^4 u_b u_c \rangle_c - 10 \langle
u_a^3
u_b^2 u_c \rangle_c \\
&& + 20 \langle u_a^3 u_b u_c u_d \rangle_c + 30 \langle u_a^2
u_b^2 u_c u_d \rangle_c - 60 \langle u_a^2 u_b u_c u_d u_e
\rangle_c + 24 \langle u_a u_b u_c u_d u_e u_f \rangle_c ) \\
&&  = - m^{-4} R^{(6)}(0)
\end{eqnarray}
where the last line results from evaluating the combination using
formula (C15-25) of Appendix C.2. of \cite{BalentsLeDoussal2005}.
All contributions of $m_6$, $s_6$, $q_4$, $R^{(4)}(0)^2$ cancel in
this calculation. This tedious and non trivial (partial) check that
$\hat R=R$ should convince the reader of the amazing efficiency of
the Legendre transform method given in the text. Furthermore formula
(\ref{derivatives}) provides a general method to express any
derivative of $\hat R$ (and thus of $R$) at zero to one linear
combination of correlation functions (others can be generated by
adding all STS identities).

\section{Legendre transform: relations between moments and $\Gamma$-moments}

\label{app:legendre}

We perform the Legendre transform in any dimension, $d=0$ is recovered
setting $g=1/m^2$ and discarding space indices.

We start from the definition of the Legendre transform (\ref{gammaud2}):
\begin{eqnarray}
&& \Gamma[u] =  \frac{1}{T} \sum_{axy} u^a_x g^{-1}_{xy} v^a_y - W[v] \label{leg1} \\
&& v^a_x = T \sum_{y} g_{xy} \frac{\delta \Gamma[u]}{\delta u^a_y}  \label{leg2}
\end{eqnarray}
where the second line is the condition $j_a^x = \frac{\delta \Gamma[u]}{\delta u_x^a}$
together with (\ref{jtov}). Taking the derivative in the $\Gamma$-cumulant
expansion (\ref{gammaud}), (\ref{leg2}) can be rewritten:
\begin{eqnarray}
&& v^a_x =  u^a_x - \frac{1}{T} \sum_{yc} g_{xy}
\frac{\delta R[u^{ac}] }{\delta u^a_y}
- \frac{1}{2 T^2} \sum_{ycd} g_{xy}
\frac{\delta S[u^a,u^c,u^d] }{\delta u^a_y}  -
\frac{1}{6 T^3} \sum_{ycde} g_{xy}
\frac{\delta Q[u^a,u^c,u^d,u^e] }{\delta u^a_y} + \text{4 replica sums} \label{deriv}
\end{eqnarray}
where we have used the symmetry of the cumulants. One can also rewrite
(\ref{leg2}) by rearranging the quadratic term and inserting the
$W$-cumulant expansion:
\begin{eqnarray}
&& \Gamma[u] = \Gamma_0 + \frac{1}{2 T} \sum_{axy} g^{-1}_{xy} u^a_x
u^a_y - \frac{1}{2 T} \sum_{axy} g^{-1}_{xy} (v-u)^a_x (v-u)^a_y
\nonumber
\\
&& - \frac{1}{2 T^2} \sum_{ab} \hat R[v^{ab}]
- \frac{1}{6 T^3} \sum_{abc} \hat S[v^{a},v^{b},v^{c}]
- \frac{1}{4! T^4} \sum_{abcd} \hat Q[v^{a},v^{b},v^{c},v^d]
 + \text{5 replica sums} \label{deriv2}
\end{eqnarray}

We can now insert (\ref{deriv}) into (\ref{deriv2}), perform Taylor expansion
in the cumulants. Remarkably, since in that operation the number of free
replica sums can only increase, it allows to identify simple relations
between cumulants by comparing with:
\begin{eqnarray}
&& \Gamma[u] = \Gamma_0 + \frac{1}{2 T} \sum_{axy} g^{-1}_{xy} u^a_x
u^a_y - \frac{1}{2 T^2} \sum_{ab} R[u^{ab}] - \frac{1}{6 T^3}
\sum_{abc} S[u^{a},u^{b},u^{c}] - \frac{1}{4! T^4} \sum_{abcd}
Q[v^{a},v^{b},v^{c},v^d]
 + \text{5 replica sums} \nonumber
\end{eqnarray}

The calculation yields the formula (\ref{corresp}) in the text as well as:
\begin{eqnarray}
&& Q[u_a,u_b,u_c,u_d] = \hat Q[u_a,u_b,u_c,u_d]
- 12 ~ \text{sym}_{abcd} \sum_{xy} g_{xy} \frac{\delta R[u^{ab}] }{\delta u^a_x}
\frac{\delta S[u_a,u_c,u_d]] }{\delta u^a_y} \\
&& - 6 ~ \text{sym}_{abcd} \sum_{xyzt} \frac{\delta^2 R[u^{ab}]
}{\delta u^a_x \delta u^b_y} g_{xz} (\frac{\delta R[u^{ac}] }{\delta
u^a_z} - \frac{\delta R[u^{bc}] }{\delta u^b_z}) g_{yt}
(\frac{\delta R[u^{ad}] }{\delta u^a_t} - \frac{\delta R[u^{bd}]
}{\delta u^b_t})
\end{eqnarray}
and using (\ref{corresp})
one finds that the same formula holds with $S \to \hat S$ and the $-6$ replaced by
$+6$.

\section{generating functionals and Legendre transform
in general, non STS case}

\label{app:general}

Since many models do not posses an exact STS symmetry we give here a
derivation valid in the general case. The only assumption is that
the expansion in number of replicas is valid. Of course the final
formula is not as nice as in the STS case, and we hence stop at the
level of second cumulant.

One defines as usual the functional $W[j]=\ln Z[j]=\ln
\overline{\prod_{a=1}^p Z_V[j_a]}$ where:
\begin{eqnarray}
Z_V[j] = \int Du e^{- \frac{1}{T} H_V[u] + \int_x j_x u_x} = e^{-
\frac{1}{T} F_V[j]}
\end{eqnarray}
where $F_V[j]$ is the free energy in presence of sources, e.g. an
external force $j_x=-f_x/T$. One writes the general expansion:
\begin{eqnarray}
W[j]=W[0] + \frac{1}{T} \sum_a \hat U[j_a] + \frac{1}{2 T^2}
\sum_{ab} \hat  R[j_a,j_b] + \frac{1}{6 T^3} \sum_{abc} \hat
S[j_a,j_b,j_c] + ..
\end{eqnarray}
in terms of fully symmetric functionals. Note that the number of
replica $p$ is arbitrary here \cite{gauge}.

One shows that:
\begin{eqnarray}
&& - \overline{F_V[j]} = \hat U[j] \\
&& \overline{F_V[j_1] F_V[j_2]}^c = \hat R[j_1,j_2] \\
&& - \overline{F_V[j_1] F_V[j_2] F_V[j_3]}^c = \hat S[j_1,j_2,j_3]
\end{eqnarray}
It is clear from the definition $W[j]=\ln \overline{\exp(-
\frac{1}{T} \sum_a F_V[j_a])} = \sum_{n=1}^\infty \frac{(-1)^n}{n!
T^n} \overline{(\sum_a F_V[j_a])^n}^c$. It can also be checked by
inserting groups of values $j_a=j_i$ for $n_i$ replicas and
expanding in multinomial powers of $n_i$.

These function(al)s can be simply measured by applying a force to
the system. For instance, at $T=0$, denoting $E_f$ the ground state
energy in one realization of disorder in presence of a force one
has:
\begin{eqnarray}
&& - \overline{E_f} = \hat U[f] \\
&& \overline{E_{f_1} E_{f_2}}^c = \hat R[f_1,f_2]
\end{eqnarray}
In some cases, STS is restored at large scales (such a a manifold on
a lattice) and these formula can be used to check that. In that case
it should depend only on $f_1-f_2$. Of course in that case a non STS
local part such as non STS gradient terms may remain, and are
typically irrelevant.

Let us now perform the Lengendre transform $\Gamma[u]+W[j]=\int_x
u^a_x j^a_x$ with $u=W'[j]$. This condition yields:
\begin{eqnarray}
&& u_a^x = \frac{1}{T} \hat U_x'[j_a] + \frac{1}{T^2} \sum_{b} \hat
R_x'[j_a,j_b] + \frac{1}{2 T^3} \sum_{bc} \hat S'_x[j_a,j_b,j_c] +
.. \label{suma}
\end{eqnarray}
where derivatives here are with respect to the first variable. Now
define $j^0[u]$ such that:
\begin{eqnarray}
&& u_x = \frac{1}{T} \hat U_x'[j^0[u]]
\end{eqnarray}
In the absence of the terms involving replica sums in (\ref{suma})
the solution would be $j^a = j^0[u^a]$. In general one writes $j^a =
j^0[u^a] + \delta j^a$ where $\delta j^a$ contains at least one
replica sum. One rewrite (\ref{suma}) as:
\begin{eqnarray}
&& u^a_x = \frac{1}{T} \hat U_x'[j^0[u^a]] + \frac{1}{T} \hat
U_{xy}''[j^0[u^a]] \delta j^a_y + \frac{1}{T^2} \sum_{b} \hat
R_x'[j^a,j^b] + \text{2 replica sums} + .. \label{sumb}
\end{eqnarray}
The two first terms cancel one gets:
\begin{eqnarray}
&& \delta j^a_y = - \frac{1}{T} \hat U''[j^0[u^a]]^{-1}_{yz}
\sum_{b} \hat R_z'[j^0[u^a],j^0[u^b]] + \text{2 replica sums} + ..
\label{sumc}
\end{eqnarray}
We can now express the effective action:
\begin{eqnarray}
&& \Gamma[u]= \sum_a \Gamma_0[u^a] + \sum_a \int_x u^a_x \delta
j^a_x -  \frac{1}{T} \hat U_x'[j^0[u^a]]  \delta j^a_x - \frac{1}{2
T^2} \sum_{a b} \hat R[j^0[u^a],j^0[u^b]] + \text{3 replica
sums} + .. \\
&& \Gamma_0[u] = \int_x u_x j_x^0[u] - \frac{1}{T} \hat U[j^0[u]]
\end{eqnarray}
where the terms $O(\delta j^2)$ yield three sums. A cancellation
occurs from the definition of $j^0$ giving the final result:
\begin{eqnarray}
&& \Gamma[u]= \sum_a \Gamma_0[u^a] - \frac{1}{2 T^2} \sum_{a b} \hat
R[j^0[u^a],j^0[u^b]] + \text{3 replica sums} + ..
\end{eqnarray}
Thus one has:
\begin{eqnarray}
&& R[u^a,u^b] = \hat R[j^0[u^a],j^0[u^b]]
\end{eqnarray}
where $j^0[u]=-f^0[u]/T$ with:
\begin{eqnarray}
&& \overline{ F'_{V x}[f^0[u]]} = u_x
\end{eqnarray}
note that $f_0[u]$ is a non random quantity depending only on the
average energy.

Finally note that if $j^0[0]=0$ then :
\begin{eqnarray}
&& \Gamma_0[0]=\overline{F_V}/T \quad , \quad
R[0,0]=\overline{F_V^2}^c
\end{eqnarray}
and $S[0,0,0]=\hat S[0,0,0]=-\overline{F_V^3}^c$, checked in the STS
case. In general one can write:
\begin{eqnarray}
&& \hat{U}[j] = \frac{T^2}{2} j G j + \hat U_i[j] \\
&& G = - \frac{\delta^2 \overline{F_V[f]}}{\delta f \delta f} \\
&& G = g \quad , \quad \text{for STS}
\end{eqnarray}
with $\hat U_i=0$ in the latter case. Thus in general:
\begin{eqnarray}
&& u_x = v^0_x[u] + \frac{1}{T} \hat U'_i[j^0[u]] \\
&& v^0 = T G j^0
\end{eqnarray}
in the STS case $v^0[u]=u$, $j^0[u]=g^{-1} u/T$.

\section{from $\Gamma$-FRG to $W$-FRG}
\label{app:WtoGERG}

We start with the second cumulant $W$-ERG:
\begin{eqnarray}
&& - m \partial_m R(v) = \frac{2 T}{m^2} R''(v) + \frac{2}{m^2} \hat
S_{110}(0,0,v)
\end{eqnarray}
and substitute using (\ref{relGW}):
\begin{eqnarray}
&& \hat S_{110}(v_{abc}) = S_{110}(v_{abc}) + \frac{1}{m^2} (-
R'''_{ab} (R'_{ac}-R'_{bc}) - R''_{ab} R''_{bc} - R''_{ac}
(R''_{ab}-R''_{cb}) )
\end{eqnarray}
and we denote $R'_{ab}=R(v_{ab})$ etc.. The limit $v_{ab} \to 0$ is
always unambiguous for $T>0$ (and here also unambiguous at $T=0$
provided $R'''$ is bounded which we assume from now on). It yields
the second cumulant $\Gamma$-ERG (\ref{Gcum2}) in the text.

Let us now consider the third cumulant $W$-ERG:
\begin{eqnarray}
 && -m \partial_m \hat S(v_{abc}) = \frac{3 T}{m^2}
[\hat S_{200}(v_{abc})] + \frac{3}{m^2} [\hat
Q_{1100}(v_{aabc})] + \frac{6}{m^2} [R'_{ab} R'_{ac}]
\end{eqnarray}
Let us write, using (\ref{relGW}):
\begin{eqnarray}
 && -m \partial_m S(v_{abc}) = -m \partial_m \hat S(v_{abc}) -
 \frac{3}{m^2} (-m \partial_m) [R'_{ab} R'_{ac}] - \frac{6}{m^2} [R'_{ab}
 R'_{ac}] \\
 && = \frac{3 T}{m^2}
[\hat S_{200}(v_{abc})] -
 \frac{3}{m^2} (-m \partial_m) [R'_{ab} R'_{ac}] + \frac{3}{m^2} [\hat
Q_{1100}(v_{aabc})] \\
&& =  \frac{3 T}{m^2} [S_{200}(v_{abc})] + \frac{9 T}{m^4} [
\partial_a^2 [R'_{ab} R'_{ac}] ]
- \frac{3}{m^2} (-m \partial_m) [R'_{ab} R'_{ac}] + \frac{3}{m^2}
[\hat Q_{1100}(v_{aabc})]
\end{eqnarray}
One has:
\begin{eqnarray}
 && [ \partial_a^2 [R'_{ab} R'_{ac}] ] = \frac{2}{3} [R''_{ab} R''_{ac}] + \frac{4}{3}
 [R'''_{ab} R'_{ac}] \\
 && -m \partial_m [R'_{ab} R'_{ac}] = 2 [ (-m \partial_m) R'_{ab} R'_{ac}]
 = \frac{4 T}{m^2} [R'''_{ab} R'_{ac}] - \frac{4}{m^2}
 [S_{111}(u_{aab})
 R'_{ac}] + \frac{8}{m^2} [(R''_{ab}-R''(0))R'''_{ab} R'_{ac}]
\end{eqnarray}
Thus we obtain:
\begin{eqnarray}
 && -m \partial_m S(v_{abc}) = \frac{3 T}{m^2} [S_{200}(v_{abc})] + \frac{6 T}{m^4}
 [R''_{ab} R''_{ac}] + \frac{12}{m^4}
 [S_{111}(v_{aab}) R'_{ac}] + \frac{3}{m^2} [\hat Q_{1100}(v_{aabc})]
 - \frac{24}{m^4} [(R''_{ab}-R''(0))R'''_{ab} R'_{ac}] \nonumber
\end{eqnarray}
One can check that:
\begin{eqnarray}
 && [\partial_a \partial_b [S_{100}(v_{abc}) R'_{ad}]|_{a=b}]
 = \frac{1}{3} [ R'_{ad} S_{111}(v_{acc}) ] + \frac{1}{3} [ R''_{ac}
 S_{110}(v_{aad})] - \frac{1}{3} [ (R''_{ac}- R''(0)) S_{110}(v_{acd})
 ] \label{contract}
\end{eqnarray}
using that $S_{111}$ is odd and using the STS relations
$S_{111}(u,u,v)+2 S_{210}(u,u,v)=0$ and $S_{200}+S_{110}+S_{101}=0$.
Thus one gets:
\begin{eqnarray}
 && -m \partial_m S(v_{abc}) = \frac{3 T}{m^2} [S_{200}(v_{abc})] + \frac{6 T}{m^4}
 [R''_{ab} R''_{ac}] + \frac{12}{m^4} (
[ R''_{ac} S_{110}(u_{aad})] - [ (R''_{ac}- R''(0))
S_{110}(u_{acd}) ]) \\
&&  + \frac{3}{m^2} [Q_{1100}(v_{aabc})]  - \frac{24}{m^4}
[(R''_{ab}-R''(0))R'''_{ab} R'_{ac}] + \frac{36}{m^4} [\partial_a
\partial_b [ R''_{ac} (R'_{ab}-R'_{cb})R'_{ad} ]|_{a=b}]
\label{inter}
\end{eqnarray}
Performing the last calculation cancellations occur and one gets the
equation (\ref{Gcum3}) in the text. Note that the $R'R'$ part of
$S-\hat S$ as well as the $R' S'$ part of $Q-\hat Q$ can be guessed
from the W-ERG equation, just solving for explicit mass dependence
in the last term (i.e. the non linear term). Finally note that the
above was derived using $R'''(0)=0$, i.e. at $T>0$. There are in
fact several terms formally proportional to $R'''(0)$ both in
(\ref{contract}) and in (\ref{inter}). One finds that these all
cancel by parity if one replaces the contractions $|_{a=b}$ by
$|_{u_a=u_b+\delta}$ and keep $\delta$ infinitesimal until the end
(i.e. the result is unambiguous).

\section{two-well droplet solution for any moment}
\label{app:dropletmoments}

\subsection{third and fourth moments}

We start with the two-well calculation of the third moment:
\begin{eqnarray}
&& \frac{\bar S_{111}(v,z,t)}{m^6} = - \frac{\overline{ \hat V'(v)
\hat V'(t)
\hat V'(z)} }{m^6} \\
&& = \overline { (- T \tilde v + \frac{u_1 X_1 + u_2 w X_2}{X_1 + w
X_2}) (- T \tilde t + \frac{u_1 Y_1 + u_2 w Y_2}{Y_1 + w Y_2})
(- T \tilde z + \frac{u_1 Z_1 + u_2 w Z_2}{Z_1 + w Z_2}) } \\
&& = - T \langle u_1^2 \rangle_P (\tilde v + \tilde t + \tilde z) +
T \langle h_{ss}(X_1,Y_1,Z_1,u_1;X_2,Y_2,Z_2,u_2) \rangle_{u_i} +
O(T^2)
\end{eqnarray}
where we have defined $X_i=e^{m^2 u_i \tilde v}$,$Y_i=e^{m^2 u_i
\tilde t}$, $Z_i=e^{m^2 u_i \tilde z}$ and:
\begin{eqnarray}
&& h(X_1,Y_1,Z_1,u_1;X_2,Y_2,Z_2,u_2) = \int_0^1 \frac{dw}{w}
\frac{u_1 X_1 + u_2 w X_2}{X_1 + w X_2} \frac{u_1 Y_1 + u_2 w
Y_2}{Y_1 + w Y_2} \frac{u_1 Z_1 + u_2 w Z_2}{Z_1 + w Z_2}
\end{eqnarray}
together with the symmetrized expression $h_{ss}$ as in
(\ref{symmetriz}). A tedious calculation then yields:
\begin{eqnarray}
&& h_{ss}(X_1,Y_1,Z_1,u_1;X_2,Y_2,Z_2,u_2) = \frac{1}{2} m^2 u_1
u_2 (u_1-u_2)^2  (\tilde v + \tilde t + \tilde z) \\
&& + \frac{1}{4} m^2 (u_1-u_2)^4 [ \frac{X_2^2 Y_1 Z_1 + X_1^2 Y_2
Z_2}{(X_2 Y_1 - X_1 Y_2) (X_2 Z_1 - X_1 Z_2)} \tilde v \\
&& + \frac{Y_2^2 X_1 Z_1 + Y_1^2 X_2 Z_2}{(X_2 Y_1 - X_1 Y_2) (-Y_2
Z_1 + Y_1 Z_2)} \tilde t + \frac{Z_2^2 X_1 Y_1 + Z_1^2 X_2 Y_2}{
(-Y_2 Z_1 + Y_1 Z_2) (- X_2 Z_1 + X_1 Z_2) } \tilde z ]
\end{eqnarray}
Here we can check that upon the shift $\tilde v \to \tilde v +
a$,$\tilde z \to \tilde z + a$,$\tilde t \to \tilde t + a$, the
change in $h_{ss}$ simplifies drastically into:
\begin{eqnarray}
&& a m^2 ( \frac{3}{2}  u_1 u_2 (u_1-u_2)^2  + \frac{1}{2}
(u_1-u_2)^4 )
\end{eqnarray}
Expanding this yields $a m^2 \langle u_1^4 - u_1^3 u_2
\rangle_{u_i}$ which cancels exactly the change from the first term
$- 3 a \langle u_1^2 \rangle_P$ due to the STS relation
$-P'(u_1)=\int du_2 (u_1-u_2) D(u_1,u_2)$. The final result is thus
invariant under translation. The first term $- T \langle u_1^2
\rangle_P$ can hence be combined with the second and third. Defining
$y=u_1-u_2$, upon a few transformations one obtains the result given
in the text.

The fourth moment can be computed in the same way:
\begin{eqnarray}
&& \frac{\bar Q_{1111}(v,u,z,t)}{m^8} = \frac{\overline{ \tilde
V'(v) \tilde V'(u) \tilde V'(t)
\tilde V'(z)} }{m^8} \\
&& = \overline { (- T \tilde v + \frac{u_1 X_1 + u_2 w X_2}{X_1 + w
X_2}) (- T \tilde u + \frac{u_1 Y_1 + u_2 w Y_2}{Y_1 + w Y_2})
 (- T \tilde z + \frac{u_1 Z_1 + u_2 w Z_2}{Z_1 + w Z_2}) (- T \tilde
 t
 + \frac{u_1 T_1 + u_2 w T_2}{T_1 + w T_2}) } \\
&& = \langle u_1^4 \rangle + T \langle
h_{ss}(X_1,Y_1,Z_1,T_1,u_1;X_2,Y_2,Z_2,T_2,u_2) \rangle_{u_i} +
O(T^2)
\end{eqnarray}
with $X_i=e^{m^2 u_i \tilde v}$,$Y_i=e^{m^2 u_i \tilde u}$,
$Z_i=e^{m^2 u_i \tilde z}$, $T_i=e^{m^2 u_i \tilde t}$. Let us
define:
\begin{eqnarray}
&& h(X_1,Y_1,Z_1,T_1,u_1;X_2,Y_2,Z_2,T_2,u_2) \\
&& = \int_0^1 \frac{dw}{w}( \frac{u_1 X_1 + u_2 w X_2}{X_1 + w X_2}
\frac{u_1 Y_1 + u_2 w Y_2}{Y_1 + w Y_2} \frac{u_1 Z_1 + u_2 w
Z_2}{Z_1 + w Z_2} \frac{u_1 T_1 + u_2 w T_2}{T_1 + w T_2} - u_1^4)
\nonumber
\end{eqnarray}
Implementing the two symmetries a nice mathematica calculation
yields:

\begin{eqnarray}
&& h_{ss}= \frac{1}{4} u_1 u_2 u_{21}^2 [ (-3 + \frac{2 T_2
X_1}{-T_2 X_1 + T_1 X_2} + \frac{2 X_2 Y_1 }{X_2 Y_1 - X_1 Y_2} +
\frac{2  X_2 Z_1 }{X_2 Z_1 - X_1 Z_2 }) \ln(X_1/X_2) + p.c. ] \\
&& + \frac{1}{4} u_{21}^4 [ \frac{X_2^3 Y_1 Z_1 T_1 + X_1^3 Y_2 Z_2
T_2 }{(- T_2 X_1 + T_1 X_2)(X_2 Y_1 - X_1 Y_2)(X_2 Z_1 - X_1 Z_2)}
\ln(X_1/X_2) + p.c. ]
\end{eqnarray}

One gets finally the result given in the text. One can check the STS symmetry (under a common shift of all the variables) noting that $sym_{u,v,z,t} F[v-u,v-t,v-z]=0$, both for circular
permutations, and full permutations, and one also finds that the second term is invariant by the shift of all variables. One finds also
the small argument behavior:
\begin{eqnarray}
m^{-8} (\bar Q_{1111}(v_1,v_2,v_3,v_4) - \langle u_1^4 \rangle_P) = -
\frac{T}{4} \big( 
12 \langle u_1 u_2 y^2 \rangle_y + O( m^2 \langle u_1 u_2 y^4 \rangle_y v_i^2) 
 + \frac{11}{3} \langle y^4 \rangle_y +  O( m^2 \langle
y^6 \rangle_y v_i^2) \big)
\end{eqnarray}
all those functions are always even in $y$, as has been imposed using the symmetries.

\subsection{solution for any moment}

We can write any odd moment as:
\begin{eqnarray}
&& \bar S^{(n)}_{1..1}(v_1,..v_n) = - T m^{2 n} \langle u_1^{n-1} \rangle_P (\sum_{i=1}^n \tilde v_i) + A_n[\tilde v_i]
+ O(T^2)
\end{eqnarray}
and every even one:
\begin{eqnarray}
&& \bar S^{(n)}_{1..1}(v_1,..v_n) = m^{2 n} \langle u_1^{n} \rangle_P + A_n[\tilde v_i] + O(T^2)
\end{eqnarray}
where:
\begin{eqnarray}
&& A_n[\tilde v_i]= m^{2n} \overline{ \prod_{i=1}^n \frac{u_2 w + u_1 a_i}{w+a_i} -
u_1^n } = T \langle h[v_i;a_i,u_1,u_2] \rangle_{u_i} \\
&& h[\tilde v_i;a_i,u_1,u_2] = m^{2n} \int_0^1 \frac{dw}{w} ( \prod_{i=1}^n \frac{u_2 w + u_1 a_i}{w+a_i} -
u_1^n )
\end{eqnarray}
and $a_i=\exp(m^2 y \tilde v_i)$, $y=u_1-u_2$. To simplify this expression it is convenient to first subtract
$u_2^n$ instead of $u_1^n$, then perform the Mathematica command Apart[,w] to extract the poles. Dividing by $w$
and applying again the command Apart[,w], one gets a term $(u_2^n - u_1^n)/w$ which can be discarded if the
initial substraction is $u_1^n$. The final result is remarkably simple:
\begin{eqnarray}
h[\tilde v_i;a_i,u_1,u_2]= m^{2n} \int_0^1 dw \sum_{i=1}^n \frac{u_{21}}{w+a_i} \prod_{j \neq i} \frac{u_2 a_i -
u_1 a_j}{a_i - a_j}
\end{eqnarray}
Now, before integrating we first symmetrize under the first
symmetry $a_i \to 1/a_i$, $u_1 \leftrightarrow u_2$. Noting that
each fraction $\frac{u_2 a_i - u_1 a_j}{a_i - a_j}$ is invariant,
one ends up to compute $\int_0^1 dw [ (w+a_i)^{-1} -
(w+1/a_i)^{-1} ]=- \ln a_i$ thus one gets simply:
\begin{eqnarray}
h_s[\tilde v_i;a_i,u_1,u_2]= \frac{y}{2} m^{2n} \sum_{i=1}^n \ln a_i \prod_{j \neq i} \frac{u_2 a_i - u_1
a_j}{a_i - a_j}
\end{eqnarray}
Hence one gets:
\begin{eqnarray}
A_n[\tilde v_i]= \frac{1}{2} m^{2 n+2} T \langle y^2 \sum_{i=1}^n \tilde v_i \prod_{j \neq i} (u_1 - y
\frac{1}{1 - e^{m^2 y (\tilde v_j-\tilde v_i)}}) \rangle_{u_1,y}
\end{eqnarray}
The last step is the symmetrization $u_i \to - u_i$. It yields:
\begin{eqnarray}
A_n[\tilde v_i]= \frac{1}{4} m^{2n+2} T \langle y^2 \sum_{i=1}^n \tilde v_i [ \prod_{j \neq i} (u_1 - y
\frac{1}{1 - e^{m^2 y (\tilde v_j-\tilde v_i)}}) + (-1)^{n-1} \prod_{j \neq i} (u_1 - y \frac{1}{1 - e^{- m^2 y
(\tilde v_j-\tilde v_i)}}) ] \rangle_{u_1,y}
\end{eqnarray}
The highest order term in $y$ is:
\begin{eqnarray}
&& A^{high}_n[\tilde v_i]= \frac{1}{4} m^{2n+2} T \langle y^{n+1} \sum_{i=1}^n \tilde v_i F[m^2 y
(\tilde v_i-\tilde v_j)] \rangle_{u_1,y} \\
&& F[\{ z_{ij} \}]= (-1)^{n-1} \prod_{j \neq i} \frac{1}{1 - e^{z_{ji}}} +
 \prod_{j \neq i} \frac{1}{1 - e^{z_{ij}}} =  \frac{1 + e^{\sum_j
z_{ij}}}{\prod_{j \neq i} (1-e^{z_{ij}})} \label{defF}
\end{eqnarray}
where $z_{ij}=z_i-z_j$,
a result consistent with the highest order term in $y$ given in the text for the third and fourth moments. To
get the other terms given in the text one must use, for the third moment, $\langle u_1^2 \rangle_P = \frac{m^2}{3}
\langle y u_1^3 \rangle_{y,u_1}$ from the STS identity (\ref{rgdrop2}) which yields $- \langle u_1^2 \rangle_P +
\frac{1}{2} \langle y^2 u_1^2 \rangle_{y,u_1} - \frac{1}{2} \langle y^3 u_1 \rangle_{y,u_1} = - \frac{1}{6}
\langle y^4 \rangle_y$, and, for the fourth moment, the identity $\tilde F(a,b,c)=3 \text{sym}_{abc}
(1-e^{-a})(1-e^{-b})^{-1}-(1-e^{a})(1-e^{b})^{-1}$ where $\tilde F$ is defined in the text.

\subsection{generalization to any $N$}

The above formula can be generalized to any $N$. We drop the tilde subscript. We need:
\begin{eqnarray}
A_n[\vec v_i]_{\alpha_1,..\alpha_n} = m^{2n} \overline{ \prod_{i=1}^n
\frac{u_{2,\alpha_i} w + u_{1,\alpha_i} a_i}{w+a_i} -
\prod_{i=1}^n u_{1,\alpha_i} } = T \langle h[\vec v_i;a_i,\vec u_1,\vec u_2]_{\alpha_1,..\alpha_n} \rangle_{\vec u_i}
\end{eqnarray}
where $a_i=\exp(m^2 \vec y \cdot \vec v_i)$, $\vec y=\vec u_1-\vec u_2$.
Again using Apart one finds:
\begin{eqnarray}
h[v_i;a_i,u_1,u_2]_{\alpha_1,..\alpha_n} =m^{2n} \int_0^1 dw \sum_{i=1}^n \frac{y_{\alpha_i}}{w+a_i}
\prod_{j \neq i} \frac{u_{2,\alpha_j} a_i - u_{1,\alpha_j} a_j}{a_i - a_j}
\end{eqnarray}
Integration and symmetrization yields:
\begin{eqnarray}
h_s[v_i;a_i,u_1,u_2]_{\alpha_1,..\alpha_n} = \frac{1}{2}m^{2n} \sum_{i=1}^n y_{\alpha_i} \ln a_i
\prod_{j \neq i} \frac{u_{2,\alpha_j} a_i - u_{1,\alpha_j} a_j}{a_i - a_j}
\end{eqnarray}
Using the symmetrization $\vec u_i \to - \vec u_i$. It yields:
\begin{eqnarray}
A_n[v_i]_{\alpha_1,..\alpha_n} = \frac{1}{4} m^{2 n+2} T \sum_{i=1}^n \langle y_{\alpha_i}
\vec y \cdot \vec v_i  [
\prod_{j \neq i} (u_{1,\alpha_j} - y_{\alpha_j} \frac{1}{1 - e^{m^2 \vec y \cdot
\vec v_{ji} }}) +
(-1)^{n-1} \prod_{j \neq i} (u_{1,\alpha_j} - y_{\alpha_j} \frac{1}{1 - e^{- m^2
\vec y \cdot
\vec v_{ji}}}) ] \rangle_{\vec u_1,\vec y}
\end{eqnarray}
The highest order term is:
\begin{eqnarray}
&& A_n^{high}[v_i]_{\alpha_1,..\alpha_n} = \frac{1}{4} m^{2 n +2} T \langle y_{\alpha_1} .. y_{\alpha_n}
\sum_{i=1}^n \vec y \cdot \vec v_i F[m^2  \vec y \cdot
\vec v_{ij}] \rangle_{\vec u_1,\vec y} 
\end{eqnarray}
where the function $F$ is defined in (\ref{defF}) and does not depend on $N$. For the second moment one obtains, restoring the temperature in the TBL variable:
\begin{eqnarray}
R''_{\alpha \beta}(\vec v)=R''_{\alpha \beta}(0)+ T m^4 \langle y_\alpha y_\beta F_2(m^2 \vec y \cdot \vec v/T) \rangle_{\vec y}
\end{eqnarray}

\section{Three loop beta function in zero dimension}

\subsection{Three loop at $T>0$ by the $\Gamma$-ERG method}

\label{app:3looperg}

We start from the Eq. (\ref{Gcum4r}) in the text. From this equation we expect
a solution for $\tilde{S}^{(4)}$ of a form similar to (\ref{2loopsolu}), schematically:
\begin{eqnarray}
&& \tilde{S}^{(4)} = \alpha' \tilde T (\tilde  R'' \tilde  R'' \tilde  R'') + \beta' (\tilde  R''
\tilde  R'' \tilde  R'' \tilde  R'')    \label{3loopsolu4}
\end{eqnarray}
To be systematic, note that any feeding term of the form $c \tilde  T^n \tilde  R^p$ in the equation for $\tilde  S^{(q)}$ yields the term $\alpha_{np}^q c \tilde  T^n \tilde  R^p$ in its solution with the explicit form for the coefficients:
\begin{eqnarray}
&& 1/\alpha_{np}^q = 2 n + 2 q - 4 + (p+1-q) \epsilon = 2 n + 4 p
- 2 q
\end{eqnarray}
Hence one has $\alpha'=4 \gamma \alpha^4_{13} =1/2$ and $\beta'=6 \gamma'
\alpha^4_{04} =3/8$ in (\ref{3loopsolu4}), and this also reproduces (\ref{alphares}).
We now write $\tilde S^{(3)}=\tilde S^{(3,0)}+\tilde S^{(3,1)}$ with, schematically:
\begin{eqnarray}
\tilde S^{(3,0)} = \frac{3}{8} \tilde T  (\tilde R'' \tilde R'') + \frac{3}{8} ( \tilde R'' \tilde R'' \tilde R'')
\end{eqnarray}
The equation for $S^{(3,1)}$ is, schematically:
\begin{eqnarray}
  && (\partial_l - 2 \epsilon + 2) \tilde S^{(3,1)} =
- \frac{3}{8} \tilde T  (2 \delta \tilde R'' \tilde R'') - \frac{3}{8} (3 \delta
\tilde R'' \tilde R'' \tilde R'') + \frac{3}{2} \tilde T  (\tilde S^{(3,0) \prime \prime}) + 3 (\tilde R''
(\tilde{S}^{(3,0)}_{110}(u_{113})-\tilde S^{(3,0) \prime \prime}) \\
&& + \frac{3}{4} \tilde T [(\tilde R'' \tilde R'' \tilde R'')''] + \frac{9}{16} [(\tilde R''
\tilde R'' \tilde R'' \tilde R'')'']
\end{eqnarray}

This leads to the various contributions to $\tilde S^{(3)}_{110}(0,0,u)$:
\begin{eqnarray}
&& \frac{9}{160} [[(\tilde R'' \tilde R'' \tilde R'' \tilde R'')'']''] =  \frac{7}{160}
(\tilde R''')^4 + \frac{3}{20} \sf R'' (\tilde
R''')^2 \tilde R''''+ \frac{1}{80} {\sf R''}^2 (\tilde R'''')^2  \\
&& \frac{3}{32} \tilde T [[(\tilde R'' \tilde R'' \tilde R'')'']''] =  \frac{1}{32}
\tilde T (\tilde R''')^2 \tilde R'''' - \frac{1}{32} \tilde T
(\tilde R''')^2
\tilde R''''(0) \\
&& \frac{3}{2} \tilde T  (\tilde S^{(3,0) \prime \prime} ) \to \frac{3}{2} \tilde T
\frac{3}{8} \alpha_{13}^3 [((R'' R'' R'')'')''] = \frac{1}{8}
\tilde T (\tilde R''')^2 \tilde R'''' - \frac{3}{32} \tilde T
(\tilde R''')^2 \tilde R''''(0) - \frac{3}{32} \tilde T \tilde
R''''(0)^2 {\sf R''} \\
&& + \frac{1}{32} \tilde T (\tilde R'''')^2 {\sf R''}+
\frac{1}{16} \tilde T  \tilde R''' \tilde R^{(5)} {\sf R''}
\\
&& \frac{3}{2} \tilde T  (S^{(3,0) \prime \prime}) \to \frac{3}{2} \tilde T^2
\frac{3}{8} \alpha_{22}^3 [((R'' R'')'')''] = \frac{1}{48} \tilde
T^2 (\tilde R'''')^2 - \frac{1}{12} \tilde T^2 \tilde R''''(0)
\tilde R'''' + \frac{1}{24} \tilde T^2 \tilde R'''  \tilde R^{(5)}
-  \frac{1}{24} \tilde T^2 {\sf R''} \tilde R^{(6)}(0) \\
&& - \frac{3}{4} \tilde T  (\delta \tilde R'' \tilde  R'') \to - \frac{3}{4}
\alpha^3_{2 2} \tilde T ^2 [((R''''-R''''(0)) R'')''] =
\frac{1}{24} \tilde T^2 \tilde R''''(0)  \tilde R'''' -
\frac{1}{24} \tilde T^2 \tilde R''' \tilde R^{(5)} + \frac{1}{24}
\tilde T^2 {\sf R''} \tilde R^{(6)}(0) \\
&& - \frac{3}{4} \tilde T  (\delta \tilde R'' \tilde R'') \to - \frac{3}{4}
\alpha^3_{1 3} \tilde T [((R'''^2+ R'' R'''') R'')''] =
\frac{1}{32} \tilde T \tilde R''''(0) (\tilde R''')^2 -
\frac{3}{32} \tilde T \tilde R'''' (\tilde R''')^2  \\
&& + \frac{3}{32} \tilde T \tilde R''''(0)^2 {\sf R''} +
\frac{1}{32} \tilde T {\sf R''} \tilde R''''(0) \tilde R'''' -
\frac{1}{32} \tilde T {\sf R''} \tilde R''' \tilde R^{(5)}
 \\
&& - \frac{9}{8} (\delta \tilde R'' \tilde R'' \tilde R'') \to - \frac{9}{8}
\alpha^3_{1 3} \tilde T [((R''''-R''''(0)) R'' R'')''] =
\frac{1}{32} \tilde T \tilde R''''(0) (\tilde R''')^2  -
\frac{1}{32} \tilde T (\tilde R''')^2 \tilde R'''' - \frac{1}{16}
\tilde T {\sf R''} \tilde R''' \tilde
R^{(5)} \nonumber \\
&& - \frac{9}{8} (\delta \tilde R'' \tilde R'' \tilde R'') \to - \frac{9}{8}
\alpha^3_{0 4} [((R'''^2+ R'' R'''') R'' R'')''] = - \frac{1}{40}
(\tilde R''')^4 -  \frac{7}{40} {\sf R''} (\tilde R''')^2 \tilde
R'''' - \frac{1}{20} {\sf R''}^2 \tilde R''' \tilde
R^{(5)} \\
&& 3 (\tilde R'' (\tilde{S}^{(3,0)}_{110}(u_{113})-\tilde S^{(3,0) \prime \prime}) \to \frac{9}{8}
\alpha^3_{13} \tilde T [(R'' (R'' R'')''_0 - R'' (R'' R'')'')''] =
- \frac{3}{32} \tilde T \tilde R''''(0) (\tilde R''')^2 +
\frac{3}{32} \tilde T \tilde R'''' (\tilde R''')^2 \nonumber \\
&& + \frac{1}{32} \tilde T {\sf R''} (\tilde R'''')^2 -
\frac{1}{32} \tilde T {\sf R''} \tilde R''''(0) \tilde R'''' +
\frac{1}{32} \tilde T {\sf R''} \tilde R''' \tilde R^{(5)} \\
&& 3 (\tilde R'' (\tilde{S}^{(3,0)}_{110}(u_{113})- \tilde S^{(3,0) \prime \prime}) \to \frac{9}{8}
\alpha^3_{04} \tilde  [(R'' (R'' R'' R'')''_0 - R'' (R'' R''
R'')'')''] = \frac{3}{40} (\tilde R''')^4 + \frac{11}{40} {\sf
R''} (\tilde R''')^2 \tilde R'''' \nonumber \\
&& + \frac{1}{20} {\sf R''}^2 (\tilde R'''')^2 + \frac{1}{20} {\sf
R''}^2 \tilde R''' \tilde R^{(5)}
\end{eqnarray}
we note that each of these terms has a vanishing second derivative
at zero (they all come from a true third cumulant, and third
cumulants start as $u^6$). Above, $\delta \tilde R''=(\partial_l +
\epsilon) {\sf R}'' = \tilde T (\tilde R''''- \tilde R''''(0)) +
(\tilde R''')^2 + {\sf R}'' \tilde R'''' + O(T R^2, R^3)$.
Remarkably, the sum of all these terms, after multiple cancellations, gives the $T>0$ beta function to three loop derived by other means in Section \ref{sec:loopexpansion}.

It is worth pointing out where each anomalous term originates from:
\begin{eqnarray}
&& \frac{3}{2} \tilde T  (\tilde S^{(3,0) \prime \prime}) \to \frac{3}{2} \tilde T
\frac{3}{8} \alpha_{13}^3 [((\tilde R'' \tilde R'' \tilde R'')'')''] = - \frac{3}{32}
\tilde T \tilde R''''(0) (\tilde R''')^2  - \frac{3}{32} \tilde T
\tilde R''''(0)^2 {\sf R''} \\
&& \frac{3}{2} \tilde T  (\tilde S^{(3,0) \prime \prime}) \to \frac{3}{2} \tilde T^2
\frac{3}{8} \alpha_{22}^3 [((\tilde R'' \tilde R'')'')''] = -  \frac{1}{24} \tilde T^2 {\sf R''} \tilde R^{(6)}(0) \\
&& - \frac{3}{4} \tilde T  (\delta \tilde R'' \tilde R'') \to - \frac{3}{4}
\alpha^3_{2 2} \tilde T ^2 [((\tilde R''''-\tilde R''''(0)) \tilde R'')''] =
\frac{1}{24}
\tilde T^2 {\sf R''} \tilde R^{(6)}(0) \\
&& - \frac{3}{4} \tilde T  (\delta \tilde R'' \tilde R'') \to - \frac{3}{4}
\alpha^3_{1 3} \tilde T [((R'''^2+ R'' R'''') R'')''] =
\frac{1}{32} \tilde T \tilde R''''(0) (\tilde R''')^2  +
\frac{3}{32} \tilde T \tilde R''''(0)^2 {\sf R''} +
\frac{1}{32} \tilde T {\sf R''} \tilde R''''(0) \tilde R'''' \nonumber \\
&& 3 (R'' (\tilde{S}^{(3,0)}_{110}(u_{113})- \tilde S^{(3,0) \prime \prime}) \to \frac{9}{8}
\alpha^3_{13} \tilde T [(\tilde R'' (\tilde R'' \tilde R'')''_0 - \tilde R'' (\tilde R'' \tilde R'')'')''] =
- \frac{3}{32} \tilde T \tilde R''''(0) (\tilde R''')^2 -
\frac{1}{32} \tilde T {\sf R''} \tilde R''''(0) \tilde R'''' \\
&&
\frac{3}{32} \tilde T [[(\tilde R'' \tilde R'' \tilde R'')'']''] =   - \frac{1}{32}
\tilde T (\tilde R''')^2
\tilde R''''(0) \\
&& - \frac{9}{8} (\delta \tilde R'' \tilde R'' \tilde R'') \to - \frac{9}{8}
\alpha^3_{1 3} \tilde T [((R''''-R''''(0)) R'' R'')''] =
\frac{1}{32} \tilde T \tilde R''''(0) (\tilde R''')^2   \nonumber \\
\nonumber
\end{eqnarray}
Cancellations of $1/\tilde T$ terms arise in the first four terms,
while the last three are more regular (with term linear in $\tilde R''$
cancelling). The cancellation of divergent terms occurs thus
between $\partial \tilde S^{(3)}$ and $\tilde T \tilde S^{(3) \prime \prime}$ in the unrescaled equation, and
presumably between $\tilde T \tilde S^{(3) \prime \prime}$ and $\zeta \tilde S^{(3)} + \zeta u
\tilde S^{(3) \prime }$ in the rescaled equation using the $\tilde R$ equation.

\subsection{Three loop at $T=0$ by the $\Gamma$-ERG method}

\label{app:3loopergT0}

Let us now recompute the $T=0$ terms from the previous Appendix, taking all limits at $0^+$:
\begin{eqnarray}
&& \frac{9}{160} [[(R'' R'' R'' R'')'']''] =  \frac{7}{160}
(\tilde R''')^4 + \frac{3}{20} \sf R'' (\tilde
R''')^2 \tilde R''''+ \frac{1}{80} {\sf R''}^2 (\tilde R'''')^2  \\
&& - \frac{3}{160} \tilde R'''(0^{++})^2 (\tilde R''')^2 -
\frac{11}{160} \tilde R'''(0^+)^2 (\tilde R''')^2 + \frac{3}{80}
{\sf R''} \tilde R'''(0^{++})^2 \tilde R''''(0^+) - \frac{3}{16}
{\sf R''} \tilde R'''(0^+)^2 \tilde R''''(0^+) \\
&& - \frac{9}{8}
(\delta R'' R'' R'') \to - \frac{9}{8} \alpha^3_{0 4} [((R'''^2+
R'' R'''') R'' R'')''] = - \frac{1}{40} (\tilde R''')^4 -
\frac{7}{40} {\sf R''} (\tilde R''')^2 \tilde
R'''' - \frac{1}{20} {\sf R''}^2 \tilde R''' \tilde R^{(5)} \\
&& + \frac{3}{80} (\tilde R''')^2 \tilde R'''(0^+)^2  +
\frac{7}{40} {\sf R''} \tilde R'''(0^+)^2 \tilde R''''(0^+) +
\frac{1}{40} {\sf R''} \tilde R'''(0^+)^2 \tilde R''''
\\
&& 3 (R'' (\tilde{S}^{(3)}_{0;110}(u_{113})-S_0'') \to \frac{9}{8}
\alpha^3_{04} \tilde  [(R'' (R'' R'' R'')''_0 - R'' (R'' R''
R'')'')''] = \frac{3}{40} (\tilde R''')^4 + \frac{11}{40} {\sf
R''} (\tilde R''')^2 \tilde R'''' \\
&& + \frac{1}{20} {\sf R''}^2 (\tilde R'''')^2 + \frac{1}{20} {\sf
R''}^2 \tilde R''' \tilde R^{(5)} \\
&& - \frac{1}{40} \tilde R'''(0^{++})^2 (\tilde R''')^2 -
\frac{1}{8} \tilde R'''(0^{+})^2 (\tilde R''')^2
 + \frac{1}{20} {\sf R}'' \tilde R'''(0^{++})^2 \tilde R''''(0^+)
- \frac{3}{10} {\sf R}'' \tilde R'''(0^{+})^2 \tilde R''''(0^+) -
\frac{1}{40} {\sf R}'' \tilde R'''(0^{+})^2 \tilde R'''' \nonumber
\end{eqnarray}
Where we have assumed that ${\sf R}''(0^+)=0$. We find that each of these expressions has no super-cusp by
itself. When two limits have to be taken the one taken first is noted $0^{++}$ (for instance this gives the
feeding of the fourth cumulant into the third). This is PLB31. It has no ambiguity. Then the second limit $0^+$
is taken, this is PBL21 or PBL211. One easily sees that there is no supercusp independently for $0^+$ and
$0^{++}$. The final result for the beta function to three loop using this procedure is:
\begin{eqnarray}
&& \partial \tilde R = (\epsilon - 4 \zeta) \tilde R + \zeta u
\tilde R' + [ \frac{1}{2} (\tilde R'')^2 - \tilde R''(0) \tilde
R'' ] + \frac{1}{4} (\tilde R''')^2 (\tilde R'' - \tilde R''(0))
- \frac{1}{4} \tilde R'''(0^+)^2  \tilde R'' \nonumber \\
&& + \frac{1}{16} (\tilde R''- \tilde R''(0))^2 (\tilde R'''')^2 +
\frac{3}{32} (\tilde R''')^4 + \frac{1}{4} (\tilde R''- \tilde
R''(0)) (\tilde R''')^2 \tilde R'''' \label{oldrg} \\
&& - \frac{5}{32} \tilde R'''(0^+)^2 (\tilde R''')^2 -
\frac{5}{16} \tilde R'''(0^+)^2 \tilde R''''(0^+) \tilde R'' -
\frac{7}{160} \tilde R'''(0^{++})^2 (\tilde R''')^2 + \frac{7}{80}
\tilde R'''(0^{++})^2 \tilde R''''(0^+) \tilde R'' \nonumber
\end{eqnarray}
Thus if one sets the anomalous $0^{++}$ terms to zero one reproduces the beta function obtained from the three
loop finite T $R$ equation using the value of $r^{(4)}(0)$ which cancels the super-cusp. Indeed the $0^{++}$
contribution has no supercusp by itself. Next, if one adds the $0^{++}$ and $0^{+}$ contributions, one still
does not obtain the full correct answer, derived in the next Section. As explained there an additional term,
called $a_1$ there, is set to zero here, see below for further explanations.


\subsection{Three loop beta function via W-ERG}

\label{app:threeloppwerg}

The simplest method to program in Mathematica, although quite memory consuming, is to use instead the
$W$-functional. One uses the W-ERG equation (\ref{Wrg}) on the {\it moments}, directly at $T=0$. It has the
symbolic form:
\begin{eqnarray} \label{hier}
&& \partial_t \bar S^{(n)} = \alpha_n (\bar S^{(n+1)})''
\end{eqnarray}
with $\alpha_n=n/2$, where in $(..)''$ denote two derivatives and their arguments set equal at the end, and
symmetrization in $n$ arguments is implicit. Similar notations were used in Appendix G of Ref.
\onlinecite{BalentsLeDoussal2005} for a different calculation, using instead the $\Gamma$-ERG equations. We
recall that $t=m^{-2}$.

One looks for a solution of the form:
\begin{eqnarray}
&& \partial_t R = \sum_{q=1}^\infty t^{2q-1} \delta_q R \\
&& \bar S^{(n)} = \sum_{p=0}^\infty t^{2p} S_p^{(n)}[R]  \quad n \quad \text{even} \\
&& \bar S^{(n)} = \sum_{p=0}^\infty t^{2p+1} S_p^{(n)}[R]  \quad n \quad \text{odd}
\end{eqnarray}
The recursion relations read:
\begin{eqnarray}
&& (2 k+1) S_k^{(n)} = \alpha_n (S_k^{(n+1)})''
- \sum_{q=1}^k S_{k-q}^{(n)}[\delta_q R,R] \quad n \quad \text{odd} \\
&& (2 k+2) S_{k+1}^{(n)} = \alpha_n (S_k^{(n+1)})''
- \sum_{q=1}^{k+1} S_{k+1-q}^{(n)}[\delta_q R,R] \quad n \quad \text{even}
\end{eqnarray}
The notation $S_{p}^{(n)}[\delta_q R,R]$ just means that one performs the derivative $\partial_t S_{p}^{(n)}[R]$
and replace each of the $p$ resulting $\partial_t R$ factors by $\delta_q R$. These are what are usually called
counterterms and here they appear automatically in the calculation.

Each $\delta_q R$ is called the $q$-loop contribution to the beta function and is homogeneous of order
$O(R^{q})$. The calculation of $\delta_1 R$, i.e. the one loop contribution to the beta function, can be
performed as follows:
\begin{eqnarray}
&& S_0^{(4)} = 3 [R R] \\
&& S_0^{(3)} = \alpha_3 (S_0^{(4)})'' \\
&& \delta_1 R = (S_0^{(3)})''
\end{eqnarray}

To obtain $\delta_2 R$, the two loop contribution to the beta function, one needs some one loop counterterms,
hence $\delta_1 R$. One does it as follows.
\begin{eqnarray}
&& S_0^{(6)} = 15 [R R R] \\
&& S_0^{(5)} = \alpha_5 (S_0^{(6)})'' \\
&& 2 S_1^{(4)} = \alpha_4 (S_0^{(5)})'' - S_0^{(4)}(\delta_1 R,R)  \\
&& 3 S_1^{(3)} = \alpha_3 (S_1^{(4)})'' - S_0^{(3)}(\delta_1 R,R) \\
&& \delta_2 R = (S_1^{(3)})''
\end{eqnarray}
Hence to a given number of loop, $q$, one performs the gaussian (Wick) truncation on the $2 q+2$ moment, and
from there solves the RG equations for all lower moments. It exactly amounts to solve the FRG hierarchy to a
given order in powers of $t$, as shown above. Note that to get $\delta_2 R$ one could instead truncate the fifth
moment into a product of the form $\bar S^{(3)} R$ and use the one-loop formula to evaluate the third moment.
One can check that it yields the same result. Note that it is convenient to to use ${\sf R}$ i.e. set $R''(0)=0$
in the calculation, since this is the object which appear in the loop corrections, but then one should not
forget when computing the beta function for ${\sf R}$, and the related counterterms, to write $\partial {\sf R}
=
\partial R - a_1 u^2/2
\partial \tilde R''(0)$, with $a_1=1$.

The result of this calculation, performed using Mathematica, reads to three loop, in the rescaled version, with
$a_1=1$:
\begin{eqnarray}
&& \partial \tilde R = (\epsilon - 4 \zeta) \tilde R + \zeta u
\tilde R' + [ \frac{1}{2} (\tilde R'')^2 - \tilde R''(0) \tilde
R'' ] + \frac{1}{4} (\tilde R''')^2 (\tilde R'' - \tilde R''(0))
- \frac{1}{4} \tilde R'''(0^+)^2  \tilde R'' \nonumber \\
&& + \frac{1}{16} (\tilde R''- \tilde R''(0))^2 (\tilde R'''')^2 +
\frac{3}{32} (\tilde R''')^4 + \frac{1}{4} (\tilde R''- \tilde
R''(0)) (\tilde R''')^2 \tilde R'''' \label{3loopapp}  \\
&& - \frac{3}{16} \tilde R'''(0^+)^2 (\tilde R''')^2 - \frac{1}{4} \tilde R'''(0^+)^2 \tilde R''''(0^+) \tilde
R''  \nonumber
\end{eqnarray}
While keeping arbitrary $a_1$ yields changes only in the last line as:
\begin{eqnarray}
&& + (\frac{a_1}{80} - \frac{1}{5}) \tilde R'''(0^+)^2 (\tilde R''')^2
- (\frac{a_1}{40} + \frac{9}{40}) \tilde R'''(0^+)^2 \tilde R''''(0^+) \tilde R''  \nonumber
\end{eqnarray}
and one can check that the result of the previous Section, Eq. (\ref{oldrg}) corresponds to setting $a_1=0$.
This is not too surprising since they were derived from a $T>0$ procedure, where no non trivial counterterms to
$\tilde R''(0)$ arises (i.e. $a_1=0$) if one justs set $T=0$. One must indeed properly take into account the
ensuing non analyticity of the function $\tilde R(u)$, as is done here. The corresponding modification in the
calculation of the previous Section does indeed lead to the correct value $a_1=1$ and the correct three loop
beta function (\ref{3loopapp}).

Note that no ambiguity ever appear in this iterative procedure. That there should be no ambiguity in each
equation (\ref{hier}) is clear. The r.h.s. involves derivatives such that, e.g. for $n=4$, $\bar
Q_{1100}(v_1,v_2,v_3,v_4) = \overline{F(v_1) F(v_2) \tilde V(v_3) \tilde V(v_4)}$. Despite the presence of
shocks, the product $F(v_1) F(v_2)$ in any correlation involving forces or anything smoother at points distinct
from $v_1,v_2$ is a continuous function and the coinciding point limit $v_1 \to v_2$ can be taken unambiguously.
This is further discussed in the text. What is less obvious is that this remains true order by order in the loop
expansion. For $N=1$, we have checked explicitly that it does, up to four loop, but we believe it holds to all
orders.

\section{properties of the fixed point functions for the Sinai landscape } 
\label{appRu}

\subsection{asymptotics of $R(v)$}

The large $v$ limit can be studied using the asymptotics (shown in Appendix of Ref.\cite{FrachebourgMartin99}):
\begin{eqnarray}
&& B(x,z_1,z_2) = \int_0^\infty e^{y z} Ai(y + z_1) Ai(y + z_2)
 \approx \frac{1}{2 \sqrt{\pi x}} e^{\frac{x^3}{12} - \frac{x}{2} (z_1+z_2)
- \frac{(z_1-z_2)^2}{4 x} }
\end{eqnarray}
Inserting this approximation which becomes exact at large $v$ yields:
\begin{eqnarray}
&& p(E) \approx \frac1{\sqrt{4 \pi v}} e^{- \frac{E^2}{4 v}} \int_{-\infty}^{+\infty} \frac{d z_1}{2 \pi i}
\int_{-\infty}^{+\infty} \frac{d z_2}{2 \pi i}
 e^{\frac{E}{ b v} (z_2-z_1) - \frac{1}{b^2 v} (z_2-z_1)^2 }
\frac{1}{Ai(z_1)^2 Ai(z_2)^2}
\end{eqnarray}
which is always normalized to unity (the part with no shock decays rapidly at large $v$) and yields in the large
$v$ limit, $p(E)  = \frac1{\sqrt{4 \pi v}} e^{- \frac{E^2}{4 v}}$ since $\int_{-\infty}^{+\infty} \frac{d z_1}{2
\pi i} \frac{1}{Ai(z_1)^2}=1$. We can now use the result of the Appendix of Ref.\cite{FrachebourgMartin99} for the integral:
\begin{eqnarray}
&& B(x,z_1,z_2) = \int_0^\infty e^{y x} Ai(y + z_1) Ai(y + z_2)
= B_1(x,z_1,z_2) (1 - \int_x^\infty dy \frac{g(y,z_1,z_2)}{B_1(y,z_1,z_2)}) \\
&& B_1(x,z_1,z_2) = \frac{1}{2 \sqrt{\pi x}} e^{\frac{x^3}{12} - \frac{x}{2} (z_1+z_2)
- \frac{(z_1-z_2)^2}{4 x} } \\
&& g(y,z_1,z_2) = (\frac{y}{4} Ai(z_1) Ai(z_2) - \frac{1}{4} (Ai'(z_1) Ai(z_2) + Ai(z_1) Ai'(z_2))
\\
&& + \frac{1}{4 y} (2 Ai'(z_1) Ai'(z_2) - (z_1+z_2) Ai(z_1) Ai(z_2) ) - \frac{1}{4 y^2} (z_2-z_1) (Ai'(z_1)
Ai(z_2) - Ai(z_1) Ai'(z_2) )]
\end{eqnarray}

This yields:

\begin{eqnarray}
&& 2 (R(0)-R(v)) = \int_{-\infty}^{+\infty} \frac{d z_1}{2 \pi i} \int_{-\infty}^{+\infty} \frac{d z_2}{2 \pi i}
(2 v + \frac{4}{b^2} (z_2-z_1)^2 )
[ \frac{1}{Ai(z_1)^2 Ai(z_2)^2} ( 1 - \int_{v/(2 a)}^\infty dy \frac{g(y,z_1,z_2)}{B_1(y,z_1,z_2)}) \\
&& + \frac{1}{v Ai(z_1) Ai(z_2)} (a b)^{-2} 2 \sqrt{\pi v} e^{- \frac{1}{48} v^3} e^{\frac{v}{2 b} (z_1+z_2) +
\frac{(z_2-z_1)^2}{b^2 v}} ]
\end{eqnarray}
where for large $v$ only the $1$ contributes yielding the formula given in the text. One can also show the alternative formula:
\begin{eqnarray}
&& 2 (R(0)-R(v)) = - a^{-2} 2 v \sqrt{\pi} e^{- \frac{1}{48} v^3} \int_{-\infty}^{+\infty} \frac{d u}{2 \pi}
\int_{-\infty}^{+\infty} \frac{d w}{2 \pi} \\
&& \times
 e^{i \frac{v}{b} u} e^{- \frac{w^2}{b^2}}
\partial_w^2 [ \frac{1}{Ai(i u + i \sqrt{v} \frac{w}{2})
Ai(i u - i \sqrt{v} \frac{w}{2})} +  v \frac{\int_0^\infty dV e^{\frac{v}{2} V } Ai(a V + i u + i \sqrt{v}
\frac{w}{2})
 Ai(a V + i u - i \sqrt{v} \frac{w}{2})}{Ai(i u + i \sqrt{v} \frac{w}{2})^2
Ai(i u - i \sqrt{v} \frac{w}{2})^2} ]
 \nonumber
\end{eqnarray}

\subsection{expansion at small $v$}

The distribution of rescaled energy can be written:
\begin{eqnarray}
&& p(\epsilon)  = e^{- \frac{1}{48} v^3 - v \frac{\epsilon^2}{4}} g(\epsilon - \frac{v}{2})
g(-\epsilon - \frac{v}{2}) \\
&& + (a b)^{-2} \frac{v}{a} e^{- \frac{1}{48} v^3
- v \frac{\epsilon^2}{4}}
\int \frac{d z_1}{2 \pi i}
\int \frac{d z_2}{2 \pi i} e^{\frac{v}{2 b} (z_1+z_2) + \frac{\epsilon}{ b}
 (z_2-z_1)} \frac{\int_0^\infty dW e^{\frac{v}{2 a} W } Ai(W + z_1)
 Ai(W+ z_2)}{Ai(z_1)^2 Ai(z_2)^2}
 \nonumber \\
\end{eqnarray}
with $\epsilon=E/v$. We use the notations of the text. Hence one has:
\begin{eqnarray}
&& p_1(\epsilon) = (a b)^{-2} \int \frac{d z_1}{2 \pi i}
\int \frac{d z_2}{2 \pi i}  e^{\frac{\epsilon}{ b} (z_2-z_1)}  \frac1{Ai(z_1) Ai(z_2)} [ \frac{1}{2 b} (z_1+z_2)
- \frac{1}{4} \epsilon^2 + \frac{1}{a} \frac{\int_0^\infty dW Ai(W + z_1)
Ai(W+ z_2)}{Ai(z_1) Ai(z_2)} ] \nonumber \\
&& = a^{-2} \int \frac{d \lambda_1}{2 \pi }
\int \frac{d \lambda_2}{2 \pi }  e^{i \epsilon (\lambda_2-\lambda_1)}  \frac1{Ai(i b \lambda_1)
Ai(i b \lambda_2)} [ \frac{i}{2} (\lambda_1 + \lambda_2)
- \frac{1}{4} \epsilon^2 + \frac{1}{a} \frac{\int_0^\infty dW Ai(W + i b \lambda_1)
Ai(W+ i b \lambda_2)}{Ai(i b \lambda_1) Ai(i b \lambda_2)} ] \nonumber
\end{eqnarray}
Thus:
\begin{eqnarray}
&& \hat p_1(\mu) =
\int d\epsilon e^{i \mu \epsilon} p_1(\epsilon)
= a^{-2} \int \frac{d \lambda_2}{2 \pi } \frac1{Ai(i b \lambda_2 + i b \mu )
Ai(i b \lambda_2)}  [ \frac{i}{2} (2 \lambda_2 + \mu)
 + \frac{1}{a} \frac{\int_0^\infty dW Ai(W + i b \lambda_2 + i b \mu)
Ai(W+ i b \lambda_2)}{Ai(i b \lambda_2) Ai(i b \lambda_2 + i b \mu)} ] \nonumber \\
&& + a^{-2} \frac{1}{4} \partial^2_\mu
\int \frac{d \lambda_2}{2 \pi } \frac1{Ai(i b \lambda_2 + i b \mu )
Ai(i b \lambda_2)} \\
&& = a^{-2} \int \frac{d \lambda_2}{2 \pi } \frac1{Ai(i b \lambda_2 + i b \mu/2 )
Ai(i b \lambda_2 - i b \mu/2)}  [ i \lambda_2
 + \frac{1}{a} \frac{\int_0^\infty dW Ai(W + i b \lambda_2 + i b \mu/2)
Ai(W+ i b \lambda_2 - i b \mu/2 )}{Ai(i b \lambda_2 + i b \mu/2) Ai(i b \lambda_2 - i b \mu/2)} ] \nonumber \\
&& + a^{-2} \frac{1}{4} \partial^2_\mu
\int \frac{d \lambda_2}{2 \pi } \frac1{Ai(i b \lambda_2 + i b \mu/2 )
Ai(i b \lambda_2 - i b \mu/2)} \\
\end{eqnarray}

The function $f(z) = Ai(z+z_1) Ai(z+z_2)$ satisfies the differential equation:
\begin{eqnarray}
&& f''''(z) - (4 z + 2 z_1 + 2 z_2) f''(z) - 6 f'(z) + (z_1-z_2)^2 f(z) = 0
\end{eqnarray}
This yields:
\begin{eqnarray}
&& \int_0^\infty dz Ai(z+z_1) Ai(z+z_2) = \frac{Ai(z_2) Ai'(z_1) - Ai(z_1)Ai'(z_2) }{z_2-z_1} \\
&& \int_0^\infty dz Ai(z+z_1)^2 = z_1 Ai(z_1)^2 - Ai'(z_1)^2
\end{eqnarray}
Let us denote
\begin{eqnarray}
&& p_{11}(\epsilon) = (a b)^{-2} \frac{1}{a} \int \frac{d z_1}{2 \pi i}
\int \frac{d z_2}{2 \pi i}  e^{\frac{\epsilon}{ b} (z_2-z_1)} \frac{\int_0^\infty dW Ai(W + z_1)
Ai(W+ z_2)}{Ai(z_1)^2 Ai(z_2)^2}  \\
&& = (a b)^{-2} \frac{1}{a} \int \frac{d z_1}{2 \pi i}
\int \frac{d z_2}{2 \pi i}  e^{\frac{\epsilon}{ b} (z_2-z_1)}
\frac{Ai(z_2) Ai'(z_1) - Ai(z_1)Ai'(z_2) }{(z_2-z_1) Ai(z_1)^2 Ai(z_2)^2}
\end{eqnarray}
One finds:
\begin{eqnarray}
&& \partial_\epsilon p_{11}(\epsilon) = - 2 a^{-1} b^{-2} \epsilon g(\epsilon) g(- \epsilon)
= - \epsilon g(\epsilon) g(- \epsilon) \\
&& p_{10}(\epsilon) = - \frac{\epsilon^2}{4} g(\epsilon) g(-\epsilon)
- \frac{1}{2} ( g'(\epsilon) g(-\epsilon) + g(\epsilon) g'(-\epsilon) )
\end{eqnarray}

Let us first check normalization $\int d\epsilon p_1(\epsilon)=0$. We need to show:
\begin{eqnarray}
&& \frac{3}{4} \int d\epsilon \epsilon^2 g(\epsilon) g(- \epsilon)
= \int d\epsilon g'(\epsilon) g(-\epsilon)
\end{eqnarray}
One has:
\begin{eqnarray}
&& \frac{3}{4} \int d\epsilon \epsilon^2 g(\epsilon) g(- \epsilon)
= \frac{3}{4} \frac{b^2}{a^2}
\int \frac{d \lambda}{2 \pi} [ \frac{- i b \lambda}{Ai(i b \lambda)^2}
+ 2 \frac{Ai'(i b \lambda)^2}{Ai(i b \lambda)^4} ] \\
&& \int d\epsilon g'(\epsilon) g(-\epsilon)
= \frac{1}{b a^2}  \int \frac{d \lambda}{2 \pi}  \frac{- i b \lambda}{Ai(i b \lambda)^2}
\end{eqnarray}
These two quantity are indeed equal, from the identity:
\begin{eqnarray}
&& 3 \int dz \frac{Ai'(z)^2}{Ai(z)^4} = \int dz \frac{z}{Ai(z)^2}
\end{eqnarray}
which can be checked by integration by part of $\int_z Ai'(z)^2/Ai(z)^4
= - \frac{1}{2} \int_z (Ai'(z)/Ai(z)) (1/Ai(z)^2)'$. This implies the expansion and derivatives given in the text.

\subsection{check of matching}

Writing:
\begin{eqnarray}
&& g(\epsilon) = \frac{1}{a b} \int_z e^{- z \epsilon/b} \frac{1}{Ai(z)} \\
&& d(y) = \frac{a}{b} \int_z e^{z \epsilon/b} \frac{Ai'(z)}{Ai(z)}
\end{eqnarray}
with $\int_z = \int dz/(2 \pi i)$. One finds:
\begin{eqnarray}
&& \int d\epsilon \epsilon^4 g(\epsilon) g(-\epsilon)
= \frac{4}{a^2} J_1 \\
&& \int d\epsilon \epsilon^2 g'(\epsilon) g(-\epsilon)
= \frac{1}{a^2} J_2 \\
&& \int du dy g(-u) y^3 d(y) g(y+u)
= \frac{2}{a^2} J_3 \\
&& \int du dy g(-u) y^2 d(y) g(y+u) = 1
\end{eqnarray}
with:
\begin{eqnarray}
&& J_1 = \int_z (\frac{1}{Ai(z)})'''' \frac{1}{Ai(z)}
= 5 I_1 - 28 I_2 + 24 I_3 = \frac{7}{15} I_1 \\
&& J_2 = \int_z (\frac{- z}{Ai(z)})'' \frac{1}{Ai(z)}
= I_1 - 2 I_2 = \frac{1}{3} I_1 \\
&& J_3 = - \int_z (\frac{Ai'(z)}{Ai(z)})''' \frac{1}{Ai(z)^2}
= 2 I_1 - 8 I_2 + 6 I_3 = \frac{8}{15} I_1 \\
\end{eqnarray}
where we have defined:
\begin{eqnarray}
&& I_1 = \int_z \frac{z^2}{Ai(z)^2} = 3 I_2 \\
&& I_2 = \int_z \frac{z Ai'(z)^2}{Ai(z)^4} = \frac{5}{3} I_3 \\
&& I_3 = \int_z \frac{Ai'(z)^4}{Ai(z)^6}
\end{eqnarray}
the above relations being derived from considering the total
derivatives $(\frac{z Ai'(z)}{Ai(z)^3})'$, $(\frac{Ai'(z)^3}{Ai(z)^5})'$
and $(\frac{1}{Ai(z)^2})'$ which integrate to zero.

The droplets predict:
\begin{eqnarray}
&& - R'''(0^+) = - \frac{1}{4} 2 \int du dy g(u) y^3 d(y) g(y+u)
= - \frac{1}{a^2} J_3 \\
&&  - R'''(0^+) =  (\frac{1}{4} \int d\epsilon \epsilon^4 g(\epsilon) g(-\epsilon)
- 3 \int d\epsilon \epsilon^2 g'(\epsilon) g(-\epsilon) )
= \frac{1}{a^2} (J_1 - 3 J_2)
\end{eqnarray}
We can see that these results agree $J_1 - 3 J_2 = - J_3 = - \frac{8}{15} I_1$.
The final result is:
\begin{eqnarray}
&& R'''(0^+) = \frac{1}{a^2} \frac{8}{15} \int_z \frac{z^2}{Ai(z)^2}
\end{eqnarray}

\subsection{calculation of  $\Delta(u)$}

Let us start from the formula given in the text. 
The first part gives, writing $F_1=F-v/2$, $F_2=F+v/2$ :
\begin{eqnarray}
&& \Delta_{ns}(v) = e^{ - \frac{1}{48} v^3} \int_{-\infty}^{+\infty} dF (F^2 - \frac{v^2}{4} )
g(F-\frac{v}{2}) g(-F-\frac{v}{2}) e^{- \frac{v}{4} F^2} \\
&& = a^{-2} e^{ - \frac{1}{48} v^3} \int_{-\infty}^{+\infty} \frac{d \lambda_1}{2 \pi} \int_{-\infty}^{+\infty}
\frac{d \lambda_2}{2 \pi}  e^{  i \frac{v}{2} (\lambda_1+\lambda_2)  } \frac{1}{Ai(i b \lambda_1) Ai(i b
\lambda_2)} \int_{-\infty}^{+\infty} dF (F^2 - \frac{v^2}{4} ) e^{ - i
(\lambda_1-\lambda_2) F - \frac{v}{4} F^2} \\
&& = \frac{1}{2} \sqrt{\pi} v^{-5/2} a^{-2} e^{ - \frac{1}{48} v^3} \int_{-\infty}^{+\infty} \frac{d
\lambda_1}{2 \pi} \int_{-\infty}^{+\infty}  \frac{d \lambda_2}{2 \pi}  e^{i \frac{v}{2} (\lambda_1+\lambda_2) -
\frac{(\lambda_1-\lambda_2)^2}{ v}  } (8 v - v^4 - 16 (\lambda_1-\lambda_2)^2) \frac{1}{Ai(i b \lambda_1) Ai(i b
\lambda_2)}
\end{eqnarray}
Using the same trick as below it can also be written:
\begin{eqnarray}
&& \Delta_{ns}(v) = - 2 \sqrt{\pi} v^{-1/2} a^{-2} b^2 e^{ - \frac{1}{48} v^3} \int_{-\infty}^{+\infty} \frac{d
\lambda_1}{2 \pi} \int_{-\infty}^{+\infty}  \frac{d \lambda_2}{2 \pi}  e^{  i \frac{v}{2} (\lambda_1+\lambda_2)
- \frac{(\lambda_1-\lambda_2)^2}{ v}   } \frac{Ai'(i b \lambda_1) Ai'(i b \lambda_2) }{Ai(i b \lambda_1)^2 Ai(i
b \lambda_2)^2}
\end{eqnarray}

The part with shocks is:

\begin{eqnarray}
&& p(F_1,F_2,v)= g(F_1) g(-F_2) dF_1 dF_2 \theta(v+F_1-F_2) e^{ - \frac{1}{48} v^3} \int_{-\infty}^{+\infty}
\frac{d \lambda_1}{2 \pi} \int_{-\infty}^{+\infty}  \frac{d \lambda_2}{2 \pi}
\int_0^\infty dV e^{\frac{1}{2} V v } \\
&& \times \int_0^{v+F_1-F_2} du e^{ - \frac{v}{4} (u -F_1 - \frac{v}{2})^2 } e^{i \lambda_2 (v+F_1-F_2) } e^{i
(\lambda_1-\lambda_2) u } \frac{Ai(a V + i b \lambda_1)
 Ai(a V + i b \lambda_2)}{Ai(i b \lambda_1)
 Ai(i b \lambda_2)} \nonumber
\end{eqnarray}

To compute $\Delta$ the best is to go back directly to:

\begin{eqnarray}
&& \int dF_1 dF_2 g_s(F_1,F_2,v) = \int du_1 du_2 du_1^* e^{ -
\frac{1}{48} v^3  - \frac{v}{4} (u_1^* - \frac{v_1+v_2}{2})^2 } \\
&&  \times \int_0^\infty dV e^{\frac{1}{2} v V }
 (v_1-u_1) g(v_1-u_1) h(u_1^*-u_1,V_1)
  h(u_2 - u_1^*,V) (v_2-u_2) g(u_2-v_2) \nonumber \\
  && = -  \int dw e^{ - \frac{1}{48} v^3  - \frac{v}{4} (w-
\frac{v}{2})^2 }\int_0^\infty dV e^{\frac{1}{2} v V } \psi(w,V) \psi(v - w,V)
\end{eqnarray}
(put bounds) where we have defined:
\begin{eqnarray}
&& \psi(w,V) = \int_0^\infty du  (u-w) g(u-w) h(u,V_1) = - \int_{-\infty}^{+\infty}  \frac{d \lambda}{2 \pi}
e^{i \lambda w} \frac{b Ai'(i b \lambda) Ai(a V + i b \lambda) }{a Ai(i b \lambda)^3}
\end{eqnarray}
with $w=u_1^*-v_1$. One has $u g(u)= - \int_{-\infty}^{+\infty} \frac{d \lambda}{2 \pi} e^{- i \lambda u}
\frac{b Ai'(i b \lambda)}{a Ai(i b \lambda)^2}$

\begin{eqnarray}
&& \Delta_s(v)= - b^2 a^{-2} e^{ - \frac{1}{48} v^3 } \int dw e^{- \frac{v}{4} (w- \frac{v}{2})^2 }
\int_{-\infty}^{+\infty} \frac{d \lambda_1}{2 \pi} \int_{-\infty}^{+\infty}  \frac{d \lambda_2}{2 \pi} e^{i
\lambda_2 v} e^{i (\lambda_1-\lambda_2) w}
\\
&& \times \int_0^\infty dV e^{\frac{1}{2} v V } \frac{ Ai'(i b \lambda_1) Ai(a V + i b \lambda_1) }{ Ai(i b
\lambda_1)^3} \frac{ Ai'(i b \lambda_2) Ai(a V + i b \lambda_2) }{ Ai(i b
\lambda_2)^3} \nonumber \\
&& = - b^2 a^{-2} e^{ - \frac{1}{48} v^3 } \int_{-\infty}^{+\infty} \frac{d \lambda_1}{2 \pi}
\int_{-\infty}^{+\infty}  \frac{d \lambda_2}{2 \pi} 2 \sqrt{\pi} v^{-1/2} e^{ -
\frac{(\lambda_1-\lambda_2)^2}{v} + i
\frac{v}{2} (\lambda_1+\lambda_2)} \\
&& \times \int_0^\infty dV e^{\frac{1}{2} v V } \frac{Ai'(i b \lambda_1) Ai(a V + i b \lambda_1) }{Ai(i b
\lambda_1)^3} \frac{Ai'(i b \lambda_2) Ai(a V + i b \lambda_2) }{ Ai(i b
\lambda_2)^3} \nonumber 
\end{eqnarray}

Putting everything together one finds the formula given in the text.

\section{Short range random potential and Kida Burgers turbulence}
\label{sec:kida} 

Consider the $d=0$ toy model
\begin{eqnarray}
 && H_{V,v}(u)= \frac{m^2}{2} (u-v)^2 + V(u)
\end{eqnarray}
where $V(u)$ has short range correlations. This is the problem studied, in the context of Burgers turbulence, by Kida \cite{kida} and is by now standard. A nice calculation using replica, and interpretation in terms of extremal statistics was given in \cite{jpkida}. The calculation of the one point function below is a slight generalization of the one in Ref. \cite{jpkida}. We recall it here and extend to the two point function and calculation of $R(u)$ for this problem. We also derive the logarithmic corrections for disorder distributions which fall in the Gumbel class. One method to define the model in the continuum is via a Poisson point process. See refs. \cite{BauerBernard99} for the solution in that case. We also use below some of the arguments given there. 

Here we will rather discretize $u$ to integers and consider the small $m$ limit where a continuum limit exists. We consider that $V(u)$ are i.i.d random variables and call $P_0(V)$ the on site probability distribution and define:
\begin{eqnarray}
 && P_{<}(V) = \int_{-\infty}^V P_0(W) dW = e^{- A(V)}
\end{eqnarray}
and $P_{>}(x)=\int_x^\infty P_0(V) dV = 1 - P_{<}(x)$. The minima statistics are controlled by the tail of $P(V)$ for small $V$. We treat here all cases in the so-called Gumbel class, and give explicit application to the case $A(V)=B |V|^\delta-\ln C$ for large negative $V$, the subcase $\delta=1$ being $P(V)=B C e^{B V}$ as $V \to -\infty$.

\subsubsection{one point function}

Let us first consider the problem for $T=0$, i.e. the inviscid limit $\nu=0$ and define $\hat V(v)=\min_u H_{V,v}(u)$. Consider first the distribution of the absolute minimum. It is equivalent to study $v=0$. Then the probability $P(u_1,V_1)$ that
the absolute minimum is at $u_1$ with energy $H=E_1=\frac {m^2}{2} u_1^2
+ V_1$ is given by:
\begin{eqnarray}
&& P(u_1,V_1) = P(V_1) \prod_{u \neq u_1}  P_{>}(V_1+\frac{1}{2} m^2 (u_1^2-u^2))
 \approx P(V_1) \exp(- \int du P_{<}(V_1 + \frac
{m^2}{2} u_1^2 - \frac {m^2}{2} u^2) ) \\
&& = -A'(V_1) e^{-A(V_1)} \exp(- \int d\tilde u \frac{1}{m} e^{- A(V_1 + \frac
{1}{2} \tilde u_1^2 - \frac {1}{2} \tilde u^2) } )
\end{eqnarray}
assuming $\ln(1-P_{<}) \approx -P_{<}$ which can be checked a posteriori to hold for the bulk of the resulting distribution. We have defined $u_1=\tilde u_1/m$ and $u=\tilde u/m$. In the limit $m \to 0$ the bulk of the probability is around $V_1=V^m$ such that:
\begin{eqnarray}
&& e^{- A(V^m)} = m         \quad, \quad A(V^m) = \ln(1/m)
\end{eqnarray}
hence $V^m = - (\ln(1/m)/B)^{1/\delta}$. Expanding $A(V_1)$ around $V^m$ to linear order, defining
$a_m=-A'(V^m)$ and performing the gaussian integral over $\tilde u$ one finds:
\begin{eqnarray}
&& P(u_1,V_1)  \approx m a_m e^{a_m (V_1-V^m)} \exp(- e^{a_m (V_1-V^m +\frac
{1}{2} \tilde u_1^2) + \frac{1}{2} \ln(2 \pi/a_m) })
\end{eqnarray}
with $a_m \approx  \delta \ln(1/m)/|V^m| = \delta \ln(1/m)^{1-\frac{1}{\delta}} B^{1/\delta}$. Hence one finds that the dependence in energy is of Gumbell type from extremal statistics of short range correlated variables, while the one point distribution for the minimum, obtained by integration over $V_1$ is a simple Gaussian (in a rescaled sense as $m \to 0$):
\begin{eqnarray}
&& P(u_1) \approx \sqrt{\frac{m^2 a_m}{2 \pi}} e^{- \frac{1}{2} m^2 a_m u_1^2}
\end{eqnarray}
hence $u_1 \sim \delta^{-1/2} B^{\frac{- 1}{2 \delta}} m^{-1} (\ln \frac{1}{m})^{\frac{1-\delta}{2 \delta}}$. This corresponds to $\zeta=1$ with logarithmic corrections.
This is also the one point distribution of the Burgers velocity field since $u(x) \leftrightarrow F(v)=(v-u(v)/t$ has the same distribution as $-u_1/t$ with $t=m^{-2}$. From this we get the kinetic energy:
\begin{eqnarray}
&& E(t) = \frac{1}{2} \overline{{\sf u}(x,t)^2} = \frac{1}{2 t^2} <u_1^2>_P = \frac{m^2}{2 a_m} = 2^{-1/\delta} \delta^{-1} B^{\frac{- 1}{ \delta}} t^{-1} (\ln t)^{\frac{1-\delta}{ \delta}}
\end{eqnarray}
and one finds $E(t) \sim 1/(t (\ln t)^{1/2})$ for the gaussian case $\delta=2$.

It is also interesting to note that the joint distribution of the renormalized potential $E=\hat V(0)=V_1+ \frac{1}{2} m^2 u_1^2$ and its derivative, the force $F=\hat V'(0)=-m^2 u_1$, at the same point is juste the product:
\begin{eqnarray}
&& P(E,F) = Q_e(E-V^m) Q_f(F) \\
&& Q_e(E) = a_m e^{a_m E} \exp(- e^{a_m E}) \\
&& Q_f(F) =  \sqrt{\frac{a_m}{2 \pi m^2}} e^{- a_m F^2/(2 m^2)} 
\end{eqnarray}
of a Gumbel by a Gaussian, hence that these two variables are statistically independent. 

\subsubsection{two point function}

Let us now consider the two point probability $P^{tot}_{v_1,v_2}(u_1,V_1,u_2,V_2)$. It has two contributions:
\begin{eqnarray}
&& P^{tot}_{v_1,v_2}(u_1,V_1,u_2,V_2) = \delta(u_2-u_1) \delta(V_2-V_1) P_{v_1,v_2}(u_1,V_1)
+ P_{v_1,v_2}(u_1,V_1,u_2,V_2)
\end{eqnarray}
according to whether there are no shock, or at least one shock in the interval $[v_1,v_2]$. We start with the second contribution, i.e. we assume that there are
shocks.  Then one
wants the two conditions to be simultaneously fulfilled:
\begin{eqnarray}
&& V(u) \geq V_1 +  \frac {m^2}{2} (u_1-v_1)^2 - \frac {m^2}{2}
(u-v_1)^2 \quad , \quad \text{equality in} \quad u_1 \\
&& V(u) \geq V_2 +  \frac {m^2}{2} (u_2-v_2)^2 - \frac {m^2}{2}
(u-v_2)^2 \quad , \quad \text{equality in} \quad u_2
\end{eqnarray}
One defines the intersection of the two parabola (the construction
is similar to Section \ref{sec:shocks}):
\begin{eqnarray}
&& u^*=\frac{v_1+v_2}{2} - \frac{1}{m^2 v_{21}} (V_{21} + \frac
{m^2}{2} ((u_2-v_2)^2 - (u_1-v_1)^2) ) = \frac{v_1+v_2}{2} - \frac{E_{21}}{m^2 v_{21}}
\end{eqnarray}
One must have $u_1<u^*<u_2$, hence the energy difference $E_{21}=\hat V(v_2)-\hat V(v_1)$ must satisfy $- \frac{1}{2} v_{21} + v_2-u_2 < \frac{E_{21}}{v_{21}} < \frac{1}{2} v_{21} + v_1-u_1$.
This is assuming there are shocks in the interval $[v_1,v_2]$. The case $u_1=u^*<u_2$ corresponds to one shock in the interval at $v=v_2$. When there are shocks between the points then:

\begin{eqnarray}
 && P_{v_1,v_2}(u_1,V_1,u_2,V_2) \approx A'(V_1) e^{A(V_1)} A'(V_2) e^{A(V_2)} \\
&& \times  \exp(-
\int_{-\infty}^{\tilde u^*} \frac{d\tilde u}{m} e^{- A(V_1 + \frac {1}{2} (\tilde u_1-\tilde v_1)^2 - \frac {1}{2} (\tilde u-\tilde v_1)^2)}
- \int^{+\infty}_{\tilde u^*} \frac{d\tilde u}{m} e^{- A(V_2 + \frac {1}{2} (\tilde u_2-\tilde v_2)^2 - \frac {1}{2} (\tilde u-\tilde v_2)^2)}) \nonumber \\
&& \approx m^2 a_m^2 e^{a_m (V_1+V_2-2V^m)} \exp(-
\int_{-\infty}^{\tilde u^*} d\tilde u e^{a_m (V_1-V^m + \frac {1}{2} (\tilde u_1-\tilde v_1)^2 - \frac {1}{2} (\tilde u-\tilde v_1)^2)} - \int^{+\infty}_{\tilde u^*} d\tilde u e^{a_m(V_2-V^m  + \frac {1}{2} (\tilde u_2-\tilde  v_2)^2 - \frac {1}{2} (\tilde u-\tilde v_2)^2)}) \nonumber
\end{eqnarray}
As above we have defined $\tilde u_i=m u_i$ and $\tilde v_i=m v_i$.
Since $\tilde u^*$ depends only on $V_{21}=V_2-V_1$ one can integrate over $(V_1+V_2)/2-V^m$ which gives:
\begin{eqnarray}
 && P_{v_1,v_2}(u_1,u_2,V_{21}) = \frac{m^2 a_m}{
[\int_{-\infty}^{\tilde u^*} d\tilde u e^{a_m (- \frac {1}{2} V_{21} + \frac {1}{2} (\tilde u_1-v_1)^2 - \frac {1}{2} (\tilde u-\tilde v_1)^2)} + \int^{+\infty}_{\tilde u^*} d\tilde u e^{a_m(\frac {1}{2} V_{21} + \frac {1}{2} (\tilde u_2-v_2)^2 - \frac {1}{2} (\tilde u-\tilde v_2)^2)}) ]^2 }
\end{eqnarray}
This yields the joint distribution, where we have restored the mass dependence:
\begin{eqnarray}
&& P_{v_1,v_2}(u_1,u_2,E_{21}) = m^2 a_m^2 ~~ p_{m a_m^{1/2} v_{21}}(m a_m^{1/2} (u_1-v_1),m a_m^{1/2} (u_2-v_2) , a_m E_{21}) \\
&& p_v(w_1,w_2,E) = \frac{e^{- \frac{1}{2} (w_1^2+w_2^2) + \frac{1}{4} v^2 + \frac{E^2}{ v^2}} } {
[ \Phi\big( \frac{v}{2} - \frac{E}{v} ) \big) + \Phi\big(\frac{v}{2} + \frac{E}{ v}) ]^2}  \theta(- \frac{v}{2}  -w_2 < \frac{E}{v} < \frac{v}{2} - w_1)
\nonumber
\end{eqnarray}
where $\Phi(w)=e^{w^2/2} \int_{-\infty}^w dz e^{-z^2/2} = \int_0^{+\infty} dz e^{-z^2/2 + w z}$ which satisfies $\Phi'(w)=1+w \Phi(w)$. 

The case with no shock corresponds to $u_1=u^*=u_2$. From the above we get:
\begin{eqnarray}
 && P_{v_1,v_2}(u_1,V_1)  \approx m a_m e^{a_m (V_1-V^m)}  \\
 && \times  \exp(- e^{a_m (V_1-V^m)} [
\int_{-\infty}^{\tilde u_1} d\tilde u e^{a_m (\frac {1}{2} (\tilde u_1-\tilde v_1)^2 - \frac {1}{2} (\tilde u-\tilde v_1)^2)} + \int^{+\infty}_{\tilde u_1} d\tilde u e^{a_m(\frac {1}{2} (\tilde u_1-\tilde  v_2)^2 - \frac {1}{2} (\tilde u-\tilde v_2)^2)} ]) \nonumber
\end{eqnarray}
Through integration over the Gumbell function of $V_1$ we get:
\begin{eqnarray}
 && P_{v_1,v_2}(u_1) = m a_m^{1/2} p_{v_{21} m a_m^{1/2}}(m a_m^{1/2} (u_1-v_1),
m a_m^{1/2} (u_1-v_2)) \\
&& p_v(w_1,w_2)= \frac{1}{ \Phi(w_1) + \Phi(-w_2) }
\end{eqnarray}

Adding the two contributions, and setting for now $a_m$ and $m$ to unity (to be restored below) we find the joint distribution for the energy difference $E= \hat V(v_2)-\hat V(v_1)$, and the forces $F_i=F(v_i)=\hat V'(v_i)=m^2(v_i-u_i)$, $i=1,2$, with $v=v_2-v_1$, as:
\begin{eqnarray}
 && P(F_1,F_2,E)=\delta(F_2-(F_1+v)) \delta(E - \frac{v}{2} (F_1+F_2))  \frac{1}{ \Phi(-F_1) + \Phi(F_2) }
\\
&&  +  \frac{e^{- \frac{1}{2} (F_1^2+F_2^2) + \frac{1}{4} v^2 + \frac{E^2}{ v^2}} } {
[ \Phi\big( \frac{v}{2} - \frac{E}{v} ) \big) + \Phi\big(\frac{v}{2} + \frac{E}{ v}) ]^2}  \theta(- \frac{v}{2}  + F_2 < \frac{E}{v} < \frac{v}{2} + F_1)
\end{eqnarray}
To restore $m$ dependence one replaces $F_i \to a_m^{1/2} F_i/m$, $E \to a_m E$ and $v \to v m a_m^{1/2}$, and correcting as needed for the probability measure to remain normalized to one.

The two point distribution of the forces is obtained by integration over $E$:
\begin{eqnarray}  \label{forceforce}
 && P(F_1,F_2)=\delta(F_2-(F_1+v))   \frac{1}{ \Phi(-F_1) + \Phi(F_2) }
\\
&&  + v e^{- \frac{1}{2} (F_1^2+F_2^2) + \frac{1}{4} v^2 } \int_{- \frac{v}{2}  + F_2}^{ \frac{v}{2} + F_1} d \epsilon \frac{e^{\epsilon^2}}{
[ \Phi\big( \frac{v}{2} - \epsilon ) \big) + \Phi\big(\frac{v}{2} + \epsilon) ]^2}  \theta(F_1-F_2+v)
\end{eqnarray}
it also yields the two point distribution for Burgers turbulence through the replacement:
\begin{eqnarray}
 && F_i \to \alpha_t^{1/2} t^{1/2} {\sf u}(x_i) \\
 && v  \to  \alpha_t^{1/2}  t^{-1/2} (x_2-x_1) 
\end{eqnarray}
with $\alpha_t=a_{m=1/\sqrt{t}}$. Hence the internal scale is $\sim (t/\alpha_t)^{1/2}$. Since the width of a shock is $\sim T t/u_{21}
\sim T t/(t/\alpha_t)^{1/2}$ the dimensionless ratio is $T \alpha_t \sim T (\ln t)^{1-\frac{1}{\delta}}$. Note that for $\delta<1$ temperature is in effect irrelevant. The case $\delta=1$ is related to the REM.

To compute $R(u)$ we instead integrate over $F_1$ and $F_2$ gives (being careful with the jacobian factor $1/v$ in the first term) and obtain the distribution of scaled energy difference $\epsilon=E/v$:
\begin{eqnarray}
 && P(\epsilon)=  \frac{1}{ \Phi\big( \frac{v}{2} - \epsilon  \big) + \Phi\big(\frac{v}{2} + \epsilon) }
 + v  \frac{ \Phi\big( \frac{v}{2} - \epsilon  \big)  \Phi\big(\frac{v}{2} + \epsilon) } {
[ \Phi\big( \frac{v}{2} - \epsilon  \big) + \Phi\big(\frac{v}{2} + \epsilon) ]^2} \\
&& =  \frac{  \Phi'\big( \frac{v}{2} - \epsilon  \big)  \Phi\big(\frac{v}{2} + \epsilon) +  \Phi \big( \frac{v}{2} - \epsilon  \big)  \Phi'\big(\frac{v}{2} + \epsilon)} {
[ \Phi\big( \frac{v}{2} - \epsilon  \big) + \Phi\big(\frac{v}{2} + \epsilon) ]^2} = \partial_{\epsilon} H(\epsilon,v) \\
&& H(\epsilon,v) =  \frac{ \Phi\big(\frac{v}{2} + \epsilon) }{ \Phi\big( \frac{v}{2} - \epsilon  \big) + \Phi\big(\frac{v}{2} + \epsilon) } 
\end{eqnarray}
using that $w \Phi(w)=\Phi'(w)-1$. Since $\Phi(w) \sim -1/w$ vanishes as $w \to -\infty$ and diverges at $\sqrt{2 \pi} e^{w^2/2}$ for  $w \to -\infty$, one has $H(-\infty)=0$ and $H(+\infty)=1$ hence $P(\epsilon)$ is correctly normalized. We now obtain:
\begin{eqnarray} \label{reskida}
 &&  R(0) - R(v) = \frac{1}{2} v^2 \int_{-\infty}^{+\infty} d\epsilon  \epsilon^2 p(\epsilon) = 2 v^2 \int_{0}^{+\infty} d\epsilon ~ \epsilon  \frac{ \Phi\big(\frac{v}{2} - \epsilon) }{ \Phi\big( \frac{v}{2} - \epsilon  \big) + \Phi\big(\frac{v}{2} + \epsilon) } \\
 && =  \frac{1}{2}  v^2  - \frac{1}{3 \sqrt \pi} v^3 + 0.0272494 v^4 -0.00114373 v^5 + O(v^6)
\end{eqnarray}
where integration by parts and symmetries have been used. One shows, using  symmetries and the differential equation for $\Phi(w)$, that the integrand can be rewritten:
\begin{eqnarray}
 && \epsilon H(-\epsilon,v)=- \partial_v \ln \big( \Phi\big( \frac{v}{2} - \epsilon  \big)+ \Phi\big(\frac{v}{2} + \epsilon) \big) + \frac{1}{2}  (\frac{v}{2} + \epsilon) + \frac{1}{ \Phi\big( \frac{v}{2} - \epsilon  \big) + \Phi\big(\frac{v}{2} + \epsilon) }
 \end{eqnarray}
 where the linear term exactly cancels the bevavior of the first term at  $\epsilon \to +\infty$. In the end one shows that:
 \begin{eqnarray}
 &&  - R'(v) = 2 v  \int_{0}^{+\infty} d\epsilon   \frac{1}{ \Phi\big( \frac{v}{2} - \epsilon  \big) + \Phi\big(\frac{v}{2} + \epsilon) } \\
 && = v - v^2/\sqrt{\pi} + 0.108998 v^3 - 0.00571863 v^4 + O(v^5)
 \end{eqnarray}
 which leads to Kida result for the two point force correlator, i.e. the velocity correlator $\Delta(u)=-R''(u)$. Here we have obtained also $R(0)$ since $R(\infty=0)=0$ one gets for large $v$:
\begin{eqnarray}
 &&  R(0) = \lim_{v \to \infty} 2 v^2 \int_0^\infty d\epsilon \epsilon \frac{1}{1 + e^{v \epsilon}}  = \frac{\pi^2}{6} 
 \end{eqnarray}
 
Restoring all $m$ factors one sees that 
\begin{eqnarray}
&&  R(v) = a_m^{-2} R_s(\sqrt{a_m} m v) 
\end{eqnarray}
where the scaled $R$, noted here $R_s(v)$ is given by formula (\ref{reskida}). It should in principle satisfy the FRG equation with $\zeta=1$ (and $\theta=0$) to leading order in $m \to 0$. Note that it should be the {\it short range} solution  for this value of $\zeta$, while the long range one (with the same value of $\zeta$) corresponds to logarithmic disorder and was studied in Appendix \ref{sec:log}. As a result the force correlator satisfies:
\begin{eqnarray}
&&  \int_0^\infty \Delta(v) dv = 0
\end{eqnarray}
and indeed one can see in Figure 10 or Ref \cite{kida} what was probably (unknowingly) the first FRG fixed point correlator (before the FRG was invented) with a nice cusp, and is reminiscent of the results for manifolds in random bond disorder measured recently in \cite{{MiddletonLeDoussalWiese2006}}. 

Consider now the small $v$ limit of the part of the two point force probability which contains at least one shock (setting $t$ to unity), from formula (\ref{forceforce}) one obtains:
\begin{eqnarray}  \label{forceforce2}
 && P_{shock}(F_1,F_2) \approx v e^{- \frac{1}{2} (F_1^2+F_2^2)} (F_1-F_2) \theta(F_1-F_2) 
 \end{eqnarray}
 Since as $v \to 0$ one can safely assume that there will be only one shock in the interval one finds that the joint distribution of shock sizes $s=u(v^+)-u(v^-)=F_1-F_2>0$ (for $t=1$) and positions $\hat u=\frac{1}{2} (u(v+)+u(v^-))$ is simply:
 \begin{eqnarray}  \label{distrib}
 && P(s,\hat u) = \frac{s}{2 \sqrt{\pi}}  e^{-s^2/4} e^{-(\hat u-v)^2} \theta(s)
 \end{eqnarray}
normalized to unity,
where $\hat u-v$ is the velocity of the shock in Burgers, and the center position of the droplet/shock, and the two variables $s$ and $\hat u-v$ are independent. From the droplet shock relation (\ref{rel0}) we find the droplet density distribution function:
\begin{eqnarray}  \label{distrib}
 && D(y) = \frac{1}{2 \sqrt{\pi}}  e^{-y^2/4} 
 \end{eqnarray}
normalized by $\int_{-\infty}^{\infty} y^2 D(y)=2$.

We will refrain from giving here results for the other classes of disorder (Weibul and Frechet) but these
are easily analyzed along the same lines. For the depinning this was done very recently in \cite{LeDoussalWiese2008}.

\section{two-well droplet calculation in higher dimension}

\label{sec:2welld} 

We start from:
\begin{eqnarray}
&& \tilde V[\{v_x\}] - \tilde V[\{v_x=0\}]  = \frac{1}{2}
\sum_{xy} g^{-1}_{xy} v_x v_y
- T \ln( p X_1 + (1-p) X_2 ) \\
&& X_i = e^{ \sum_{x} \tilde v_x (\sum_{y} g^{-1}_{xy} u_{i,y}) }
\end{eqnarray}
with $p=1/(1+w)$. As before calculation is simpler if one
considers the force:
\begin{eqnarray}
&& - {\cal F}[v]_x = \frac{\delta \tilde V[\{v_z\}]}{\delta v_x} =
g^{-1}_{xy} ( T \tilde v_y - \langle u_y \rangle ) \nonumber 
\\
&& = g^{-1}_{xy} ( T \tilde v_y - \frac{u_{1,y} X_1 + u_{2,y} X_2
w  }{X_1 + w X_2} )
\end{eqnarray}
One defines the moments:
\begin{eqnarray}
&& \bar S^{11..1}_{x_1..x_n} = \overline{ {\cal F}[v_1]_{x_1} .. {\cal F}[v_n]_{x_n} } =
g^{-1}_{x_1 z_1} .. g^{-1}_{x_n z_n}  C_n[v_1,..v_n]_{z_1,..z_n} \\
&& C_n[v_1,..v_n]_{x_1,..x_n}
= \overline{ \prod_{i=1}^n \frac{u_{1,x_i} a_i + u_{2,x_i} w  }{a_i + w } } \\
&& C_{2 n} [v_1,..v_n]_{x_1,..x_n} = \langle \prod_{i} u_{1,x_i}
\rangle_{u_1}
+ A_{2 n} [v_1,..v_n]_{x_1,..x_n} \\
&& C_{2 n+1}[v_i]_{z_1,..z_n} = A_{2 n+1}[v_i]_{z_1,..z_n}
- T (\sum_i \tilde v_{i,z_i} \langle \prod_{j \neq i} u_{1,z_j}
\rangle_{u_1} )
\end{eqnarray}
with $a_i=X_{1,i}/X_{2,i}$. The calculation is similar to Appendix \ref{app:dropletmoments}.
The result is:

\begin{eqnarray}
&& A_n[v_i]_{x_1,..x_n} = \frac{1}{4}  T \sum_{i=1}^n \langle
Y_{x_i} (Y g^{-1} v_i) [
\prod_{j \neq i} (u_{1,x_j} - Y_{x_j} \frac{1}{1 - e^{(Y g^{-1} v_{ji}) }}) \\
&& + (-1)^{n-1} \prod_{j \neq i} (u_{1,x_j} - Y_{x_j} \frac{1}{1 -
e^{- (Y g^{-1} v_{ji})}}) ] \rangle_{u_1,Y}
\end{eqnarray}
where $Y_{x} = u_{1,x} - u_{2,x}$ and $(Y g^{-1} v) =\sum_{xy} Y_x
g^{-1}_{xy} v_y$. The two symmetries used are $u_{i,x} \to -
u_{i,x}$ and $u_{1,x} \leftrightarrow u_{2,x}$. The highest order
term is:
\begin{eqnarray}
&& A^{high}_n[v_i]_{x_1,..x_n} = \frac{1}{4} T \langle Y_{x_1} .. Y_{x_n}
\sum_{i=1}^n (Y g^{-1} v_i) F_n[ (Y g^{-1} v_{ij}) ] \rangle_{u_1, Y} 
\end{eqnarray}
where the function $F_n$ was given in (\ref{defF}).
For the second and third moments we obtain the result given in the text.

\section{STS and ERG identities for correlations in higher $d$ and droplets}
\label{app:sts}

\subsection{functional form of STS and ERG identities}

The general STS identity is:
\begin{eqnarray}
&& T \sum_c \langle \frac{\delta O[u]}{\delta u_c^x  }  \rangle = g^{-1}_{xy}  \sum_f \langle  O[u] u_f^y
\rangle
\end{eqnarray}
Specializing to observables $O[u]$ depending only on a single replica one finds:
\begin{eqnarray}
&&  T \overline{ \langle \frac{\delta O[u]}{\delta u_x  }  \rangle } = g^{-1}_{xy}  ( \overline{ \langle
O[u] u_y \rangle } - \overline{ \langle O[u]  \rangle \langle u_y \rangle } )
\end{eqnarray}
In the zero $T$ limit it yields:
\begin{eqnarray}
&&  \langle \frac{\delta O[u_1]}{\delta u_{1x} } \rangle_{P[u_1]} = g^{-1}_{xy}  \langle (u_{1x} - u_{2x})
O[u_1] \rangle_{D[u_1,u_2]}
\end{eqnarray}
which yields the functional equation given in the text. Similarly the ERG equation is:
\begin{eqnarray}
&& \partial \langle O[u] \rangle = - \frac{1}{2 T} \sum_f (u_f
\partial g^{-1} u_f) O[u] \rangle
\end{eqnarray}
Specializing to $g$ independent observables depending on a single replica one gets
the functional equation given in the text.

\subsection{droplet calculation of correlation functions}

In the case of independent well separated droplets one evaluates the thermal and disorder averages as follows. The mean
of any observable is:
\begin{eqnarray}
&& \frac{A(u_1)+w_i A(u_1+u_{21}^{(i)})+ \sum_{i \neq j} w_i w_j A(u_1+u_{21}^{(i)}+u_{21}^{(j)}) + .. }{1 +
\sum_i w_i + \sum_{i \neq j} w_i w_j + .. }
\end{eqnarray}
Since any disorder average containing more than one distinct $w_i$ will be higher order in $T$ upon computing
the above average one finds:
\begin{eqnarray}
&& \overline{ \langle O[u] u \rangle } - \overline{ \langle O[u] \rangle \langle u \rangle } = \overline{ \sum_i
\frac{w_i}{(1+ w_i)^2} (O(u_1) - O(u_2^{(i)}) (u_1 - u_2^{(i)}) } + O(T^2)
\end{eqnarray}

We now define $u_{1x}=\hat u_x - \frac{1}{2} u^{21}_x$, $u_{2x}=\hat u_x + \frac{1}{2} u^{21}_x$, then one has
$\langle u_x \rangle = \hat u_x + (\frac{1}{2} -p) u^{21}_x$. Only correlations with an even number of $\hat u$
and an even number of $u_{21}$ are non zero and satisfy the symmetries. This
yields for the two point functions:
\begin{eqnarray}
&& G_{xy} = \langle u_{1x} u_{1y} \rangle - \frac{T}{2} \sum_i \langle
u^{(i)}_{21x} u^{(i)}_{21y} \rangle \\
&& \tilde G_{xy} = \langle u_{1x} u_{1y} \rangle
\end{eqnarray}

We now want to compute the four point functions. We need to relate $R''''_{xyzt}[0]$ and $Q^{1111}_{xyzt}[0]$ to droplets since all correlations
can be obtained from them. We will also check the consistency, i.e. that, as for $d=0$ all three STS relations
amount to one droplet identity, and similarly that the two ERG relations yield only one droplet identity.
The only four point correlation which are non zero in the droplet calculation are $\langle \hat u_{x} \hat u_{y} u_{21z}
u_{21t} \rangle$ and $\langle u_{21x} u_{21y} u_{21z} u_{21t} \rangle$. We now show that they are related:

Consider the identity (from the above):
\begin{eqnarray}
&& \langle u^a_x u^a_y u^b_z u^c_t \rangle - \langle u^a_x u^b_y u^c_z u^d_t \rangle \equiv \overline{ (\langle
u_x u_y \rangle - \langle u_x \rangle \langle u_y \rangle )  \langle u_z \rangle \langle u_t \rangle } = T
g_{xy} G_{zt}
\end{eqnarray}
Let us compute within droplets:
\begin{eqnarray}
&& \overline{ (\langle u_x u_y \rangle - \langle u_x \rangle \langle u_y \rangle )  \langle u_z \rangle \langle
u_t \rangle } = T \overline{p(1-p)} \langle u_{21x} u_{21y} \hat u_{z} \hat u_{t} \rangle + T \overline{p(1-p)
(\frac{1}{2}-p)^2} \langle u_{21x} u_{21y} u_{21z} u_{21t} \rangle
\end{eqnarray}
This yields the relation:
\begin{eqnarray}
&&  \langle u_{21x} u_{21y} \hat u_{z} \hat u_{t} \rangle + \frac{1}{12} \langle u_{21x} u_{21y} u_{21z} u_{21t}
\rangle = 2 g_{xy} G_{zt}
\end{eqnarray}
the right hand side being presumably the disconnected part of the first correlation on the l.h.s.

The droplet calculation of correlations yields:

\begin{eqnarray}
&& \langle u^a_x u^a_y u^a_z u^a_t \rangle = \langle u_{1x} u_{1y}
u_{1z} u_{1t} \rangle \\
&& \langle u^a_x u^a_y u^a_z u^b_t \rangle = \langle u_{1x} u_{1y} u_{1z} u_{1t} \rangle - \frac{1}{8}  T
\langle u_{21x} u_{21y}
u_{21z} u_{21t} \rangle \\
&& - \frac{1}{2} T ( \langle u_{21z} u_{21t} \hat u_{21x} \hat u_{21y} \rangle + \langle u_{21x} u_{21t} \hat
u_{21z} \hat u_{21y} \rangle + \langle u_{21y} u_{21t} \hat u_{21x} \hat
u_{21z} \rangle ) \nonumber \\
&& = \langle u_{1x} u_{1y} u_{1z} u_{1t} \rangle - T g_{zt} G_{xy}
- T g_{yt} G_{xz} - T g_{xt} G_{yz} \nonumber \\
&& \langle u^a_x u^a_y u^b_z u^b_t \rangle =  \langle u_{1x} u_{1y} u_{1z} u_{1t} \rangle - \frac{1}{2} T (
\langle u_{21y}
u_{21z} \hat u_{21x} \hat u_{21t} \rangle \\
&& + \langle u_{21x} u_{21z} \hat u_{21y} \hat u_{21t} \rangle + \langle u_{21y} u_{21t} \hat u_{21x} \hat
u_{21z} \rangle + \langle u_{21x}
u_{21t} \hat u_{21y} \hat u_{21z} \rangle ) \nonumber \\
&& = \langle u_{1x} u_{1y} u_{1z} u_{1t} \rangle + \frac{1}{6}  T \langle u_{21x} u_{21y} u_{21z} u_{21t}
\rangle - T (g_{yz} G_{xt} + g_{xz} G_{yt} + g_{yt} G_{xz} + g_{xt} G_{yz}) \nonumber
\\
 && \langle u^a_x u^a_y u^b_z u^c_t \rangle = \langle u_{1x}
u_{1y} u_{1z} u_{1t} \rangle - \frac{1}{8}  T \langle u_{21x} u_{21y} u_{21z} u_{21t} \rangle - \frac{1}{2} T (
\langle u_{21y} u_{21z}
\hat u_{t} \hat u_{x} \rangle + 4 \text{perm} )\\
&& = \langle u_{1x} u_{1y} u_{1z} u_{1t} \rangle - \frac{1}{8} T \langle u_{21x} u_{21y} u_{21z} u_{21t} \rangle
- \frac{1}{2} T ( \langle u_{21y} u_{21z} \hat u_{t} \hat u_{x} \rangle + 5 \text{perm} ) - \langle u_{21x}
u_{21y} \hat u_{z} \hat u_{t}
\rangle \nonumber \\
&& = \langle u_{1x} u_{1y} u_{1z} u_{1t} \rangle + \frac{1}{12}  T \langle u_{21x} u_{21y} u_{21z} u_{21t}
\rangle - T ( g_{xy} G_{zt} + 5 \text{perm}  ) + T g_{xy} G_{zt} \nonumber
\\
 && \langle u^a_x u^b_y u^c_z u^d_t \rangle = \langle u_{1x}
u_{1y} u_{1z} u_{1t} \rangle - \frac{1}{6}  T \langle u_{21x} u_{21y} u_{21z} u_{21t} \rangle - \frac{1}{2} T (
\langle u_{21x} u_{21y}
\hat u_{z} \hat u_{t} \rangle + 5 \text{perm} ) \\
&& = \langle u_{1x} u_{1y} u_{1z} u_{1t} \rangle + \frac{1}{12}  T \langle u_{21x} u_{21y} u_{21z} u_{21t}
\rangle - T ( g_{xy} G_{zt} + 5 \text{perm}  ) \nonumber
\end{eqnarray}
where in the last steps we use the above equality. Thus everything is consistent with:
\begin{eqnarray}
&& g_{xx'} g_{yy'}g_{zz'} g_{tt'} T R''''_{x'y'z't'}[0] =
\frac{1}{12}   \langle u_{21x} u_{21y} u_{21z} u_{21t} \rangle \\
&&  g_{xx'} g_{yy'}g_{zz'} g_{tt'} ( Q^{1111}_{x'y'z't'}[0] - T^2
R''''_{x'y'z't'}[0] ) = \langle u_{1x} u_{1y} u_{1z} u_{1t} \rangle_c \\
\end{eqnarray}
and
we must check whether these identities are compatible with our result for these cumulants. Finally the ERG equation yields, using the above:
\begin{eqnarray}
&& \partial G_{zt} = \frac{1}{12} \langle (u_{21} \partial g^{-1} u_{21}) u_{21z} u_{21t}  \rangle - \partial
g^{-1}_{x'y'} (g_{x'z} G_{y't} + g_{x't} G_{y'z})
\end{eqnarray}
consistent with the equations given in the text.

\section{One loop ERG using multilocal expansion}

\label{app:multilocal} 

The multilocal expansion was studied in Ref. \cite{ergchauve,schehr_co_pre}. Let us just indicate here the rules for derivatives and fix notations.

\subsection{multilocal expansion and derivatives}

One writes:
\begin{eqnarray}
&& R[v] = \int_x R(v_x) +  \frac{1}{2} \int_{xy} R(v_x,v_y,x-y) + \frac{1}{6} \int_{xyz} R(v_x,v_y,v_z,x,y,z) +..
\end{eqnarray}
up to trilocal part. At this stage we do not try to define these components unambiguously, apart from the local part which has been defined in the text. The results for the derivatives are:
\begin{eqnarray}
&& R'_x[v] = R'(v_x) +  \int_{y} R_{10}(v_x,v_y,x-y) +  \frac{1}{2} \int_{yz} R_{100}(v_x,v_y,v_z,x,y,z) +.. \\
&& R''_{xy}[v]= R''(v_x) \delta_{xy} + R_{11}(v_x,v_y,x-y)  + \delta_{xy} \int_{z} R_{20}(v_x,v_z,x-z) +  \int_{z} R_{110}(v_x,v_y,v_z,x,y,z) \\
&& +  \frac{1}{2} \delta_{xy}
\int_{zt} R_{200}(v_x,v_z,v_t,x,z,t) \nonumber \\
&& R'''_{xyz}[v]= R'''(v_x) \delta_{xyz} + \delta_{xz} R_{21}(v_x,v_y,x-y) + \delta_{yz} R_{12}(v_x,v_y,x-y) + \delta_{xy}
R_{21}(v_x,v_z,x-z) \\
&& +
\delta_{xyz} \int_{t} R_{30}(v_x,v_t,x-t) + R_{111}(v_x,v_y,v_z,x,y,z) + \delta_{xz} \int_{t} R_{210}(v_x,v_y,v_t,x,y,t) + \delta_{yz}
\int_{t}
R_{120}(v_x,v_y,v_t,x,y,t) \nonumber \\
&& + \delta_{xy} \int_{t} R_{210}(v_x,v_z,v_t,x,z,t) + \frac{1}{2} \delta_{xyz}
\int_{zt} R_{300}(v_x,v_z,v_t,x,z,t) \nonumber
\end{eqnarray}
and so on. For the third moment one has:
\begin{eqnarray}
&& S[v_{123}] = \int_x S(v_{123x}) + \frac{1}{2} \int_{xy} S(v_{123x},v_{123y},x-y) + ..\\
&& S^{100}_x[v_{123}] = S_{100}(v_{123x}) + \int_{y} S_{100;000}(v_{123x},v_{123y},x-y) + ..\\
&& S^{110}_{xy}[v_{123}] = S_{110}(v_{123x}) \delta_{xy} +S_{100;010}(v_{123x},v_{123y},x-y) + \delta_{xy} \int_{z}
S_{110;000}(v_{123x},v_{123z},x-z)
\end{eqnarray}
and so on. Note that for uniform configurations:
\begin{eqnarray} \label{unif} 
&& R[v]|_{v_x=v} = L^d R(v) \\
&& R_{xy}[v]|_{v_x=v} = \delta_{xy} R''(v) + R_{11}(v,v,x-y) \\
&& R_{xyz}[v]|_{v_x=v} = \delta_{xyz} R'''(v) + (\delta_{xy}
R_{11}(v,v,z-x) + 2 p.c.) + R_{111}(v,v,v,x,y,z) \\
&& R_{xyzt}[v]|_{v_x=v} = \delta_{xyzt} R''''(v) + (\delta_{xyz} R_{31}(v,v,t-x) + 3 p.c.) +  (\delta_{xy}
\delta_{zt}
R_{22}(v,v,t-x) + 2 p.c.) \\
&& + (\delta_{xy} R_{211}(v,v,v,x,z,t) + 5 p.c.) + R_{1111}(v,v,v,v,x,y,z,t) \nonumber 
\end{eqnarray}

\subsection{one loop analysis using multilocal expansion}

Let us apply the first W-ERG and $\Gamma$-ERG equations to a uniform $v$. One finds the exact equations:
\begin{eqnarray}
&& \partial R(v) = T \partial g_{x=0} R''(v) + T \int_x \partial g_{x} R_{11}(v,v,x) + \partial g_{x=0} \bar
S_{110}(0,0,v) + \int_x \partial g_{x} \bar S_{100;010}(0,0,v;0,0,v;x)
\end{eqnarray}
and
\begin{eqnarray}
&& \partial R(v) = T \partial g_{x=0} R''(v) + \int_x \partial
g_{x} g_{x} {\sf R}''(v)^2 + \partial g_{x=0} S_{110}(0,0,v) \\
&& + T \int_x \partial g_{x} R_{11}(v,v,x) + \int_x \partial g_{x}
S_{100;010}(0,0,v;0,0,v;x) \nonumber  \\
&& + 2 \int_{xy} \partial g_{y} g_{xy} {\sf R}_{11}(v,v,-x) {\sf R}''(v) + \int_{xyz} \partial g_{z} g_{xy}
{\sf R}_{11}(v,v,x-z) {\sf R}_{11}(v,v,y) \nonumber
\end{eqnarray}
In the TBL regime these equations are, by construction, exactly obeyed by the droplet solution described in the text. 
\begin{eqnarray}
&& R''(v) = T m^4 \int_x \langle u_{210} u_{21x} F_2( \tilde v \int_z
u_{21z}) \rangle \\
&& R_{11}(v,v,x) = T ( (m^2 - \nabla^2_x)^2 \langle u_{210} u_{21x} F_2( \tilde v \int_z u_{21z}) \rangle -
\delta(x) m^4 \int_y \langle u_{210} u_{21y}
F_2( \tilde v \int_z u_{21z}) \rangle ) \\
&& S_{110}(0,0,v) = T^2 m^4 \int_x \langle u_{210} u_{21x} G_3(
\tilde v \int_z u_{21z}) \rangle \\
&& S_{100;010}(0,0,v;0,0,v;x) = T^2 ( (m^2 - \nabla^2_x)^2 \langle u_{210} u_{21x} G_3( \tilde v \int_z u_{21z})
- \delta(x) m^4 \int_y \langle u_{210} u_{21y} G_3( \tilde v \int_z u_{21z}) \rangle )
\end{eqnarray}

Let us now obtain a RG equation for the bilocal part. The simplest, unambiguous way to proceed is to obtain the
flow of $R_{11}(v,v,x)$. One starts from the exact equation:
\begin{eqnarray}
&& \partial R_{xy}[v] = T \partial g_{zt} R''''_{xyzt}[v] +
\partial
g_{zz'} \tilde S^{112}_{zz'xy}[0,0,v] \\
&& = T \partial g_{zt} R''''_{ztxy}[v] + 2 \partial g_{zz'} g_{tt'} ( {\sf R}''''_{ztxy}[v] {\sf R}''_{z't'}[v]
+ {\sf R}'''_{ztx}[v] {\sf R}'''_{z't'y}[v] ) +
\partial g_{zz'} S^{112}_{zz'xy}[0,0,v]
\end{eqnarray}
We now take a uniform configuration and use formula such as (\ref{unif}) to express 
the flow in terms of the multilocal components. In the resulting exact equation, we now get rid of the explicit $\delta_{xy}$ parts (which yield the FRG equation for the local part) using a projector $(1-P)$ which projects out the local part:
\begin{eqnarray}
&& \partial R_{11}(v,v,x-y)  =  (1-P)[  2 T
\partial g_{x=0} R_{31}(v,v,x-y) + 2 T
\partial g_{xy} R_{22}(v,v,x-y) + 4 T \int_t \partial g_{xt} R_{211}(v,v,v,x,y,t) \nonumber   \\
&& + 4 \int_z \partial g_{z} g_{-z} {\sf R}''(v) R_{31}(v,v,x-y) + 4 \int_{z} \partial g_{xz} g_{zy} {\sf
R}''(v) R_{22}(v,v,x-y)
+ 8 \int_{z t}  \partial g_{xz} g_{zt} {\sf R}''(v) R_{211}(v,v,v,x,y,t) \nonumber \\
&& + 4 \int_{zt} \partial g_{xz} g_{xt} {\sf R}_{11}(v,v,z-t) R_{31}(v,v,x-y) + 4 \int_{zt} \partial g_{xz}
g_{yt} {\sf
R}_{11}(v,v,z-t) R_{22}(v,v,x-y) \nonumber \\
&& + 8 \int_{z'tt'} \partial g_{xz'} g_{tt'} {\sf
R}_{11}(v,v,z'-t') ) R_{211}(v,v,v,x,y,t)  + 2 \partial g_{xy} g_{xy} R'''(v)^2 \\
&& + 4 \partial g_{xy} g_{xz} R'''(v) R_{11}(v,v,z-y) +  4
\partial g_{xz} g_{xy} R'''(v) R_{11}(v,v,z-y) + 4 \partial g_{xz} g_{xz} R'''(v) R_{11}(v,v,y-z) \nonumber
\\
&& + 4 \int_{zt} \partial g_{xz} g_{xt} R'''(v) R_{111}(v,v,y,z,t) + 2 \int_{ztz't'} \partial g_{zz'} g_{tt'}
R_{111}(v,v,v,x,z,t)
R_{111}(v,v,v,y,z',t') \nonumber \\
&& + 2 \int_{zt} ( \partial (g_{xy} g_{zt}) + 2
\partial (g_{yz} g_{zt}) + 2 \partial g_{xt} g_{yz}
+ \partial g_{zt} g_{zt} ) R_{11}(v,v,x-z) R_{11}(v,v,y-t) \nonumber \\
&& + 4 \int_{z z't} ( \partial g_{xz'} g_{zt} +  \partial g_{zz'} g_{xt} +
\partial g_{zz'} g_{zt} ) R_{11}(v,v,x-z) R_{111}(v,v,y,z',t') +
\partial g_{zz'} S^{112}_{zz'xy}[0,0,v] ]\nonumber
\end{eqnarray}
where symmetrization w.r.t. $xy$ is often implicit, and one should project out the local part, as is indicated
by the projector $1-P$. We have not decomposed the third cumulant up to its tri-local part as it should appear.
This equation is exact, and the droplet solution thus obeys it automatically all through the TBL.

Applying the $\epsilon$ expansion counting, one finds that the equation reduces to its $T=0$ limit and to lowest order reduces to:
\begin{eqnarray}
&& \partial R_{11}(v,v,x)  = 2 (\partial g_{x} g_{x}- \delta(x) \int_y \partial g_{y} g_{y}) R'''(v)^2
\end{eqnarray}
The discussion of this result was given in the text.

\section{W-ERG}
\label{app:werg} 

To one loop we need:

\begin{eqnarray}
&& \partial R[v] = \partial g_{zz'} \tilde S^{110}_{zz'}[0,0,v] \\
&& \partial \bar S[v_{123}] =  \frac{3}{2} \text{sym}_{123}
\partial g_{xy} \bar Q^{1100}_{xy}[v_{1123}]
\end{eqnarray}

We write the cumulant truncation:
\begin{eqnarray}
&& \bar Q[v_{1234}] = R[v_{12}] R[v_{34}] + R[v_{13}] R[v_{24}] + R[v_{14}] R[v_{23}] \\
&& \bar Q^{1100}_{xy}[v_{1234}] = - R''_{xy}[v_{12}] R[v_{34}] + R'_x[v_{13}] R'_y[v_{24}] + R'_x[v_{14}]
R'_y[v_{23}]
\end{eqnarray}
Hence cumulant truncation+weak (resp strong) continuity of $R''_{xy}[v_{12}]$ implies weak (resp strong)
continuity of $\bar Q$ to one loop with the result:
\begin{eqnarray}
&& \partial g_{xy} \bar Q^{1100}_{xy}[v_{1123}] = 2 \partial g_{xy} R'_x[v_{12}] R'_y[v_{13}]
\end{eqnarray}
up to the gauge term $- R''_{xy}[0] R[v_{23}] $. Hence:
\begin{eqnarray}
&& \partial \bar S[v_{123}] = \partial g_{xy} ( R'_x[v_{12}] R'_y[v_{13}] + R'_x[v_{21}] R'_y[v_{23}] +
R'_x[v_{31}] R'_y[v_{32}])
\end{eqnarray}
To this order it integrates into:
\begin{eqnarray}
&& \bar S[v_{123}] = g_{xy} ( R'_x[v_{12}] R'_y[v_{13}] + R'_x[v_{21}] R'_y[v_{23}] + R'_x[v_{31}] R'_y[v_{32}])
\end{eqnarray}
since $\partial R$ is higher order. Let us now evaluate:
\begin{eqnarray}
&& \bar S^{110}_{zt}[v_{123}] = g_{xy} ( R'''_{xzt}[v_{12}] (R'_y[v_{23}] - R'_y[v_{13}]) - R''_{xt}[v_{12}]
R''_{yz}[v_{13}] - R''_{xz}[v_{21}] R''_{yt}[v_{23}] + R''_{xz}[v_{31}] R''_{yt}[v_{32}]) )
\end{eqnarray}
Assuming that:
\begin{eqnarray}
&& \lim_{v_{12} \to 0} g_{xy} ( R'''_{xzt}[v_{12}] (R'_y[v_{23}] - R'_y[v_{13}]) = 0
\end{eqnarray}
then one obtains the equations in the text. One has then strong/weak continuity for $\bar S$ if $R$ satisfies
the same.

\section{details of two loop $\Gamma$-ERG calculation in $\epsilon$ expansion}
\label{app:details}

Let us denote $M_{12}={\sf R}''_{yr}[v_{12}]$ and rewrite Eq. (\ref{solu3mom}) in the text as:
\begin{eqnarray}
&& S[v_{123}] = \frac{1}{2}  A \\
&& A = tr[g M_{12} g M_{12} (g M_{13} + g M_{23} )] + tr[g M_{23}
g M_{23} (g M_{12} + g M_{13} )] \\
&& + tr[g M_{13} g M_{13} (g M_{12} + g M_{23} )]  - 2 tr[g M_{12} g M_{23} g M_{13}]
\end{eqnarray}
On can rewrite:
\begin{eqnarray}
&& A = A_1 + A_2 + A_3 \\
&& A_1 = tr[g M_{12} g M_{12} (g M_{13} + g M_{23} )] \\
&& A_2 = tr[g M_{12} g (M_{13} - M_{23}) g (M_{13} - M_{23})] \\
&& A_3 = \frac{1}{3} tr[g (M_{13} + M_{23}) g (M_{13} + M_{23}) g (M_{13} + M_{23}) ]
\end{eqnarray}
up to gauge terms, where we have used cyclic properties of the trace, as well as identity under transposition and that
all matrices are symmetric. The second term is of order $v_{12}^3$. To perform the expansion in the last term we use
symmetric expansion of a three replica functional as explained in Appendix of \cite{BalentsLeDoussal2005} . Let $f[v_{13},v_{23}]$ the symmetric functional. One can
either expand:
\begin{eqnarray}
&& f[v_{13},v_{13}-v_{12}] = f[v_{13},v_{13}] - v_{12x} f^{01}_x[v_{13},v_{13}] + \frac{1}{2} v_{12x} v_{12y}
f^{02}_{xy}[v_{13},v_{13}] + O(v_{12}^3) \\
&& f[v_{23}+v_{12},v_{23}] = f[v_{23},v_{23}] + v_{12x} f^{10}_x[v_{23},v_{23}] + \frac{1}{2} v_{12x} v_{12y}
f^{20}_{xy}[v_{23},v_{23}]   + O(v_{12}^3)  \\
&& = f[v_{23},v_{23}] + v_{12x} ( f^{10}_x[v_{13},v_{13}] - v_{12y} ( f^{20}_{xy}[v_{13},v_{13}]+
f^{11}_{xy}[v_{13},v_{13}]) + \frac{1}{2} v_{12x} v_{12y} f^{20}_{xy}[v_{13},v_{13}]  + O(v_{12}^3)
\end{eqnarray}
performing the half sum and discarding the zero-th order terms which are gauge, one gets:
\begin{eqnarray}
&& f[v_{13},v_{23}] = - \frac{1}{2} v_{12x} v_{12y} f^{11}_{xy}[v_{13},v_{13}] + O(v_{12}^3)
\end{eqnarray}
We obtain:
\begin{eqnarray}
&& A_3 = - 2 v_{12x} v_{12y} tr[ g R'''_x[v_{13}] g R'''_y[v_{13}] g R''[v_{13}] ] + O(v_{12}^3)
\end{eqnarray}
The first term gives:
\begin{eqnarray}
&& A_1 = 2 tr[ g {\sf R}''[v_{12}] g {\sf R}''[v_{12}] g R''[v_{13}] ] + O(v_{12}^3)
\end{eqnarray}
replacing  ${\sf R}_{xy}''[v_{12}]
=\delta_{xy} R(v_{12x})$ it becomes:
\begin{eqnarray}
&& A_1 = 2 \int_{xyz} g_{xy} {\sf R}''(v_{12y}) g_{yz} {\sf R}''(v_{12z}) g_{zx} {\sf R}''(v_{13x})
\end{eqnarray}

\section{periodic case and mean field limit}
\label{app:last} 

Here we give some partial results about two solvable cases. The random periodic class appears simpler in any $d$, and in $d=0$ it is useful as it provides an exact solution of the FRG hierarchy.  It then raises questions about shocks which can be answered in another solvable limit, large $d$ or the fully connected model. 

\subsection{random periodic class in $d=0$}

Consider in $d=0$, for $N=1$, a random potential $V(u)$ periodic of period one. 
Then $\hat V(v)$ is also periodic of period one. Consider the $T=0$ problem such that
$\hat V(v) = \min_{u} ( V(u) + (u-v)^2/(2 t) )$. It is easy to compute the $T=0$ fixed point of the FRG in the limit $t=m^{-2} \to \infty$. Denote
$u_1$ the absolute minimum of $V(u)$ on a period, say $-1/2 \leq u_1<1/2$. Since for infinite $t$ the curvature of the quadratic well
goes to zero, $u_1(v)=u_1$ for $u_1-1/2< v<u_1+1/2$, $u_1(v)=u_1+1$ for $u_1+1/2< v<u_1+3/2$ etc..
Hence there is a single shock in the unit cell $[0,1[$, at position $0 \leq v_s=u_1 + 1/2 < 1$, and on the real axis there is a periodic array of shocks at $v^{(i)}_s = u_1 + (2 i+1)/2$. The force $t F(v)=v - u_1(v)$ is then for $0 \leq v < 1$:
\begin{eqnarray}
&& t F(v) = v - v_s + \frac{1}{2} \quad , \quad v < v_s \\
&& t F(v) = v - v_s - \frac{1}{2} \quad , \quad v > v_s
\end{eqnarray}
The velocity of the shock is zero, because
of the constraint that the integral of the force over a period is zero. Note that this solution indeed satisfies $\int_0^1 dv
\tilde F(v)=0$.

Because of statistical translational invariance $u_1$ is uniformly distributed in $[-1/2,1/2[$ and so is $v_s$ in $[0,1[$. One then obtains the second and third moments:
\begin{eqnarray} \label{resRP}
&& - t^2 R''(v) = t^2 \overline{F(0) F(v)} = \int_0^1 dv_s (\frac{1}{2} - v_s)
(v - v_s - \frac{1}{2} \epsilon(v-v_s)) = \frac{1}{12} - \frac{1}{2} v (1-v) \\
&& t^3 \bar S_{111}(v_1,v_2,v_3) = - t^3 \overline{F(v_1) F(v_2) F(v_3)} 
=  \frac{1}{6} (1+v_1+v_2-2 v_3)(1+2 v_1- v_2- v_3)(v_1- 2 v_2 +  v_3)
\end{eqnarray}
for $0 \leq v \leq 1$ and for $0 \leq v_1 < v_2 < v_3  \leq 1$, all other cases can be obtained by symmetry or periodicity. We can now compute the limits:
\begin{eqnarray}
&& t^3 \tilde S_{111}(0,0^+,v) =  \frac{1}{6} v (1 - 2 v)(1-v) \\
&&  t^3 \tilde S_{112}(0,0^+,v) =  \frac{1}{6} - v (1-v)
\end{eqnarray}
The first FRG equation reads:
\begin{eqnarray}
&&  \partial_t R''(v) = T R''''(v)+  \bar S_{112}(0,0,v)
\end{eqnarray}
and one easily checks from the above that its $T=0$ version is obeyed (i.e. setting $T=0$ and 
replacing $\tilde S_{112}(0,0,v) \to \tilde S_{112}(0,0^+,v)$). One also checks that $\langle u_1^2 \rangle = - R''(0) t^2 = \frac{1}{12}$ since for $v=0$ the minimum is necessarily in the interval $[-1/2,1/2[$ and
essentially uniformly distributed (note that it crosses over to $\langle u_1^2 \rangle \sim t T$ at for large $t T$ - see below). 

Expressions for higher moments become rapidly complicated, but some generating functions can be computed, e.g (setting $t=1$): 
\begin{eqnarray}
&& \langle e^{w_1 F(v_1) + w_2 F(v_2)} \rangle
= \frac{1}{w_1+w_2}
e^{- \frac{1}{2} (w_1+w_2)} [e^{ w_1 (1-v) + w_2} - e^{ w_1 (1-v)}
+ e^{ w_1 + v w_2} - e^{ v w_2} ]
\end{eqnarray}
with $0< v=v_2-v_1$. 

Although we will not do it here in details, it is easy to study the case $T>0$. A shock is broadened as:
\begin{eqnarray}
&& t F(v) = v - v_s - \frac{1}{2} \tanh(\frac{1}{2 t T} (v-v_s) + O(T t))
\end{eqnarray}
since $u_{21}=1$, and one can redo the previous calculation using this form and get results consistent with the general ones obtained in the text for the shape of the TBL, specializing to a very simple droplet distribution $D(y)=\delta(|y|-1)$. However, this form works only for $t T \ll 1$. Since temperature is relevant here (with $\zeta=0$ and $\theta=-2$) the shock width grows as $T t$, while their spacing is unity, hence at large time they overlap.  Exact solutions are easily written using the diffusion equation for $Z=e^{-\hat V/T}$, one interesting example being the solution :
\begin{eqnarray}
&& \hat V(v) = - T \ln \sum_{n=-\infty}^{n=\infty} \frac{1}{\sqrt{2 \pi T t}} e^{-(v-v_s+n)^2/(2 T t)}
\end{eqnarray}
which represents a solution of decaying Burgers with an initial condition a periodic set of (zero temperature) shocks at locations $v_s+n$ (this solution arised in the study of SLE on a cylinder \cite{ch}. Averaging over $v_s$ yields a general solution at any $T$ for the FRG hierarchy, which describes the flow away from the zero $T$ fixed point when temperature is turned to a non zero value. Finally one can check directly on the $T>0$ solution that $\lim_{T \to 0} \tilde S_{111}(0,0,v)=\tilde S_{111}(0,0^+,v)$ the continuity property discussed in the text.

Although the $d=0$ random periodic (RP) model may appear trivial, it is an interesting limit of the RP class in higher $d$. Apart from temperature becoming relevant for $d<2$, we do not expect any bifurcation between $d=4$ and $d=0$ in the $T=0$ fixed point itself. Let us indeed compare with the result from the $\epsilon=4-d$ expansion (to two loop) of Ref. \cite{twolooplong}:
\begin{eqnarray}
\Delta(u) = m^\epsilon \frac{K^{d/2}}{\epsilon \tilde J_2} \big(
\frac{\epsilon}{36} + \frac{\epsilon^2}{54} - (\frac{\epsilon}{6} + \frac{\epsilon^2}{9}) u (1-u) + O(\epsilon^3) \big )
\end{eqnarray}
with $\epsilon \tilde J_2=2 (4 \pi)^{-d/2} \Gamma(3-\frac{d}{2})$. Note that from the shear fact that it contains only $u(1-u)$ the universal number $r_1=a \Delta''(0^+)/ |\Delta'(0^+)|=2$  and next, from the fact that it integrates to zero $r_2=a \Delta'(0^+)/ |\Delta(0)| = 6$ (where $a$ is the lattice period, $a=1$ here). Note that for depinning $\epsilon^2/54$ is changed into $\epsilon^2/108$ and the universal number becomes $r_2=6 + 2 \epsilon + O(\epsilon^2)$ as the constraint that the integral of the force is zero is relaxed. Finally, there is a global amplitude, $r= m^{-\epsilon} \Delta(0)/K^{d/2}$, with $r = 1/36 + \epsilon/54$ in the $\epsilon$ expansion, which comes not too far from the exact result $r=1/12$ in $d=0$. So we see that despite its simplicity, this model does not fare too badly as compared to $d>0$.

\subsection{large $d$ limit, fully connected model}

The model can be studied in the large $d$ limit of a hypercubic lattice, as was done in  the Appendix H2 of Ref. \cite{BalentsLeDoussal2005}. Here we study a simpler, but closely related, fully connected model. It is simple enough to allow for easy understanding of the results obtained in Ref. \cite{BalentsLeDoussal2005}, seen here under a different perspective (the definition of the renormalized correlators is slightly different there). One starts from:
\begin{eqnarray}
&& H_V[\{u_i\},v] = H_V[\{u_i\},v,u_0 = \frac{1}{N} \sum_i u_i] \\
&& H_V[\{u_i\},v,u_0] = K \sum_i (u_i - u_0)^2 + \frac{1}{2 t} \sum_i (u_i -v)^2 + V_i(u_i) 
\end{eqnarray}
with $N$ continuous variables $u_i$, $i=1,..N$. The first term can also be written $\frac{K}{2 N} \sum_{ij} (u_i-u_j)^2$. Since $u_0$ is defined by a minimization condition on the first line, one has:
\begin{eqnarray}
&& \hat V(v) = min_{u_i} H_V[\{u_i\},v] = min_{u_i, u_0}  H_V[\{u_i\},v,u_0]
\end{eqnarray}
Regrouping terms one easily sees that:
\begin{eqnarray}
&& \hat V(v) = N \min_{u_0} [ \frac{1}{N} \sum_i \hat V_i(v'=\frac{v + 2 t K u_0}{1 + 2 t K})
+ \frac{K}{1+ 2 K t} (u_0-v)^2 ]  \label{effr} \\
&& \hat V_i(v) =  \min_{u} [ \frac{1+2 t K }{2 t} (u -v)^2 + V_i(u)]  
\end{eqnarray}
and note that the force is:
\begin{eqnarray}
&& F(v)= \hat V'(v) =  N (v-u_0)/t
\end{eqnarray}
as can be checked comparing derivatives w.r.t. $v$ and to $u_0$ of the first line. Hence we are back to a $d=0$ model for the center of mass $u_0$, but it is feeling an effective disorder which is an average over a sum of a large number $N$ of independent random potentials $W(v')=\frac{1}{N} \sum_i \hat V_i(v')$. Hence it should be small and also almost gaussian in distribution. Note however that each of them has shocks if the parameter $t/(1+2 K t) > t_c$ corresponding to a Larkin scale, hence the correlator will have a cusp. For $t \to \infty$ this corresponds to a critical $K_c$ such that for $K<K_c$ the system is in a strong disorder phase with a cusp, while for $K>K_c$ there is no cusp in each layer. 

Since layers are uncorrelated the correlator of the effective force reads:
\begin{eqnarray}
&& R_W''(v) = - \overline{W'(0) W'(v)} = \frac{1}{N} R_{lay}''(v)   \label{rescd} 
\end{eqnarray}
where $R_{lay}(v)$ is the renormalized correlator of a $d=0$ model $R_{lay}''(v)=-\overline{\hat V_i'(0) \hat V_i'(v)}$ assumed identical for any $i$ (we have also assumed the $V_i$ uncorrelated). Similarly the three point cumulant of $W$ is related to $1/N^2 \bar S$ where $\bar S$ is the three point correlator in each layer
and we see that indeed the effect of higher correlators is reduced by the $N$ counting.

There are various behaviors in this model depending on the way one scales the parameters. The simplest one is to keep $t$ and $K$ fixed and $N \to \infty$ first. Then since disorder is reduced one has $u_0=v$ and $\hat V(v) = W(v)$ and (\ref{rescd}) gives the final result $R(v)=R_W(v)$. In the cusp phase of each layer, the resulting potential of the full system $\hat V$ will exhibit many independent shocks. For the random periodic class then one can then use the result (\ref{resRP}) for $R_{lay}$. On the other hand one could consider $t \to \infty$ at fixed $N$ and $K$. Then if $K>K_c$ although there are no shocks in each layer, there are shocks from the center of mass $u_0(v)$ since the curvature of the term $u_0-v$ in (\ref{effr}) goes to zero. These are {\it global shocks} quite different from the {\it local shocks} which occur independently in each layer in the other limit. In general both occur, with an interesting scaling behaviour as a function of $N/t^2$.

\subsection{periodic case, many shocks} 

Let us close on a remark about the form of the correlator ubiquitously found for the random periodic class in the $\epsilon$ expansion and in recent numerics \cite{MiddletonLeDoussalWiese2006} in $d=3,2$, i.e.  $\Delta''(v)$ independent of $v$ ($v$ not integer). Consider $n$ shocks in the interval $]0,1]$:
\begin{eqnarray}
F(v)= \sum_{i=0}^{n} (v - u_i) \theta(v_i < v < v_{i+1})
\end{eqnarray}
with $v_0=0$, $v_{n+1}=1$ and the condition $u_n=u_0+1$, so that $F(0)=F(1)$. We set $t=1$.
Then one has for $0<v \neq v' <1$
\begin{eqnarray}
\overline{ F'(v) F'(v') }= -1 + \sum_{i \neq j =1}^{n} \overline{  \alpha_i \alpha_j \delta(v-v_i)
\delta(v'-v_j)}
\end{eqnarray}
with $\alpha_i=u_{i-1}-u_{i}$, hence $\sum_{i=1}^{n} \alpha_i =1$ and we have used that $\overline{ F'(v)} = 0$. For $v \neq v'$ the term $i=j$ can be discarded
since it produces only a $\delta$ function. For two shocks it yields:
\begin{eqnarray}
\overline{ F'(v) F'(v') }= -1 + 2 \int d\alpha \alpha (1- \alpha) P(\alpha,v-v')
\end{eqnarray}
where $P(\alpha,v_1,v_2)=P(\alpha,v_1-v_2)$ is the probability of shock positions and amplitude.
It can only depend on the difference from translational invariance. The form $-R''(v) = \Delta(v) \sim v(1-v)$ (periodized on the real axis) thus can only occur if
the shock distance distribution is uniform i.e independently located shocks.
In that case the number:
\begin{eqnarray}
\overline{ F'(v) F'(v') }= -1 + 2 \overline{ \alpha (1- \alpha) } = - 2 \overline{ \alpha^2 }
\end{eqnarray}
by symmetry. This is more general one has for $n$ independent shocks:
\begin{eqnarray}
\overline{ F'(v) F'(v') } = -1 + \sum_{i \neq j =1}^{n} \overline{  \alpha_i \alpha_j }
= -  \sum_i \overline{ \alpha_i^2 } = - n \overline{ \alpha^2 }
\end{eqnarray}
Hence $\Delta''(v) = n \overline{ \alpha^2 }$ and we recall that $\overline{\alpha}=1/n$.
This model can be generalized in various ways, such as $n$ not fixed, but it illustrates simply 
what features of the shock distribution lead to the usual form of the random periodic fixed point.

\end{widetext}

\tableofcontents


\begin{thebibliography}{32}
\bibitem{gauge}
Note that there was a degree of arbitrary in the definition of these
functions (a gauge invariance). One can perform for instance the
simultaneous changes (in the more general notations of Appendix
\ref{app:general}): $\hat R[j_a,j_b] \to \hat R[j_a,j_b] + T
\phi[j_a] + T \phi[j_b]$, $\hat U[j_a] \to \hat U[j_a] - p
\phi[j_a]$. This gauge choice is fixed by the sample dependent
definition of $\hat V$ chosen here.
\end{thebibliography}
\end{document}